\documentclass[11pt,twoside]{article}

\usepackage[dvipdfmx]{graphicx,color}
\usepackage[dvipsnames]{xcolor}
\usepackage[font={small},labelsep=space]{caption}
\usepackage{graphbox,subcaption}
\usepackage{amsmath,amssymb,mathrsfs,bm,bold-extra}
\usepackage{booktabs,multirow,dcolumn,bigdelim}
\usepackage{fancybox,arydshln,mdframed}
\usepackage{enumitem,sectsty,titlesec}
\usepackage[hang,flushmargin]{footmisc}
\usepackage{pdflscape,afterpage,capt-of}
\usepackage{endnotes,filecontents,etoolbox}
\usepackage{here,stackengine,empheq,cases,braket}
\usepackage[T1]{fontenc}
\usepackage{fancyhdr}
\usepackage[makeroom]{cancel}
\usepackage{mwe,ulem}
\usepackage{newtxtext,newtxmath}
\usepackage{setspace}

\usepackage[compress,authoryear,round]{natbib}
\setcitestyle{notesep={;},aysep={,},yysep={;}}

\usepackage[hyphens]{xurl}
\usepackage[pdfencoding=auto,linkbordercolor=CornflowerBlue,citecolor=tb,linkcolor=tb,pdfpagelabels=false]{hyperref}
\hypersetup{
  colorlinks=true,
  allbordercolors=CornflowerBlue
}

\DeclareCaptionLabelFormat{vert}{\textbf{{#1} {#2}~|}}
\usepackage[modulo]{lineno}

\newcommand{\f}[2]{\frac{#1}{#2}}

\newcommand{\abs}[1]{\left| #1 \right|}
\newcommand{\vev}[1]{\left\langle #1 \right\rangle}
\newcommand{\bmt}[1]{{{\mbox{\boldmath$ #1 $}}}}
\newcommand{\q}[1]{`#1'}

\newcommand{\disc}[2]{\href{https://explore.openalex.org/concepts/#1}{\textcolor{BlueViolet}{#2}}}

\renewcommand*{\thepage}{\footnotesize\arabic{page}}
\newcolumntype{d}{D{.}{.}{2.4}}

\newcommand{\rn}[1]{\uppercase\expandafter{\romannumeral #1\relax}}
\newcommand{\period}[2]{\color{Cerulean}{\textrm{\textbf{\underline{\rn{#1}.\quad {#2}}}}}}

\def\ep{\epsilon}
\def\X{\mathrm{X}}
\def\Y{\mathrm{Y}}
\def\cS{\mathcal{S}}
\def\cC{\mathcal{C}}
\def\xsize{0.615}
\def\csize{0.35}
\def\dsize{0.4}
\def\chordsize{0.32}
\def\marrow{\LARGE\color{Cerulean}{$\blacktriangle$}}

\def\OpenAlex{\href{https://docs.openalex.org}{OpenAlex}}
\def\ai{\disc{C154945302}{Artificial Intelligence}}
\def\quantum{\disc{C62520636}{Quantum Science}}
\def\bio{\disc{C150903083}{Biotechnology}}
\def\nano{\disc{C171250308}{Nanotechnology}}
\def\agri{\disc{C88463610}{Agricultural Engineering}}
\def\particle{\disc{C109214941}{Particle Physics}}
\def\aerospace{\disc{C146978453}{Aerospace Engineering}}
\def\nuclear{\disc{C116915560}{Nuclear Engineering}}
\def\marine{\disc{c199104240}{Marine Engineering}}
\def\neuro{\disc{c169760540}{Neuroscience}}
\def\condensed{\disc{C26873012}{Condensed Matter Physics}}
\def\envi{\disc{C87717796}{Environmental Engineering}}
\def\earth{\disc{c1965285}{Earth Science}}
\def\astro{\disc{c1276947}{Astronomy}}
\def\math{\disc{C202444582}{Pure Mathematics}}

\def\mtitle{A half-century of global collaboration in science and the `Shrinking World'}

\newif\ifabbreviation
\pretocmd{\thebibliography}{\abbreviationfalse}{}{}
\AtBeginDocument{\abbreviationtrue}
\DeclareRobustCommand\acroauthor[2]{%
  \ifabbreviation #2\else #1\fi}

\def\bibfont{\small}
\makeatletter
\newcommand*{\myfnsymbol}[1]{\ensuremath{%
\ifcase#1 \or \ast \or \dagger \or \spadesuit \or \diamondsuit \or \clubsuit \or \heartsuit \else \@ctrerr \fi}}
\makeatother

\definecolor{tb}{rgb}{0.24, 0.43, 0.91}
\renewcommand*{\cite}[1]{\textcolor{tb}{\citep{#1}}}
\newcommand*{\citek}[1]{\textcolor{tb}{\citet{#1}}}

\begin{document}

\quad\vspace{-1.4cm}
\begin{flushright}
August 2023 [v2]
\end{flushright}

\vspace{0.5cm}

\begin{center}
\fontsize{15pt}{16pt}\selectfont\bfseries
\mtitle
\end{center}

\renewcommand*{\thefootnote}{\textcolor{black}{\myfnsymbol{\value{footnote}}}}

\vspace*{0.8cm}
\centerline{%
{Keisuke Okamura}\,\footnote{\,{\tt okamura@ifi.u-tokyo.ac.jp}}${}^{,}$%
\footnote{\,\href{https://orcid.org/0000-0002-0988-6392}{\tt \textcolor{black}{orcid.org/0000-0002-0988-6392}}}${}^{;\,1,\,2}$}

\vspace*{0.6cm}
\small{
\centerline{\textit{
${}^{1}$Institute for Future Initiatives (IFI), The University of Tokyo,\footnote{\,\href{https://ror.org/057zh3y96}{\tt \textcolor{black}{ror.org/057zh3y96}}}}}
\centerline{\textit{
7-3-1 Hongo, Bunkyo-ku, Tokyo 113-0033, Japan.}}
\vspace*{3mm}\centerline{\textit{
${}^{2}$SciREX Center, National Graduate Institute for Policy Studies (GRIPS),\footnote{\,\href{https://ror.org/03gz5xp73}{\tt \textcolor{black}{ror.org/03gz5xp73}}}}}
\centerline{\textit{
7-22-1 Roppongi, Minato-ku, Tokyo 106-8677, Japan.}}
}
\vspace*{0.5cm}

\vspace{1.5cm}
\noindent\textbf{Abstract.}
\quad
Recent decades have witnessed a dramatic shift in the cross-border collaboration mode of researchers, with countries increasingly cooperating and competing with one another. 
It is crucial for leaders in academia and policy to understand the full extent of international research collaboration, their country's position within it, and its evolution over time. 
However, evidence for such world-scale dynamism is still scarce. 
This paper provides unique evidence of how international collaboration clusters have formed and evolved over the past 50 years across various scientific publications, using data from OpenAlex, a large-scale Open Bibliometrics platform launched in 2022. 
We first examine how the global presence of top-tier countries has changed in 15 natural science disciplines over time, as measured by publication volumes and international collaboration rates.
Notably, we observe that the US and China have been rapidly moving closer together for decades but began moving apart after 2019. 
We then perform a hierarchical clustering to analyse and visualise the international collaboration clusters for each discipline and period. 
Finally, we provide quantitative evidence of a `Shrinking World' of research collaboration at a global scale over the past half-century. 
Our results provide valuable insights into the big picture of past, present and future international collaboration.

\vspace{0.8cm}


\vfill


\thispagestyle{empty}
\setcounter{page}{1}
\setcounter{footnote}{0}
\setcounter{figure}{0}
\setcounter{table}{0}
\setcounter{equation}{0}

\setlength{\skip\footins}{10mm}
\setlength{\footnotesep}{4mm}

\vspace{-1.6cm}

\newpage
\renewcommand{\thefootnote}{\arabic{footnote}}

\setlength{\skip\footins}{10mm}
\setlength{\footnotesep}{4mm}

\urlstyle{sf}

\let\oldheadrule\headrule
\renewcommand{\headrule}{\color{VioletRed}\oldheadrule}

\pagestyle{fancy}
\fancyhead[LE,RO]{\textcolor{VioletRed}{\footnotesize{\textsf{\leftmark}}}}
\fancyhead[RE,LO]{}
\fancyfoot[RE,LO]{\color[rgb]{0.04, 0.73, 0.71}{}}
\fancyfoot[LE,RO]{\scriptsize{\textbf{\textsf{\thepage}}}}
\fancyfoot[C]{}

\thispagestyle{empty}

\newpage
\tableofcontents


\vfill

\begin{mdframed}[linecolor=Bittersweet]
\begin{tabular}{l}
\begin{minipage}{0.09\hsize}
\hspace{-3mm}\includegraphics[width=1.5cm,clip]{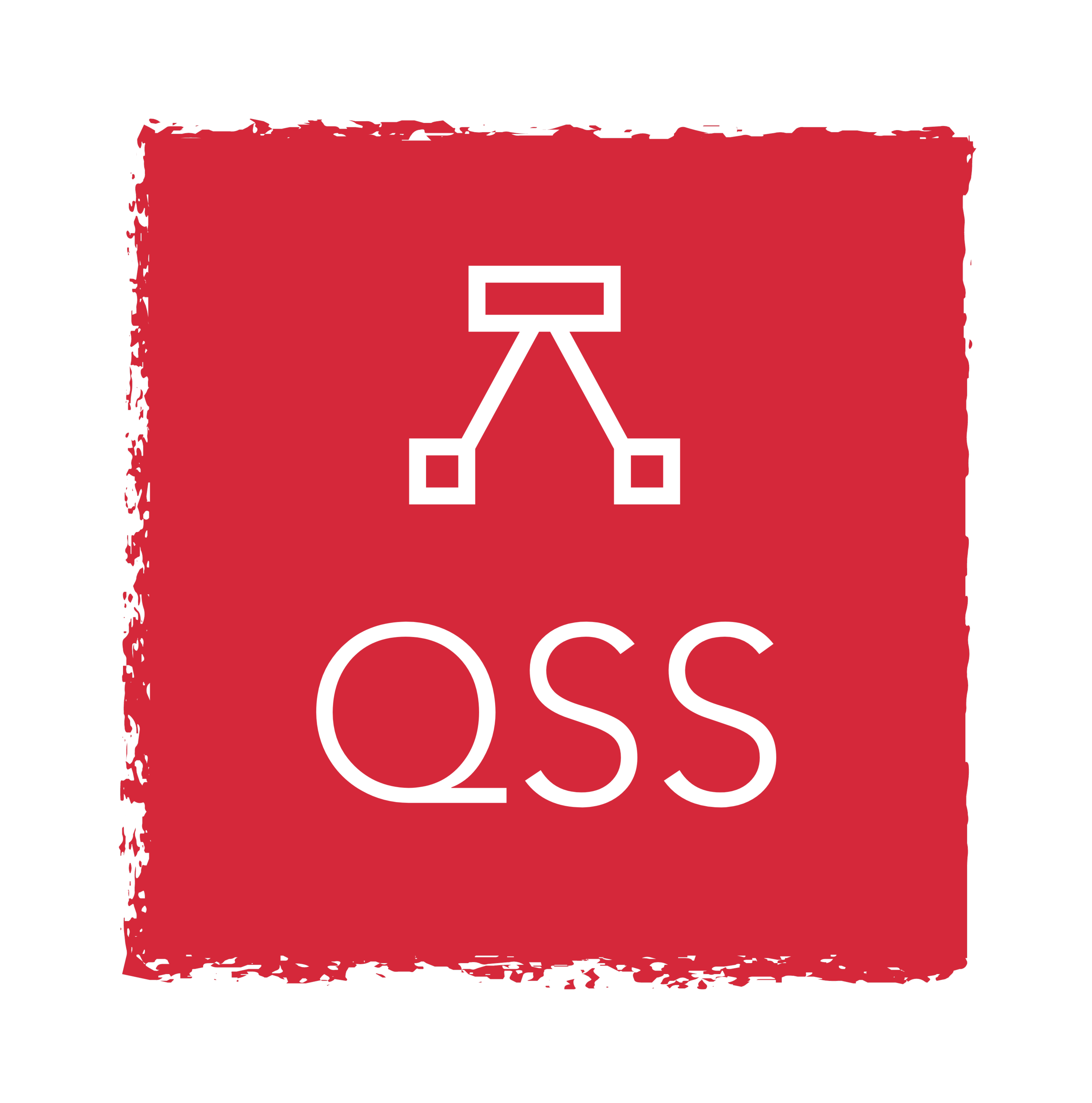}
\end{minipage}
\begin{minipage}{0.88\hsize}
\setstretch{1.1}
\textcolor{Bittersweet}{\footnotesize\textsf{%
This version is the accepted manuscript to appear in \href{https://direct.mit.edu/qss}{\textit{Quantitative Science Studies}}.
The Online Appendix accompanying the published version of the paper has been seamlessly integrated into this single file as \q{Supplementary Materials}.}}
\end{minipage}
\end{tabular}
\end{mdframed}

\newpage
\section{Introduction: Landscape of international collaboration revisited\label{sec:Introduction}}

With the rapid evolution of advanced digital communication platforms, the world has entered a new era of intense competition for knowledge.
The volume of information being produced, the degree to which it is integrated and utilised, and the range and depth of the community involved at the interface between Open Science and highly digitised society are all growing exponentially \cite{Miedema22,Beck22,Wittenburg21,Burgelman19,Dong17}.
There is no doubt that the power of science and technology (S\&T) to explore the knowledge frontier is the key to innovation, driving national growth and international competitiveness.
As a result, policymakers in many countries have begun to pay significant attention to the \q{Science of Science} \cite{Fortunato18}, which quantifies and investigates all activities associated with S\&T, providing valuable insights for policymaking.
Bibliometricians with expertise in describing and evaluating the research communities' activities through S\&T-related metrics are increasingly collaborating with policymakers and institutional practitioners, influencing the policymaking process \cite{NSB-NSF21,Cabezas-Clavijo21,Wagner17,Doria17,OECD16,Wilsdon15,Hicks15,Ismail12,RF21,OECDnd}.

The bibliometric method provides a systematic, quantitative and objective overview of information about researchers, research institutions and venues of research results publication (such as journals, conferences, institutional repositories, preprint servers and book chapters).
It also provides numerous metrics and evaluation indicators derived from the information. 
Indeed, bibliometric approaches have various conceptual and methodological difficulties and limitations, and their use requires careful attention in practical application \cite{Waltman16,Wilsdon15,Hicks15,Haustein15}.
However, to understand things at some macro level, bibliometrics can provide a unique lens through which people can see the lively ecosystem of scholarly communications.
This observation is even more true today when accessing and utilising big data through various open platforms is increasingly possible.

Policymakers in many countries recognise the significance of bibliometrics in making decisions on research and development (R\&D) investment portfolios at the state or institutional level.
An increasingly important issue of global interest in the policy arena is how to address international research collaboration \cite{Okamura23,Kwiek22,Dusdal21,Chen19,NASEM18,Wagner17,Adams12}.
On the one hand, collaboration across countries is essential for large-scale academic R\&D projects such as accelerator science and earth and planetary science, which require extended R\&D costs and periods.
It is also indispensable in addressing world-scale issues such as the United Nations Sustainable Development Goals or responding to global crises such as the COVID-19 pandemic \cite{Maher21,NSF20}.
On the other hand, geopolitical aspects such as economic security and defence-related R\&D set other primary boundary conditions for proceeding with international collaboration. 
Therefore, the policy environment surrounding the frontiers of international research collaboration has become increasingly complex and challenging in recent years, both scientifically and geopolitically.

Amid these multi-layered dimensions of cooperation and competition in international research collaboration, policymakers are keen to understand their country's collaboration partners and the clusters of research collaborations around the world \cite{Kwiek22,Adams18,Yuan18,OECD17,OECD16,He09,Mattsson08,Adams07}.
Knowing how the composition of international collaboration clusters of interest has changed over time and which policies are effective drivers of that change, whether for the better or the worse, is undoubtedly advantageous for a country.
However, it is a challenging task for various reasons.
A significant challenge from the policymaker's perspective is data availability and accessibility, which involves at least three aspects.
First, data coverage matters.
Commercial databases that focus on journal articles often do not cover enough data in today's rapidly changing, highly digitised R\&D world, where considerable scholarly communication occurs outside journals \citep[Suppl.~Fig.~S2 in][]{Okamura22}.
Furthermore, there is a field-dependent bias concerning data coverage.
For instance, in computer science, conferences and preprint servers have been more common venues for publishing research results than journals \cite{Kim19}.
If there are substantial biases or inconsistencies in data coverage, the credibility of the results of analyses based on such data will be flawed.

The second challenge concerns the timing of data availability.
For policymakers and practitioners, planning and executing R\&D investments at the right time is of utmost importance.
However, the results of bibliometric analyses based on a database that only covers journal articles may provide information too late for policymakers to reflect on their policies.
It is known that for many publications, it takes several years from when research results are generated to when they are published in peer-reviewed journals \cite{Okamura22,Lariviere14,Aman13}.
Consequently, if metrics based on such databases are generated, evaluated and then used to inform the policymaking process, by the time those metrics are used to make decisions on R\&D investments, the situation and trends in R\&D may have already changed by the time decisions on R\&D investments are made.

The third hurdle regarding accessibility, related to the first and second points, is data acquisition and utilisation autonomy.
In many cases, data on academic publications are held by major publishers or companies providing subscription-based services, such as Elsevier (with \href{https://www.elsevier.com/solutions/scopus}{Scopus}) and Clarivate Analytics (with \href{https://clarivate.com/webofsciencegroup/solutions/web-of-science/}{Web of Science}), who provide various commercial services for a fee.
The added value, including the data quality, must be fully appreciated \cite{Visser21}.
The data they provide is also valuable for bibliometricians who wish to conduct detailed quantitative analyses, where precision and comprehensiveness of data are essential.
However, not all policymakers and institutional practitioners necessarily require such precision at a fee and with restrictions for usage every time.
More critical for them could be to access data as and when they need it, even if at a different level of quality than commercial services.
Here, comparing the quality of the service of commercial and open ones in itself is an interesting issue that requires validation; as different perspectives and degrees of quality are required for different types of use, it may not always be assumed that open service is inferior to commercial data.

Metrics used in policymaking must enable practical and transparent policy accountability \cite{Wilsdon15,Hicks15}.
Therefore, for bibliometric analysis to be timely and valuable for policy considerations, the bibliometrics platform should be open, accessible, large-scale, systematic and continuously updated and operated.
{\OpenAlex} \cite{Priem22} is a promising candidate for such an Open Bibliometrics platform.
This paper employs data from {OpenAlex} to present the results of our preliminary analyses of how international collaboration clusters have formed and evolved over the past half-century for a broad set of scientific publications.
These results reflect the underlying trend of purely academia-driven or/and various state-driven cooperations, providing valuable insights for all stakeholders involved in international research collaboration.

In summary, this paper stands out for utilising OpenAlex's open data instead of commercial data to investigate 50 years of research across up to 15 distinct natural science disciplines. 
The research encompasses various types of publications, including journal articles and non-journal outputs. 
Additionally, the paper employs a hierarchical clustering technique to visualise and clarify the international collaborative relationships between countries worldwide. 
This approach provides a more nuanced perspective than a simplistic network structure analysis that focus solely on the number of coauthored papers.

The rest of this paper is organised as follows.
In Section \ref{sec:Methods}, we describe the {OpenAlex} data used in this study and the R\&D disciplines focused on.
Section \ref{sec:Results} presents the results of our analyses for each discipline, including changes over time in publication volume and international collaboration rate for each country.
We highlight that the US and China have rapidly moved closer together over the decades but started moving apart after 2019.
Furthermore, we analyse and visualise the international collaboration clusters of top-tier countries in each discipline and their evolution over time.
We also provide quantitative evidence for the \textit{\q{Shrinking World}} phenomenon of the past half-century's research collaboration. 
Finally, Section \ref{sec:Discussion} is devoted to summary, discussion and concluding remarks.
\ref{app:method} supplements the technical details of the data analysis conducted in this paper.
\ref{app:add.examples} presents the analysis results for the disciplines not fully presented in the main body of the paper, along with other complementary analysis results.

\section{Methods: Through the lens of Open Bibliometrics\label{sec:Methods}}

This section provides an overview of the data used in this study; Appendix \ref{app:data} provides additional information on the data analysis and visualisation platforms.
First, we explain the method used to acquire the data and describe the data's characteristics and practical applications.
Subsequently, we clarify the concept of the \q{nationality} of a scientific publication adopted in this paper.
In addition, we provide a description of the R\&D disciplines focused on in this paper.

\subsection{The OpenAlex data\label{subsec:data}}

The data utilised in this paper was obtained through the \href{https://docs.openalex.org/api/}{OpenAlex API}\endnote{%
OpenAlex API, {\url{https://docs.openalex.org/api/}}. Accessed 31st October 2022.}, which is a fully open catalogue of global research systems \cite{Priem22}.
It was launched to replace \href{https://www.microsoft.com/en-us/research/project/microsoft-academic-graph/}{Microsoft Academic Graph (MAG)} \cite{Sinha15}, which retired at the beginning of 2022.
OpenAlex collects information on scientific publications, including journal articles, non-journal articles, preprints, conference papers, books, and datasets---hereinafter collectively referred to as \q{works}---from various sources such as \href{https://www.crossref.org}{Crossref}, \href{https://orcid.org}{ORCID} (Open Researcher and Contributor ID), \href{https://ror.org}{ROR} (Research Organization Registry), \href{https://pubmed.ncbi.nlm.nih.gov}{PubMed}, preprint servers such as \href{https://arxiv.org}{arXiv}, and institutional or disciplinary repositories such as \href{https://zenodo.org}{Zenodo}.
OpenAlex indexes about 239 million works, with approximately 50,000 new works added daily \cite{Priem22}. 
The preprint version of this study, submitted to arXiv on 8th November 2022, is based on data obtained on 25th, 29th and 30th October and 7th November 2022, containing data published until 2021.
This study's speed was made possible by several Open Science/Bibliometrics platforms, including OpenAlex as an open data source, \href{https://docs.openalex.org/how-to-use-the-api/api-overview}{OpenAlex API} as an open standard API, and \href{https://arxiv.org}{arXiv} as an open preprint server.
In addition, R and Python, open-source programming languages updated and enhanced daily by the open data community, were used for data acquisition, analysis and visualisation.
Further, the datasets generated and/or analysed during this study can be found on \href{https://zenodo.org}{Zenodo} (see the \q{Data Availability} statement), an open dissemination research data repository.

The advantages of using OpenAlex data as a data source are summarised below, with particular emphasis on its usefulness for our analysis.
First, it provides extensive, if not exhaustive, coverage of meta-information on works, including those not published in journals.
This feature is advantageous because it can more accurately supplement the volume of R\&D activities and their associated outputs without underestimating it, even in disciplines where journals are not the primary venue for publishing research results, such as computer science.
This approach also enables the capture of outputs in preprint format, which may exist for a certain period ranging from months to years, or indefinitely, without ever becoming journal articles, as well as other data formats. 
This is particularly significant given the increasing importance of such outputs in certain disciplines in recent years \cite{Okamura22,Lariviere14}; see Appendix \ref{app:nonjrate} and Suppl.~Fig.~\ref{fig:nonjrate}.
Therefore, our approach provides a more comprehensive measure of scholarly outputs produced by each country during a certain period, including those beyond journal articles.
Although some preprint servers provide their house APIs (such as \href{https://arxiv.org/help/api/}{arXiv API}, \href{https://api.biorxiv.org/}{bioRxiv API} and \href{https://api.medrxiv.org/}{medRxiv API}), there have been no other freely accessible platforms than OpenAlex that covers all disciplines, from natural sciences to humanities and social sciences, on such a large scale.

It is important to acknowledge that there are potential disadvantages to not distinguishing research outputs in different formats, such as treating an article the same way as a dataset or a book when counting outputs.
However, this issue is not unique to this study's approach. 
Equating different journal articles could also have the same issue when counting outputs due to differences in content, length, and quality, even within the same discipline.
The primary aim of this paper is to quantify the \q{momentum} of scholarly knowledge production outputs by different countries, regardless of format, and the potential disadvantages mentioned above are not the primary concern. 
Properly identifying disciplines, isolating fields with a homogeneous publishing culture, and comparing countries within that homogeneity can mitigate these potential disadvantages, as this study does.

The second advantage of using OpenAlex is that all data is organised at the micro level, allowing users to selectively acquire and reorganise data according to their needs with a relatively high degree of flexibility.
For instance, users can selectively extract metadata about journal articles, as also demonstrated in the present study.
Third, OpenAlex is entirely open to the public and freely accessible, allowing a wide range of individuals, including data scientists, bibliometricians and other interested parties, to utilise the data and ensure the transparency and reproducibility of analysis results.
These advantages establish OpenAlex as one of the standard infrastructures supporting bibliometrics, in line with the growing momentum of Open Science and Data Science \cite{Wittenburg21,Dong17}, which we refer to as Open Bibliometrics in this paper.

\subsubsection*{R\&D disciplines}

Policy documents that discuss international research collaboration often provide an overall assessment of trends across all R\&D fields, sometimes with a field weighting.
However, significant differences exist across fields regarding their characteristics, including the resources required for R\&D, time scale and collaboration methods.
Consequently, such a generalised or averaged picture of the R\&D field is often of limited practical use.
Therefore, it is crucial to identify and adopt an appropriate classification scheme for various R\&D fields to derive meaningful policy implications for international research collaboration.
In this regard, OpenAlex has an attribution called \q{concept} assigned to each work, equivalent to a well-defined set of R\&D fields.
More than 87\% of the works on OpenAlex have been associated with one or more concepts, i.e.\ specific research areas or technologies \cite{Priem22}. 
The concepts have various levels of granularity, with 19 concepts at the coarsest (primitive) level 0; 284 at a slightly more specific level 1; followed by levels 3, 4 and 5, for a total of 65,026 concepts at 6 different levels.
OpenAlex's concept tree is a version of that used in MAG \cite{Shen18,Sinha15}, improved with a new algorithm unique to the OpenAlex API.

In this study, we specifically focus on 15 level-1 concepts from the OpenAlex classification: {\ai}, {\quantum}, {\bio}, {\nano}, {\agri}, {\particle}, {\aerospace}, {\nuclear}, {\marine}, {\neuro}, {\condensed}, {\envi}, {\earth}, {\astro} and {\math}.
We leverage the fact that OpenAlex assigns accompanying \q{related concepts} to each concept, which can be more refined or coarser than the concept's level.
For example, the level-1 concept of {\ai} is associated with level-2 subconcepts such as \q{\href{https://explore.openalex.org/concepts/C50644808}{Artificial Neural Network}} and \q{\href{https://explore.openalex.org/concepts/C108583219}{Deep Learning}}, as well as level-0 concepts such as \q{\href{https://explore.openalex.org/concepts/C41008148}{Computer Science}} and \q{\href{https://explore.openalex.org/concepts/C33923547}{Mathematics}}.
To construct an enhanced notion of R\&D discipline, we include all associated subconcepts of level 2 or higher for each of the above 15 level-1 concepts.
For instance, our defined discipline of {\ai} includes OpenAlex's level-2 concepts of \q{\href{https://explore.openalex.org/concepts/C50644808}{Artificial Neural Network}} and \q{\href{https://explore.openalex.org/concepts/C108583219}{Deep Learning}}, but not the level-0 concepts of \q{\href{https://explore.openalex.org/concepts/C41008148}{Computer Science}} or \q{\href{https://explore.openalex.org/concepts/C33923547}{Mathematics}}.

While we construct our R\&D disciplines based on level-1 concepts, disciplines constructed based on higher concept levels could provide even more practical suggestions and implications depending on the situation.
For example, using \q{\href{https://explore.openalex.org/concepts/c58053490}{Quantum Computer}} (level 3) instead of {\quantum} or \q{\href{https://explore.openalex.org/concepts/C2781047461}{Biopharmaceuticals}} (level 2) instead of {\bio} could provide more concrete implications for a country's R\&D activities or international collaborations.
We will defer such specific analyses to future work and focus on analysing the broad level-1 discipline in this paper, where the volume of work can accumulate to the scale of hundreds of thousands to millions.

\subsubsection*{\q{Nationality} of works and the counting method}

While there are various intrinsic difficulties to bibliometric methods, one of the most challenging aspects to capture for individual works is information about the research institutions to which the contributors belong \cite{Lammey20}.
To investigate the R\&D activities' status at the state or international level, we need information about the countries where the research institutions are located at the time of publication.
Indeed, the literature has demonstrated that internationally coauthored publications are a reliable proxy for research collaboration \cite{Leydesdorff08,Glanzel05,Glanzel01,Melin96,Luukkonen92}.
However, in many cases, such information is unknown or unavailable in a database.
Even if it is available, accurately analysing the metadata can be difficult due to the identification or aggregation of institution names.
This problem is typical of any bibliographic database, including commercial databases, preprint servers and repositories.\endnote{%
As an illustration, arXiv does not require institutional affiliation information on a submitted eprint, resulting in a situation where obtaining the corresponding metadata via the arXiv API is challenging.}

To alleviate, if not resolve, this issue, we employed information recorded in OpenAlex's data on \q{institutions}.
OpenAlex indexes about 109,000 institutions, around 94\% of which have the ROR ID \cite{Lammey20} as the canonical identifier \cite{Priem22}.
The data about institutions are derived from metadata found in Crossref, PubMed, ROR, MAG and publisher websites, liked to individual works through its unique algorithm.
By calling necessary data filtered on the OpenAlex API according to the appropriate conditions, data on the number of works matching the conditions and associated metadata can be obtained in a broken-down format by country.
For simplicity, we refer to the work produced by contributors from institutions in Country X as a \q{work of nationality X}.
Work produced through the collaboration of two contributors, one from an institute in Country X and the other from an institute in Country Y, has dual nationality of X and Y, counted both as a work of nationality X and a work of nationality Y.
According to the terminology introduced here, work generally has multiple nationalities.
If the country of the institution to which all contributors belong is unknown, the work is called a \q{work of unknown nationality}.

The results of the analysis on the nationality status of works by discipline and year are presented in Suppl.~Fig.~\ref{fig:unknownrate}.
Although there were yearly fluctuations, the percentage of works with unknown nationality consistently decreased in all disciplines over the past 50 years.
{\agri} had the highest percentage of works with unknown nationality, at about 60--80\% over the past few decades.
In contrast, {\nano} and {\condensed} had a lower percentage in recent years, at about 30--40\%.
We excluded works with unknown nationality from the analysis due to data unavailability, even though the proportion of such works is large.
We conducted, visualised and interpreted the analysis assuming that the excluded works have trends similar to those of works with a known nationality.

It is worth noting our method for counting works. 
As this paper aims to analyse the international presence of countries and how it has changed over time, we quantify the presence of each country by using a binary indicator (0 or 1) based on whether or not the country's name appears in the affiliation of one of the authors, indicating their involvement.
Let us consider an example of an article coauthored by three individuals, two of whom are affiliated with an institute in Country X and one of whom is affiliated with an institute in Country Y.
If the standard full counting method based on authorship is used, each author is assigned a weight of 1, and the sum of the weights is the number of coauthors, which in this case is 3.
If the fractional counting method based on authorship is used, each author is assigned a weight of one-third, and the sum of the weights is always 1, regardless of the number of coauthors.
In contrast, the binary counting method based on nationality used in this paper assigns a weight of 1 to each country for each article, regardless of the number of authors from that country. 
Therefore, in the example above, the article is counted with a weight of 1 for each Country X and Y, and Country X is not assigned a weight of 2, but rather a weight of 1, since as long as there are nonzero authors from a country, a weight of 1 is assigned to that country.

Let us take another example to see how this specific counting method would work better for the purpose of the present study. 
Consider an article with 10 coauthors from each of 10 different countries, making a total of 100 coauthors. 
Indeed, it is not uncommon for the number of coauthors to exceed 100 (or even 1,000) in the case of extensive collaborative studies \cite{Chawla19,Nogrady23}.
If the full counting method based on authorship were adopted, this article would assign a weight of 10 to each country. 
However, this could lead to an overestimation of each country's contribution to a single scientific result or a significant variation in the value implied by the weights for each article. 
On the other hand, if the fractional counting method based on authorship were adopted, the weights assigned to each country would be 0.1 each. 
However, this could lead to an underestimation of the presence of each country and would only be given a minimum weight in terms of nationality. 
These situations are not ideal for quantifying international presence in every scientific work.
The counting method adopted in this paper, i.e.\ the binary counting method based on nationality presence/absence, assigns a weight of 1 to each country, mitigating the effects of bias that arise from the full and fractional counting methods based on authorship. 
In other words, the involvement or non-involvement of each country in each work can be assessed more appropriately.

\subsection{Clustering of countries\label{subsec:clustering}}

Identifying international collaboration clusters requires grouping countries that cooperate closely. 
However, before we can do this, we must first find a way to quantify the distance between two countries in a reasonable manner.
This requires careful consideration, as simply conceptualising proximity between two countries as the number of collaborative works produced would result in an ill-defined notion of distance.
We require that the distance used for clustering to satisfy the triangle inequality, which means that if Country X and Country Y are close and Country Y and Country Z are close, then Country X and Country Z must also be close to each other.
Notice that even if there are many collaborative works in X and Y and many in Y and Z, this does not necessarily mean that the number of collaborative works in X and Z is large.
This clearly illustrates that the number of collaborative works between two countries cannot simply be associated with the proximity between them.

To ensure conceptual soundness and mathematical well-definedness, we first establish the \textit{affinity} between X and Y, denoted as $A_{\X,\Y}$.
We define this as the number of works with nationalities of both X and Y divided by the total number of works with nationalities of at least one of X and Y.
Mathematically, this is expressed as:
$$
A_{\X,\Y}
\coloneqq\f{\abs{S_{\X}\cap S_{\Y}}}{\abs{S_{\X}\cup S_{\Y}}}
=\f{n_{\X,\Y}}{n_{\X}+n_{\Y}-n_{\X,\Y}}\,,
$$
where $S_{\X}$ denotes the set of works of nationality X in a given period, $n_{\X}\coloneqq \abs{S_{\X}}$ denotes its size, and the same for $S_{\Y}$ and $n_{\Y}$.
Subsequently, we define the \textit{distance} between X and Y as $D_{\X,\Y}\coloneqq1-A_{\X,\Y}$.
See Appendix \ref{app:distance} for further details.
The resulting distance measure ranges between 0 and 1 for arbitrary country pair $\{\X,\Y\}$.
If $D_{\X,\Y}=0$, X and Y always collaborate on all works, indicating that no works of nationality Y are without nationality X, and vice versa.
If $D_{\X,\Y}=1$, then X and Y never collaborate on works, meaning that there are no works with both X and Y nationalities.
This set-theoretic distance metric satisfies the triangle inequality and can be used to calculate the distance between arbitrary country pairs in each period, as well as how the international collaborative clusters have changed over time.

Finally, a hierarchical cluster analysis (HCA) can be performed for the above-defined distance matrix $D$; see Appendix \ref{app:clustering} for the technical details.
HCA is a widely used family of unsupervised statistical methods for classifying a set of items into some hierarchy of clusters (groups) according to the distances among the items. 
This method can provide a new way of looking at the international collaboration sphere when applied to the current context. Specifically, it informs us of which countries are close to each other and to what extent, and as a result, which countries can be considered to form an international research collaboration cluster at what threshold for the closeness.

\section{Results: Kaleidoscopes of international collaboration\label{sec:Results}}

This section presents our main results on work production and the state of international collaboration as viewed through the lens of Open Bibliometrics using the OpenAlex data.

\subsection{Number of works\label{subsec:nworks}}

We begin by presenting the results of our analysis on the number of works produced over the past half-century.
To count works for each country, we adopted the binary counting method based on nationality, as introduced in the previous section.
To reiterate the rule, if a work has nationalities of X and Y, it is counted as 1 work output for each country.
Even if there are multiple contributors from X, the work is only counted as 1 in the production volume for X.
If no country information is known for all the contributors to a given work, the work is counted as \q{unknown}.

\begin{figure}[!htp]
\centering
\vspace{-0.5cm}
    \begin{tabular}{c}
{\small\textrm{\textbf{(a)~ \ai}}}\\
	\end{tabular}
	\begin{tabular}{c}
\begin{minipage}{0.5\hsize}
\begin{center}
\raisebox{-\height}{\includegraphics[align=c, scale=\xsize, vmargin=0mm]{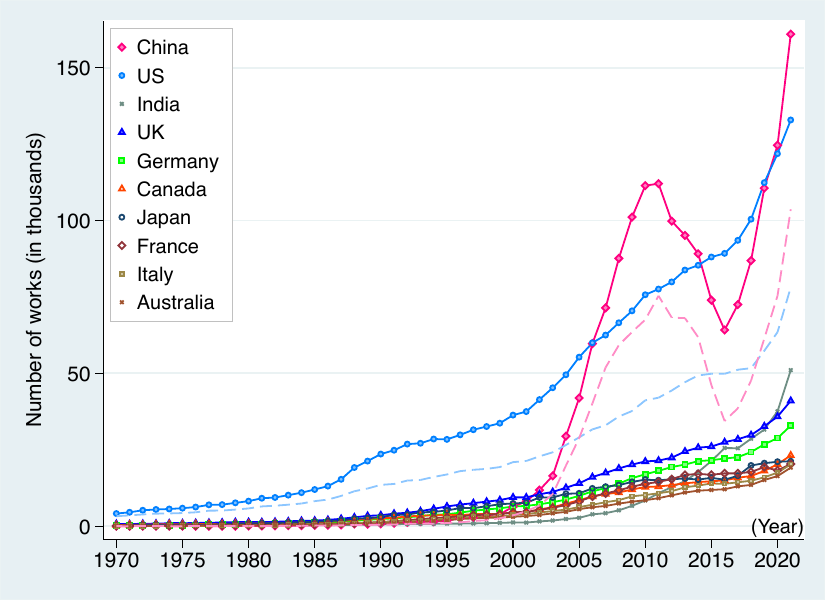}}
\end{center}
	\end{minipage}
\begin{minipage}{0.5\hsize}
\begin{center}
\raisebox{-\height}{\includegraphics[align=c, scale=\xsize, vmargin=0mm]{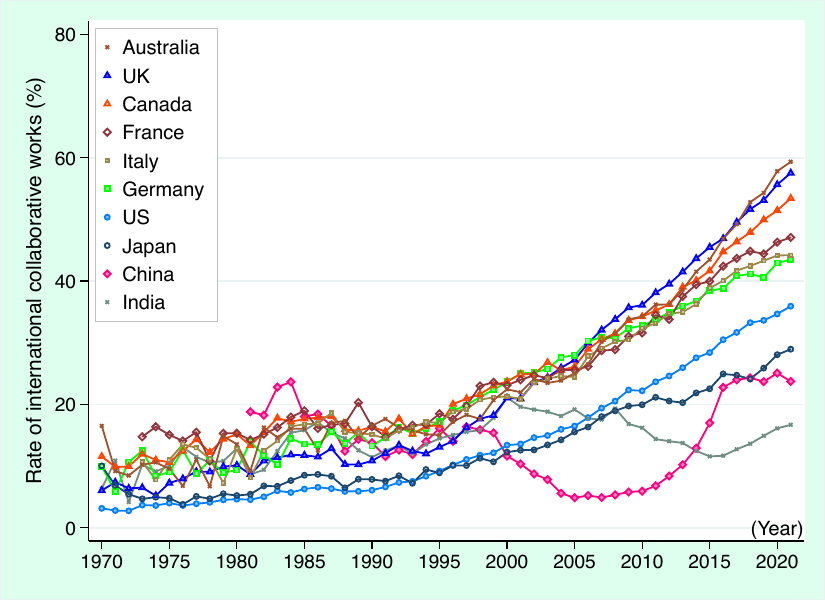}}
\end{center}
	\end{minipage}
    \end{tabular}
\vspace{3mm}\\
    \begin{tabular}{c}
{\small\textrm{\textbf{(b)~ \quantum}}}\\
	\end{tabular}
	\begin{tabular}{c}
\begin{minipage}{0.5\hsize}
\begin{center}
\raisebox{-\height}{\includegraphics[align=c, scale=\xsize, vmargin=0mm]{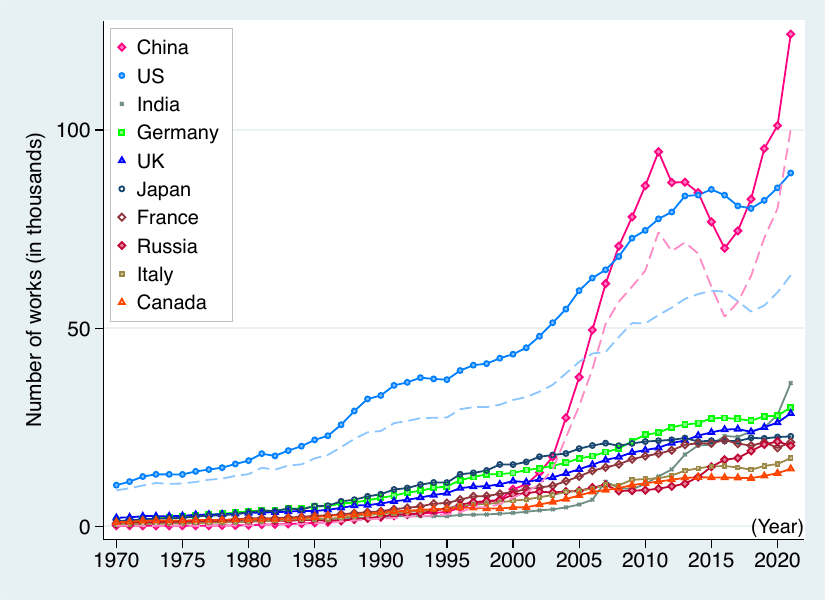}}
\end{center}
	\end{minipage}
\begin{minipage}{0.5\hsize}
\begin{center}
\raisebox{-\height}{\includegraphics[align=c, scale=\xsize, vmargin=0mm]{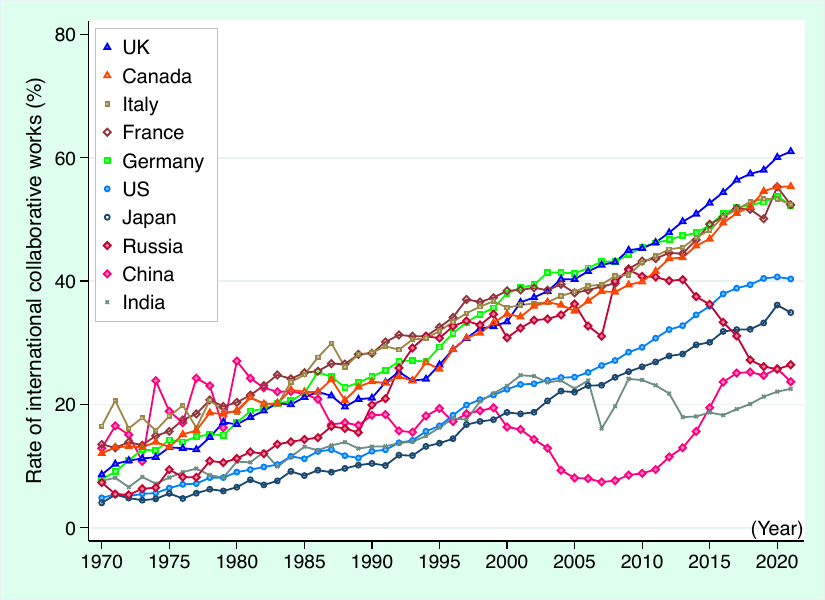}}
\end{center}
	\end{minipage}
    \end{tabular}
\vspace{3mm}\\
    \begin{tabular}{c}
{\small\textrm{\textbf{(c)~ \bio}}}\\
	\end{tabular}
	\begin{tabular}{c}
\begin{minipage}{0.5\hsize}
\begin{center}
\raisebox{-\height}{\includegraphics[align=c, scale=\xsize, vmargin=0mm]{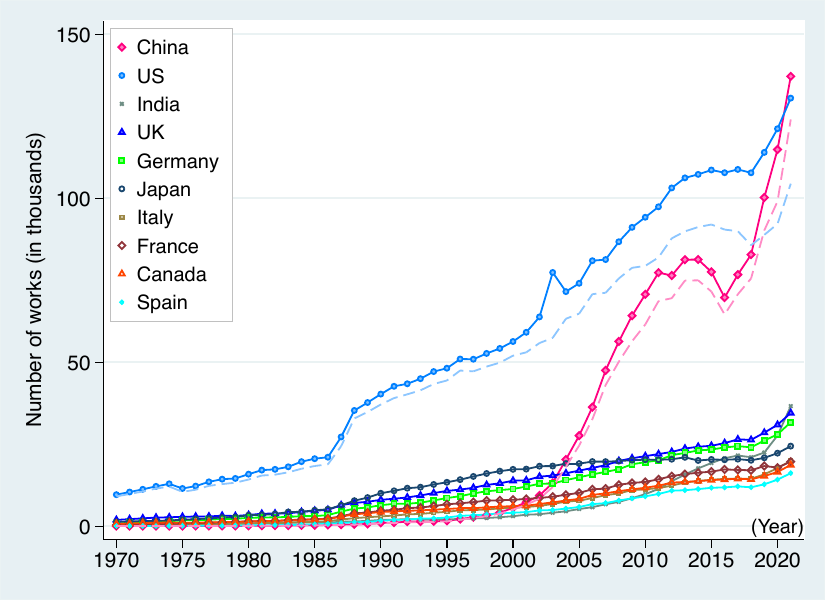}}
\end{center}
	\end{minipage}
\begin{minipage}{0.5\hsize}
\begin{center}
\raisebox{-\height}{\includegraphics[align=c, scale=\xsize, vmargin=0mm]{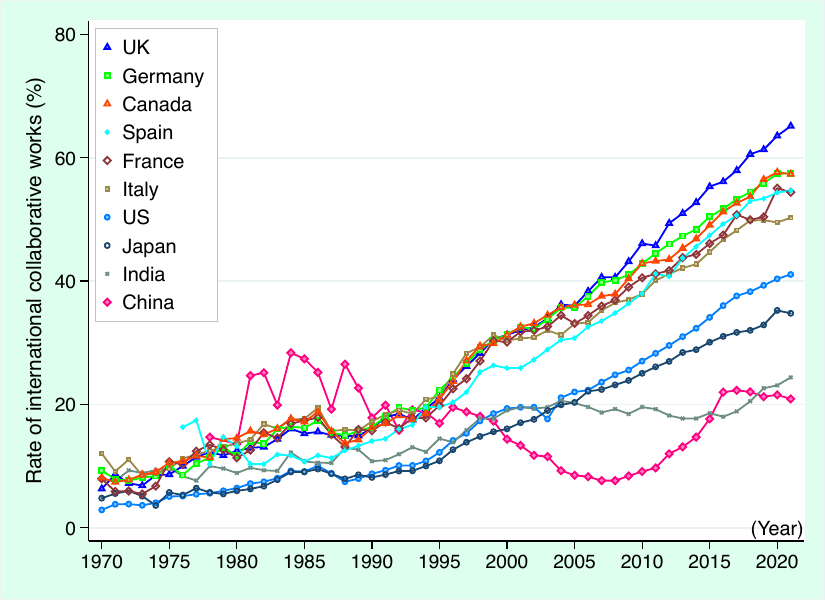}}
\end{center}
	\end{minipage}
    \end{tabular}
\vspace{3mm}
\caption{\textbf{Trends in the number of works (scientific publications) (left) and the international collaboration rate (right).}
The number of works is displayed in thousands.}
\label{fig:line_npaper_intlrate_1}
\end{figure}

The left-hand side graphs of Fig.~\ref{fig:line_npaper_intlrate_1} and Suppl.~Fig.~\ref{fig:line_npaper_intlrate_2} show the trend in the number of works for the top 10 countries in work production in each discipline in 2001--2020.
To save space, we present the results for 3 disciplines---{\ai}, {\quantum} and {\bio}---in Fig.~\ref{fig:line_npaper_intlrate_1} and those for the rest 12 disciplines in Suppl.~Fig.~\ref{fig:line_npaper_intlrate_2} in \ref{app:add.examples}.
In all of these disciplines, it is noticeable that China has shown dramatic growth over the past two decades \cite{Okamura23,NSB-NSF21,Yuan18,He09,RF21}.
For example, in {\ai}, China surpassed the US around 2020, producing more than 150,000 works in 2021 (Fig.~\ref{fig:line_npaper_intlrate_1}a).
It also overtook the US in {\quantum} (Fig.~\ref{fig:line_npaper_intlrate_1}b) and {\bio} (Fig.~\ref{fig:line_npaper_intlrate_1}c) by 2021.
Thus, what was an era of the US single-power a few decades ago has transitioned to a new era of two powerhouses, i.e.\ the US and China.
For reference, the graphs for the US and China are also shown for the case where only journal papers are counted (dashed lines in weaker colours).
It can be seen that the trends over time are generally similar between the case where all works are counted and the case where only journal papers are counted.
However, the volume of works differs remarkably depending on the discipline, indicating how significant the contributions of works in forms other than journal papers can be for some disciplines \cite{Okamura22,Kim19,Lariviere14}.\endnote{%
See \citek{Okamura22b} for the results for all the top 30 countries and the 15 disciplines, where the same conclusion can be confirmed.}

While we do not delve into detailed analyses of individual curve profiles in this paper, it is important to consider the underlying reasons for each observation. 
For instance, in China, many disciplines exhibit an \q{N-shaped} curve with a peak around 2011, followed by a sharp decline, and then a sharp rise from around 2016. 
A major factor contributing to this trend could be the number of researchers. 
From 2000 to 2020, the total number of researchers in China has been increasing overall, but there was a significant drop in 2008 and 2009. According to data from the \citek{OECD23}, the number of researchers per 1,000 employed decreased from 2.11 in 2008 to 1.52 in 2009. 
Similarly, data from the \citek{UIS23} shows a decrease in researchers in R\&D (per million people) from 1,176 in 2008 to 847 in 2009. 
Although the reason for this decline in the number of researchers is still unclear, it is possible that its impact is reflected in a decrease in the number of scientific works after about two years. 
We will revisit this and other individual curves in future studies.

\subsection{International collaboration rate\label{subsec:intlcolrate}}

Next, the right-hand side graphs of Fig.~\ref{fig:line_npaper_intlrate_1} and Suppl.~Fig.~\ref{fig:line_npaper_intlrate_2} depict the trend of the international collaboration rate by discipline and country over the past half-century.
The international collaboration rate for a specific year and country is the yearly number of international collaborative works divided by the total number of works produced.
Only cases in which the yearly number of works produced is 100 or more are shown for each discipline and country.
As a result, data are missing around 1970--1990 for some disciplines and countries.
The countries selected for display are the top 10 countries in work production in each discipline during 2001--2020, which are the same as the corresponding left-hand side graphs.\endnote{%
This approach does not reveal countries that have high international cooperation rates but are outside the top 10 in terms of work production. 
To complement the analysis results, Suppl.~Fig.~\ref{fig:line_intlrate2} presents an additional set of diagrams.}

In all disciplines, there has been a steadily increasing trend in the international collaboration rate, in line with the previous studies' findings based on commercial data \cite{Kwiek22,Leydesdorff08}.\endnote{%
This trend does not have a direct causal relationship with the upward trend in the rate of works with unknown nationality seen in Suppl.~Fig.~\ref{fig:unknownrate} because the international collaboration rate discussed here only refers to works with a known nationality.}
Over the past two decades, the UK has been among the highest in many disciplines, while European countries such as Germany, France and Italy have maintained high levels across the board.
By contrast, despite the US's rising trend \cite{NSB-NSF21,Adams18}, its international collaboration rate is generally lower than the top-tier countries mentioned above in all 15 disciplines.
For example, in 2021, it is around 40\% for the 3 disciplines displayed in Fig.~\ref{fig:line_npaper_intlrate_1}.
The international collaboration rate observed for China and India is notably lower \cite{NSB-NSF21,OECD17,OECD16}, which is consistent with prior studies based on commercial data.
It appears that a relatively high percentage of work is completed only with R\&D resources within their own countries, although it is unclear whether this is due to the policies of R\&D institutions or a natural consequence of their large researcher populations.
Moreover, some disciplines, such as {\particle}, {\aerospace}, {\nuclear} and {\astro} (Suppl.~Fig.~\ref{fig:line_npaper_intlrate_2}c, d, e and k, respectively), have a marked downward trend in Russia's international collaboration rate over the past decade \citep[c.f.][]{Kwiek22}.
In recent years, other eye-catching features include Canada's high rate of international collaboration in many disciplines, including {\ai}, {\quantum}, {\bio}, {\neuro} and {\earth} (Fig.~\ref{fig:line_npaper_intlrate_1}a, b, c; Suppl.~Fig.~\ref{fig:line_npaper_intlrate_2}g and j, respectively), Germany's and Australia's in {\agri} (Suppl.~Fig.~\ref{fig:line_npaper_intlrate_2}b), Spain's in {\particle} and {\astro} (Suppl.~Fig.~\ref{fig:line_npaper_intlrate_2}c and k, respectively), France's and Canada's in {\aerospace} (Suppl.~Fig.~\ref{fig:line_npaper_intlrate_2}d), Italy's in {\nuclear} (Suppl.~Fig.~\ref{fig:line_npaper_intlrate_2}e) and Australia's in {\envi} (Suppl.~Fig.~\ref{fig:line_npaper_intlrate_2}i).\endnote{%
Switzerland is ranked 12th in works production in {\particle} and is not shown in Suppl.~Fig.~\ref{fig:line_npaper_intlrate_2}c.
However, if it were, it would consistently be ranked the highest in terms of the international collaboration rate, reflecting the influence of CERN, the European Organization for Nuclear Research (as shown in Suppl.~Fig.~\ref{fig:line_intlrate2}).}

Additionally, comparisons across disciplines are also implicative, as shown in Suppl.~Fig.~\ref{fig:line_intlrate}.
International collaboration is particularly indispensable in large-scale academic R\&D disciplines such as {\particle} and {\astro}, which usually require extended time and high cost, gathering many contributors from many countries.
The international collaboration rates have been consistently high throughout the past half-century.
Other disciplines, such as {\condensed} and {\earth}, have also shown relatively high rates of international collaboration in recent decades, and {\quantum} has recently shown a comparatively high rate as well.

\subsection{Bilateral collaborative relationships\label{subsec:bilateral}}

\begin{figure}[htp]
\centering
\vspace{-0.5cm}
    \begin{tabular}{c}
    \begin{minipage}{0.03\hsize}
\begin{flushleft}
    \hspace{-0.7cm}\rotatebox{90}{\period{4}{2011--2020}}
\end{flushleft}
	\end{minipage}
	\begin{minipage}{0.33\hsize}
\begin{flushleft}
\raisebox{-0.0cm}{\hspace{-6mm}\small\textrm{\textbf{(a)~ \ai}}}\\[6mm]
\raisebox{\height}{\includegraphics[trim=2.0cm 1.8cm 0cm 1.5cm, align=c, scale=\chordsize, vmargin=0mm]{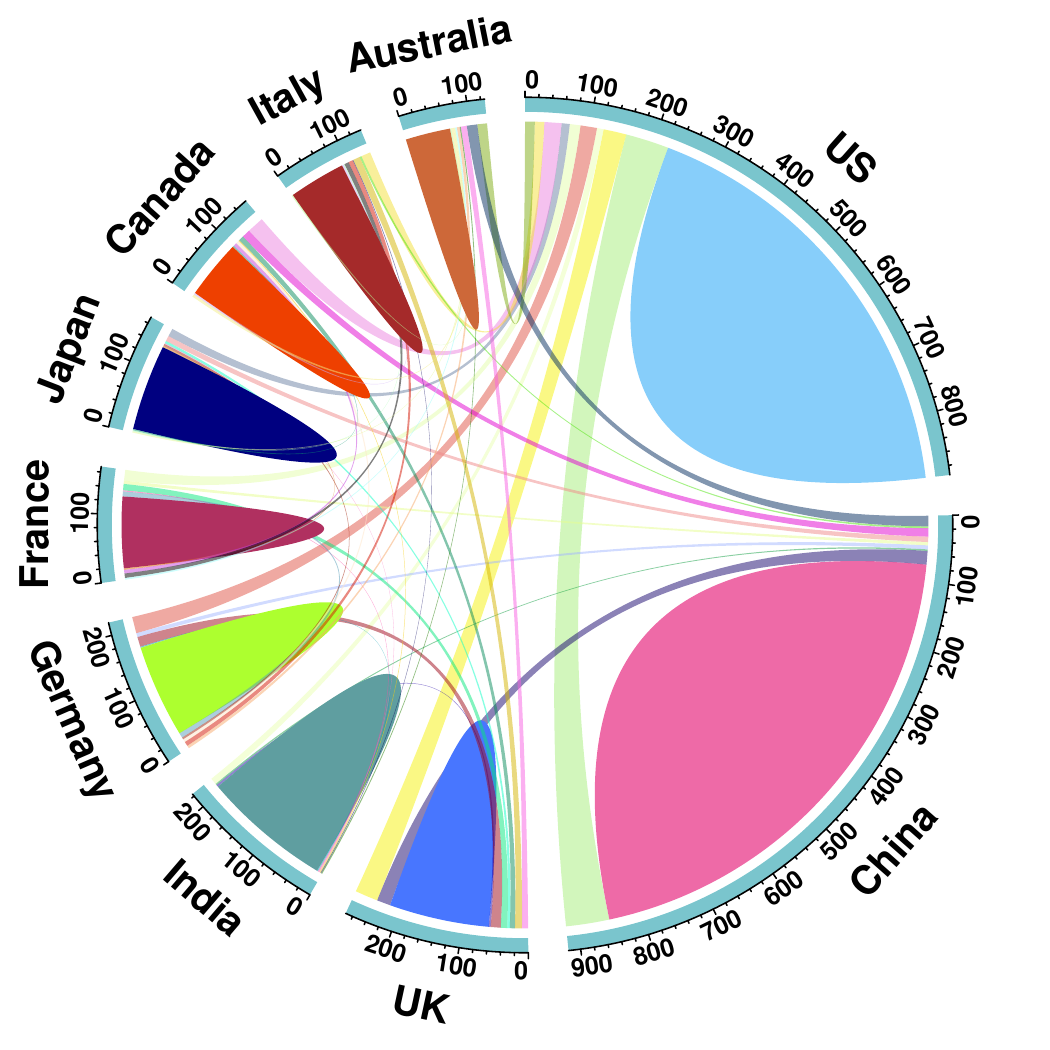}}
\end{flushleft}
    \end{minipage}
	\begin{minipage}{0.33\hsize}
\begin{flushleft}
\raisebox{-0.0cm}{\hspace{-6mm}\small\textrm{\textbf{(b)~ \quantum}}}\\[6mm]
\raisebox{\height}{\includegraphics[trim=2.0cm 1.8cm 0cm 1.5cm, align=c, scale=\chordsize, vmargin=0mm]{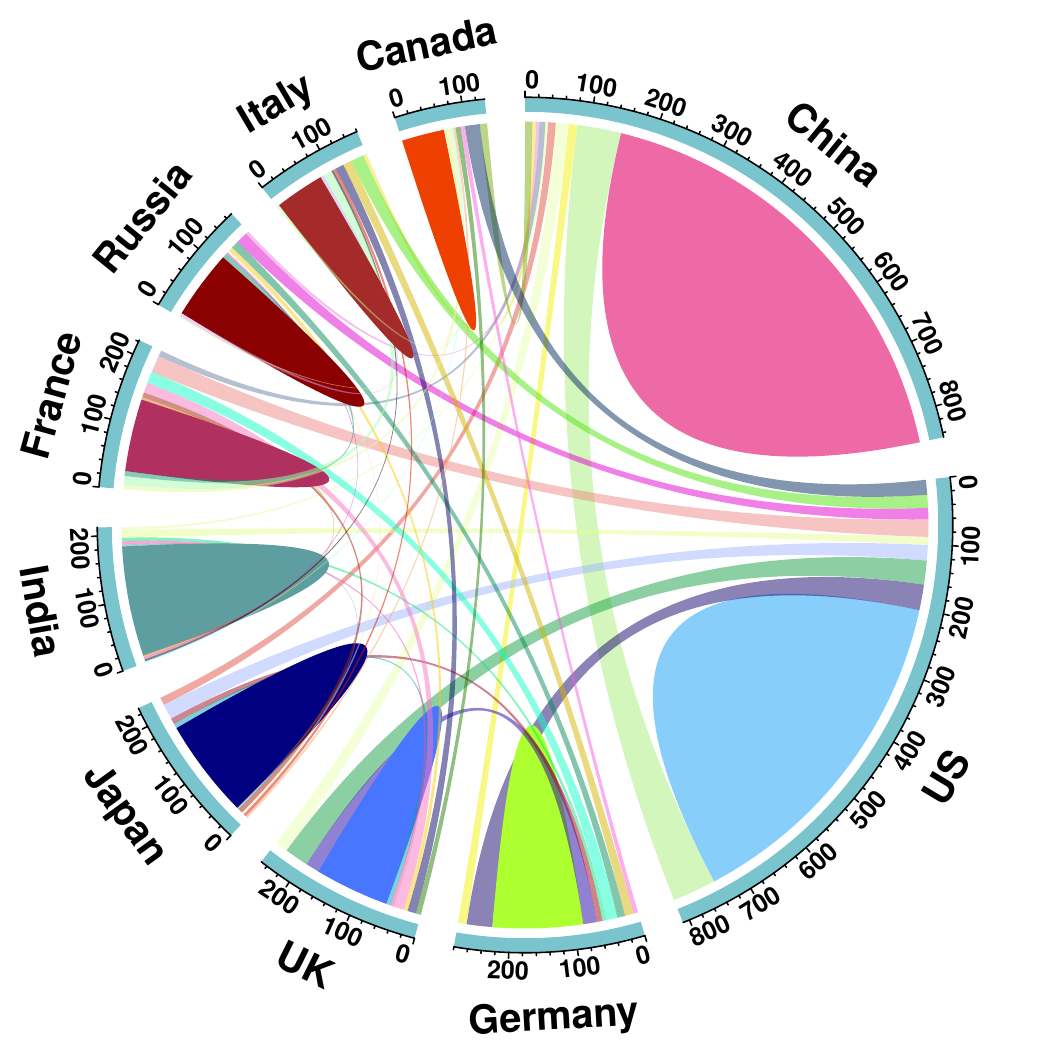}}
\end{flushleft}
	\end{minipage}
	\begin{minipage}{0.33\hsize}
\begin{flushleft}
\raisebox{-0.0cm}{\hspace{-6mm}\small\textrm{\textbf{(c)~ \bio}}}\\[6mm]
\raisebox{\height}{\includegraphics[trim=2.0cm 1.8cm 0cm 1.5cm, align=c, scale=\chordsize, vmargin=0mm]{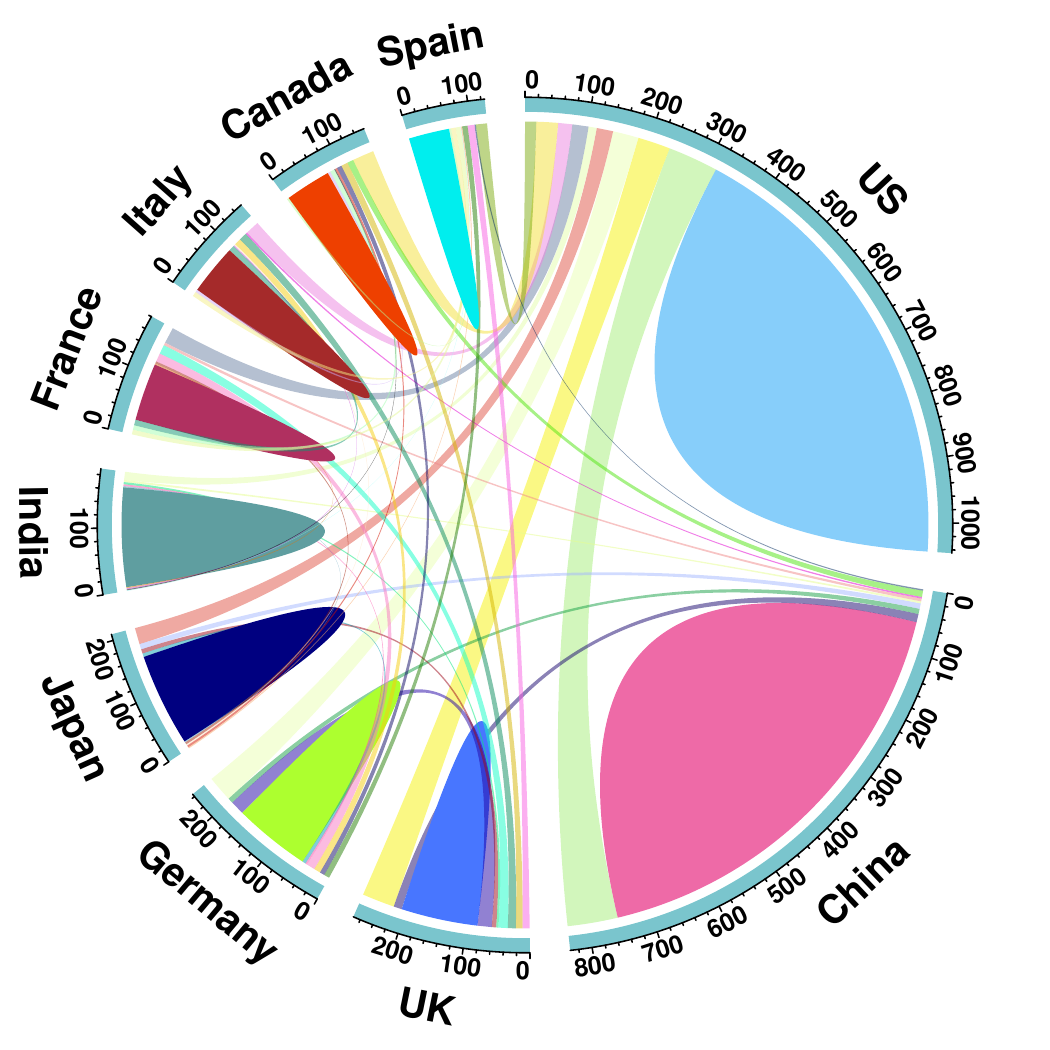}}
\end{flushleft}
	\end{minipage}
    \end{tabular}

\vspace{1mm}
\hspace{-0.5cm}{\marrow}\quad\dotfill
\vspace{4mm}

    \begin{tabular}{c}
    \begin{minipage}{0.03\hsize}
\begin{flushleft}
    \hspace{-0.7cm}\rotatebox{90}{\period{3}{2001--2010}}
\end{flushleft}
	\end{minipage}
	\begin{minipage}{0.33\hsize}
\begin{flushleft}
\raisebox{\height}{\includegraphics[trim=2.0cm 1.8cm 0cm 1.5cm, align=c, scale=\chordsize, vmargin=0mm]{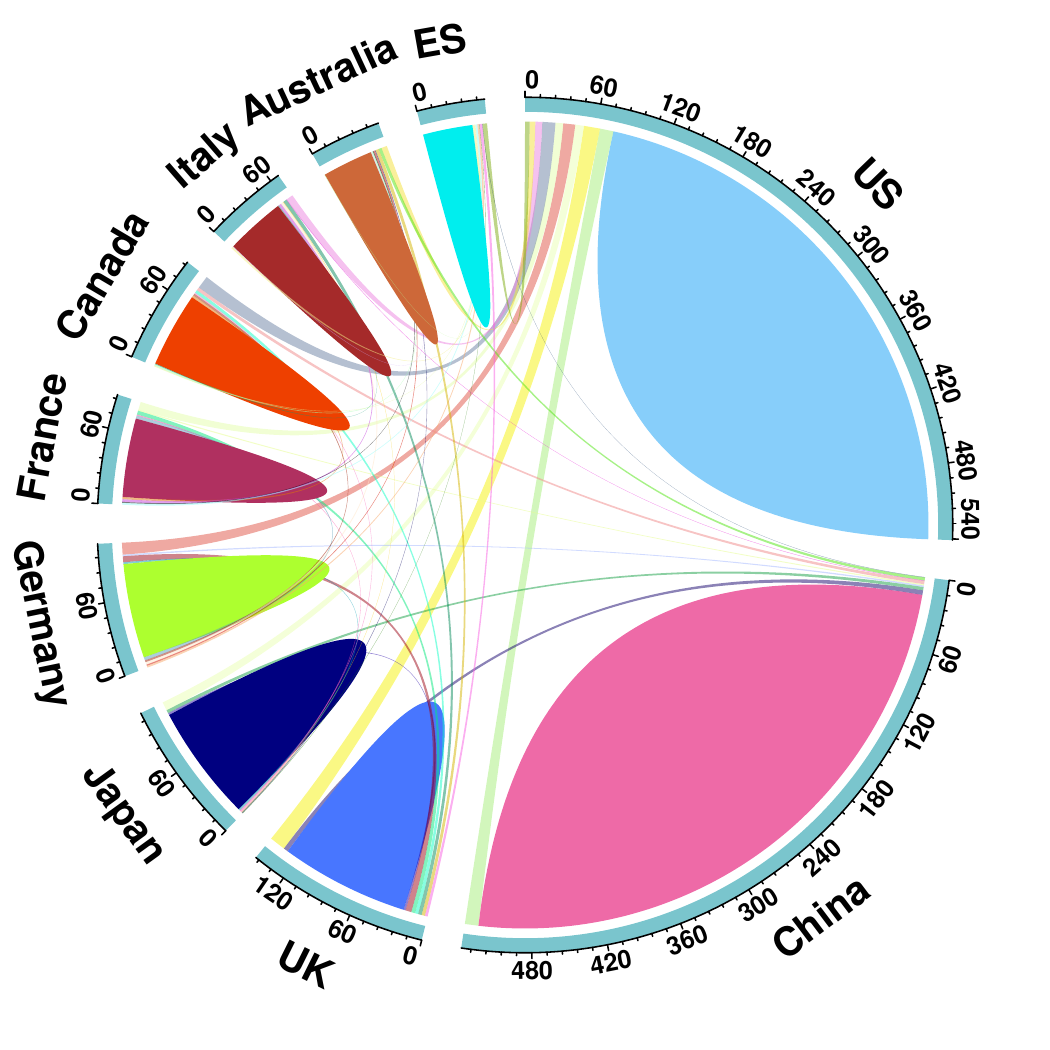}}
\end{flushleft}
    \end{minipage}
	\begin{minipage}{0.33\hsize}
\begin{flushleft}
\raisebox{\height}{\includegraphics[trim=2.0cm 1.8cm 0cm 1.5cm, align=c, scale=\chordsize, vmargin=0mm]{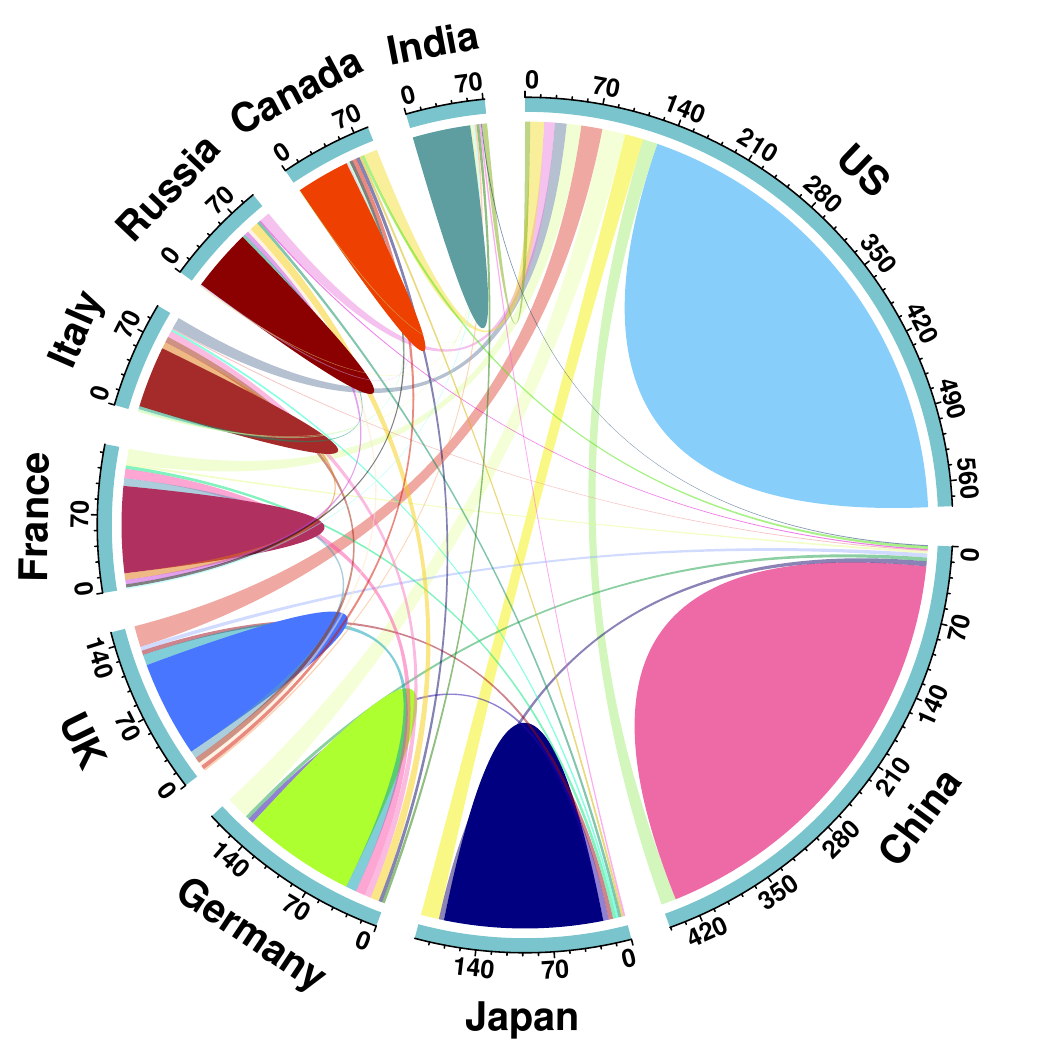}}
\end{flushleft}
	\end{minipage}
	\begin{minipage}{0.33\hsize}
\begin{flushleft}
\raisebox{\height}{\includegraphics[trim=2.0cm 1.8cm 0cm 1.5cm, align=c, scale=\chordsize, vmargin=0mm]{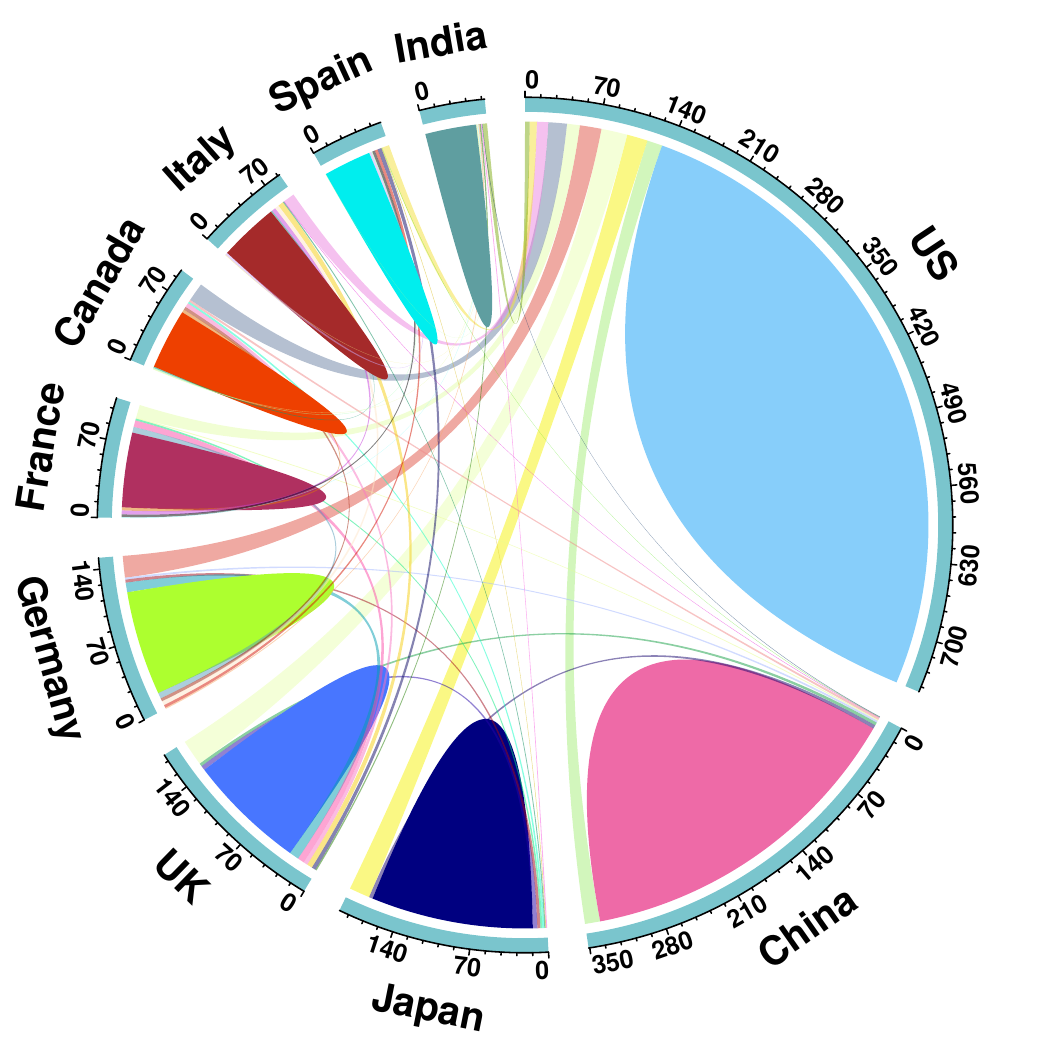}}
\end{flushleft}
	\end{minipage}
    \end{tabular}

\vspace{1mm}
\hspace{-0.5cm}{\marrow}\quad\dotfill
\vspace{4mm}

    \begin{tabular}{c}
    \begin{minipage}{0.03\hsize}
\begin{flushleft}
    \hspace{-0.7cm}\rotatebox{90}{\period{2}{1991--2000}}
\end{flushleft}
	\end{minipage}
	\begin{minipage}{0.33\hsize}
\begin{flushleft}
\raisebox{\height}{\includegraphics[trim=2.0cm 1.8cm 0cm 1.5cm, align=c, scale=\chordsize, vmargin=0mm]{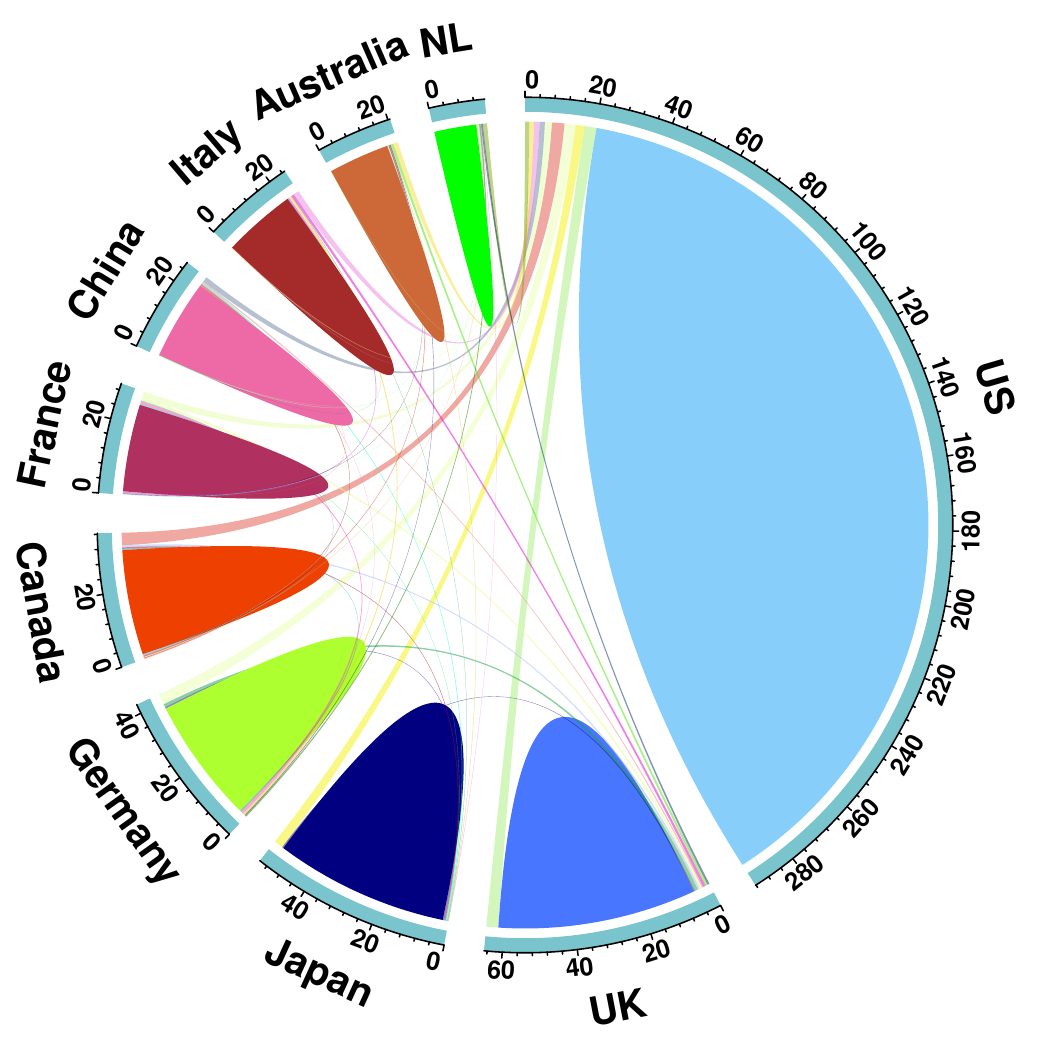}}
\end{flushleft}
    \end{minipage}
	\begin{minipage}{0.33\hsize}
\begin{flushleft}
\raisebox{\height}{\includegraphics[trim=2.0cm 1.8cm 0cm 1.5cm, align=c, scale=\chordsize, vmargin=0mm]{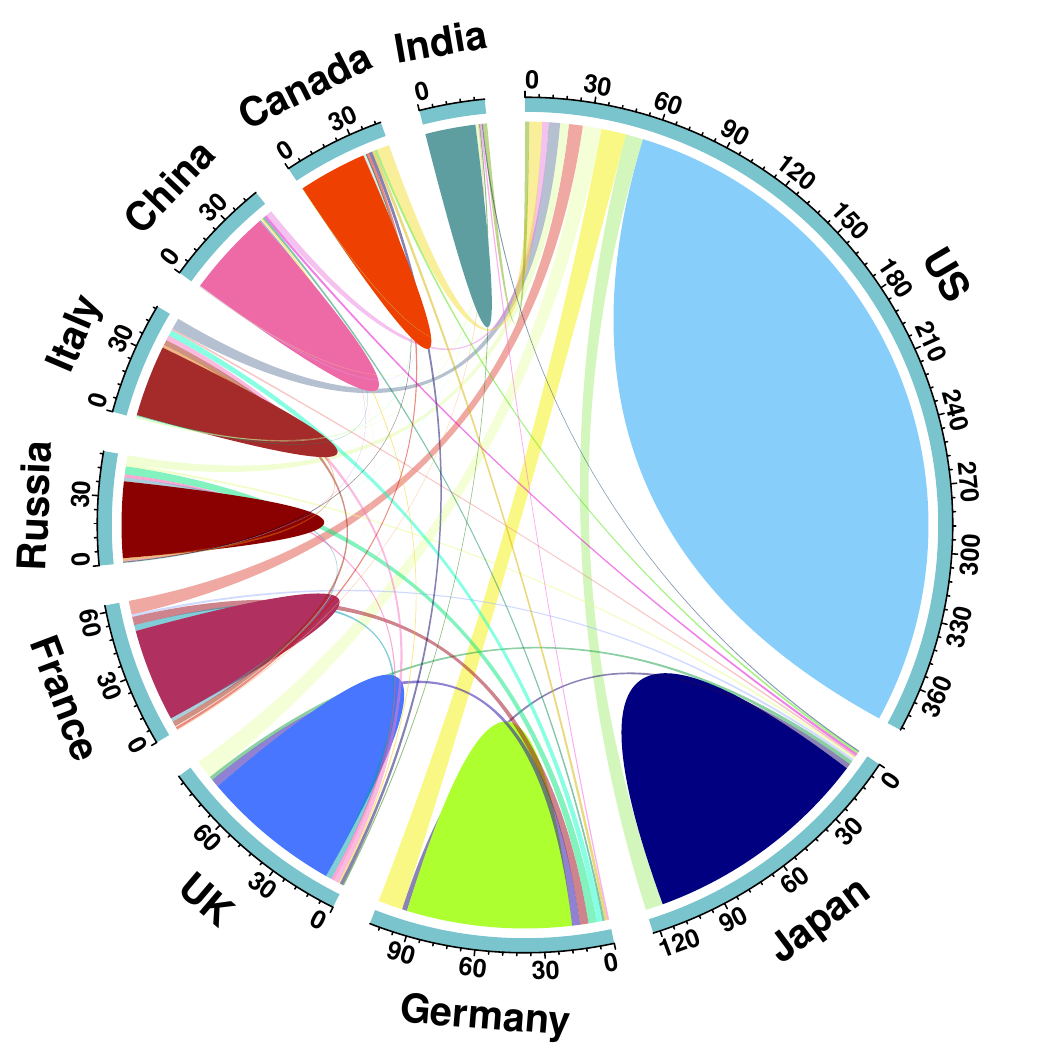}}
\end{flushleft}
	\end{minipage}
	\begin{minipage}{0.33\hsize}
\begin{flushleft}
\raisebox{\height}{\includegraphics[trim=2.0cm 1.8cm 0cm 1.5cm, align=c, scale=\chordsize, vmargin=0mm]{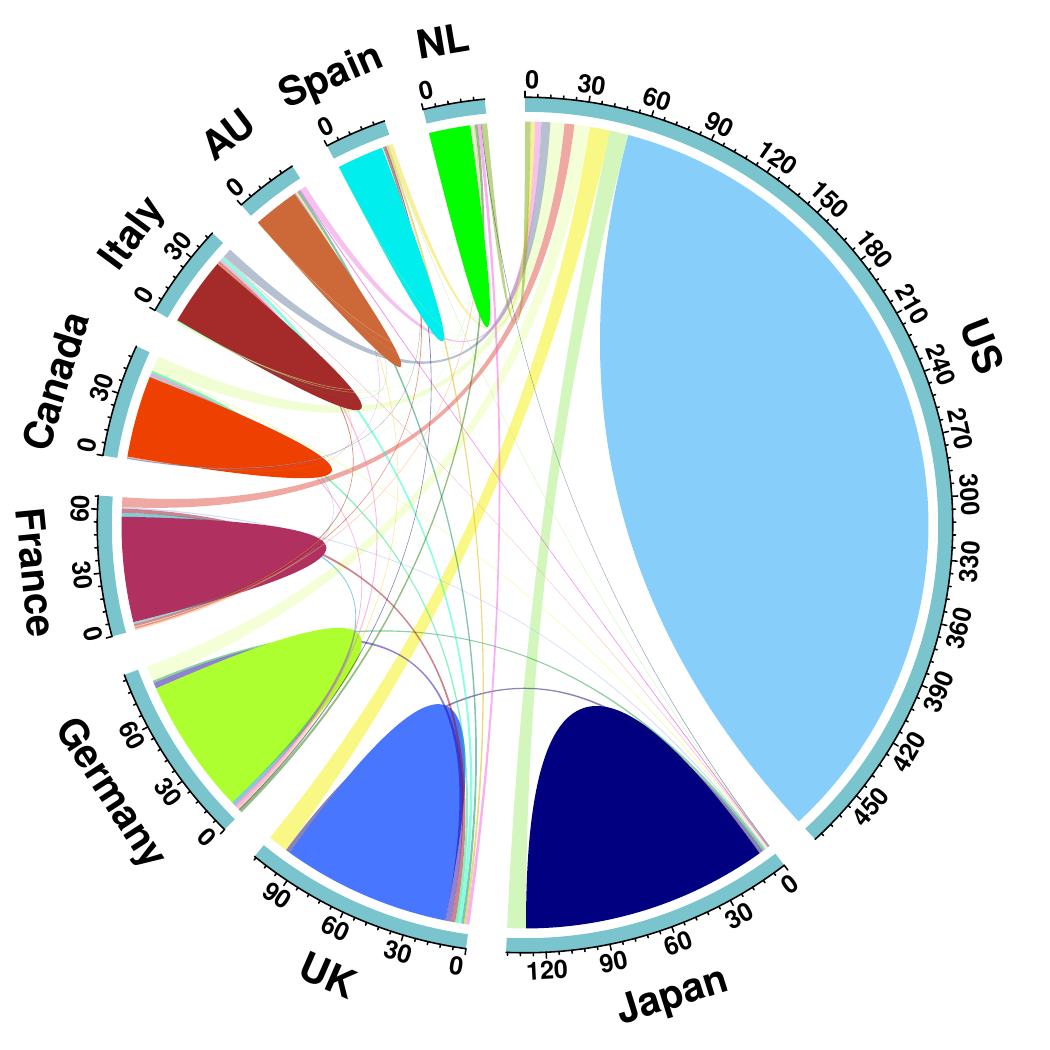}}
\end{flushleft}
	\end{minipage}
    \end{tabular}

\vspace{1mm}
\hspace{-0.5cm}{\marrow}\quad\dotfill
\vspace{4mm}

    \begin{tabular}{c}
    \begin{minipage}{0.03\hsize}
\begin{flushleft}
    \hspace{-0.7cm}\rotatebox{90}{\period{1}{1971--1990}}
\end{flushleft}
	\end{minipage}
	\begin{minipage}{0.33\hsize}
\begin{flushleft}
\raisebox{\height}{\includegraphics[trim=2.0cm 1.8cm 0cm 1.5cm, align=c, scale=\chordsize, vmargin=0mm]{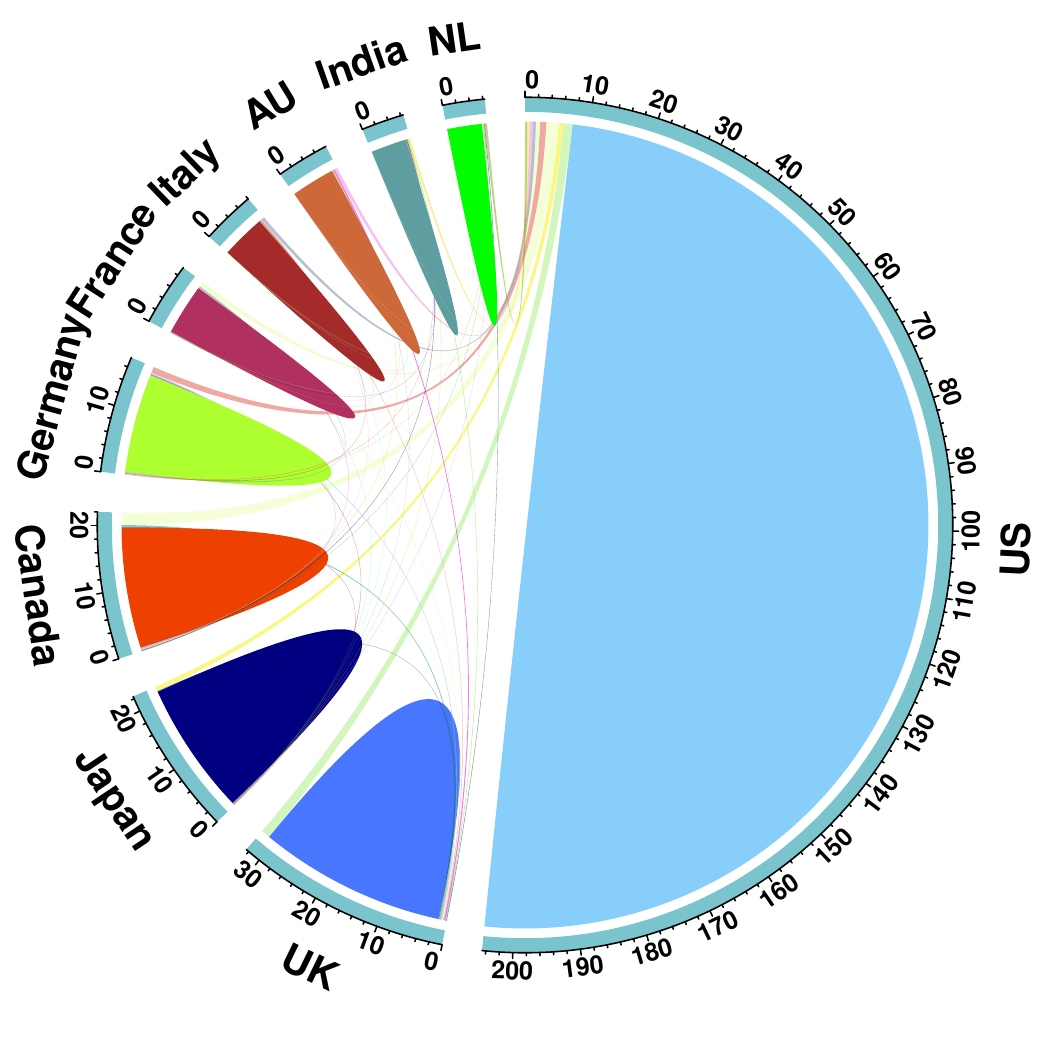}}
\end{flushleft}
    \end{minipage}
	\begin{minipage}{0.33\hsize}
\begin{flushleft}
\raisebox{\height}{\includegraphics[trim=2.0cm 1.8cm 0cm 1.5cm, align=c, scale=\chordsize, vmargin=0mm]{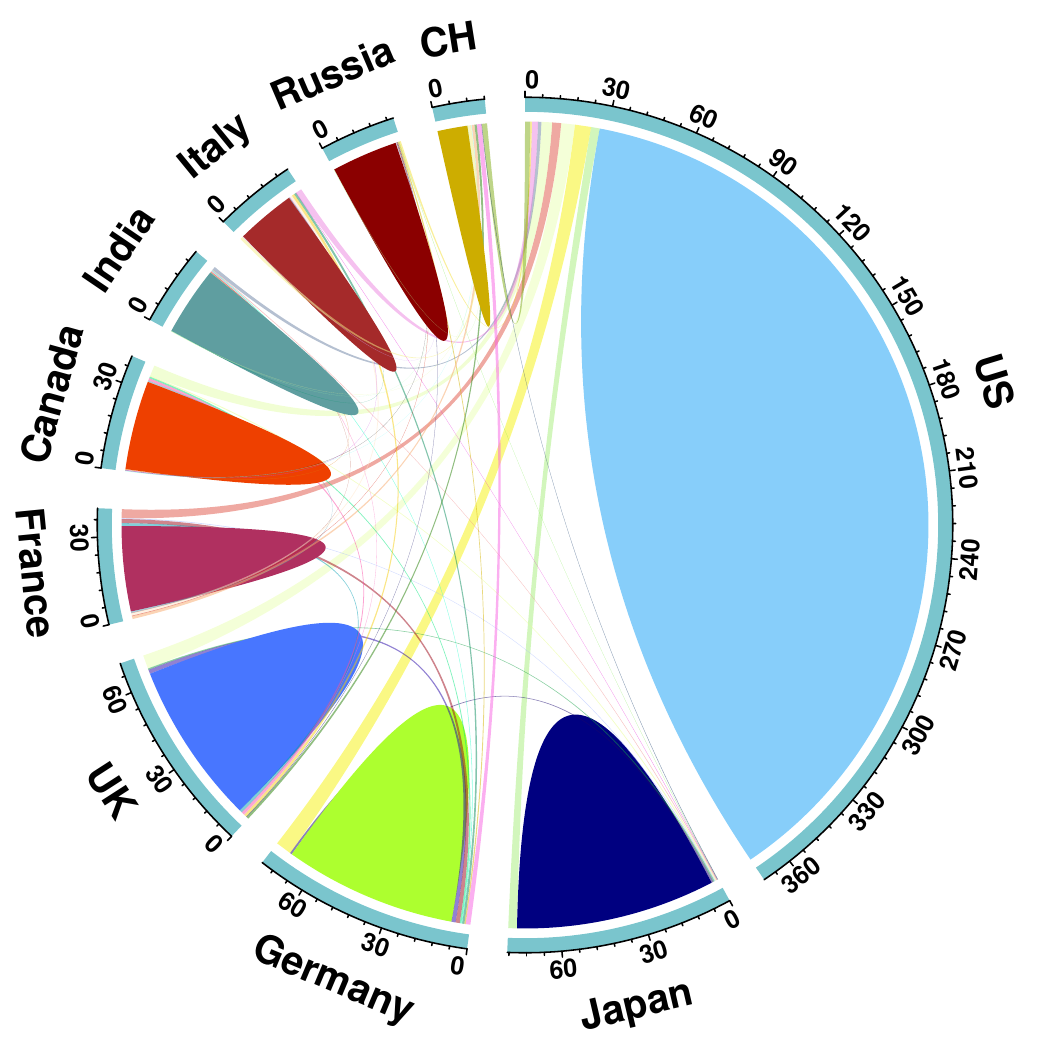}}
\end{flushleft}
	\end{minipage}
	\begin{minipage}{0.33\hsize}
\begin{flushleft}
\raisebox{\height}{\includegraphics[trim=2.0cm 1.8cm 0cm 1.5cm, align=c, scale=\chordsize, vmargin=0mm]{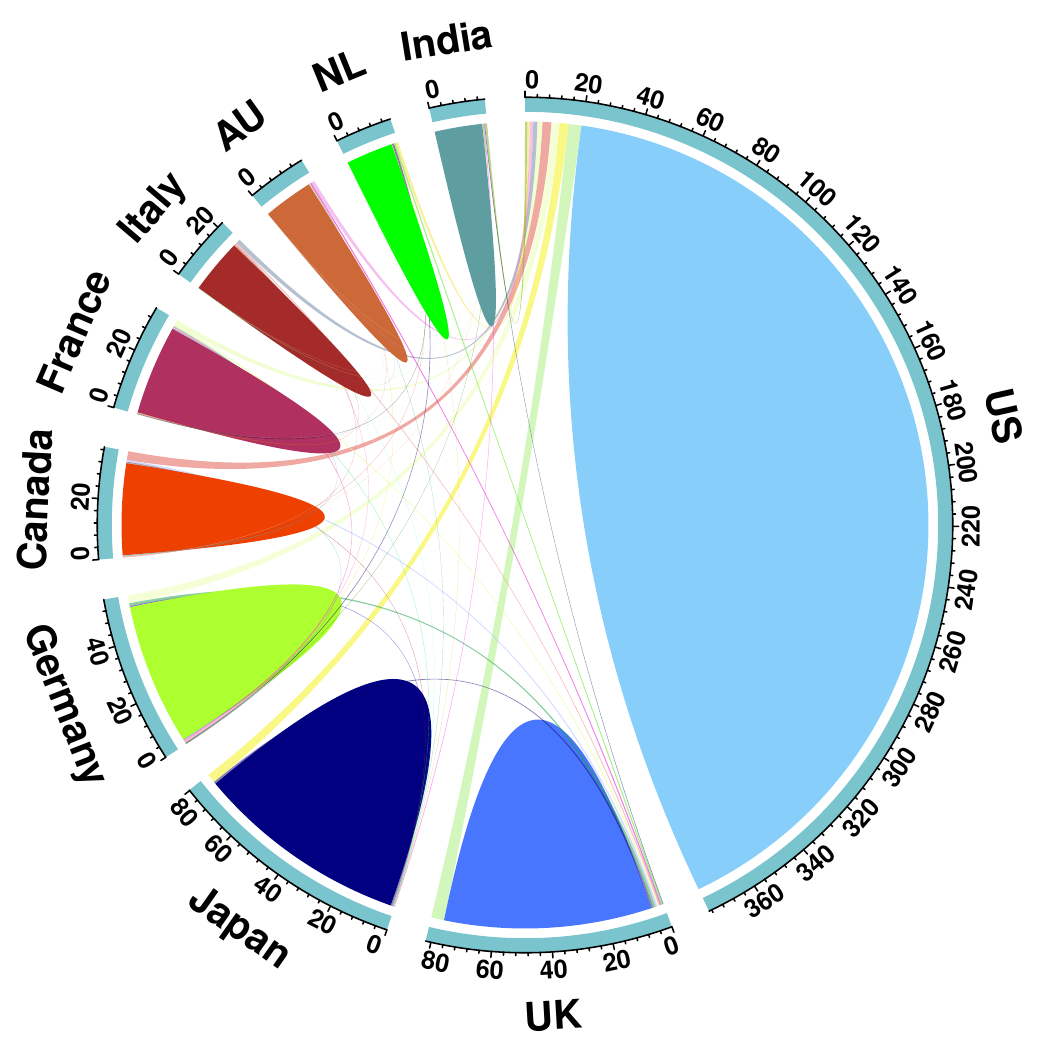}}
\end{flushleft}
	\end{minipage}
    \end{tabular}
\vspace{2.5mm}
\caption{\textbf{Change in bilateral relationships over time.}
The number of works is displayed in thousands.}
\label{fig:chord_1}
\end{figure}

Many previous studies have reported on trends and patterns of international collaboration by country \cite{Kwiek22,NSB-NSF21,Adams18,Yuan18,OECD17,OECD16,Adams12,He09,Mattsson08,Adams07,Glanzel01,Luukkonen92}.
Below we present the results of our analysis of the breakdown of specific collaborative partner countries based on the OpenAlex data for each of the 15 disciplines.
Fig.~\ref{fig:chord_1} and Suppl.~Fig.~\ref{fig:chord_2} divide the half-century from 1971 to 2020 into four periods: Period \rn{1} (1971--1990), Period \rn{2} (1991--2000), Period \rn{3} (2001--2010) and Period \rn{4} (2011--2020).\endnote{%
The four periods defined here are simply a mechanical division of a 50-year time span into four for the sake of calendar convenience.
They do not take into account any economic, social or geopolitical changes that may have occurred during this time. 
Therefore, it should be noted that trends such as the \q{N-shaped} trend in the number of works in China noted in Section \ref{subsec:nworks} would be masked when illustrated as snapshots (Fig.~\ref{fig:chord_1}) in integral values for each period. 
This equally applies to the later Section \ref{subsec:clusters} (Fig.~\ref{fig:cdend_1}).}
The chord diagrams visualise the status of bilateral collaborative relationships for each discipline and period.
The countries selected for display are the top 10 countries in work production in each discipline and period.
The scale along the circumference edge indicates the number of produced works (in thousands), and the width of the band connecting the two country's arcs is proportional to the number of works collaboratively produced by them during each period.
Some country names are abbreviated by two-letter country codes (\href{https://en.wikipedia.org/wiki/ISO_3166-1_alpha-2}{ISO 3166-1 alpha-2}) to make the diagrams easier to read.

A common trend among the disciplines is, again, China's remarkable progress that began this century \cite{Yuan18,He09,RF21}, accompanied by a decline in the relative positions of the US and other major countries.
For example, in the {\bio} discipline (Fig.~\ref{fig:chord_1}c), the US accounted for just under half of global output in Period \rn{1}, and its international collaboration rate was low, at around 5\%.
Over time, the US presence has declined significantly in relative terms.
By Period \rn{4}, its presence on the chord diagram of these top 10 countries had dropped to about a quarter of the total.
The reason for this is China's major breakthrough since Period \rn{3}.
When viewed on a 10-year period-integrated basis, the US still holds the top position in Period \rn{4}, but when viewed on an annual basis, China has already overtaken the US in the top position by 2021 (see Fig.~\ref{fig:line_npaper_intlrate_1}c).

Also evident is the revitalisation of diverse international collaboration.
The growing mutual presence of the US and China can be seen from the expanding width of the band connecting the two countries.
The international collaboration rate in the US has been on an upward trend with China and other countries, which is consistent with the findings from previous studies based on commercial data \cite{NSB-NSF21,Adams18}, resulting in the declining share of solely produced works (a hump-shaped part) on the US arc, from approximately 95\% to 70\%.
These chord diagrams indicate that over the past half-century, many disciplines have moved away from an era of single power (i.e.\ the US) and towards an era of collaboration among a diverse range of countries.
This feature is particularly evident in {\particle} and {\astro} (Suppl.~Fig.~\ref{fig:chord_2}c and k), where the chord diagram becomes more colourful and balanced as we move towards Period \rn{4}.

In the following, we particularly focus on the collaborative relationship between the US and China.
As noted above, the two countries have indeed deepened the relationship as an overall trend over the past half-century.
However, given the strained US--China relationship in recent years in the policy arena, we aim to examine whether geopolitical aspects have impacted international research collaboration through our bibliometric analysis. 
Looking solely at the increase or decrease in the number of coauthored papers between the two countries is inadequate in providing a complete picture of this impact.
This is because an increase in the number of coauthored papers does not necessarily indicate a deepening relationship if the number of papers is increasing worldwide as a global trend.
To effectively measure the degree of bilateral collaborative relationships, we must scale the absolute number of works per the trend of the times.
As an appropriate indicator for this purpose, we adopt the affinity measure introduced in Section \ref{subsec:clustering}.

\begin{figure}[!tp]
\centering
\vspace{-0.5cm}
\noindent
\includegraphics[align=c, scale=0.94, vmargin=1mm]{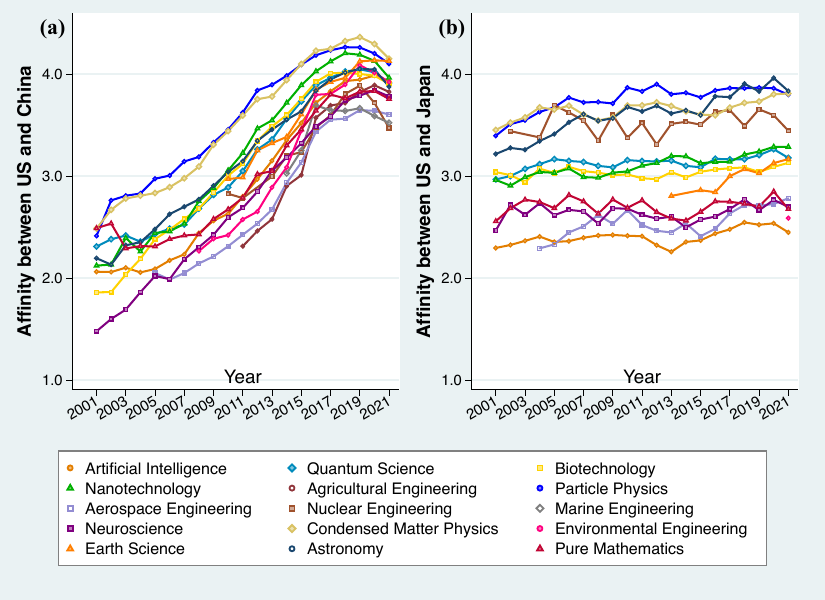}
\caption{\textbf{Trends in the affinity between the US and China (a) and the US and Japan (b).}
}
\label{fig:US-CN(-JP)}
\end{figure}

Figure \ref{fig:US-CN(-JP)}a depicts the affinity between the US and China over the past two decades, considering only data in which the yearly number of collaborative works between the two countries is 100 or more.
For clarity, the affinity measure is rescaled; specifically, the rescaling is achieved as $A\mapsto \tilde{A}\coloneqq \ln(A/\ep)$ with $\ep=0.001$.
Note that while the absolute value of this rescaled affinity measure ($\tilde{A}$) does not have direct physical significance, its relative comparison across disciplines or periods and increase/decrease over time do.
Regarding the comparison across disciplines, on the one hand, the affinity between the two countries has been relatively small in massive and heavy R\&D fields such as {\aerospace}, {\nuclear} and {\marine} in recent years. 
On the other hand, a relatively large affinity between the two countries is observed for {\condensed} and {\particle}.
It is noteworthy that the affinity between the US and China has been remarkably growing in all disciplines in common.
As a reference, Fig.~\ref{fig:US-CN(-JP)}b uses the same analytical method to show the affinity between the US and Japan, also located in Asia.
Despite the fact that the number of coauthorships between the US and Japan has increased during the past decade, the affinity between the two countries has remained almost the same in all disciplines with slight fluctuations, albeit at different levels per discipline. 
Therefore, it cannot be concluded that the relationship between the two countries has significantly deepened during the period. 
By contrast, the overall increasing affinity trend in Fig.~\ref{fig:US-CN(-JP)}a suggests that the relationship between the US and China is actually deepening.

Another noteworthy point about the US--China relationship in Fig.~\ref{fig:US-CN(-JP)}a is that the affinity between the two countries started decreasing again around 2019 for most disciplines.
Although the implications of this phenomenon require careful examination, it could reflect the \q{chilling effect} stemming from the measures taken in the US around 2018 to prevent technology outflows to China.
The observed recent repulsive trend aligns with earlier research based on commercial data, which found a sharp decrease in the number of researchers with affiliations in both the US and China in 2021 \cite{VanNoorden22}.
It is also consistent with the recent report that Chinese-origin scientists conducting research in the US have been distancing themselves from the US \cite{Xie23}.

\subsection{International research collaboration clusters\label{subsec:clusters}}

\begin{figure}[htp]
\centering
\vspace{-0.5cm}
    \begin{tabular}{c}
    \begin{minipage}{0.03\hsize}
\begin{flushleft}
    \hspace{-0.7cm}\rotatebox{90}{\period{4}{2011--2020}}
\end{flushleft}
	\end{minipage}
	\begin{minipage}{0.33\hsize}
\begin{flushleft}
\raisebox{-0.0cm}{\small\textrm{\textbf{(a)~ \ai}}}\\[6mm]
\raisebox{\height}{\includegraphics[trim=2.0cm 1.8cm 0cm 1.5cm, align=c, scale=\csize, vmargin=0mm]{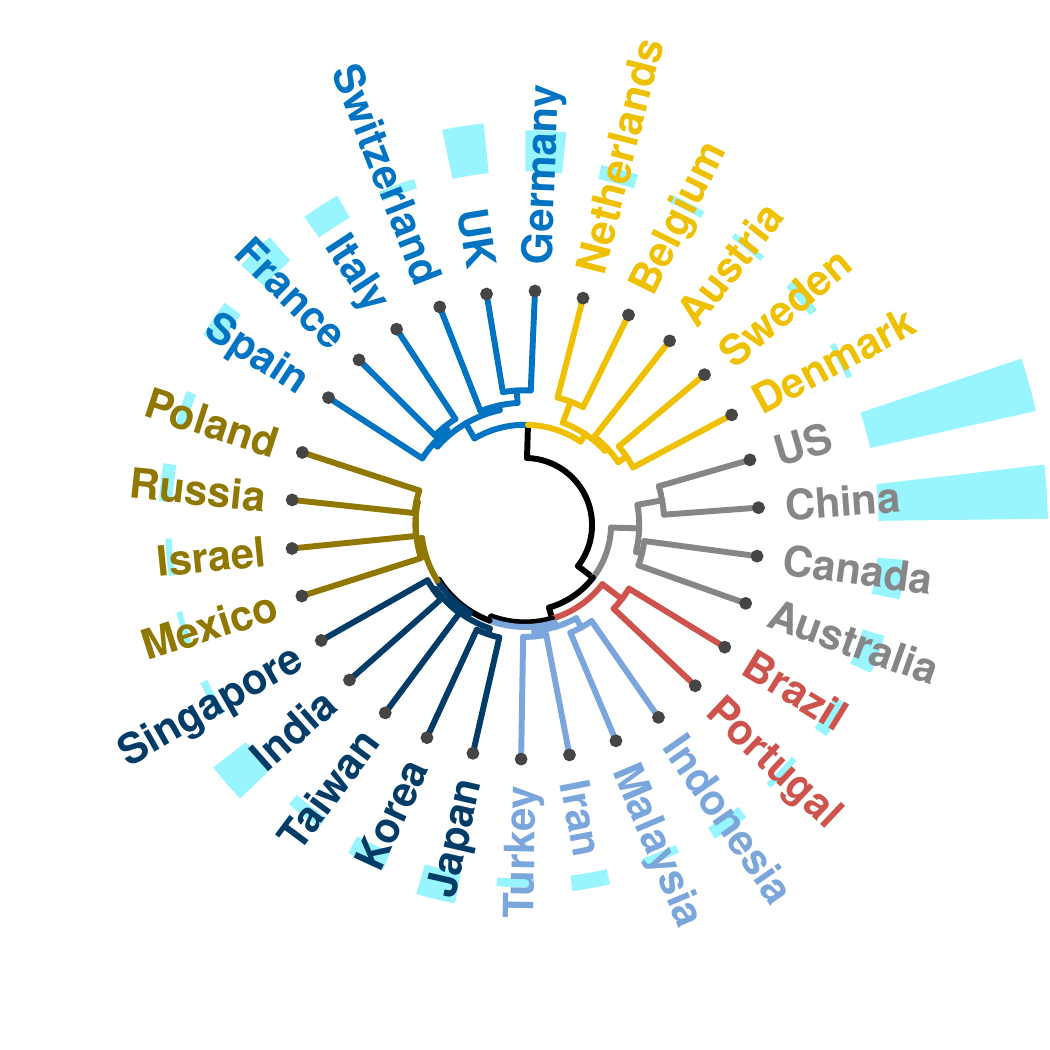}}
\end{flushleft}
    \end{minipage}
	\begin{minipage}{0.33\hsize}
\begin{flushleft}
\raisebox{-0.0cm}{\small\textrm{\textbf{(b)~ \quantum}}}\\[6mm]
\raisebox{\height}{\includegraphics[trim=2.0cm 1.8cm 0cm 1.5cm, align=c, scale=\csize, vmargin=0mm]{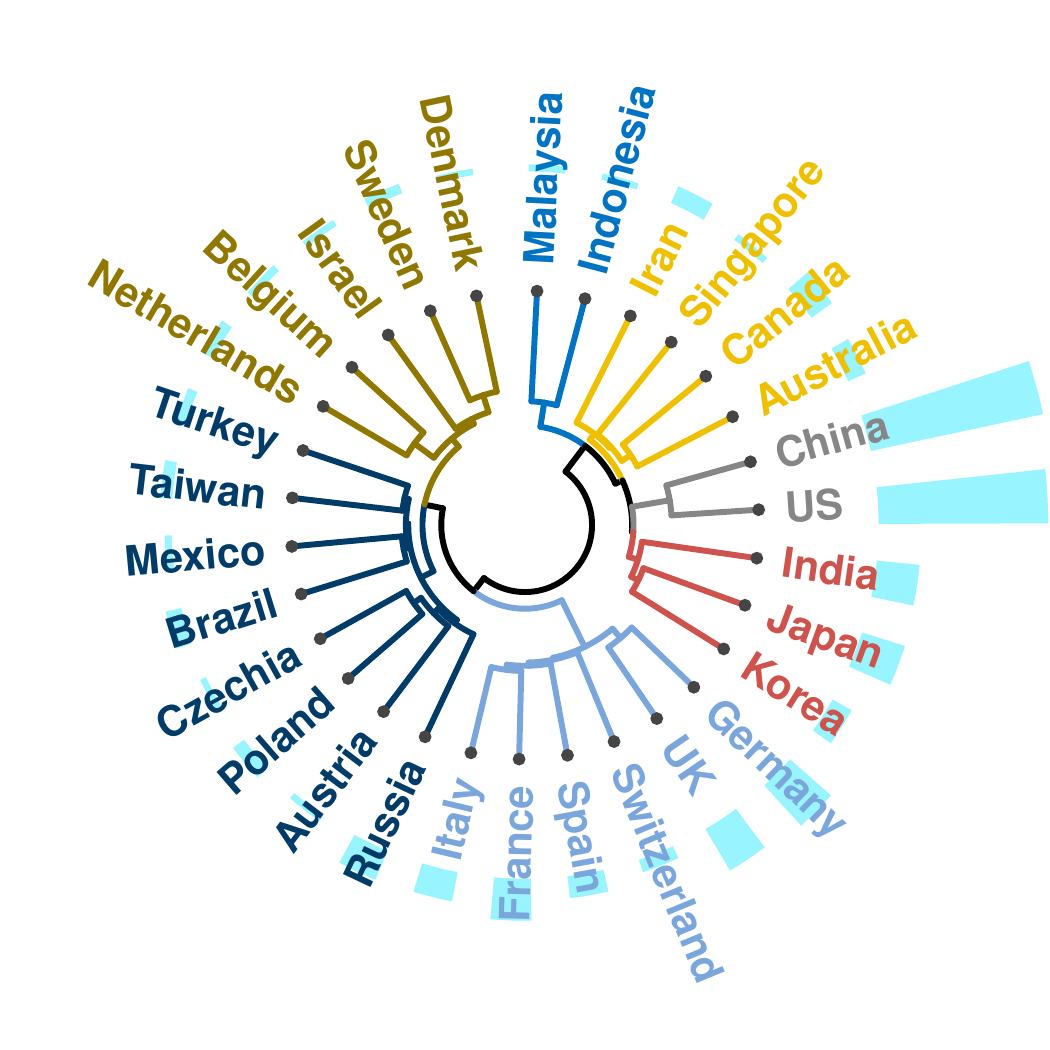}}
\end{flushleft}
	\end{minipage}
	\begin{minipage}{0.33\hsize}
\begin{flushleft}
\raisebox{-0.0cm}{\small\textrm{\textbf{(c)~ \bio}}}\\[6mm]
\raisebox{\height}{\includegraphics[trim=2.0cm 1.8cm 0cm 1.5cm, align=c, scale=\csize, vmargin=0mm]{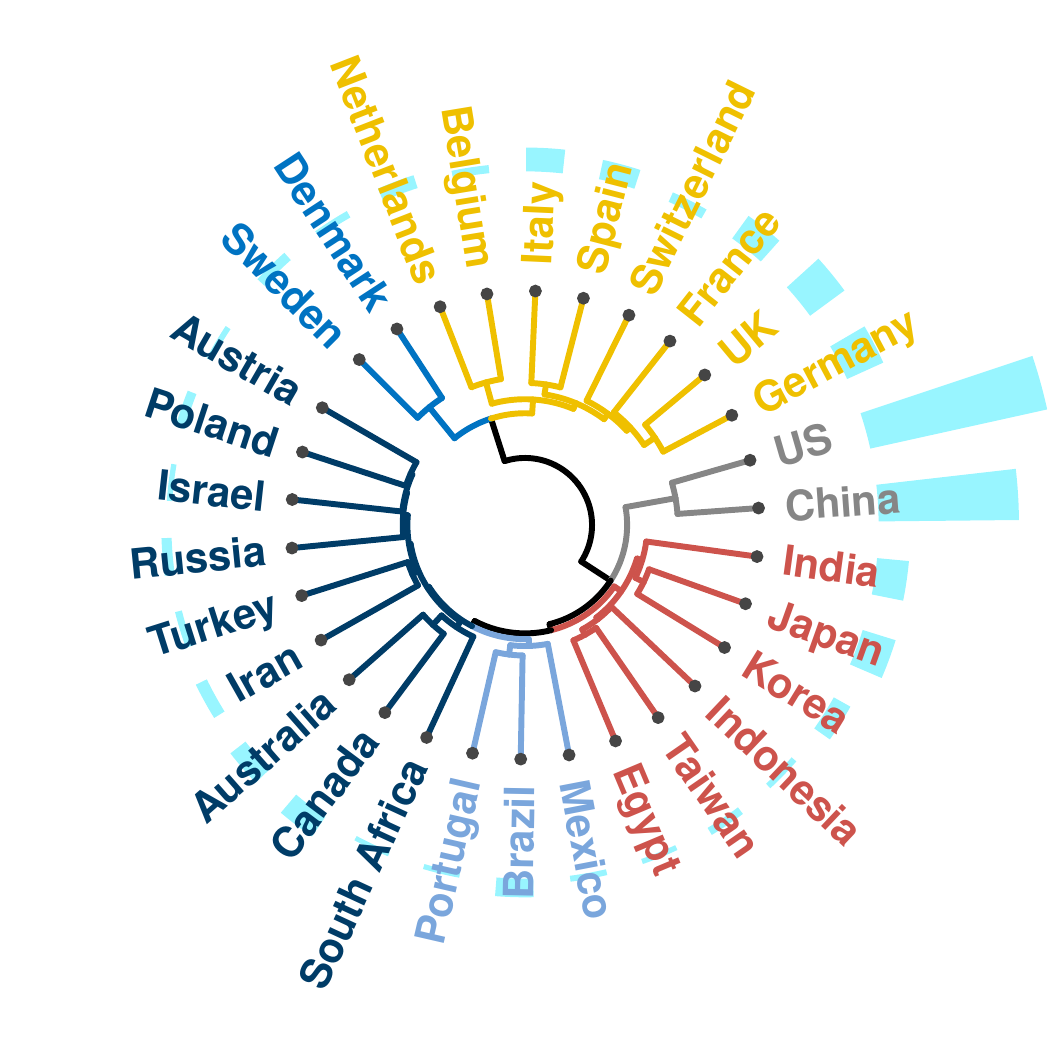}}
\end{flushleft}
	\end{minipage}
    \end{tabular}

\hspace{-0.5cm}{\marrow}\quad\dotfill

    \begin{tabular}{c}
    \begin{minipage}{0.03\hsize}
\begin{flushleft}
    \hspace{-0.7cm}\rotatebox{90}{\period{3}{2001--2010}}
\end{flushleft}
	\end{minipage}
	\begin{minipage}{0.33\hsize}
\begin{flushleft}
\raisebox{\height}{\includegraphics[trim=2.0cm 1.8cm 0cm 1.5cm, align=c, scale=\csize, vmargin=0mm]{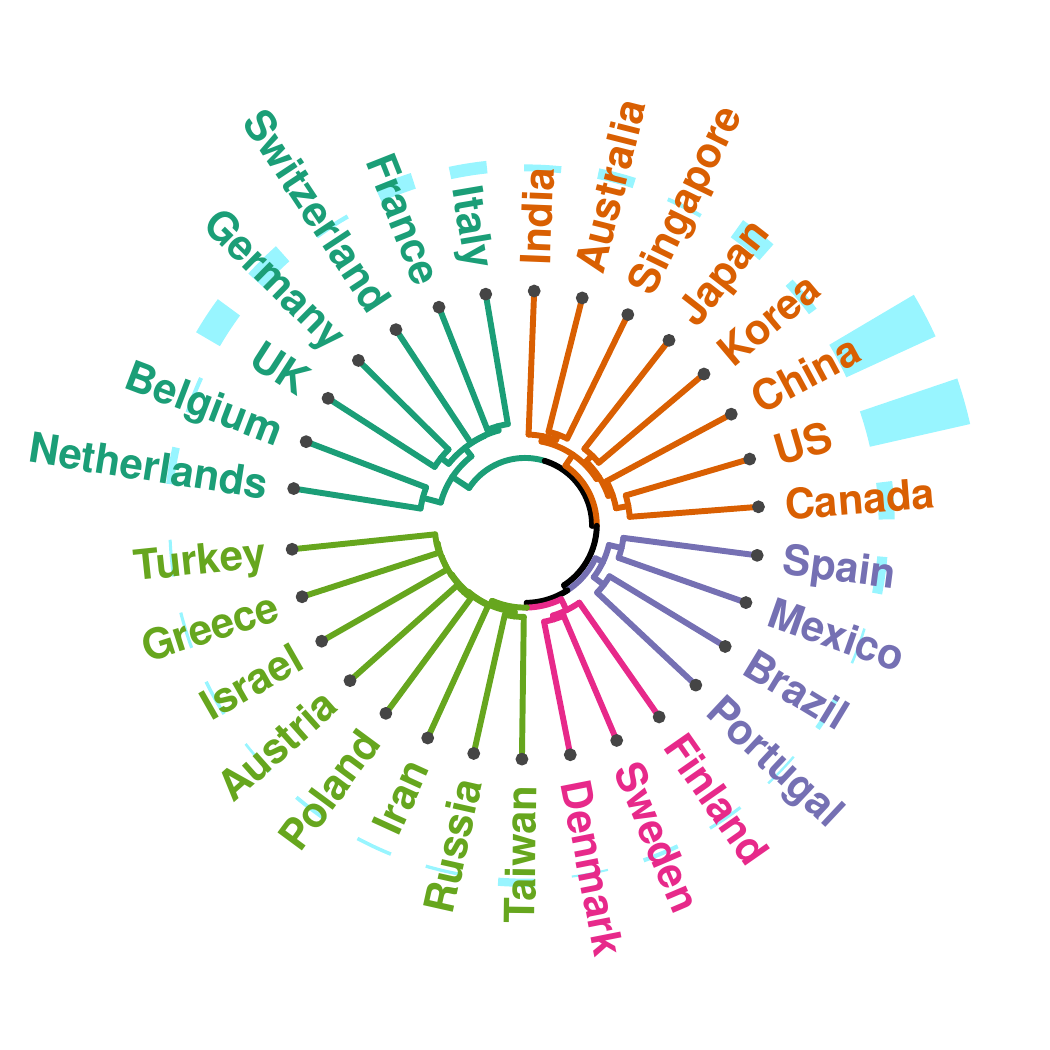}}
\end{flushleft}
    \end{minipage}
	\begin{minipage}{0.33\hsize}
\begin{flushleft}
\raisebox{\height}{\includegraphics[trim=2.0cm 1.8cm 0cm 1.5cm, align=c, scale=\csize, vmargin=0mm]{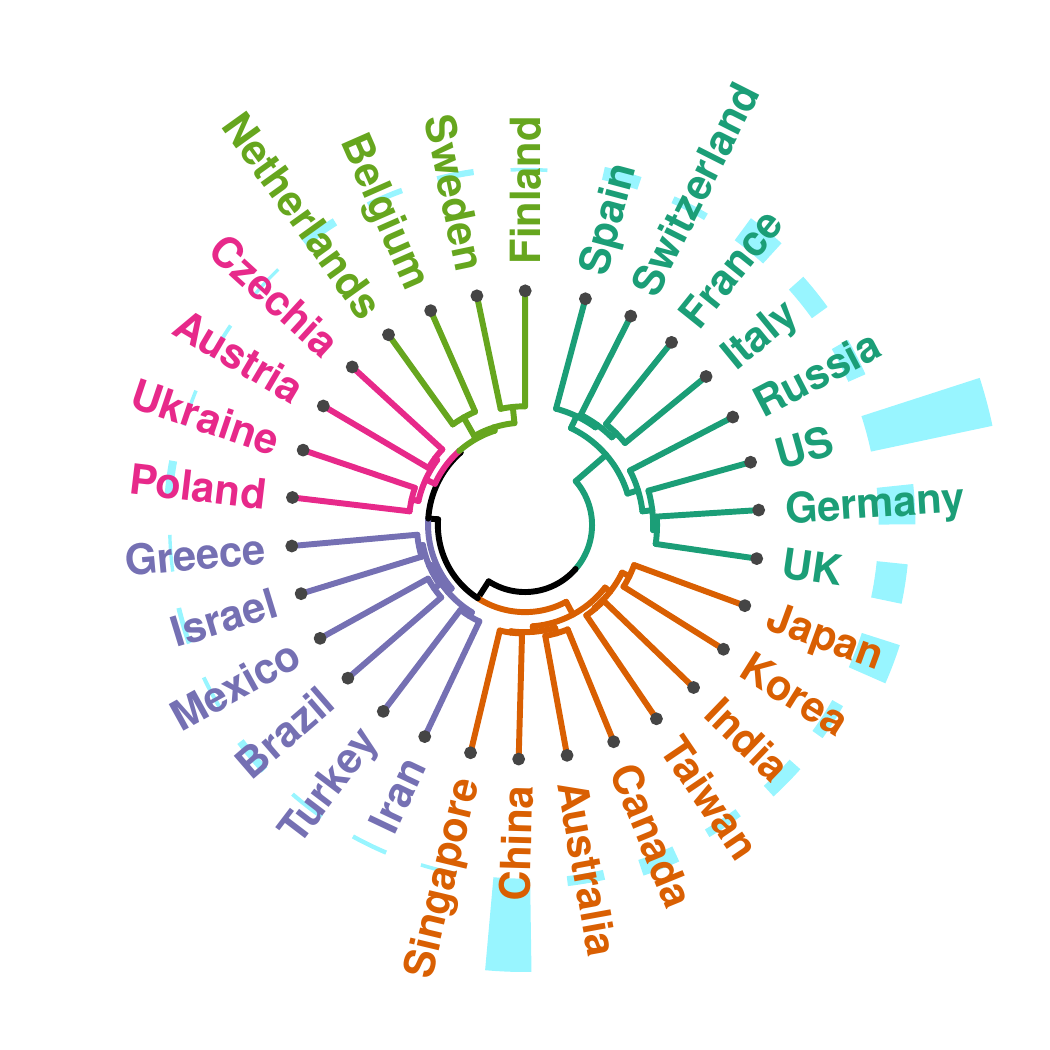}}
\end{flushleft}
	\end{minipage}
	\begin{minipage}{0.33\hsize}
\begin{flushleft}
\raisebox{\height}{\includegraphics[trim=2.0cm 1.8cm 0cm 1.5cm, align=c, scale=\csize, vmargin=0mm]{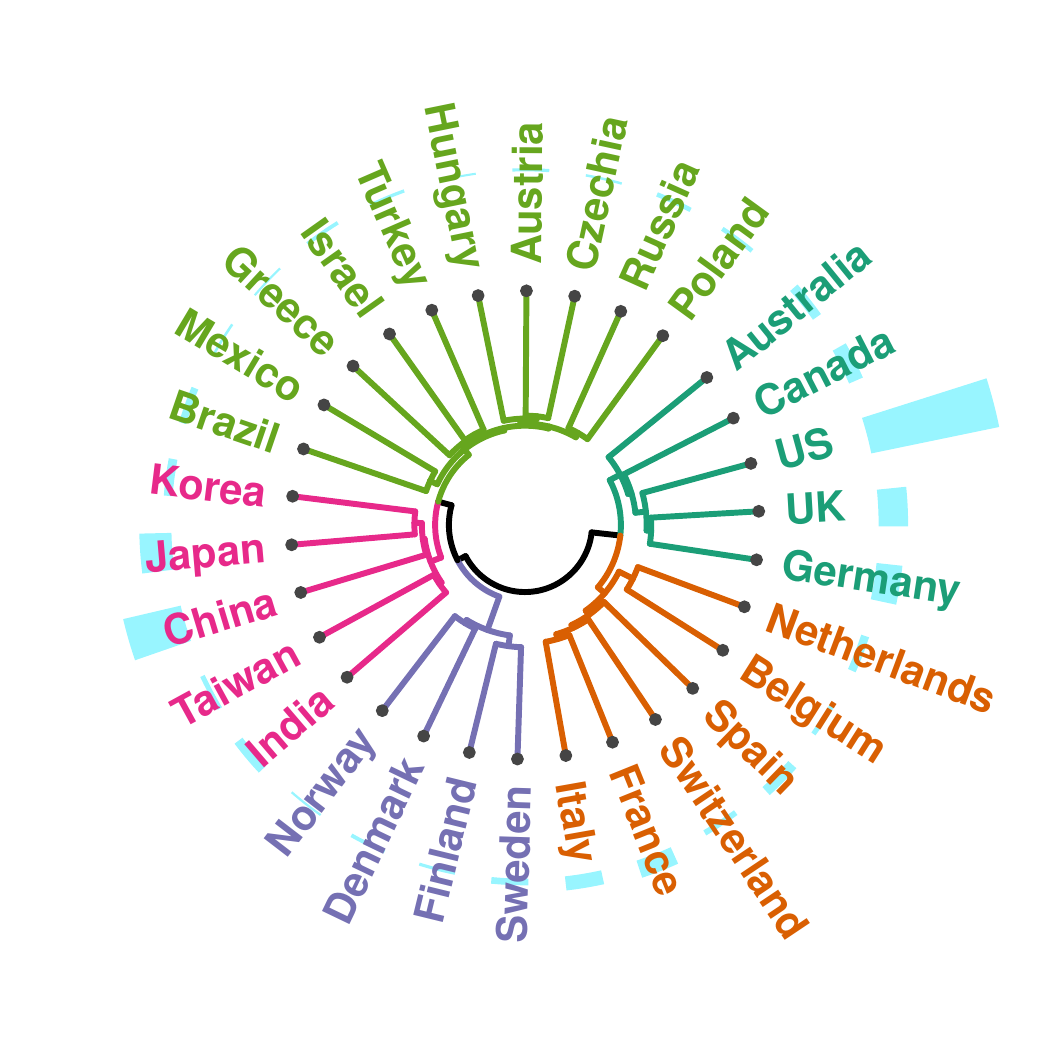}}
\end{flushleft}
	\end{minipage}
    \end{tabular}

\hspace{-0.5cm}{\marrow}\quad\dotfill

    \begin{tabular}{c}
    \begin{minipage}{0.03\hsize}
\begin{flushleft}
    \hspace{-0.7cm}\rotatebox{90}{\period{2}{1991--2000}}
\end{flushleft}
	\end{minipage}
	\begin{minipage}{0.33\hsize}
\begin{flushleft}
\raisebox{\height}{\includegraphics[trim=2.0cm 1.8cm 0cm 1.5cm, align=c, scale=\csize, vmargin=0mm]{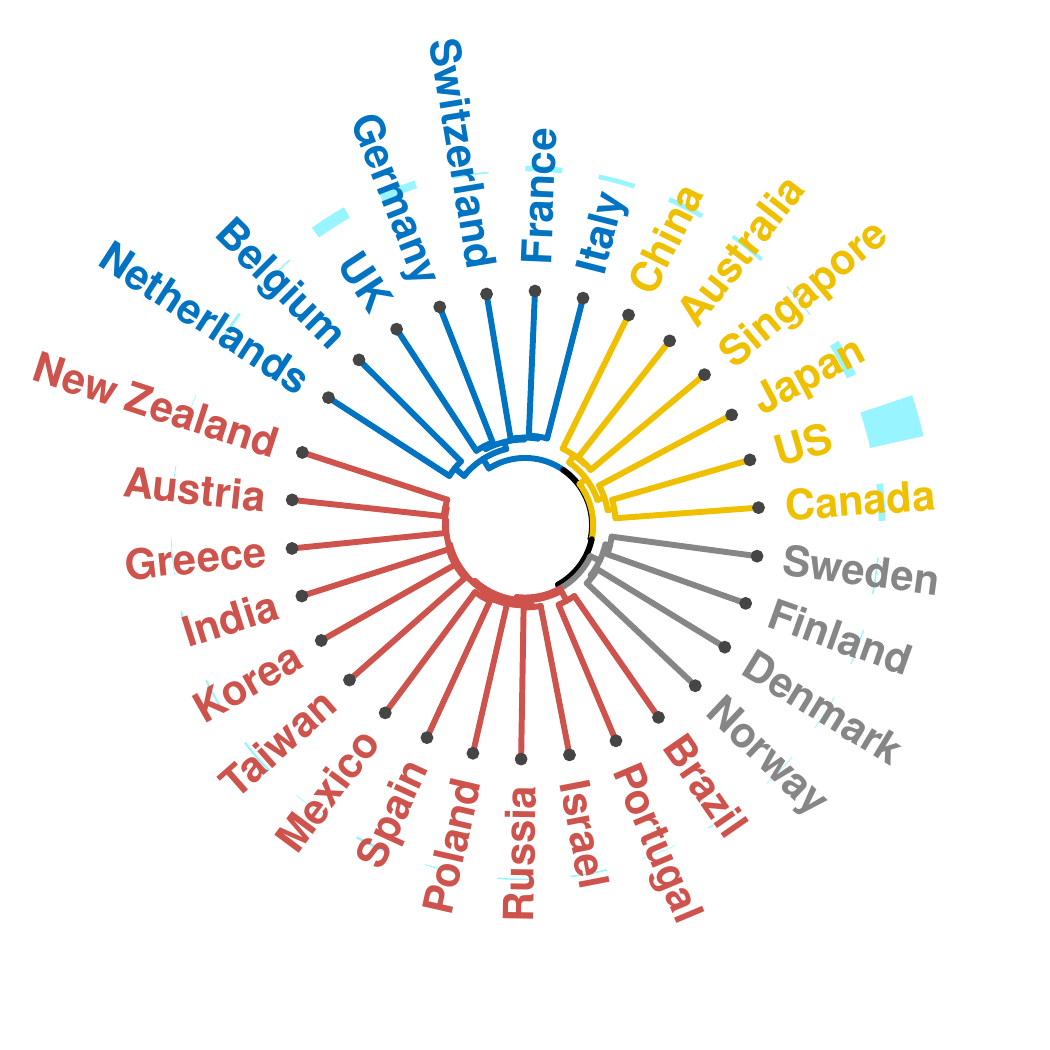}}
\end{flushleft}
    \end{minipage}
	\begin{minipage}{0.33\hsize}
\begin{flushleft}
\raisebox{\height}{\includegraphics[trim=2.0cm 1.8cm 0cm 1.5cm, align=c, scale=\csize, vmargin=0mm]{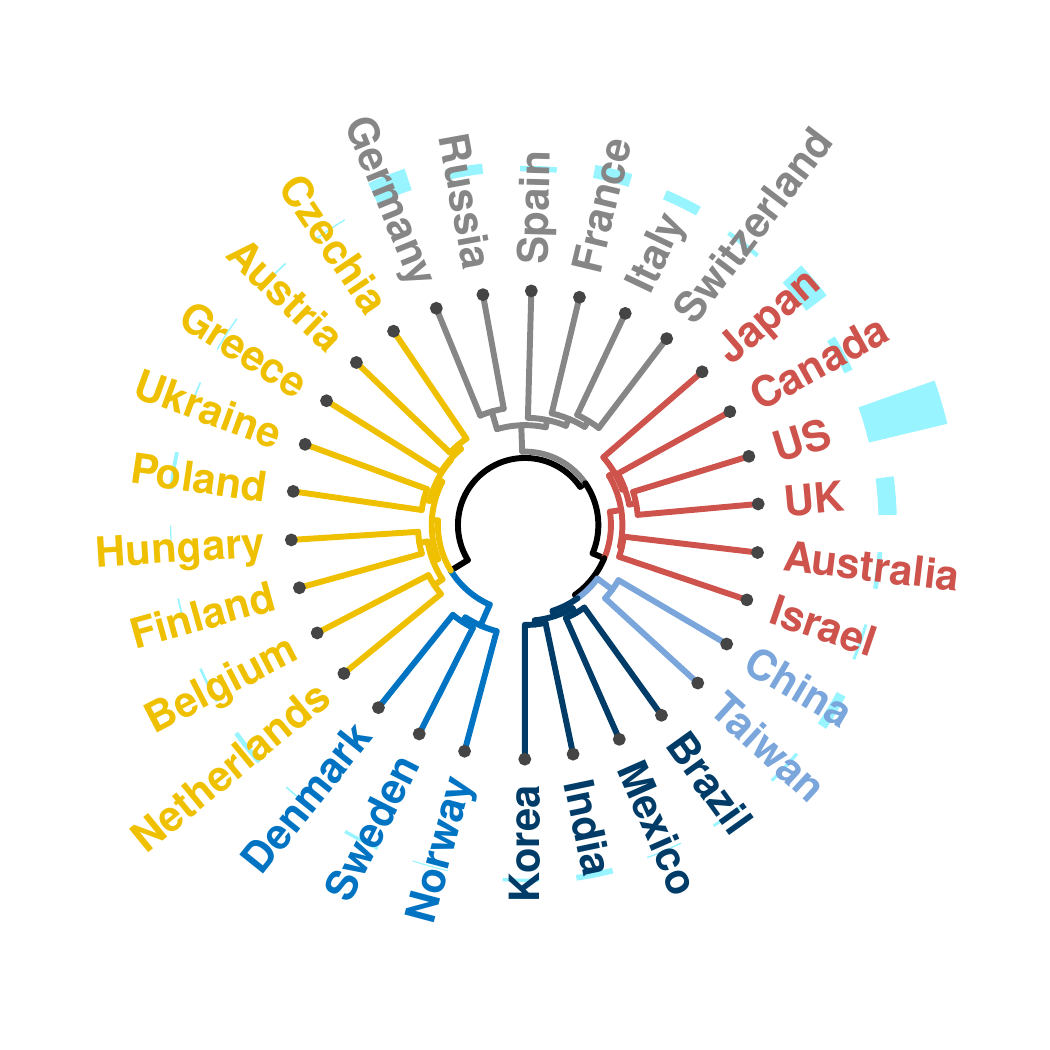}}
\end{flushleft}
	\end{minipage}
	\begin{minipage}{0.33\hsize}
\begin{flushleft}
\raisebox{\height}{\includegraphics[trim=2.0cm 1.8cm 0cm 1.5cm, align=c, scale=\csize, vmargin=0mm]{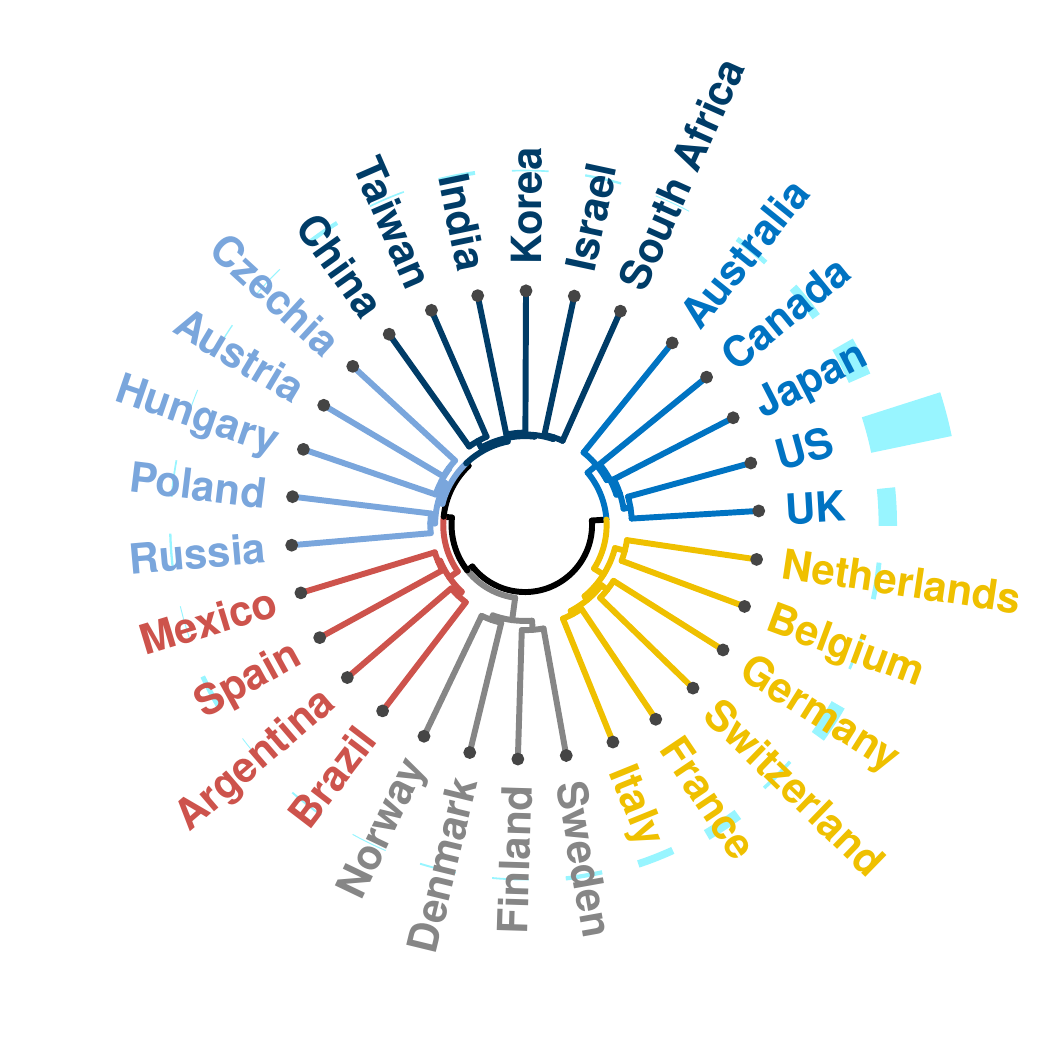}}
\end{flushleft}
	\end{minipage}
    \end{tabular}

\hspace{-0.5cm}{\marrow}\quad\dotfill

    \begin{tabular}{c}
    \begin{minipage}{0.03\hsize}
\begin{flushleft}
    \hspace{-0.7cm}\rotatebox{90}{\period{1}{1971--1990}}
\end{flushleft}
	\end{minipage}
	\begin{minipage}{0.33\hsize}
\begin{flushleft}
\raisebox{\height}{\includegraphics[trim=2.0cm 1.8cm 0cm 1.5cm, align=c, scale=\csize, vmargin=0mm]{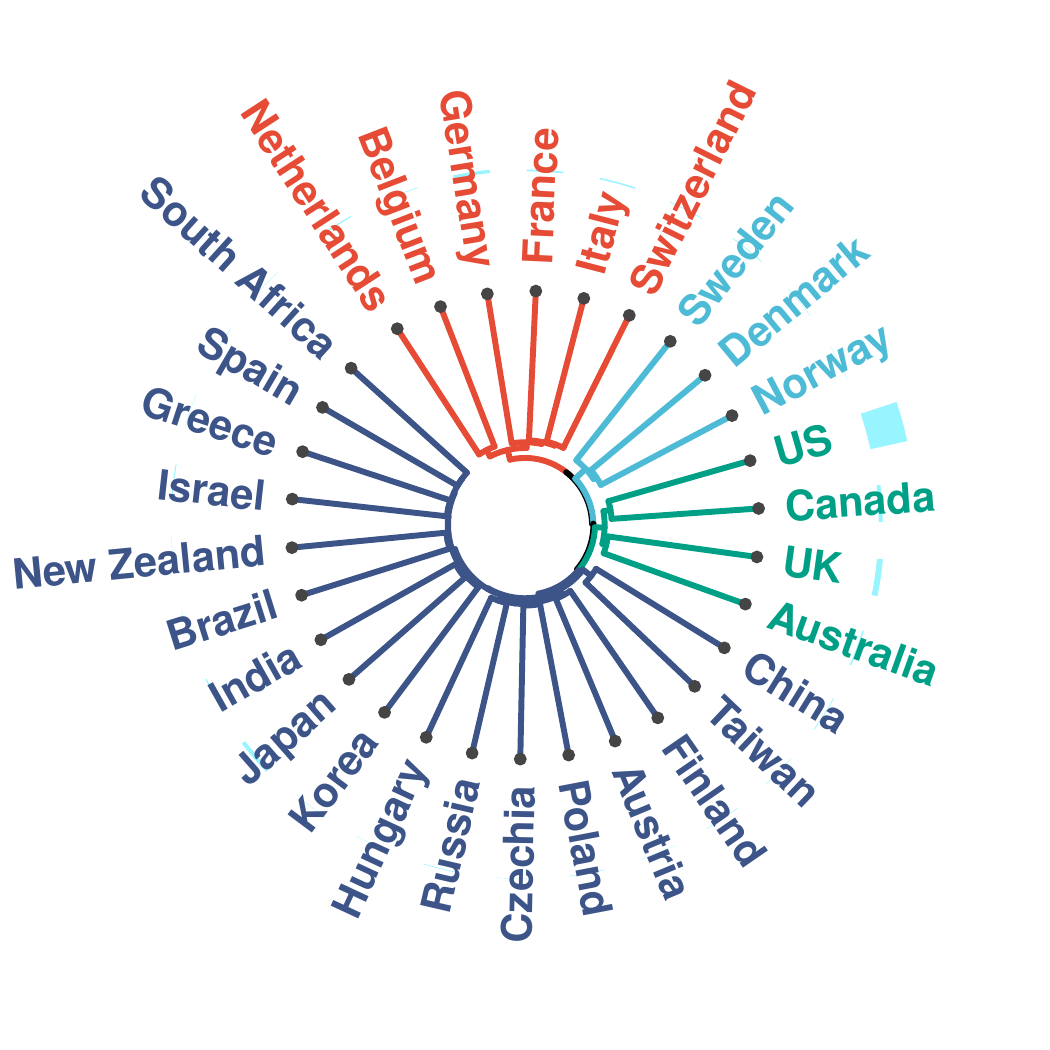}}
\end{flushleft}
    \end{minipage}
	\begin{minipage}{0.33\hsize}
\begin{flushleft}
\raisebox{\height}{\includegraphics[trim=2.0cm 1.8cm 0cm 1.5cm, align=c, scale=\csize, vmargin=0mm]{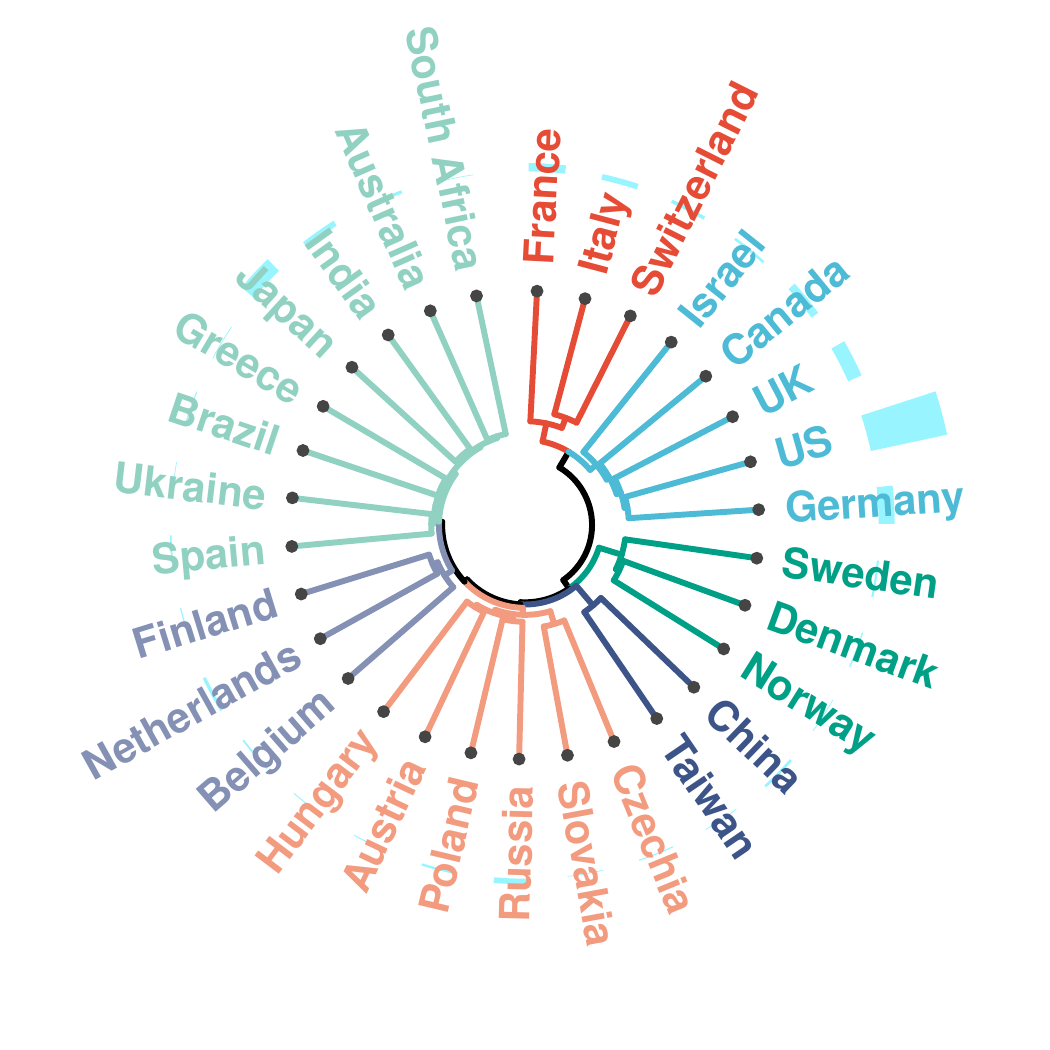}}
\end{flushleft}
	\end{minipage}
	\begin{minipage}{0.33\hsize}
\begin{flushleft}
\raisebox{\height}{\includegraphics[trim=2.0cm 1.8cm 0cm 1.5cm, align=c, scale=\csize, vmargin=0mm]{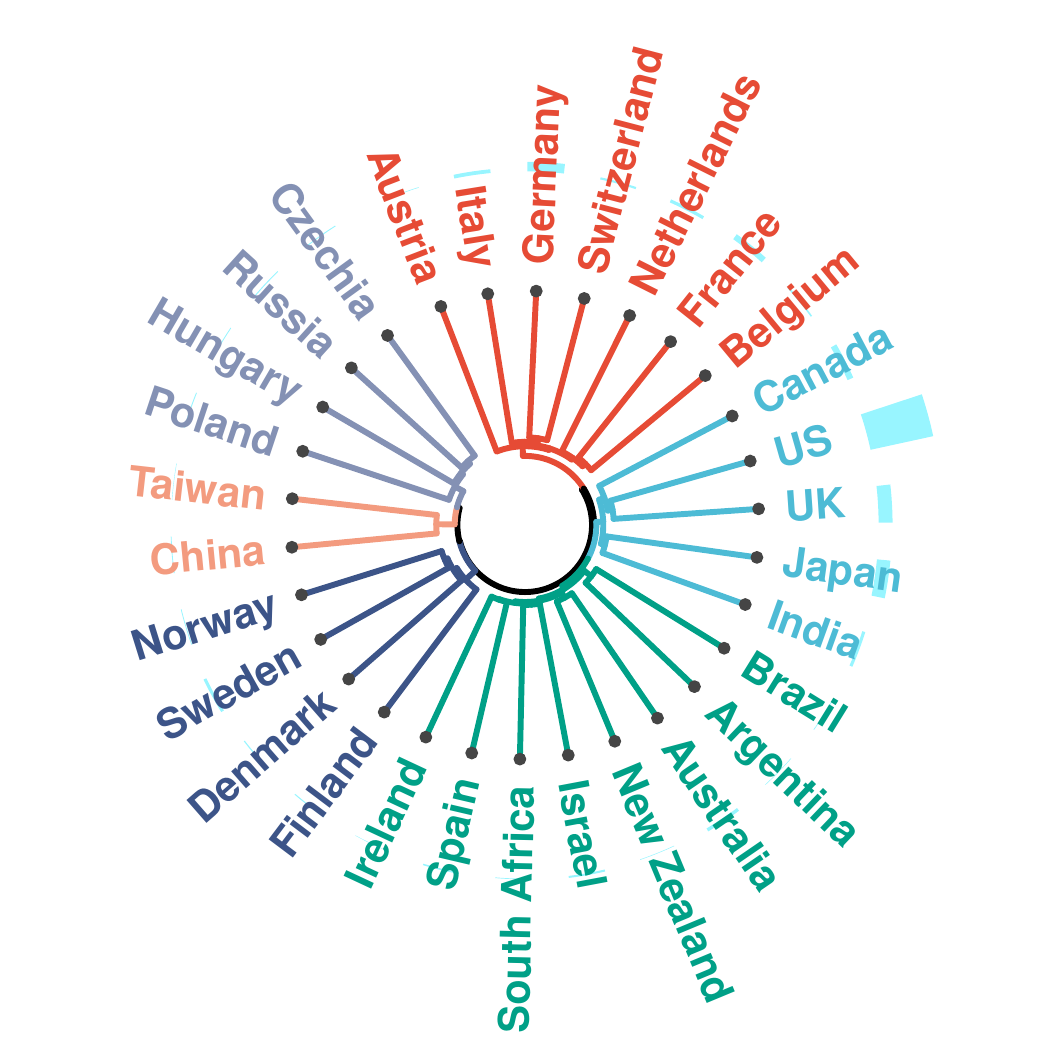}}
\end{flushleft}
	\end{minipage}
    \end{tabular}
\vspace{-0.0cm}
\caption{\textbf{Evolution of international research collaboration clusters.}}
\label{fig:cdend_1}
\end{figure}

Figure \ref{fig:cdend_1} and Suppl.~Fig.~\ref{fig:cdend_2} show the analysis results for the formation of international research collaboration clusters in each discipline for the same four periods (I--IV) as before, spanning the half-century from 1971 to 2020.
The visualisation is based on a series of circular dendrograms that represent the results of HCA for the distance matrices defined in Section \ref{subsec:clustering}.
A dendrogram is a branching diagram based on the distances among a group of entities.
In the case of the circular dendrogram employed here, the countries or clusters closer to each other are combined earlier as one moves from the outer edge of the circle towards its centre.
The height of the branching points, measured from the circumference and referred to as the coupling height, indicates how distant the countries or clusters of countries are from each other; the greater the coupling height, the farther away their relationships are.

The countries selected for display are the top 30 countries in work production in each discipline and period.
The circular bar graph in the outer region of the circular dendrogram shows the number of works produced by each country during each period.\endnote{%
The circular bar graphs are comparable within the same discipline across different periods, but not between different disciplines.}
The number of clusters that are colour-coded was calculated based on the preset threshold value for the coupling height.
As Suppl.~Table \ref{tab:ncluster} shows, the number of clusters does not necessarily increase or decrease with time; it highly depends on the discipline-wise situations and the preset coupling height threshold (Appendix \ref{app:clustering}).
In this regard, the distribution of coupling heights is more useful than the number of clusters to characterise the structure of international research collaboration clusters and compare it across disciplines and periods.
If all the coupling heights are maximally high, then the circular dendrogram would look like a shape in which lines run parallel from equally spaced points on the circumference to the centre of the circle, only to join together at a certain minimal radius all at once.
If the coupling heights are all relatively low, the branches soon couple with each other as they move from a point on the circumference to the centre of the circle, quickly forming clusters, with the so-formed clusters also coupling one after another well before the minimal radius.

It is useful first to have a big picture of overall trends.
For all R\&D disciplines illustrated in Fig.~\ref{fig:cdend_1} and Suppl.~Fig.~\ref{fig:cdend_5}, we can observe that over time, the open space in the centre of the circle tends to expand, like a tightly closed bud opening, and the structure of the branching tree becomes easier to see.
This observation suggests that countries worldwide are increasing their collaborative tendencies.
Within this overall trend, different trends can be observed in the status of connections within the dendrogram, i.e.\ the formation of clusters, depending on the period and the disciplines.
For example, in the 3 disciplines shown in Fig.~\ref{fig:cdend_1} (and also in the other disciplines indicated in Suppl.~Fig.~\ref{fig:cdend_5}), the UK and Germany have always been the first pair to connect since this century. 
To that pair, France, Italy, Switzerland and Spain have attached to form a European subcluster.
Belgium, the Netherlands, Denmark and Sweden often connect first and tend to form another set of subclusters. 
The distance between these subclusters varies according to disciplines and period, even within the same Europe.
Furthermore, in many disciplines, the UK in the last century was more deeply tied to the US, Canada and Japan than European countries.
Thus, various combinations and recombinations over time have formed a snapshot of the international research collaboration clusters of the time.

There may be various circumstances behind the formation of a particular cluster and its change over time \cite{Vieira22,Hou21,Luukkonen92}.
As discussed, the geographical proximity of the countries involved, such as in Europe or Asia, is often the most significant factor affecting the international collaboration status \cite{Fitzgerald21,Doria17,Katz94}.
For example, the European cluster can be identified as the direction from 10 to 2 o'clock of the {\ai} diagram (Fig.~\ref{fig:cdend_1}a) in Period \rn{4}.
It is noteworthy that as time passes, in many R\&D disciplines, China has moved out of what may be regarded as the Asian circle and moved into the prominent top-tier group per the works produced.
Alternatively, the formation of clusters may be related to geopolitical and historical perspectives \cite{Maher21,Luukkonen92}.
There may also have been movements among universities or research institutions where researcher exchanges flourish due to policy support under state-led scientific agreements or economic cooperation.
Thus, the background to forming the international research collaboration clusters involves a variety of national and international policies and changes in the R\&D environment surrounding the academic arena, resulting in each cluster snapshot. 
In other words, they are relatively determined rather than determined by the policies of a single country, and it is challenging to decipher them convincingly.
Although this paper will not go into contextual interpretations of individual observations, a deeper contextual discussion, complementing expert knowledge that cannot be obtained from the bibliometric approach, will provide more implications for the structure and the formation dynamics of collaboration clusters.

\subsection{{The `Shrinking World'}\label{subsec:shrinking}}

To analyse the trend of increasing collaboration discussed earlier in a more quantitative manner, we divided the half-century from 1971 to 2020 into 10 periods of 5 years each and performed cluster analysis in each discipline and period.
We then rescaled the set of coupling heights with an appropriate monotonically increasing function to make the graph easier to read (Appendix \ref{app:clustering}).
By comparing the mean of the rescaled coupling height distribution---hereinafter referred to as the (mean) International Coupling Distance (ICD)\endnote{%
The suggested view of the \textit{\q{Shrinking World}} of research collaboration would remain the same even if we used the median, instead of the mean, of the rescaled coupling height distribution.}---across periods, we can determine whether countries are getting closer or further away from each other on average over time.

\begin{figure}[!tp]
\centering
\vspace{-0.5cm}
\noindent
\includegraphics[align=c, scale=0.95, vmargin=1mm]{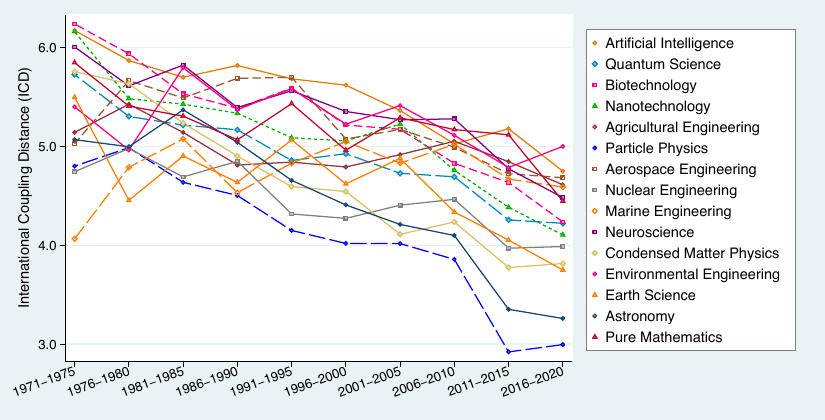}
\caption{\textbf{\textit{\q{Shrinking World}}: Change in the level of International Coupling Distance (ICD).}
}
\label{fig:line_ICD}
\vspace{5mm}
\end{figure}

Figure \ref{fig:line_ICD} displays the trends in the ICD index calculated for each discipline over time, showing variations in ICDs among disciplines.
{\ai} exhibits a relatively high level of ICD, possibly because individual researchers in the field tend to work independently without needing to cross borders.
The low level of ICD for {\particle} and {\astro} is a good reflection of reality; single countries cannot accomplish the mission in such large-scale academic disciplines, and international collaboration is indispensable, consistent with the previous observation in Suppl.~Fig.~\ref{fig:line_intlrate}.
The level of ICD is also low for {\nuclear}, which may reflect that it is a broad engineering discipline that spans many fields, from nuclear physics to materials science and applied chemistry, requiring combining technical knowledge and expertise from various countries and sectors.

While having its ups and downs at different times, the overall trend is that ICD has fallen over the past half-century, indicating a \textit{\q{Shrinking World}} of research collaboration.
Suppl.~Figure \ref{fig:kdensity_ICD} shows the result of the kernel density estimation of ICD for each period.
As time passes, the density curve's peak position shifts to a smaller value of ICD, supporting the observation at the discipline-aggregate level.
These results suggest that the S\&T world has been consistently getting smaller, and research collaboration has been more active across borders to better address expanding knowledge.
This phenomenon might be attributed to the fact that the scale of social issues targeted by S\&T has expanded to the global scale, and technological advances have made it possible to address such global-scale issues with improved international connectivity among researchers.
Additionally, policymakers, aware that S\&T is the source of national strength and industrial competitiveness and key to economic security, have launched strategic international collaborative projects from the political arena.\endnote{%
The evidence obtained for the repelling force between the US and China at the end of Section \ref{subsec:bilateral}, along with the observation in Section \ref{subsec:shrinking} of a shrinking trend in research collaboration, suggests a picture of a \textit{\q{Shrinking-and-Polarising World}}, provided these trends continue.}

\section{Summary and discussion: Across borders, disciplines and generations\label{sec:Discussion}}

The past few decades have witnessed the development and implementation of various digital technologies in society, drastically changing the relationship between people, S\&T and society.
The development of digital communication tools and platforms in the advanced information society has significantly updated how scientists and engineers interact and transfer knowledge, resulting in a smaller world (Fig.~\ref{fig:line_ICD}).
Driven by the rapid and irreversible movement of Open Science \cite{Miedema22,Burgelman19}, all forms of scientific publication, including non-journal articles, preprints, databases and social networking services, have become indispensable tools and platforms in scholarly communication today. 
It would be fair to say that a new initiative of Open Bibliometrics is replacing the traditional bibliometrics that relies solely on commercial databases centred on journal articles.
This initiative will undoubtedly play an essential role in forming future S\&T and innovation policies, evaluating and publicising them, and accelerating interdisciplinary approaches \cite{Yanai20,Mol20,Okamura19,Ledford15}.

With this philosophy in mind, the present study provided unique evidence of how international collaboration clusters have formed and evolved over the past half-century for a broad set of scientific publications based on the OpenAlex dataset.
The study first reviewed the global presence change of top-tier countries for each research discipline, as measured by publication volumes and international collaboration rates.
Notably, the US and China were shown to have rapidly moved closer together for decades but started moving apart after 2019.
Subsequently, the study analysed and visualised the international collaboration clusters for each discipline and period based on a hierarchical clustering method. 
Finally, the study provided global-scale quantitative evidence for a \textit{`Shrinking World'} of the past half-century's research collaboration.
These results provide valuable insights into the big picture of past, present and future international collaboration.

Several methodological innovations were developed and demonstrated in this study, making these dynamic quantitative analyses possible.
Specifically, the first and most novel device was formulating the distance between two countries as a simple set-theoretic distance, the so-called Jaccard distance.
This approach highlighted the dynamic distance relationship between countries and groups of countries, which had not been possible before. 
The second device was applying the Jaccard distance function to the hierarchal clustering of countries with Ward's method to identify international research collaboration clusters each period for each R\&D discipline. 
Furthermore, the visualisation method was also novel; the circular dendrogram is frequently used in papers on genetic phylogenetics, but this was the first time it was utilised in scientometrics/bibliometrics.

There are several directions in which the results of this study can be further developed.
One important area for future research is the contextual understanding of cluster formation dynamics.
A more in-depth discussion of national and international S\&T-related policies that impact global R\&D collaboration would provide additional implications for the study's findings.
Instead of simply investigating bilateral relationships using the \q{symmetric} measure of distance (\q{$D_{\X,\Y}$}) or affinity (\q{$A_{\X,\Y}$}) utilised in this paper, analysis based on an \q{asymmetric} measure that distinguishes between the mobility of researchers from Country X to Country Y and vice versa would also be informative. 
Exploring this approach, in combination with the study's results, would offer a complementary understanding of and deeper insights into the dynamics of international collaboration.
Moreover, while this study focused on the top 10 (Figs.~\ref{fig:line_npaper_intlrate_1}, \ref{fig:chord_1} and Suppl.~Figs.~\ref{fig:line_npaper_intlrate_2}, \ref{fig:line_intlrate2}, \ref{fig:chord_2}) or top 30 (Fig.~\ref{fig:cdend_1} and Suppl.~Fig.~\ref{fig:cdend_2}) countries in work production in each discipline and period, it is also relevant to consider countries outside the top 10 or top 30 for certain policy purposes.
Additionally, it would be interesting to analyse not only the absolute value of the number of works but also the relative value divided by other R\&D-related indicators defined at the state level, including researcher population, total R\&D budget and GDP.

Again, we must be fully aware of the limitations of bibliometrics or, more broadly, scientometrics.
It should always be kept in mind that due to various methodological difficulties in scientometrics, their policy implications are inherently limited \cite{Waltman16,Wilsdon15,Hicks15}; see also \citek{Okamura22,Okamura19}.
The method we proposed and implemented in this paper is no exception.
In addition to the various limitations discussed in the previous sections, including the metadata availability of OpenAlex, our results are also likely highly dependent on the R\&D field classification scheme.
Different specifications of disciplines could have been applied, leading to quantitatively different implications on each country's international presence and the landscape of international research collaborations.
Furthermore, it should be noted that a considerable volume of R\&D outputs is still not published in papers or data, including classified research results related to national security and defence, which are not recorded in open databases.
The insights from bibliometric analysis, including those of the present paper, can only represent a part of the unclassified, open world.

Despite its limitation, this paper provides valuable knowledge and new insights into the macro trends in international research collaboration and its current situation.
When looked at through the same bibliometric lens in several years from now, the results will show a new landscape that reflects the integrated impact of all major global issues underway, including the COVID-19 pandemic, various geopolitical issues and highly digitised and diversified scholarly communication modes.
The new landscape offered by the \q{Science of Science} \cite{Fortunato18} will continue to expand, where scientometric methods will be used in increasingly sophisticated and exciting ways.
A new generation of scientometricians will create new values for the times exposed to new data platforms and a highly digitised society.
They can change the angle to see the world, adjust the resolution, transcend across disciplines and gain unique perspectives on the S\&T ecosystem.
Hopefully, these new generations will bring practical hints and actions that resonate and sympathise with many people on how to proceed with our international research collaboration for a better society.

\textit{\q{Science has no borders, but scientists have their homelands}}, said Louis Pasteur (\citeauthor{Dubos50}, \citeyear{Dubos50}, pp.~84--85; \citeauthor{Vallery-Radot15}, \citeyear{Vallery-Radot15}, p.~399).
How many scientists and engineers over the decades and generations have had their minds blown by these words, only to be confronted with the gap between their ideals and reality?
We know that, by definition, science has no borders.
At the same time, we must accept that \emph{accessible} science has borders in reality.
Now that the power and use of S\&T determine the course of the world, all stakeholders must revisit what the borders and homelands mean and redefine them in the contemporary context of responsible R\&D.
How can we embody the value of S\&T literally without borders?
To this end, how can we create a policy environment that maximises the social value of borderless academia, and how can we pass that on to the next generation?
The challenge to answer these questions confronts all policymakers and stakeholders of S\&T today.
Amid this unprecedentedly complex and unpredictable international situation, we hope this paper sheds light on some essential nature of global R\&D cooperation for those seeking to open up new horizons at the interface of S\&T and society.


\vspace{0mm}
\paragraph{\textbf{Acknowledgements.}}
The author would like to thank the two anonymous reviewers of \textit{Quantitative Science Studies} for their valuable comments.
The views and conclusions contained herein are those of the author and should not be interpreted as necessarily representing the official policies or endorsements, either expressed or implied, of any of the organisations to which the author is affiliated.

\vspace{0mm}
\paragraph{\textbf{Author Contributions.}}
KO: 
Conceptualisation, Methodology, Software, Validation, Formal analysis, Investigation, Data curation, Writing (Original Draft, Review \& Editing), Visualisation, Project administration.

\vspace{0mm}
\paragraph{\textbf{Competing Interests.}}
The author has no competing interests.

\vspace{0mm}
\paragraph{\textbf{Funding Information.}}
The author did not receive any funding for this research.

\vspace{0mm}
\paragraph{\textbf{Data Availability.}}
The datasets and figures generated and/or analysed during this study can be found in the Zenodo repository at \url{https://doi.org/10.5281/zenodo.7297122}.
For reference for interested readers, the package also contains data and figures based on bibliometric data from only journal articles.

\vspace{-0cm}
\theendnotes
\addcontentsline{toc}{section}{Notes}


\bibliographystyle{apalike-imp}
\addcontentsline{toc}{section}{References}

\setlength{\bibsep}{0\baselineskip plus 0.2\baselineskip}
\renewcommand*{\bibfont}{\footnotesize}

\clearpage
\newpage

\renewcommand{\thesubsection}{Appendix\,~\Alph{subsection}}
\renewcommand{\thesubsubsection}{\Alph{subsection}.\,\arabic{subsubsection}}

\pagestyle{fancy}
\fancyhead[LE,RO]{\textcolor{orange}{\footnotesize{\textsf{SUPPLEMENTARY MATERIALS}}}}
\fancyhead[RE,LO]{}
\fancyfoot[RE,LO]{\color[rgb]{0.04, 0.73, 0.71}{}}
\fancyfoot[LE,RO]{\scriptsize{\textbf{\textsf{\thepage}}}}
\fancyfoot[C]{}

\addcontentsline{toc}{section}{Supplementary Materials}

\quad
\vspace{-0.5cm}
\begin{center}
\fontsize{15pt}{16pt}\selectfont\bfseries
Supplementary Materials
\end{center}

\vspace{-0.2cm}
\begin{center}
{for \textit{\q{\mtitle}} by K.~Okamura (2023).}
\end{center}

\vspace{1.0cm}

\renewcommand{\figurename}{Suppl.~Figure}
\renewcommand{\tablename}{Suppl.~Table}
\renewcommand{\thefigure}{S\arabic{figure}}
\renewcommand{\thetable}{S\arabic{table}}
\renewcommand{\theequation}{S\arabic{equation}}
\setcounter{section}{0}
\setcounter{figure}{0}
\setcounter{table}{0}
\setcounter{equation}{0}

\renewcommand{\headrule}{\color{orange}\oldheadrule}

\subsection[{\hspace{3.5eM}} Methodological details]{Methodological details\label{app:method}}

This appendix supplements the details of the analysis and visualisation methods utilised in the main text.

\subsubsection{Data analysis and visualisation\label{app:data}}

In this paper, all data were obtained through \href{https://docs.openalex.org/api/}{OpenAlex API} \cite{Priem22} and analysed using STATA/IC software (version 13; StataCorp LP, Texas, USA) and R software (version 4.2.1; R Core Team).
Line plots (Figs.~\ref{fig:line_npaper_intlrate_1}, \ref{fig:US-CN(-JP)}, \ref{fig:line_ICD}; Suppl.~Figs.~\ref{fig:nonjrate}, \ref{fig:unknownrate}, \ref{fig:line_npaper_intlrate_2}, \ref{fig:line_intlrate2} and \ref{fig:line_intlrate}) and kernel density plots (Suppl.~Fig.~\ref{fig:kdensity_ICD}) were created with STATA/IC software.
Chord diagrams (Fig.~\ref{fig:chord_1}; Suppl.~Figs.~\ref{fig:chord_2} and \ref{fig:chord_cdend_allf} (left)) were generated using the \href{https://www.rdocumentation.org/packages/circlize/versions/0.4.15/topics/chordDiagram}{\textsf{chordDiagram}} function from the \textsl{circlize} package \cite{Gu14} in R software.
Hierarchical cluster analysis (HCA) was performed with the \href{https://www.rdocumentation.org/packages/stats/versions/3.6.2/topics/hclust
}{\textsf{hclust}} function implemented in R software, and circular dendrograms (Fig.~\ref{fig:cdend_1}; Suppl.~Figs.~\ref{fig:cdend_2} and \ref{fig:chord_cdend_allf} (right)) were visualised using the \textsl{circlize} \cite{Gu14} and \textsl{dendextend} \cite{Galili15} packages in R software.
The method of dendrogram visualisation used in this paper is based on an example described on the \textsl{dendextend} website \cite{web:dendextend}.
The datasets and figures generated and/or analysed during this study are available in the Zenodo repository \cite{Okamura22b}.

\subsubsection{Quantifying distances among countries\label{app:distance}}

Let $S_{\X|\alpha}$ denote the set of works of nationality X published in discipline $\alpha$ during a given period, and let $n_{\X|\alpha}\coloneqq \abs{S_{\X|\alpha}}$ denote its size.
Similarly, let $S_{\X,\Y|\alpha}$ denote the set of works with nationalities X and Y published in the same discipline $\alpha$ during the same period, and let $n_{\X,\Y|\alpha}\coloneqq \abs{S_{\X,\Y|\alpha}}$ denote its size.
Then, the affinity $A_{\alpha}(\X,\Y)$ between X and Y for the period is defined by
\begin{equation}
A_{\alpha}(\X,\Y)
=\f{\abs{S_{\X|\alpha}\cap S_{\Y|\alpha}}}{\abs{S_{\X|\alpha}\cup S_{\Y|\alpha}}}
=\f{n_{\X,\Y|\alpha}}{n_{\X|\alpha}+n_{\Y|\alpha}-n_{\X,\Y|\alpha}}\,.
\end{equation}
This index is known as the Jaccard index or Jaccard similarity coefficient, which measures the size of the intersection of two sets divided by the size of their union.
The distance $D_{\alpha}(\X,\Y)$ between X and Y is then defined by
\begin{equation}
D_{\alpha}(\X,\Y) = 1-A_{\alpha}(\X,\Y)\,.
\end{equation}
This distance metric is called the Jaccard distance, which ranges from 0 and 1, with 0 indicating that X and Y are identical and 1 indicating that they are entirely distinct.
The Jaccard distance satisfies the mathematical definition of distance in the set-theoretic sense.
A similar idea was used by \citek{Okamura19} to define the distance between research disciplines.

\subsubsection{Hierarchical clustering analysis\label{app:clustering}}

\noindent\textbf{The Ward's method.}
~
The distance function defined in the previous section allows us to obtain an $n\times n$ distance matrix for the country set $\cC=\{\X_{i}\,|\,i\in\cS\}$, where $\cS=\{1,\,\dots,\,n\}$.
All cases discussed in Section \ref{subsec:clusters} were associated with $n=30$, but this size can be arbitrarily.
The distance $D_{ij}\coloneqq D(\X_{i},\X_{j})$ satisfies the properties $0\leq D_{ij}\leq 1$, $D_{ij}=D_{ji}$, $D_{ii}=0$ and $D_{ij}\leq D_{jk}+D_{ki}$ (the triangle inequality) for any $i,\,j,\,k\in\cS$.

HCA was performed on this distance matrix using the \href{https://www.rdocumentation.org/packages/stats/versions/3.6.2/topics/hclust}{\textsf{hclust}} function implemented in R software with the option \q{\textsf{ward.D2}} (i.e.\ the original Ward’s method) \cite{Murtagh14} specified.
It can be shown that our distance matrix $\bmt{D}$ is Euclidean, in the sense that there exists a Euclidean space $(\mathbb{R}^{n},\vev{\cdot,\cdot})$ and $n$ points $\{\bmt{x}_{i}\,|\,i\in\cS\}\subset \mathbb{R}^{n}$ such that $D_{ij}=\|\bmt{x}_{i}-\bmt{x}_{j}\|$ for all $i,j\in\cS$, with $\|\cdot\|$ the norm induced by the inner product $\vev{\cdot,\cdot}$ on $\mathbb{R}^{n}$.
Let us define an auxiliary $n\times n$ matrix $\bmt{Q}$ by $Q_{ij}=\big(D_{1j}^{2}+D_{i1}^{2}-D_{ij}^{2}\big)\big/2$, and decompose it as $\bmt{Q}=\bmt{P}\bmt{\mathit{\Lambda}}\bmt{P}^\intercal$, where $\bmt{P}$ is the $n\times n$ matrix whose $i$th column is the eigenvector $\bmt{p}_{i}$ of $\bmt{Q}$, and $\bmt{\mathit{\Lambda}}$ is the $n\times n$ diagonal matrix whose diagonal elements are the corresponding eigenvalues, $\mathit{\Lambda}_{ii}=\lambda_{i}$.
Then, each row vector of the matrix $\bmt{P}\sqrt{\bmt{\mathit{\Lambda}}}$, with $(\sqrt{\mathit{\Lambda}})_{ii}=\sqrt{{\lambda}_{i}}$, represents the coordinate of $\bmt{x}_{i}$ associated with Country $\X_{i}\in\cC$.
In this way, it is possible to obtain the set of coordinates $\{\bmt{x}_{i}\}$ corresponding to each country in $\cC$ in each of the four periods (Period \rn{1}: 1971--1990, Period \rn{2}: 1991--2000, Period \rn{3}: 2001--2010, Period \rn{4}: 2011--2020) used in Fig.~\ref{fig:cdend_1} and Suppl.~Fig.~\ref{fig:cdend_2}, or the ten periods of five years (1971--1975, 1976--1980, $\dots$, 2016--2020) used in Fig.~\ref{fig:line_ICD} and Suppl.~Fig.~\ref{fig:kdensity_ICD}, ensuring that the Ward's method can be readily applied to our distance matrix $\bmt{D}$.
Note that the methods proposed and employed in this paper should still be evaluated for both methodological soundness and tthe plausibility of the derived results.

\vspace{4mm}
\noindent\textbf{Number of clusters.}
~
We automatically and systematically determine the number of clusters ($N_{*}$) to be colour-coded and displayed in the circular dendrogram.
First, coupling heights $\{h_{k}\}_{k=1,\,2,\,\dots}$ are obtained for each discipline and period through HCA.
We set a common threshold value ($h_{*}$) for the coupling height across all disciplines and periods.
The number of clusters is then determined by the number of coupling heights that are larger than the threshold value, plus one, i.e.\
\begin{equation}
N_{*}=\abs{\{h_{k}\,|\,h_{k}\geq h_{*}\}}+1\,.
\end{equation}
If $h_{*}$ is sufficiently large, i.e.\ $h_{*}>h_{\max}\coloneqq\max_{k}\{h_{k}\}$, then $N_{*}=1$, and all $n$ (here, 30) countries are grouped into a single cluster.
Alternatively, if $h_{*}$ is sufficiently small, then $N_{*}=n$, the number of all fundamental entities.
In this paper, we set $h_{*}=1.005$, and the resulting number of clusters is summarised in Suppl.~Table \ref{tab:ncluster}.

\vspace{4mm}
\noindent\textbf{International Coupling Distance (ICD).}
~
To better understand whether countries have been collaborating more or less over time in a quantitative manner, we rescaled the original set of coupling heights to obtain the following rescaled coupling heights:
\begin{equation}\label{eq:scaling}
h_{k}\mapsto\tilde{h}_{k}\coloneqq -\ln(h_{0}-h_{k})\,.
\end{equation}
Here, $h_{0}$ is a constant (slightly) greater than $h_{\max}$, which we set to $h_{0}=1$ in this study.
In the main text, the mean value of the distribution of $\{\tilde{h}_{k}\}$ was called the (mean) International Coupling Distance (ICD), defined for each discipline and period.
Note that this scaling was introduced to facilitate the visualisation of the ICD (Fig.~\ref{fig:line_ICD} and Suppl.~Fig.~\ref{fig:kdensity_ICD}); any other monotonically increasing functions with an appropriate domain and range could be used as an alternative.

\subsection[{\hspace{3.5eM}} Additional analysis results]{Additional analysis results\label{app:add.examples}}

\subsubsection{Rate of non-journal article works\label{app:nonjrate}}

While many commercial services primarily focus on journal articles, it is important to recognise the increasing importance of other scholarly outputs, such as datasets, preprints and books, which are included in the \q{Works} category in the OpenAlex data. 
To highlight this point, we included Suppl.~Fig.~\ref{fig:nonjrate}a, which illustrates the proportion of non-journal article works as a percentage of total scholarly output published each year for different disciplines. 
This figure shows that non-journal article works account for a significant portion of published output in fields such as {\ai} and {\math}, with approximately 40\% in recent years. 
Even in {\bio}, where the rate of non-journal papers historically has been low, it has exceeded about 20\% in recent years. 
By limiting our analysis to journal articles alone, we would miss out on a significant weight of scientific activity and underestimate the momentum of knowledge production each year \cite{Okamura22}.

Suppl.~Fig.~\ref{fig:nonjrate}b, c and d also provide a country-specific analysis for the US, China and Japan, respectively. 
These figures show that while the low proportion of {\bio} publications is still common across these countries, there are interesting country-specific characteristics, such as the high proportion of non-journal article works in {\marine} and {\aerospace} in the US.

\subsubsection{Visualising the remaining 12 disciplines\label{app:rest_twelve}}

In the main body of this paper, Figs.~\ref{fig:line_npaper_intlrate_1}, \ref{fig:chord_1} and \ref{fig:cdend_1} only presented the diagrams for {\ai}, {\quantum} and {\bio} to save space. 
This appendix presents those for the remaining 12 disciplines: {\nano}, {\agri}, {\particle}, {\aerospace}, {\nuclear}, {\marine}, {\neuro}, {\condensed}, {\envi}, {\earth}, {\astro} and {\math}.
The corresponding figures are Suppl.~Figs.~\ref{fig:line_npaper_intlrate_2}, \ref{fig:chord_2} and \ref{fig:cdend_2}, respectively. 
The procedures and points to note in obtaining, analysing and visualising the data are the same as those described in the main body of the paper and the technical explanations in the previous appendix.
We welcome comments on and contextual interpretations of the presented results from experts in the respective fields and interested readers.

\subsubsection{The aggregate category: `Natural Sciences'\label{app:allf}}

When examining the state of international collaborative research, it is crucial to take into account the unique characteristics inherent in different research disciplines. 
Without properly defining disciplinary units, it becomes challenging to draw meaningful conclusions about the state of international research collaboration. 
Additionally, combining fields with varying disciplinary characteristics may even lead to misleading results and implications. 
That is why the main body of this paper did not include analysis results that aggregated all fields without distinguishing between them.

However, if a study aims to understand and compare the volume of scientific knowledge produced in various forms, such as papers or datasets, an analysis that aggregates all fields can still provide some informative insights. 
With this in mind, Suppl.~Fig.~\ref{fig:chord_cdend_allf} includes the results of the analysis of bilateral collaborative relationships (corresponding to Fig.~\ref{fig:chord_1} and Suppl.~Fig.~\ref{fig:chord_2}) and the international collaboration clusters (corresponding to Fig.~\ref{fig:cdend_1} and Suppl.~Fig.~\ref{fig:cdend_2}) for the entire category of \q{Natural Sciences}.
This category represents the aggregation of all 15 level-1 R\&D fields that are the focus of this paper.

\afterpage{\clearpage%
\begin{figure}[tp]
\centering
\vspace{-0.5cm}
\noindent
\includegraphics[align=c, scale=1.0, vmargin=1mm]{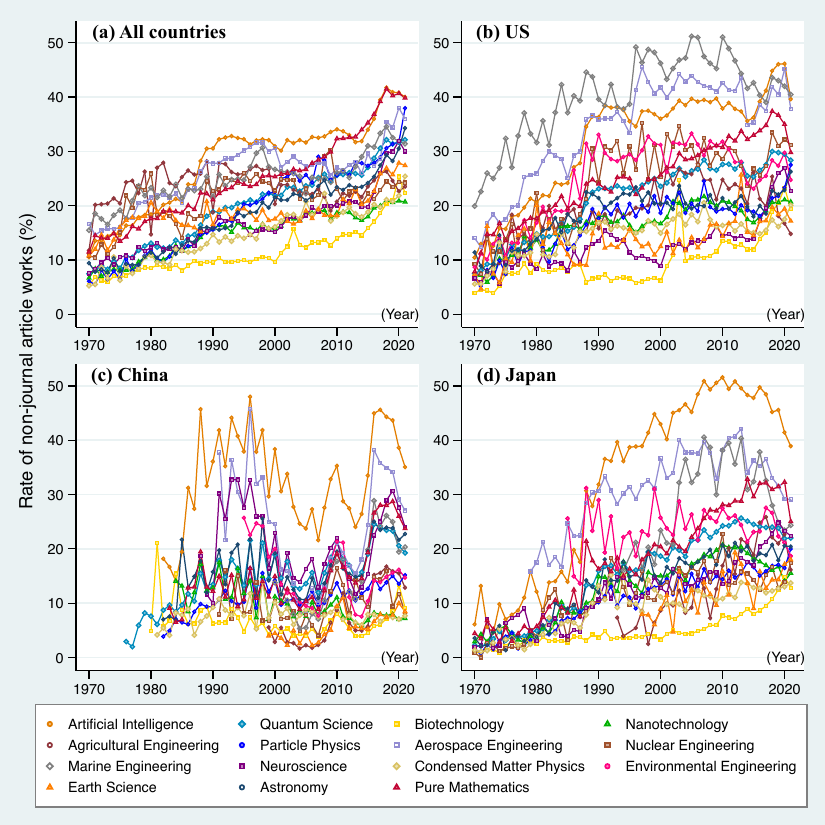}
\caption{\textbf{Trend in the rate of non-journal article works.}
(a) All countries, (b) the US, (c) China and (d) Japan.}
\label{fig:nonjrate}
\vspace{5mm}
\end{figure}
\quad\\
\vfill
}

\afterpage{\clearpage%
\begin{figure}[tp]
\centering
\vspace{-0.5cm}
\noindent
\includegraphics[align=c, scale=0.85, vmargin=1mm]{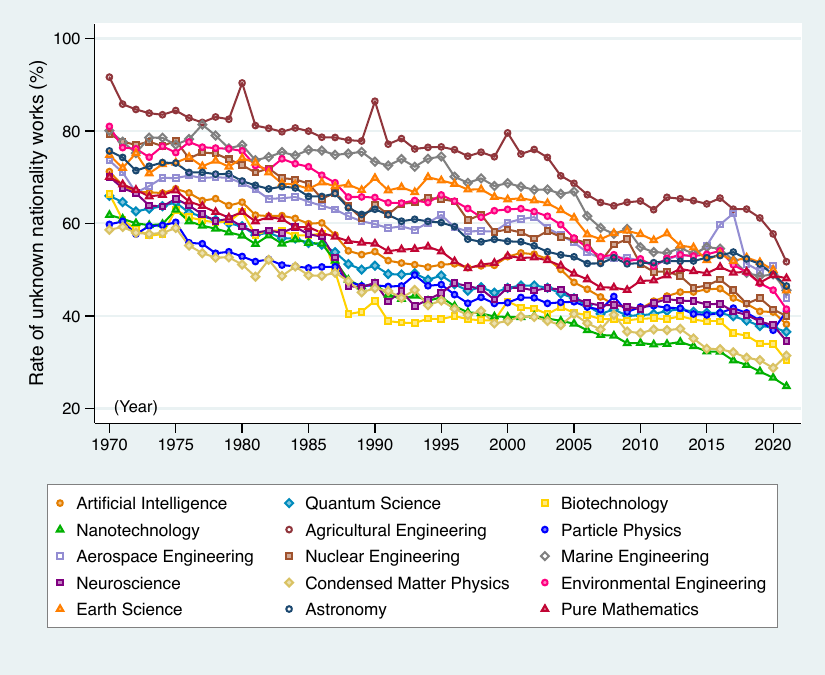}
\caption{\textbf{Trend in the rate of unknown nationality works.}
}
\label{fig:unknownrate}
\vspace{5mm}
\end{figure}
\quad\\
\vfill
}

\afterpage{\clearpage%
\begin{figure}[!tp]
\centering
\begin{subfigure}{1.0\textwidth}
\vspace{-0.5cm}
    \begin{tabular}{c}
{\small\textrm{\textbf{(a)~ \nano}}}\\
\begin{minipage}{0.5\hsize}
\raisebox{-\height}{\includegraphics[align=c, scale=\xsize, vmargin=0mm]{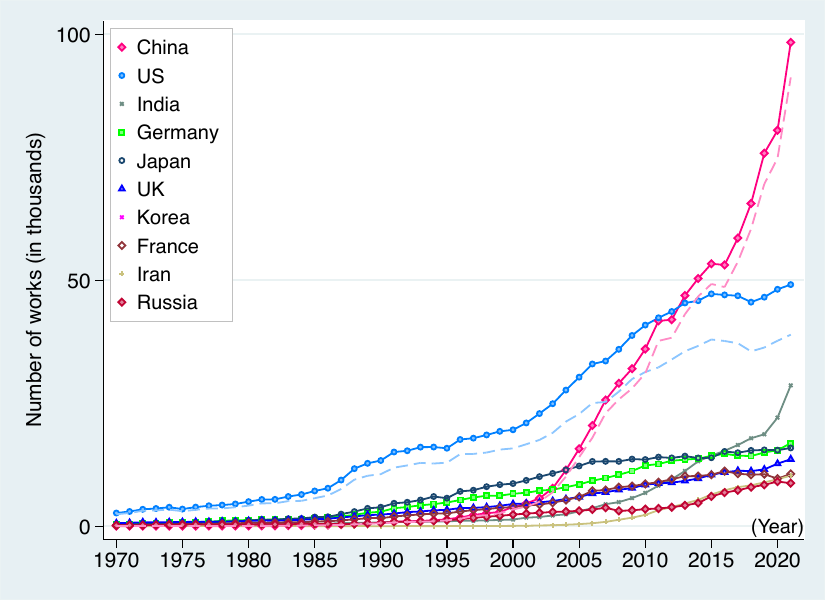}}
	\end{minipage}
\begin{minipage}{0.5\hsize}
\raisebox{-\height}{\includegraphics[align=c, scale=\xsize, vmargin=0mm]{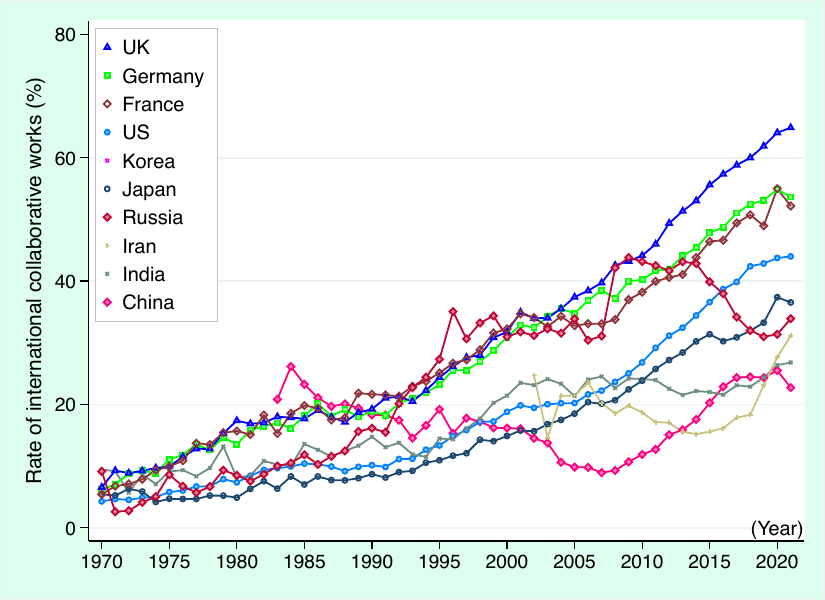}}
	\end{minipage}
    \end{tabular}
\vspace{3mm}\\
    \begin{tabular}{c}
{\small\textrm{\textbf{(b)~ \agri}}}\\
\begin{minipage}{0.5\hsize}
\raisebox{-\height}{\includegraphics[align=c, scale=\xsize, vmargin=0mm]{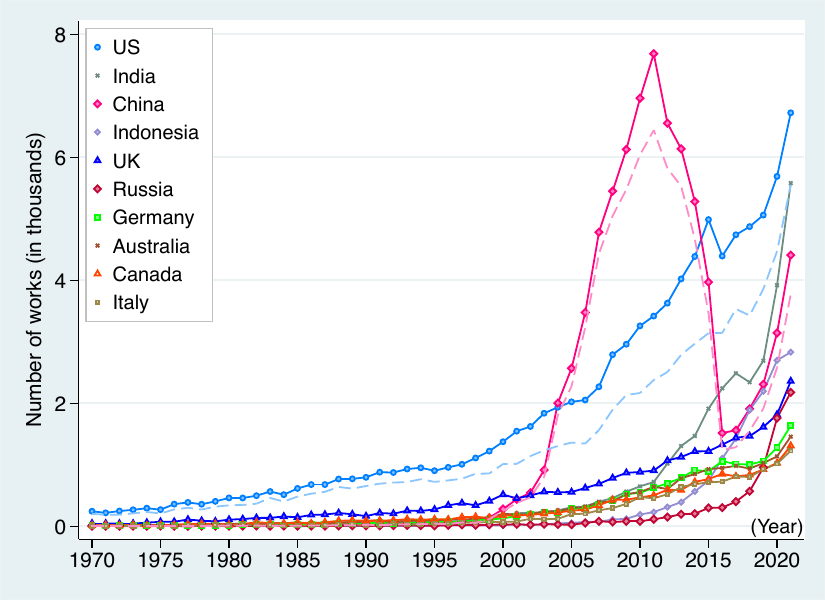}}
	\end{minipage}
\begin{minipage}{0.5\hsize}
\raisebox{-\height}{\includegraphics[align=c, scale=\xsize, vmargin=0mm]{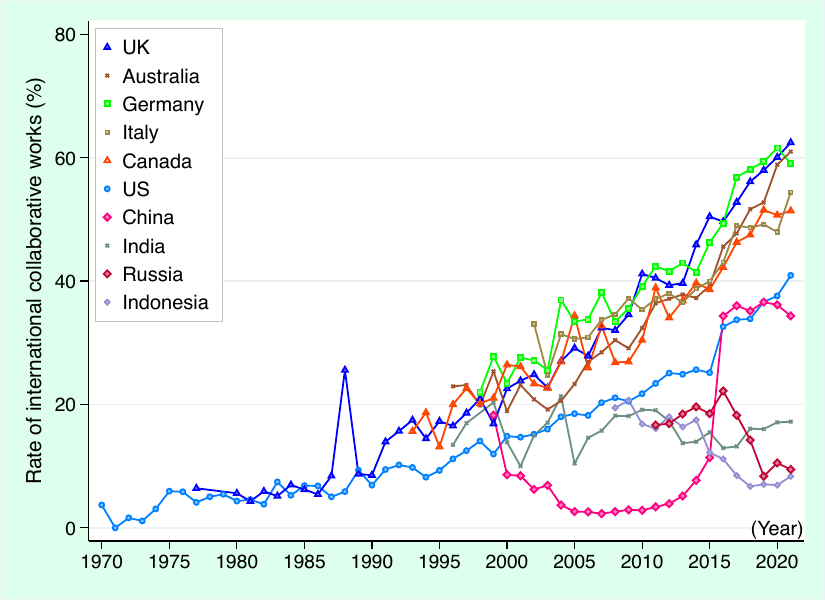}}
	\end{minipage}
    \end{tabular}
\vspace{3mm}\\
    \begin{tabular}{c}
{\small\textrm{\textbf{(c)~ \particle}}}\\
\begin{minipage}{0.5\hsize}
\raisebox{-\height}{\includegraphics[align=c, scale=\xsize, vmargin=0mm]{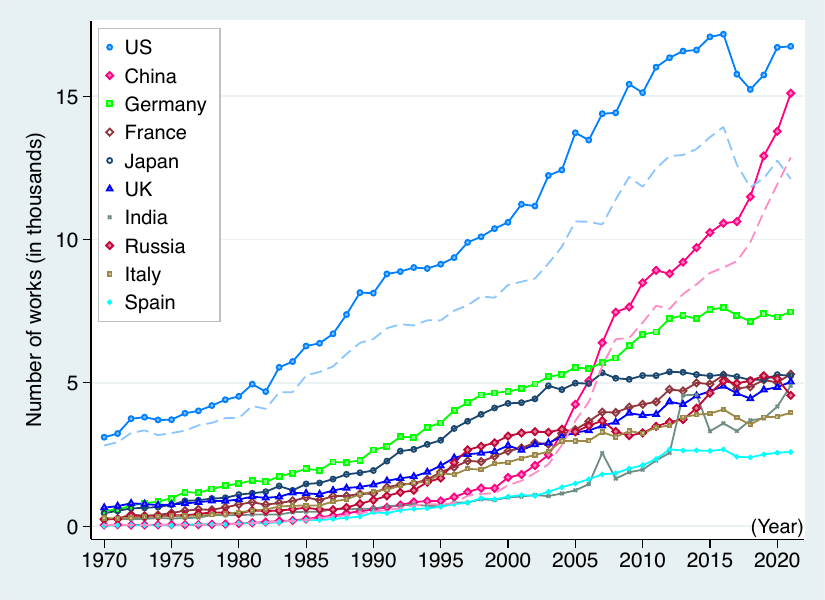}}
	\end{minipage}
\begin{minipage}{0.5\hsize}
\raisebox{-\height}{\includegraphics[align=c, scale=\xsize, vmargin=0mm]{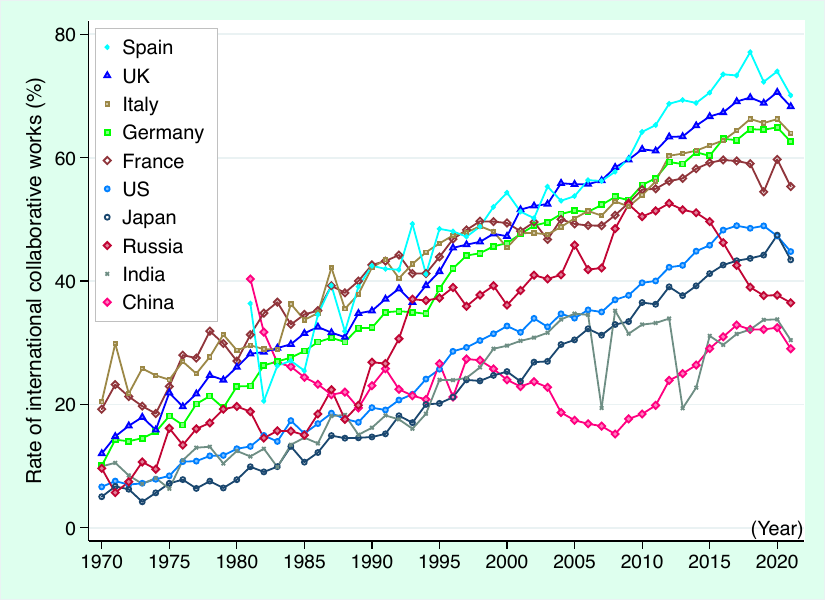}}
	\end{minipage}
    \end{tabular}
\end{subfigure}
\vspace{3mm}
\caption{\textbf{Trends in the number of works (scientific publications) (left) and the international collaboration rate (right).}
The number of works is displayed in thousands.}
\label{fig:line_npaper_intlrate_2}
\end{figure}
}
\afterpage{\clearpage%
\begin{figure}[!tp]\ContinuedFloat
\centering
\begin{subfigure}{1.0\textwidth}
\vspace{-0.5cm}
    \begin{tabular}{c}
{\small\textrm{\textbf{(d)~ \aerospace}}}\\
\begin{minipage}{0.5\hsize}
\raisebox{-\height}{\includegraphics[align=c, scale=\xsize, vmargin=0mm]{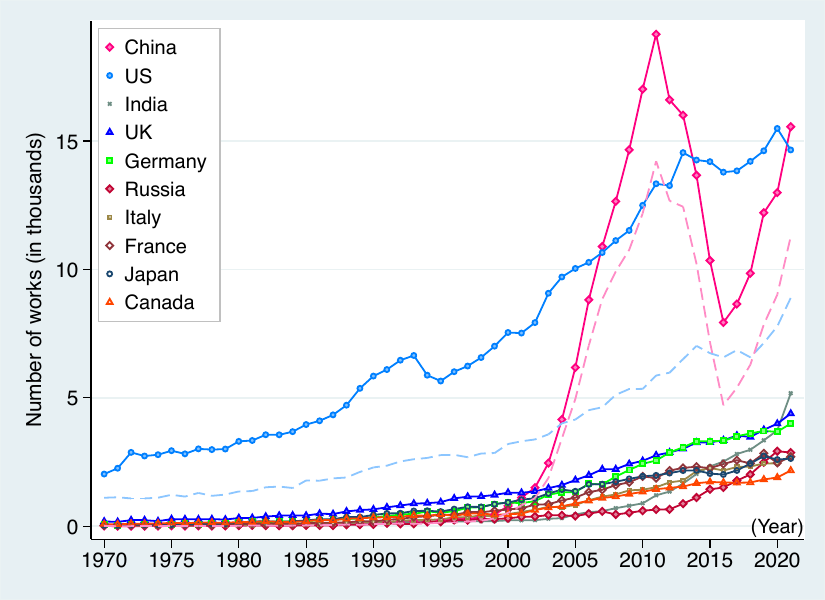}}
	\end{minipage}
\begin{minipage}{0.5\hsize}
\raisebox{-\height}{\includegraphics[align=c, scale=\xsize, vmargin=0mm]{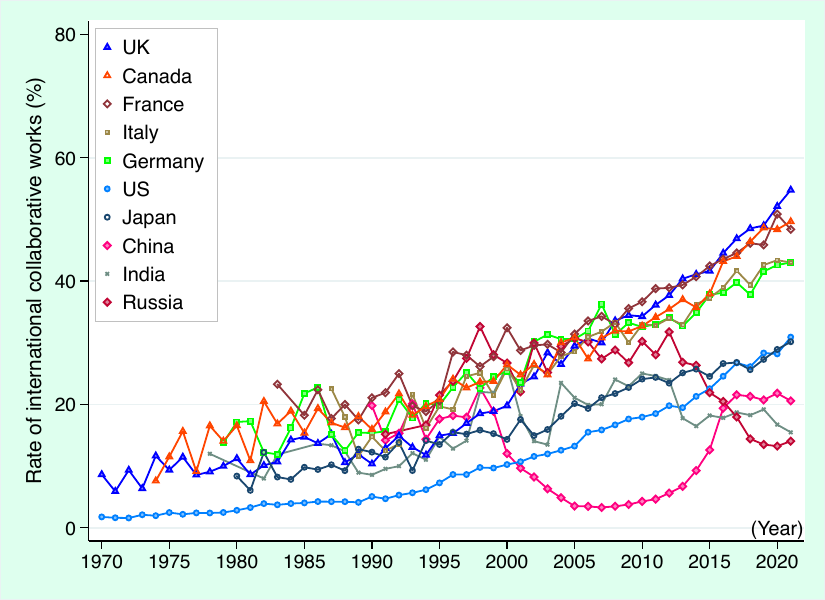}}
	\end{minipage}
    \end{tabular}
\vspace{3mm}\\
    \begin{tabular}{c}
{\small\textrm{\textbf{(e)~ \nuclear}}}\\
\begin{minipage}{0.5\hsize}
\raisebox{-\height}{\includegraphics[align=c, scale=\xsize, vmargin=0mm]{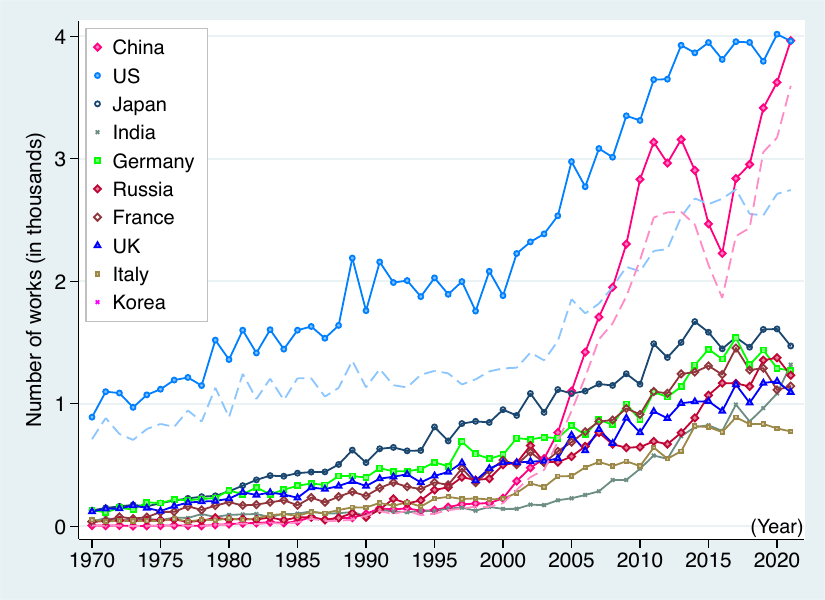}}
	\end{minipage}
\begin{minipage}{0.5\hsize}
\raisebox{-\height}{\includegraphics[align=c, scale=\xsize, vmargin=0mm]{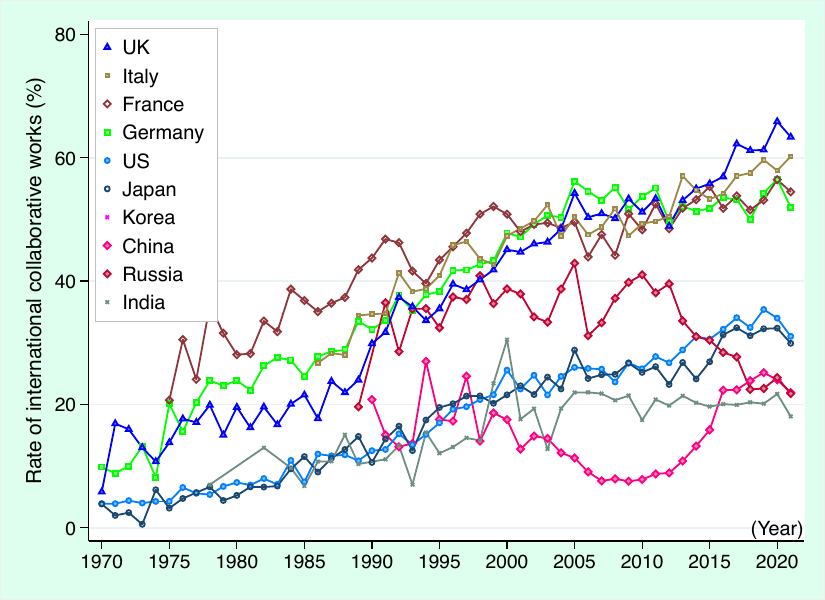}}
	\end{minipage}
    \end{tabular}
\vspace{3mm}\\
    \begin{tabular}{c}
{\small\textrm{\textbf{(f)~ \marine}}}\\
\begin{minipage}{0.5\hsize}
\raisebox{-\height}{\includegraphics[align=c, scale=\xsize, vmargin=0mm]{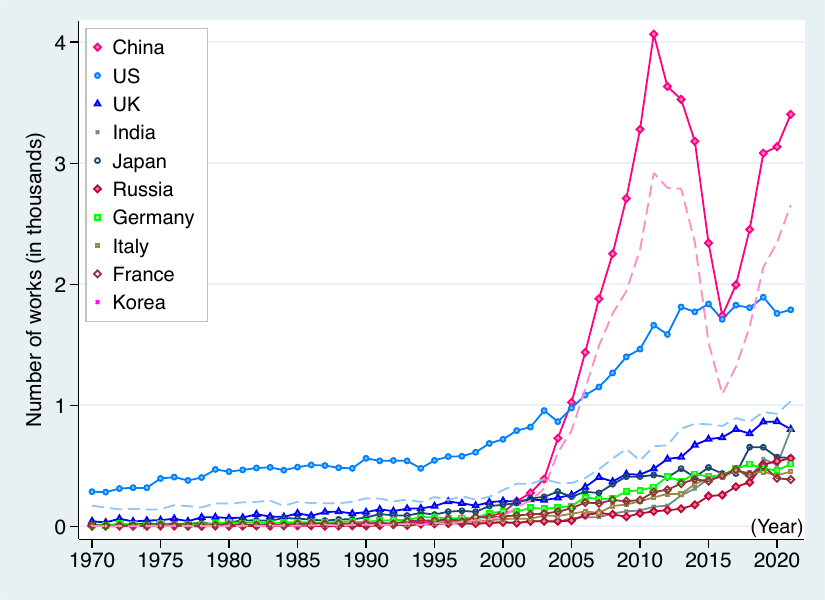}}
	\end{minipage}
\begin{minipage}{0.5\hsize}
\raisebox{-\height}{\includegraphics[align=c, scale=\xsize, vmargin=0mm]{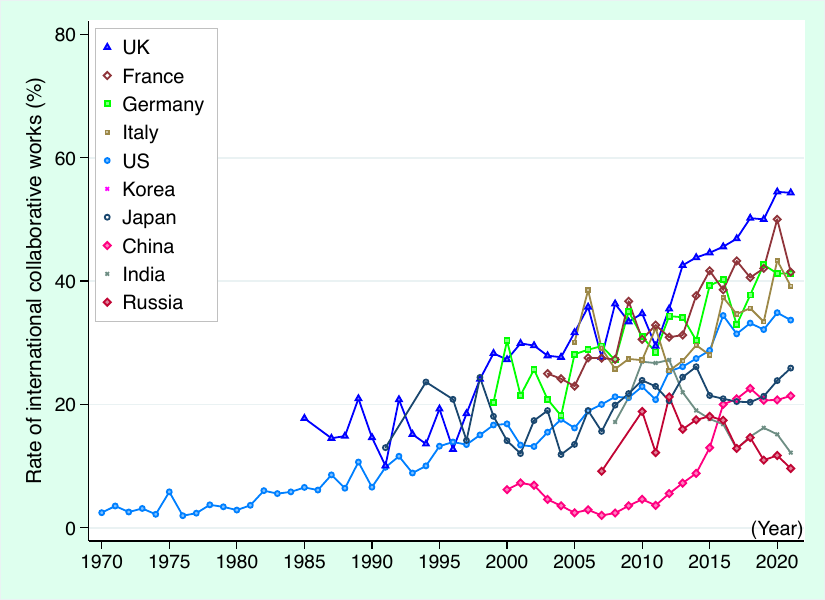}}
	\end{minipage}
    \end{tabular}
\end{subfigure}
\vspace{3mm}
\caption{\textbf{Trends in the number of works (scientific publications) (left) and the international collaboration rate (right). \emph{(Cont.)}}
The number of works is displayed in thousands.}
\label{fig:line_npaper_intlrate_3}
\end{figure}
}
\afterpage{\clearpage%
\begin{figure}[!tp]\ContinuedFloat
\centering
\begin{subfigure}{1.0\textwidth}
\vspace{-0.5cm}
    \begin{tabular}{c}
{\small\textrm{\textbf{(g)~ \neuro}}}\\
\begin{minipage}{0.5\hsize}
\raisebox{-\height}{\includegraphics[align=c, scale=\xsize, vmargin=0mm]{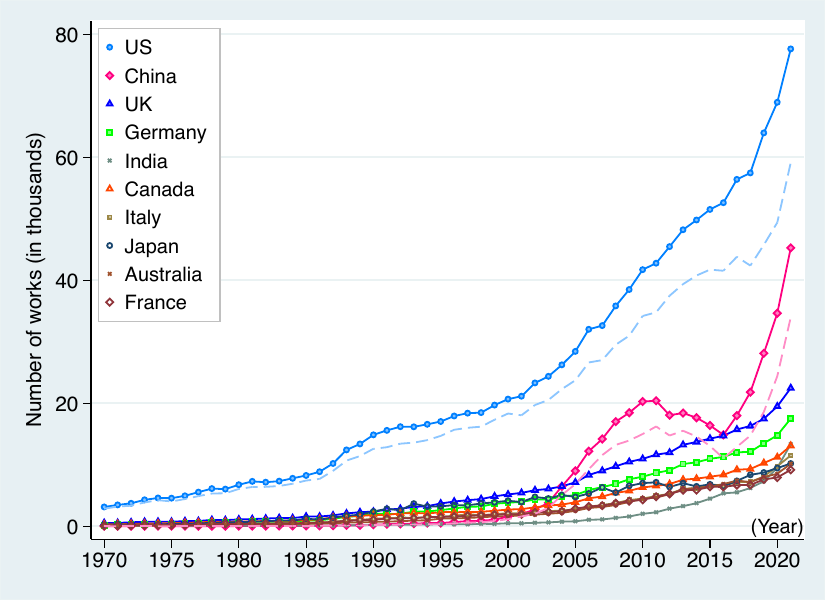}}
	\end{minipage}
\begin{minipage}{0.5\hsize}
\raisebox{-\height}{\includegraphics[align=c, scale=\xsize, vmargin=0mm]{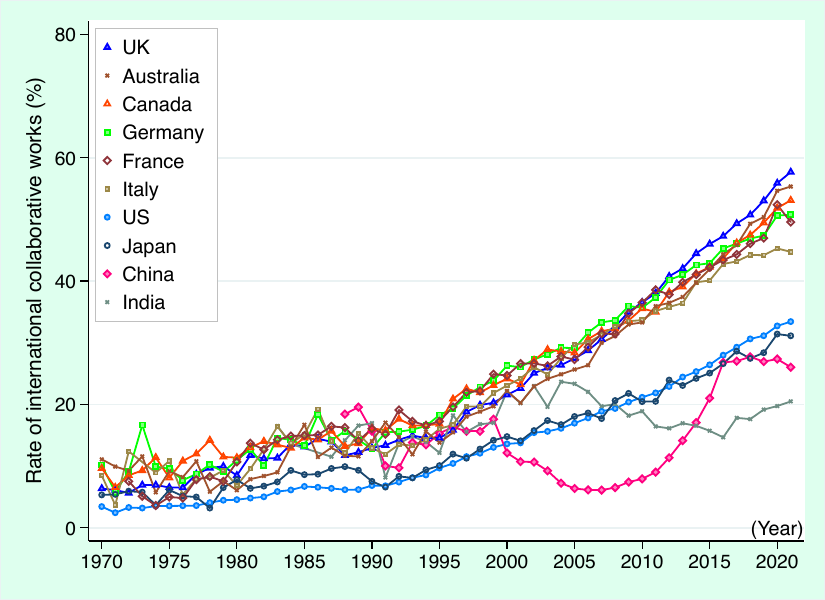}}
	\end{minipage}
    \end{tabular}
\vspace{3mm}\\
    \begin{tabular}{c}
{\small\textrm{\textbf{(h)~ \condensed}}}\\
\begin{minipage}{0.5\hsize}
\raisebox{-\height}{\includegraphics[align=c, scale=\xsize, vmargin=0mm]{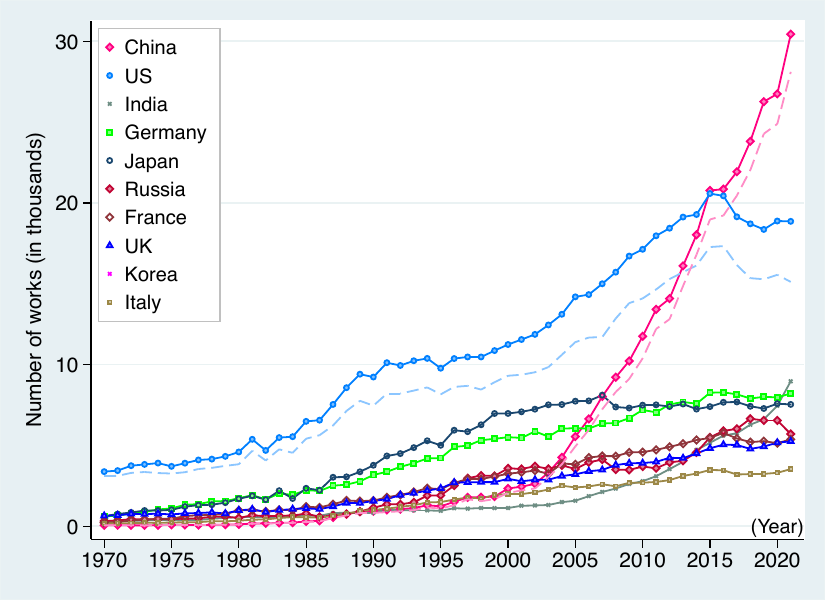}}
	\end{minipage}
\begin{minipage}{0.5\hsize}
\raisebox{-\height}{\includegraphics[align=c, scale=\xsize, vmargin=0mm]{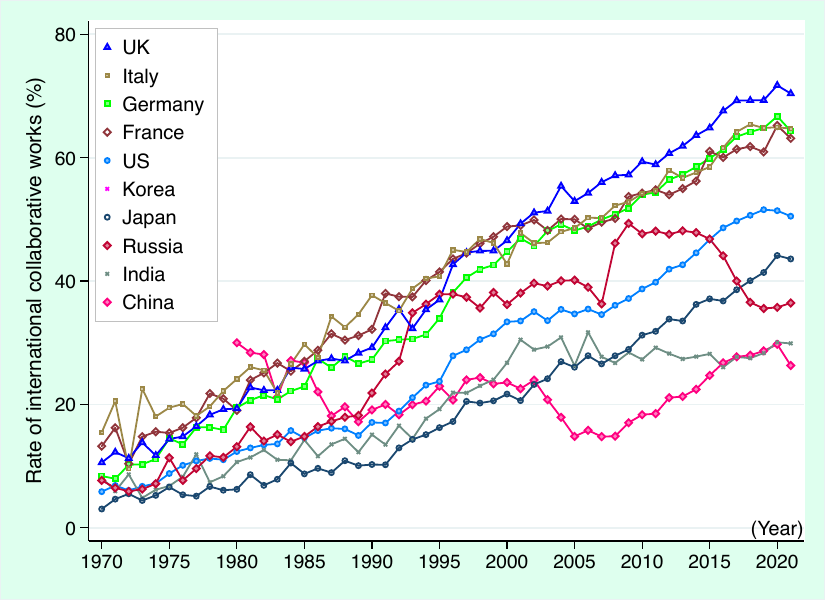}}
	\end{minipage}
    \end{tabular}
\vspace{3mm}\\
    \begin{tabular}{c}
{\small\textrm{\textbf{(i)~ \envi}}}\\
\begin{minipage}{0.5\hsize}
\raisebox{-\height}{\includegraphics[align=c, scale=\xsize, vmargin=0mm]{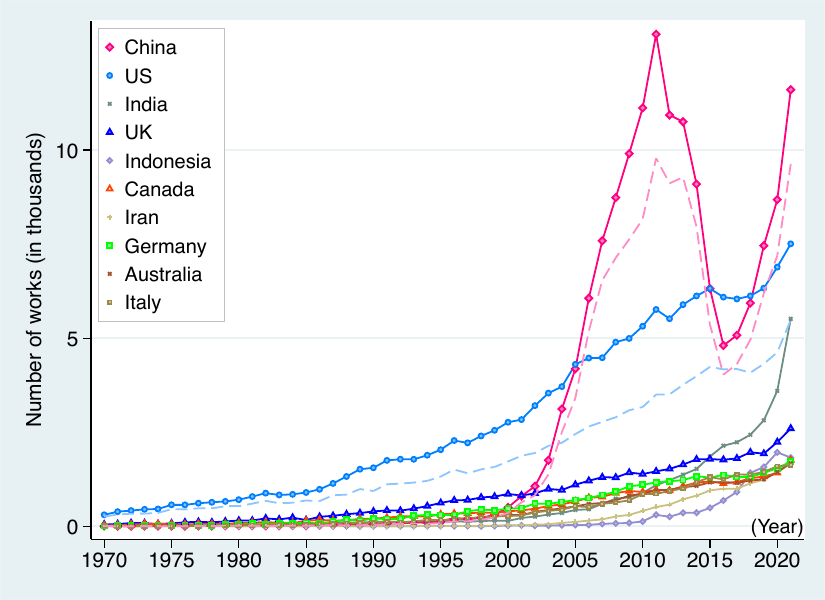}}
	\end{minipage}
\begin{minipage}{0.5\hsize}
\raisebox{-\height}{\includegraphics[align=c, scale=\xsize, vmargin=0mm]{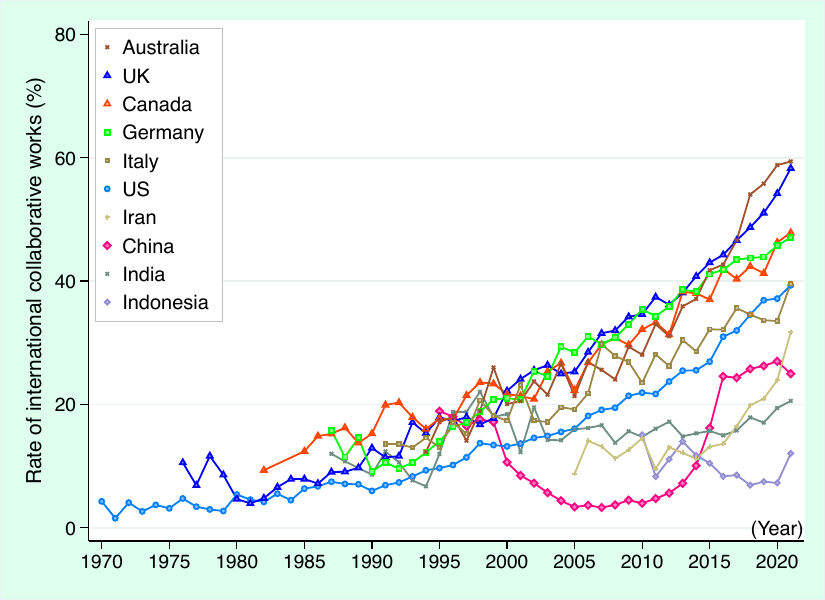}}
	\end{minipage}
    \end{tabular}
\end{subfigure}
\vspace{3mm}
\caption{\textbf{Trends in the number of works (scientific publications) (left) and the international collaboration rate (right). \emph{(Cont.)}}
The number of works is displayed in thousands.}
\label{fig:line_npaper_intlrate_4}
\end{figure}
}
\afterpage{\clearpage%
\begin{figure}[!tp]\ContinuedFloat
\centering
\begin{subfigure}{1.0\textwidth}
\vspace{-0.5cm}
    \begin{tabular}{c}
{\small\textrm{\textbf{(j)~ \earth}}}\\
\begin{minipage}{0.5\hsize}
\raisebox{-\height}{\includegraphics[align=c, scale=\xsize, vmargin=0mm]{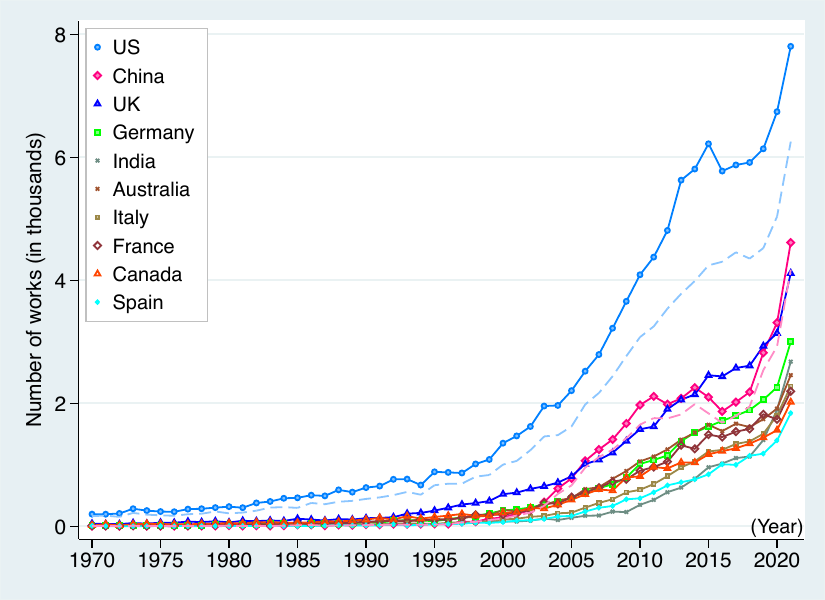}}
	\end{minipage}
\begin{minipage}{0.5\hsize}
\raisebox{-\height}{\includegraphics[align=c, scale=\xsize, vmargin=0mm]{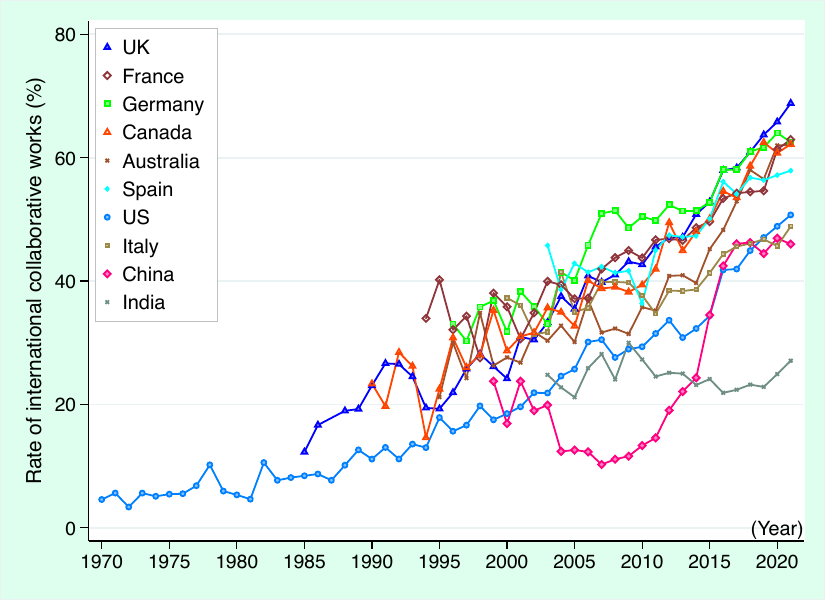}}
	\end{minipage}
    \end{tabular}
\vspace{3mm}\\
    \begin{tabular}{c}
{\small\textrm{\textbf{(k)~ \astro}}}\\
\begin{minipage}{0.5\hsize}
\raisebox{-\height}{\includegraphics[align=c, scale=\xsize, vmargin=0mm]{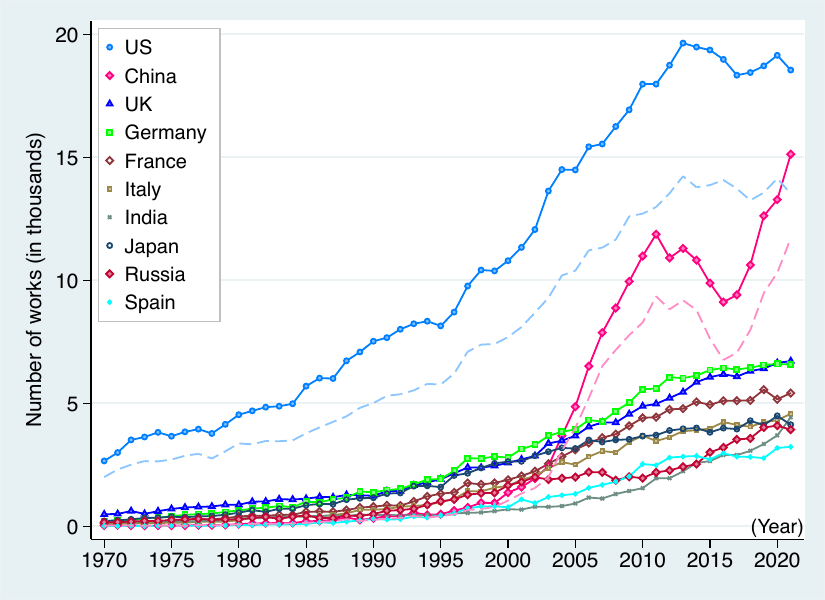}}
	\end{minipage}
\begin{minipage}{0.5\hsize}
\raisebox{-\height}{\includegraphics[align=c, scale=\xsize, vmargin=0mm]{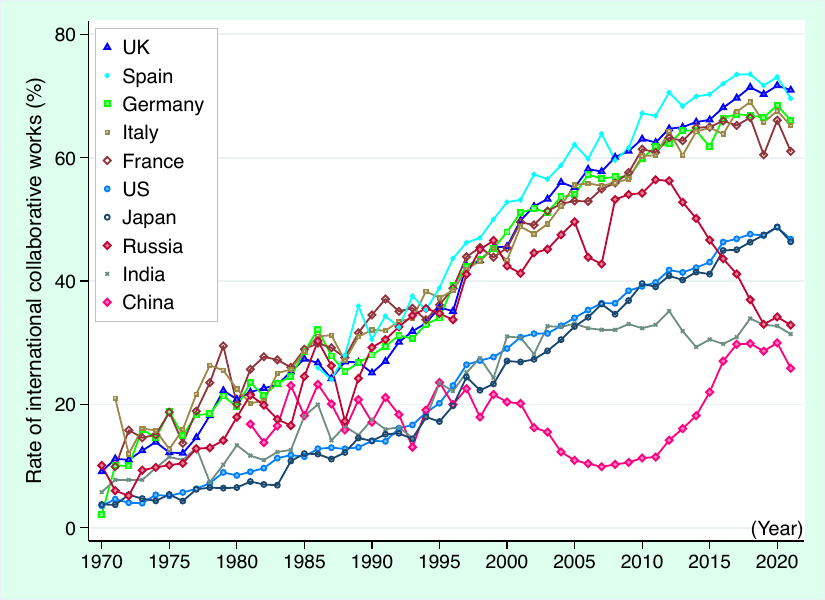}}
	\end{minipage}
    \end{tabular}
\vspace{3mm}\\
    \begin{tabular}{c}
{\small\textrm{\textbf{(l)~ \math}}}\\
\begin{minipage}{0.5\hsize}
\raisebox{-\height}{\includegraphics[align=c, scale=\xsize, vmargin=0mm]{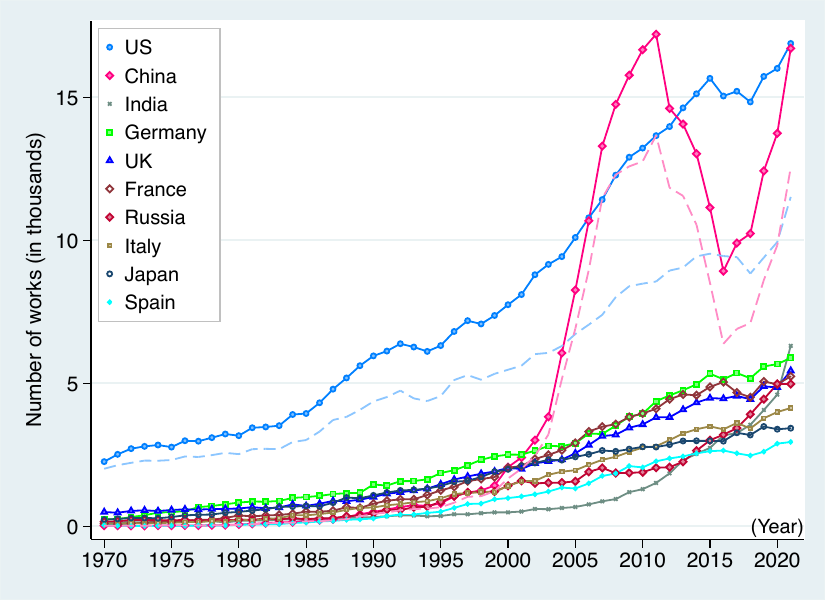}}
	\end{minipage}
\begin{minipage}{0.5\hsize}
\raisebox{-\height}{\includegraphics[align=c, scale=\xsize, vmargin=0mm]{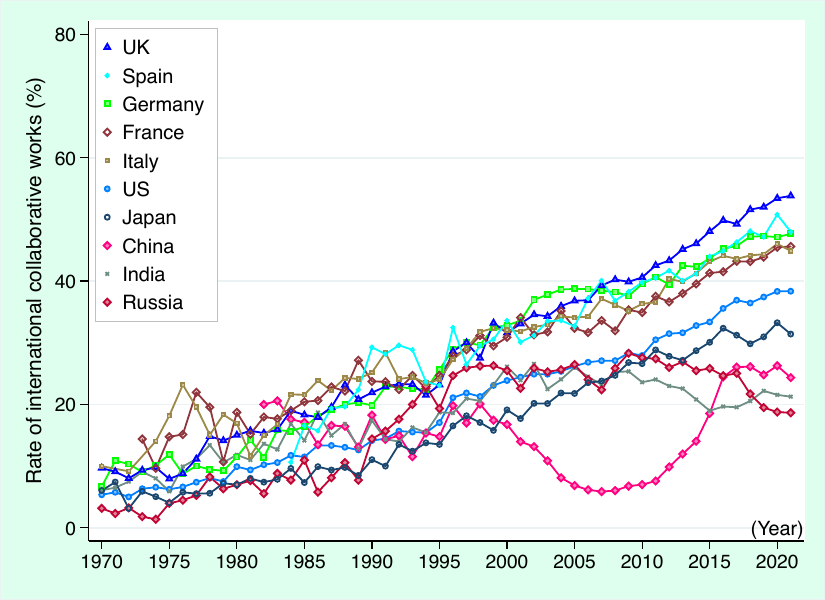}}
	\end{minipage}
    \end{tabular}
\end{subfigure}
\vspace{3mm}
\caption{\textbf{Trends in the number of works (scientific publications) (left) and the international collaboration rate (right). \emph{(Cont.)}}
The number of works is displayed in thousands.}
\label{fig:line_npaper_intlrate_5}
\end{figure}
}

\afterpage{\clearpage%
\begin{figure}[!htp]
\centering
\vspace{-0.5cm}
	\begin{tabular}{c}
\begin{minipage}{0.33\hsize}
\begin{center}
\raisebox{-\height}{\includegraphics[align=c, scale=\dsize, vmargin=0mm]{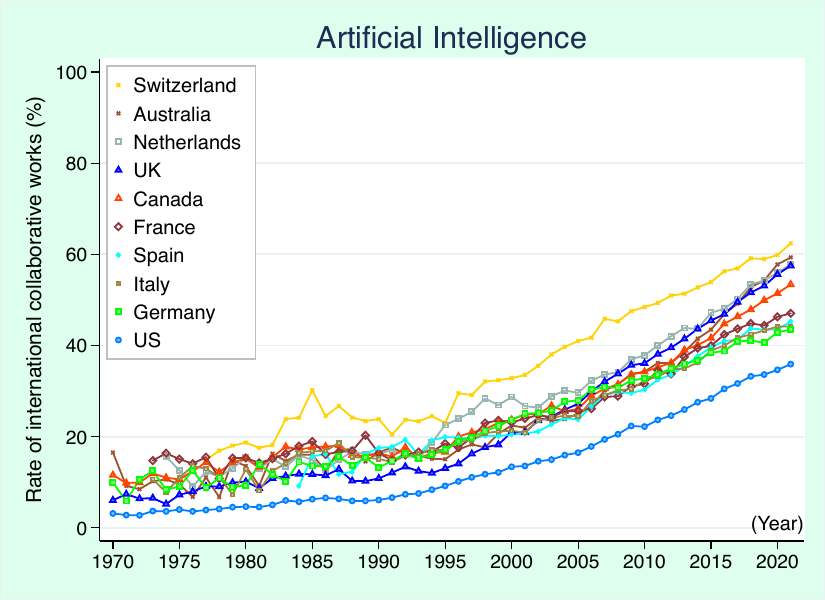}}
\end{center}
	\end{minipage}
\begin{minipage}{0.33\hsize}
\begin{center}
\raisebox{-\height}{\includegraphics[align=c, scale=\dsize, vmargin=0mm]{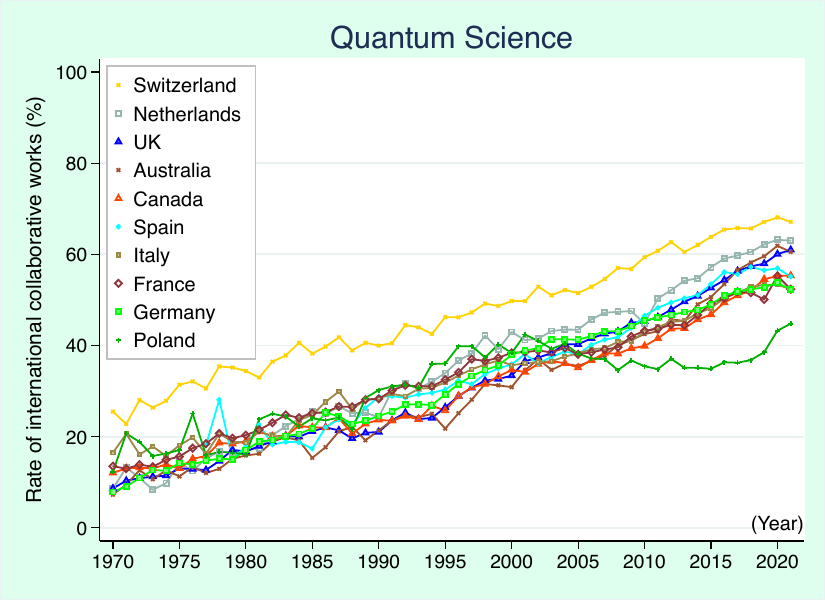}}
\end{center}
	\end{minipage}
\begin{minipage}{0.33\hsize}
\begin{center}
\raisebox{-\height}{\includegraphics[align=c, scale=\dsize, vmargin=0mm]{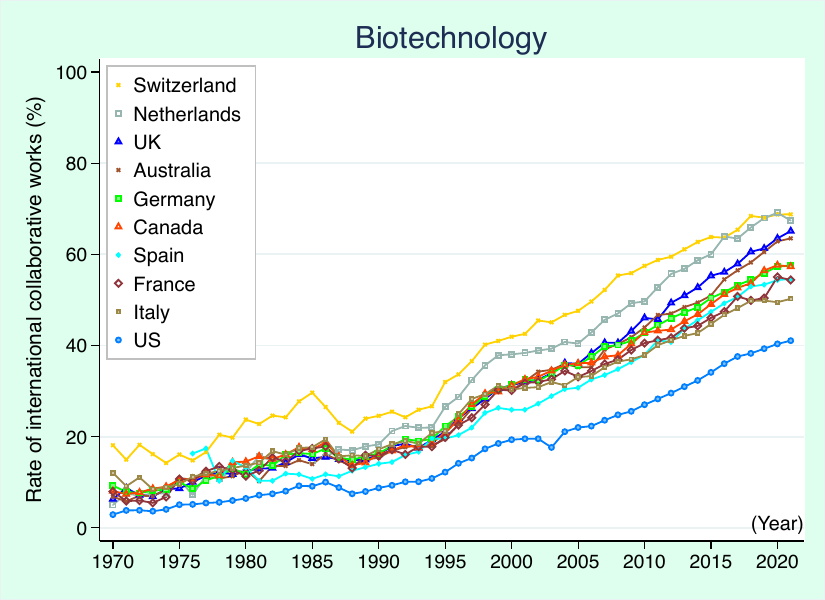}}
\end{center}
	\end{minipage}
    \end{tabular}
\vspace{1mm}\\
	\begin{tabular}{c}
\begin{minipage}{0.33\hsize}
\begin{center}
\raisebox{-\height}{\includegraphics[align=c, scale=\dsize, vmargin=0mm]{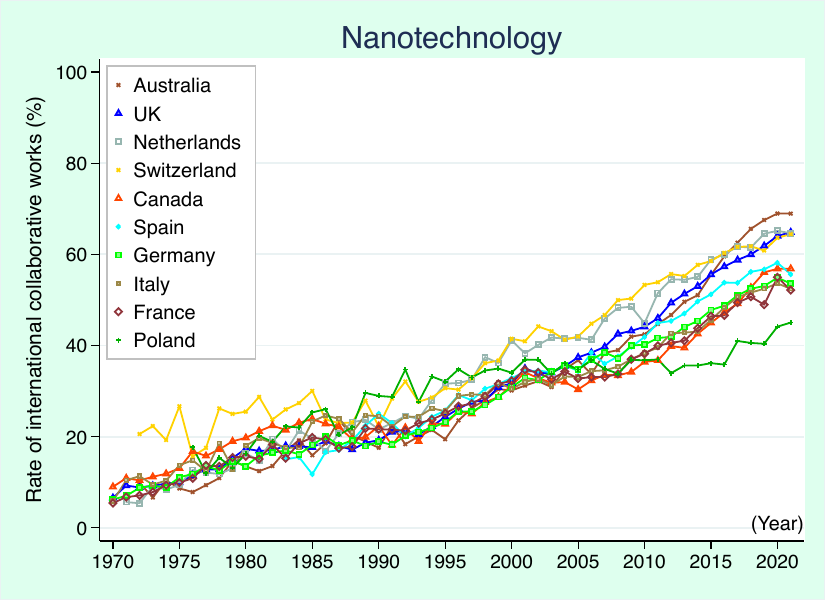}}
\end{center}
	\end{minipage}
\begin{minipage}{0.33\hsize}
\begin{center}
\raisebox{-\height}{\includegraphics[align=c, scale=\dsize, vmargin=0mm]{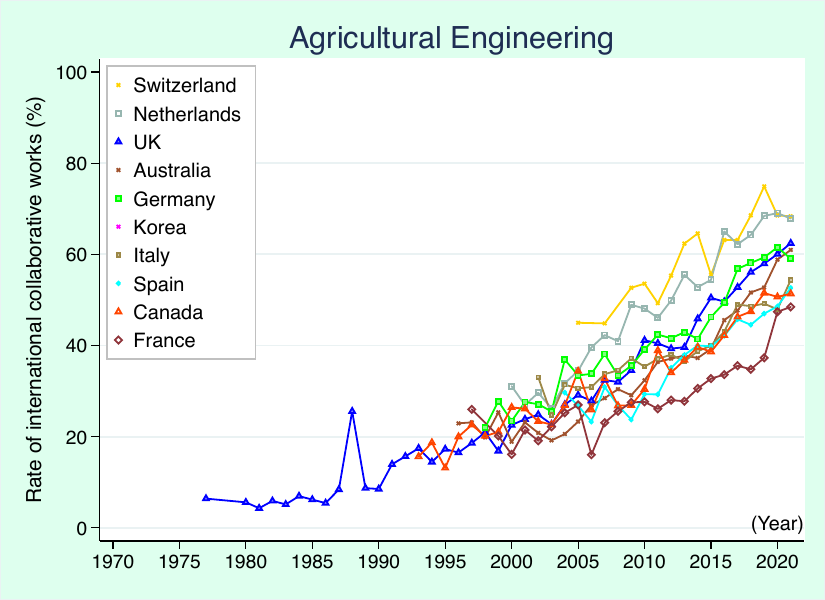}}
\end{center}
	\end{minipage}
\begin{minipage}{0.33\hsize}
\begin{center}
\raisebox{-\height}{\includegraphics[align=c, scale=\dsize, vmargin=0mm]{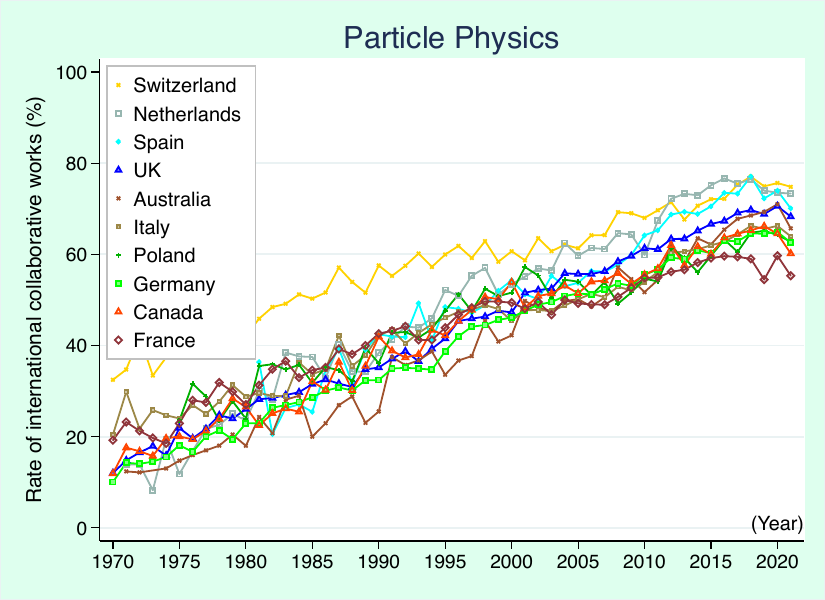}}
\end{center}
	\end{minipage}
    \end{tabular}
\vspace{1mm}\\
	\begin{tabular}{c}
\begin{minipage}{0.33\hsize}
\begin{center}
\raisebox{-\height}{\includegraphics[align=c, scale=\dsize, vmargin=0mm]{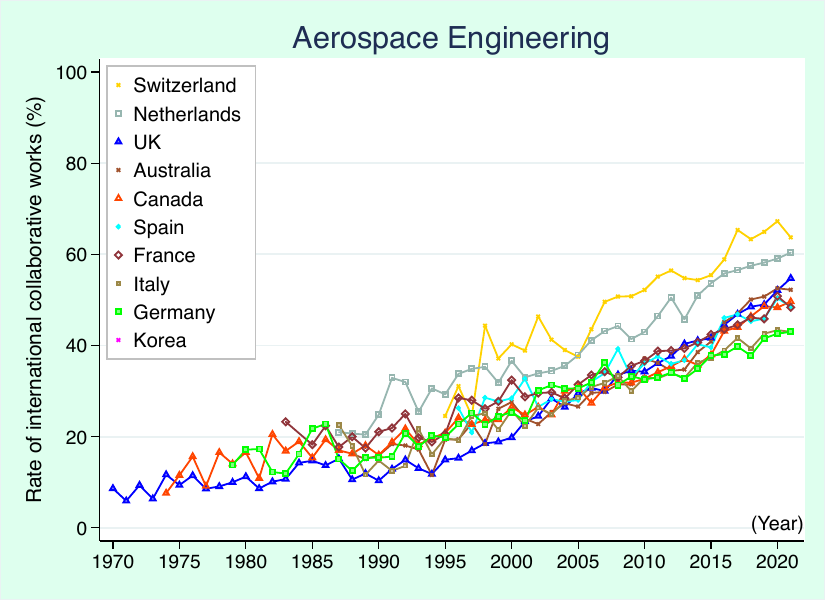}}
\end{center}
	\end{minipage}
\begin{minipage}{0.33\hsize}
\begin{center}
\raisebox{-\height}{\includegraphics[align=c, scale=\dsize, vmargin=0mm]{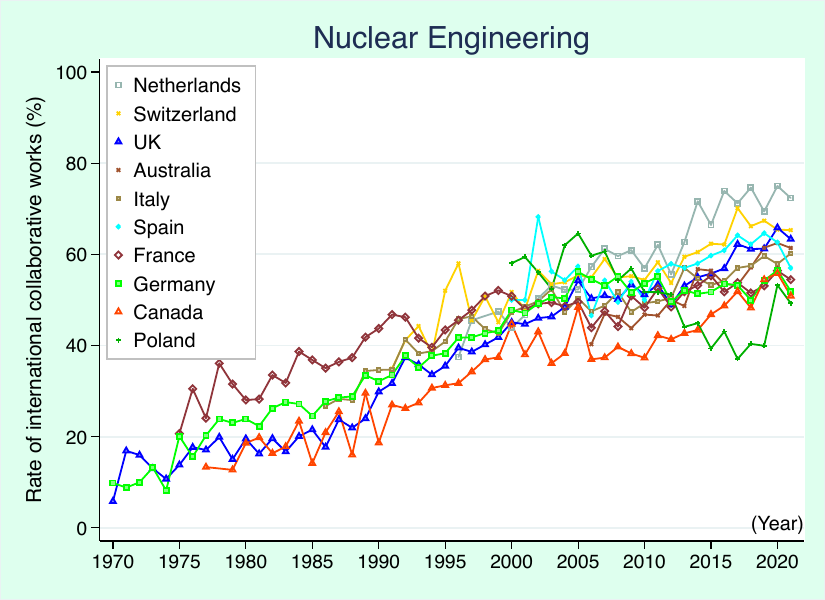}}
\end{center}
	\end{minipage}
\begin{minipage}{0.33\hsize}
\begin{center}
\raisebox{-\height}{\includegraphics[align=c, scale=\dsize, vmargin=0mm]{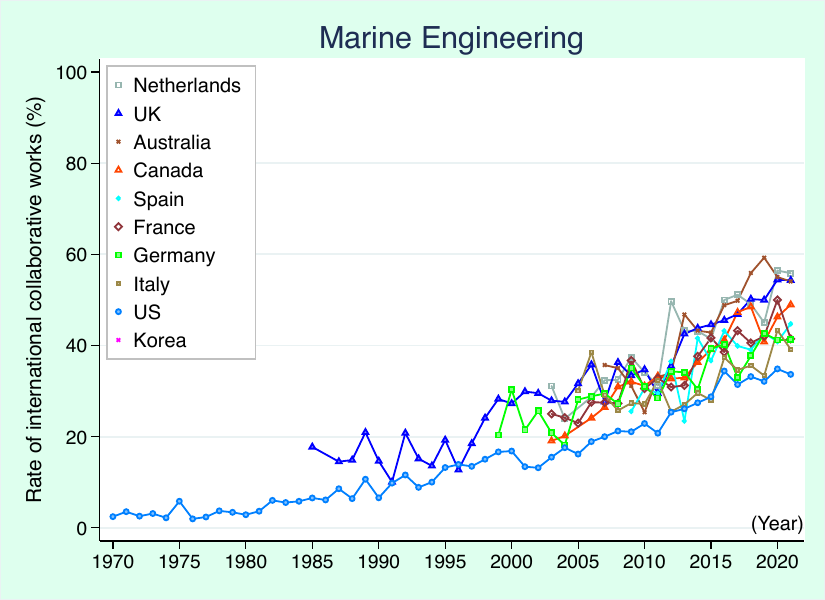}}
\end{center}
	\end{minipage}
    \end{tabular}
\vspace{1mm}\\
	\begin{tabular}{c}
\begin{minipage}{0.33\hsize}
\begin{center}
\raisebox{-\height}{\includegraphics[align=c, scale=\dsize, vmargin=0mm]{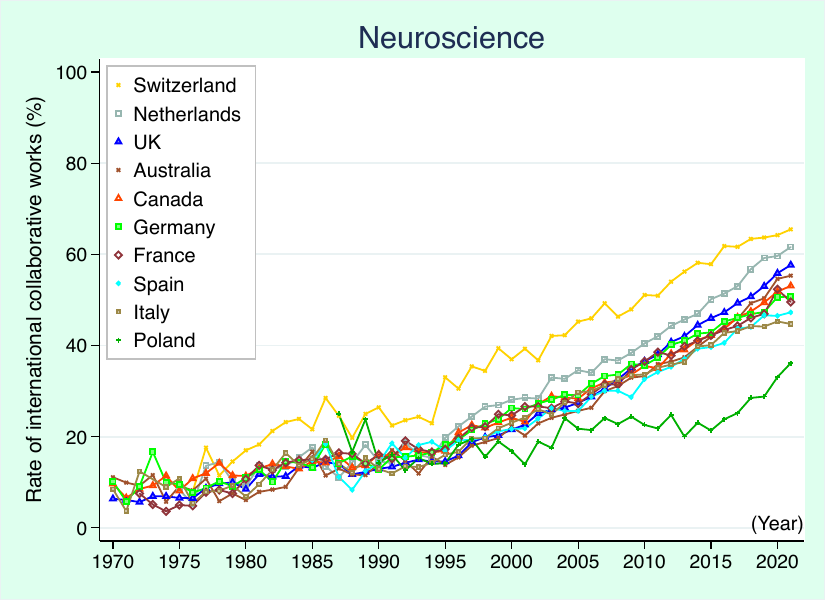}}
\end{center}
	\end{minipage}
\begin{minipage}{0.33\hsize}
\begin{center}
\raisebox{-\height}{\includegraphics[align=c, scale=\dsize, vmargin=0mm]{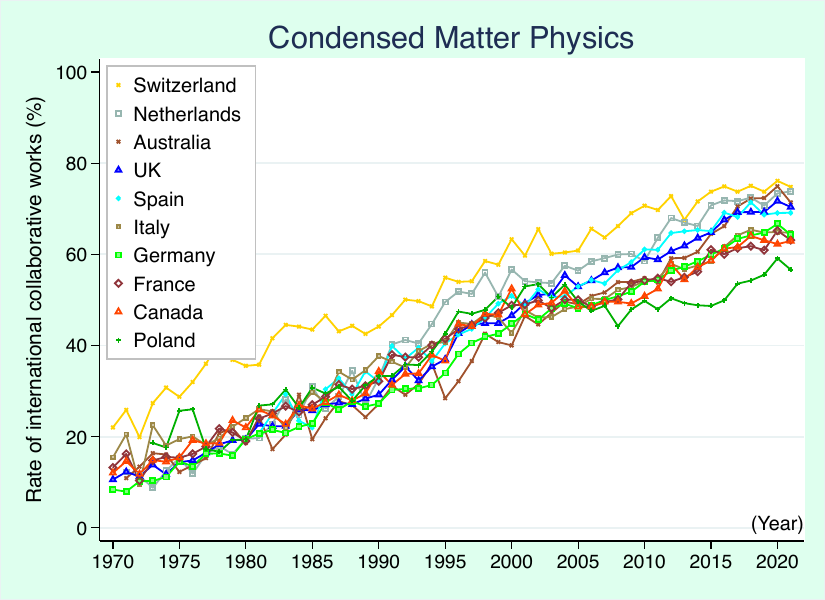}}
\end{center}
	\end{minipage}
\begin{minipage}{0.33\hsize}
\begin{center}
\raisebox{-\height}{\includegraphics[align=c, scale=\dsize, vmargin=0mm]{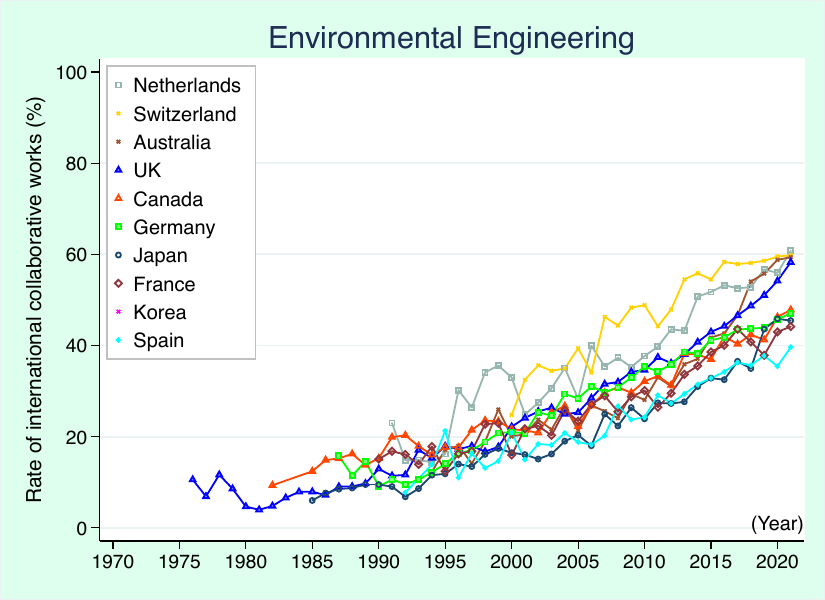}}
\end{center}
	\end{minipage}
    \end{tabular}
\vspace{1mm}\\
	\begin{tabular}{c}
\begin{minipage}{0.33\hsize}
\begin{center}
\raisebox{-\height}{\includegraphics[align=c, scale=\dsize, vmargin=0mm]{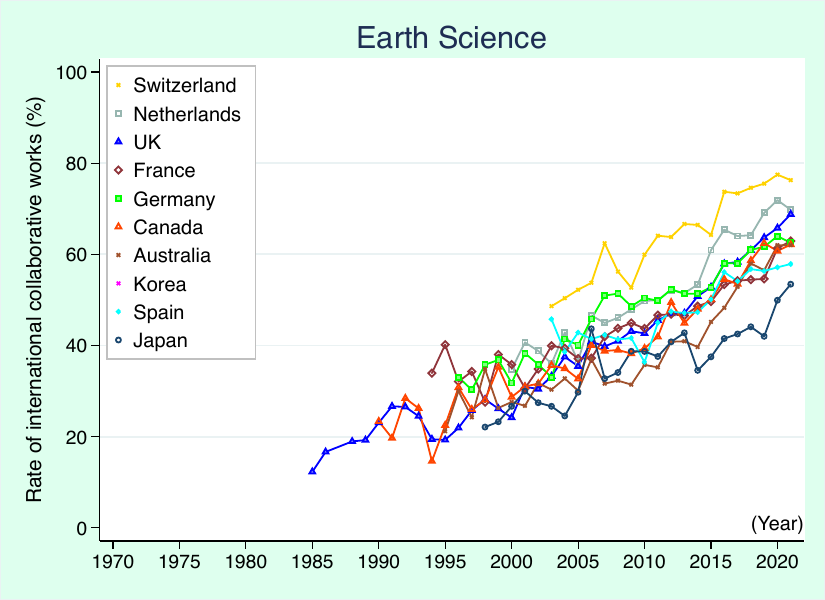}}
\end{center}
	\end{minipage}
\begin{minipage}{0.33\hsize}
\begin{center}
\raisebox{-\height}{\includegraphics[align=c, scale=\dsize, vmargin=0mm]{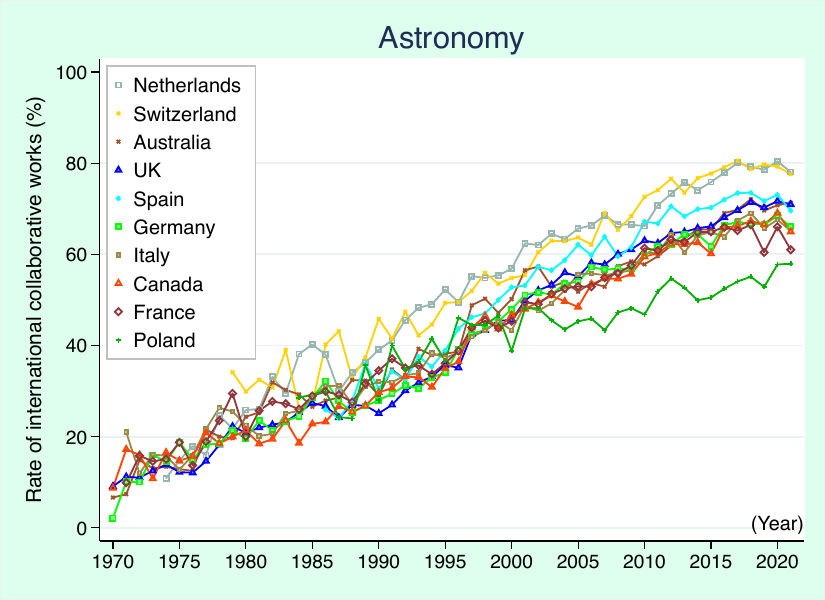}}
\end{center}
	\end{minipage}
\begin{minipage}{0.33\hsize}
\begin{center}
\raisebox{-\height}{\includegraphics[align=c, scale=\dsize, vmargin=0mm]{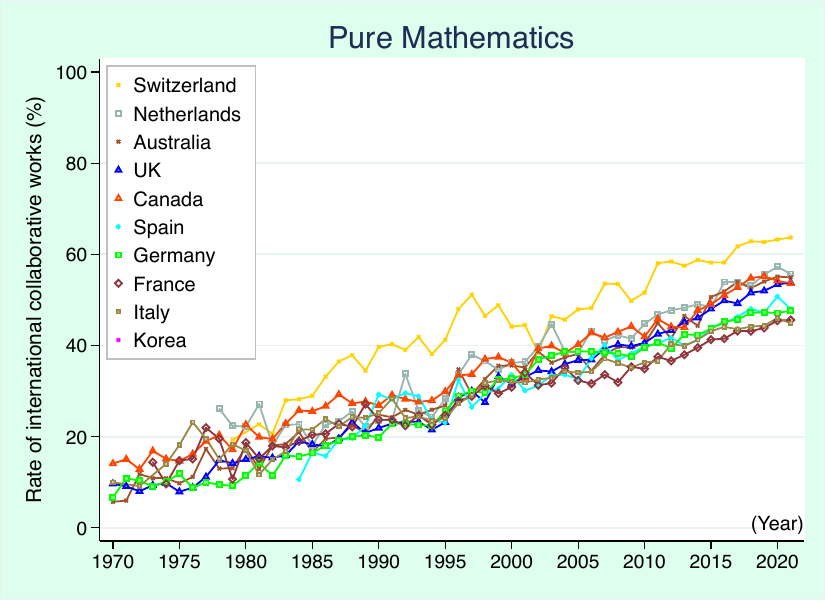}}
\end{center}
	\end{minipage}
    \end{tabular}

\vspace{3mm}
\caption{\textbf{Trends in the international collaboration rate.}
The top 20 countries in work production in 2001--2020 were first identified for each discipline, from which the top 10 countries in international collaboration rate are displayed.}
\label{fig:line_intlrate2}
\end{figure}
}

\afterpage{\clearpage%
\begin{figure}[!tp]
\centering
\vspace{-0.5cm}
\noindent
\includegraphics[align=c, scale=0.85, vmargin=1mm]{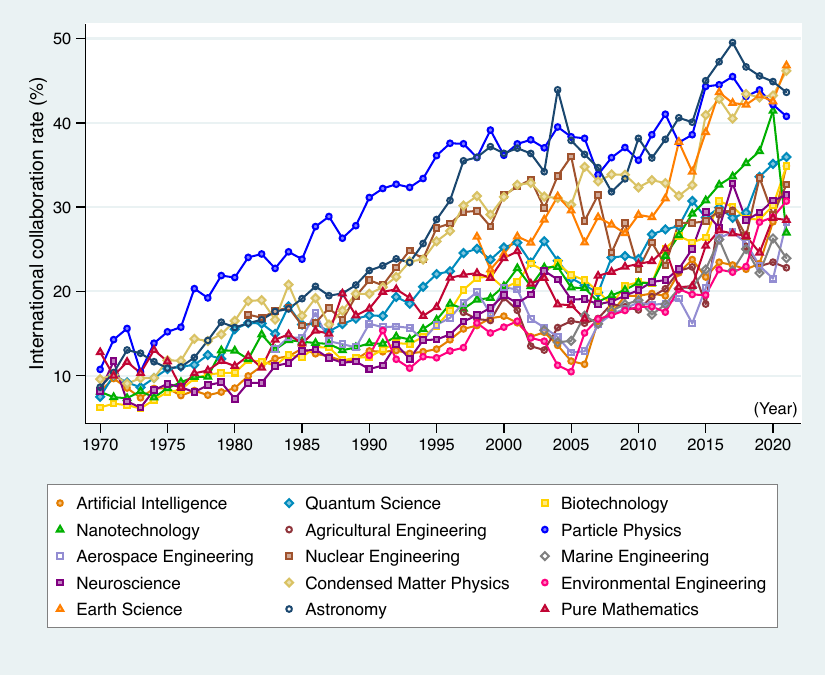}
\caption{\textbf{Trends in the international collaboration rate by discipline.}
}
\label{fig:line_intlrate}
\vspace{5mm}
\end{figure}
\quad\\
\vfill
}

\afterpage{\clearpage%
\begin{figure}[htp]
\centering
\begin{subfigure}{1.0\textwidth}
\vspace{-0.5cm}
    \begin{tabular}{c}
    \begin{minipage}{0.03\hsize}
\begin{flushleft}
    \hspace{-0.7cm}\rotatebox{90}{\period{4}{2011--2020}}
\end{flushleft}
	\end{minipage}
	\begin{minipage}{0.33\hsize}
\begin{flushleft}
\raisebox{-0.0cm}{\hspace{-6mm}\small\textrm{\textbf{(a)~ \nano}}}\\[6mm]
\raisebox{\height}{\includegraphics[trim=2.0cm 1.8cm 0cm 1.5cm, align=c, scale=\chordsize, vmargin=0mm]{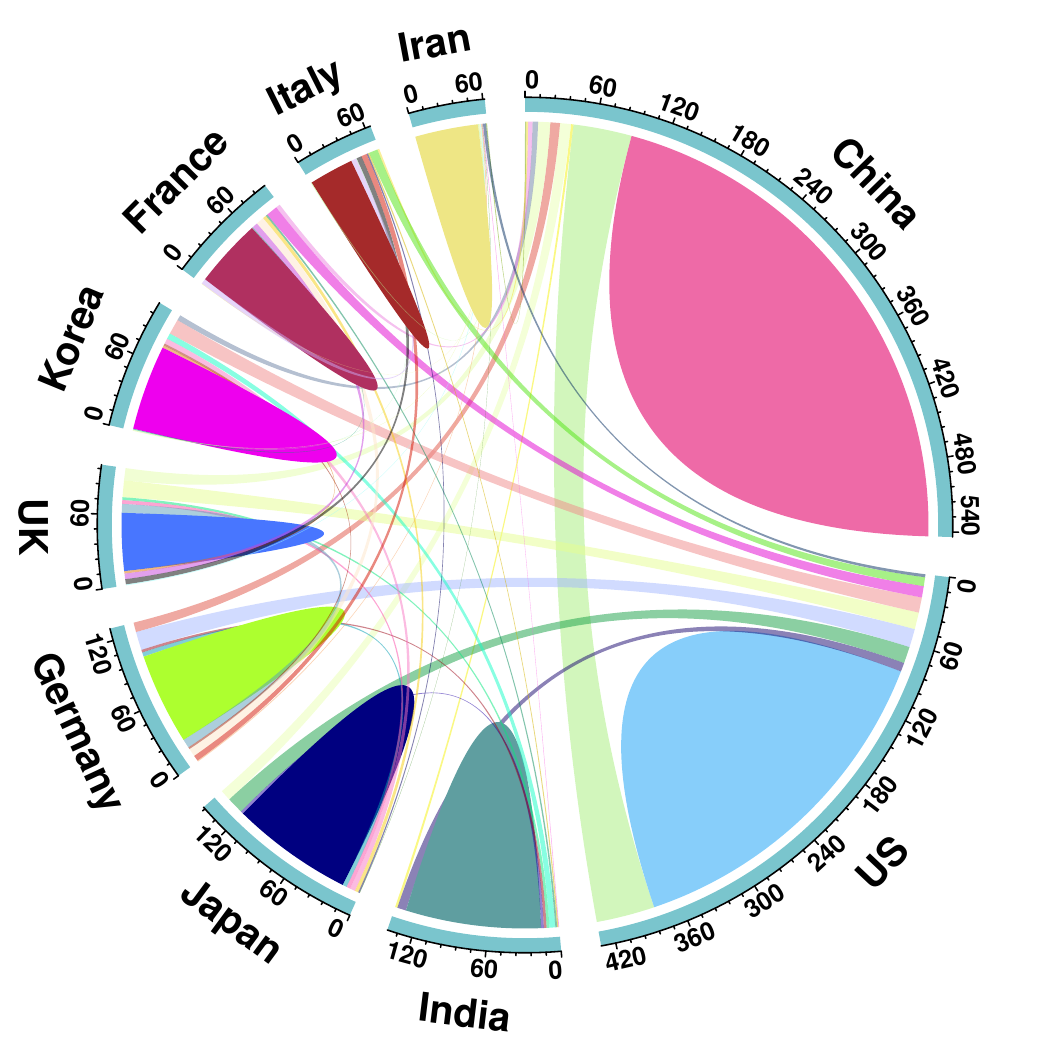}}
\end{flushleft}
    \end{minipage}
	\begin{minipage}{0.33\hsize}
\begin{flushleft}
\raisebox{-0.0cm}{\hspace{-6mm}\small\textrm{\textbf{(b)~ \agri}}}\\[6mm]
\raisebox{\height}{\includegraphics[trim=2.0cm 1.8cm 0cm 1.5cm, align=c, scale=\chordsize, vmargin=0mm]{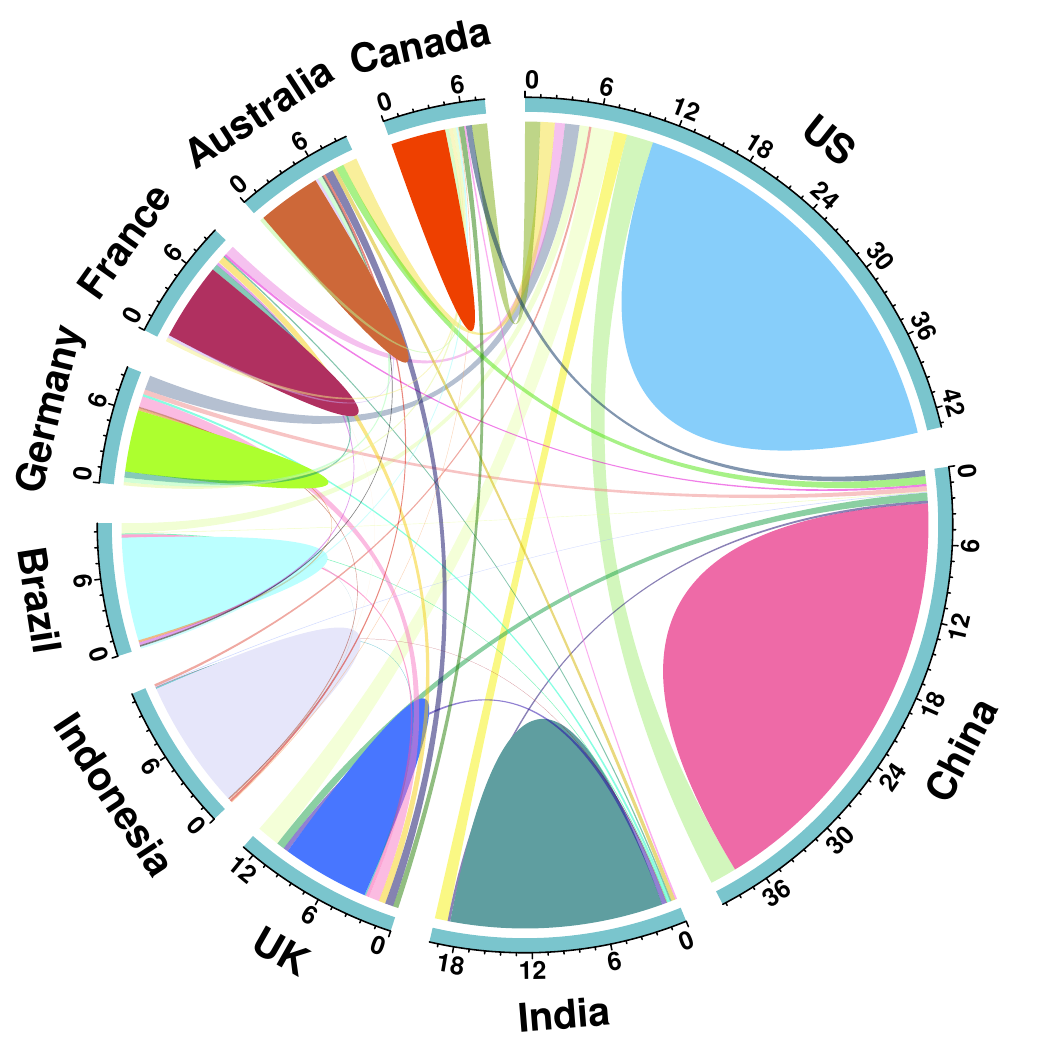}}
\end{flushleft}
	\end{minipage}
	\begin{minipage}{0.33\hsize}
\begin{flushleft}
\raisebox{-0.0cm}{\hspace{-6mm}\small\textrm{\textbf{(c)~ \particle}}}\\[6mm]
\raisebox{\height}{\includegraphics[trim=2.0cm 1.8cm 0cm 1.5cm, align=c, scale=\chordsize, vmargin=0mm]{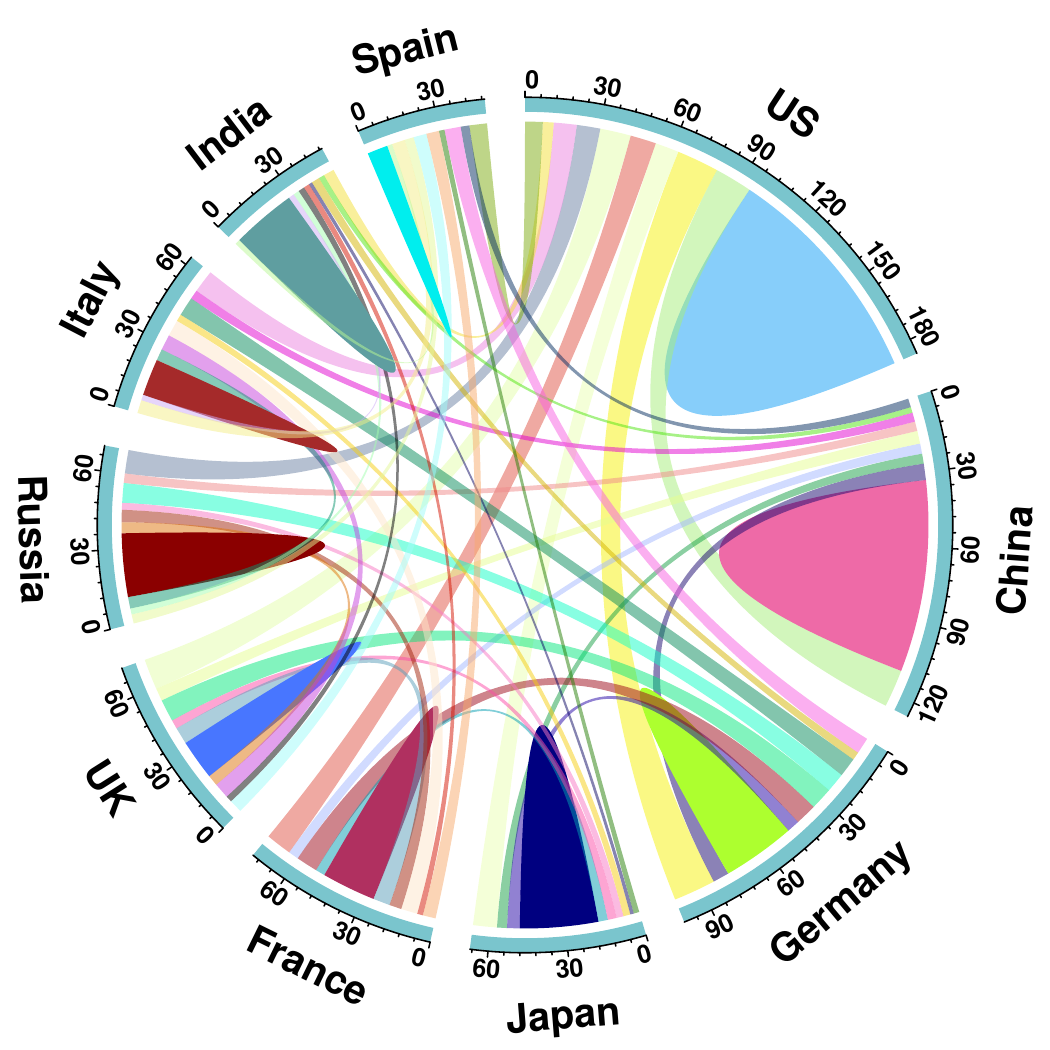}}
\end{flushleft}
	\end{minipage}
    \end{tabular}

\vspace{1mm}
\hspace{-0.5cm}{\marrow}\quad\dotfill
\vspace{4mm}

    \begin{tabular}{c}
    \begin{minipage}{0.03\hsize}
\begin{flushleft}
    \hspace{-0.7cm}\rotatebox{90}{\period{3}{2001--2010}}
\end{flushleft}
	\end{minipage}
	\begin{minipage}{0.33\hsize}
\begin{flushleft}
\raisebox{\height}{\includegraphics[trim=2.0cm 1.8cm 0cm 1.5cm, align=c, scale=\chordsize, vmargin=0mm]{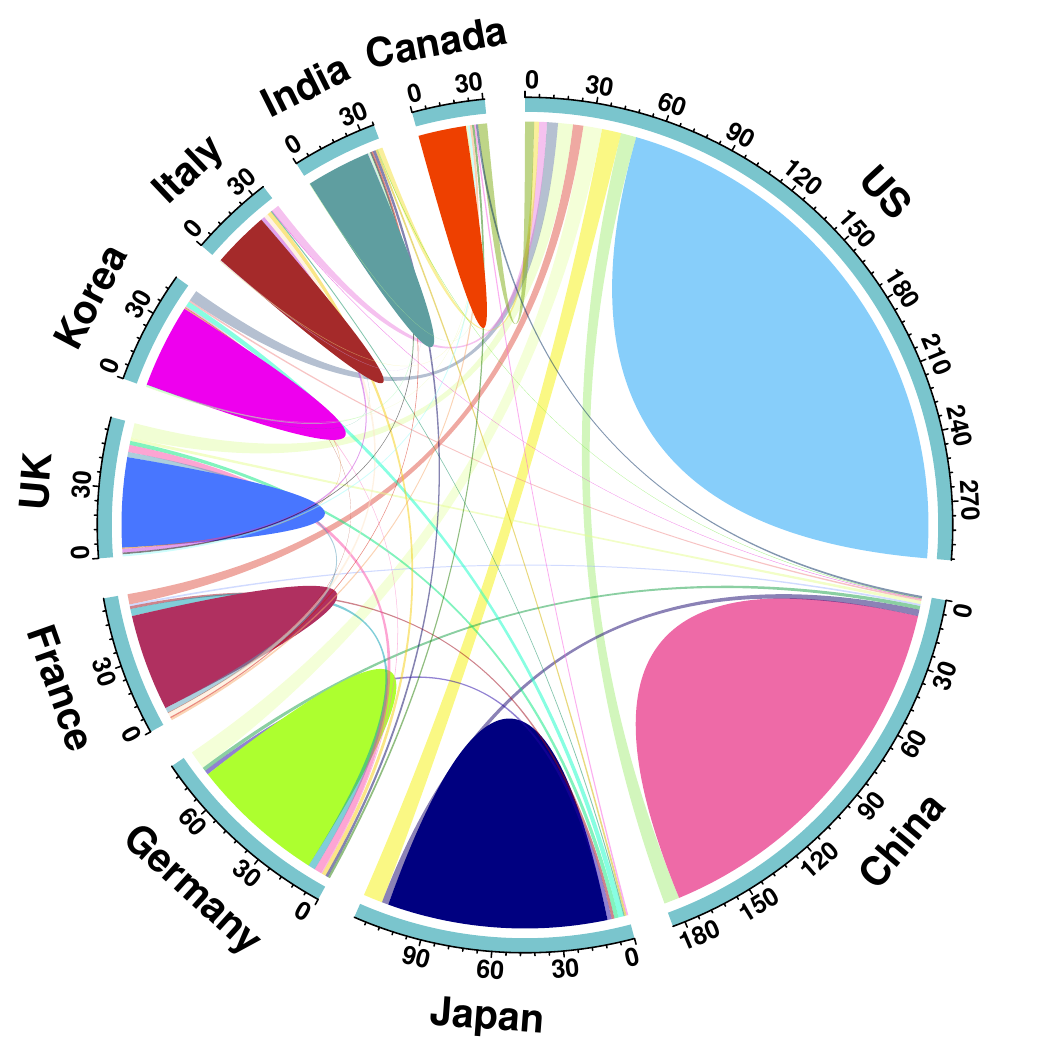}}
\end{flushleft}
    \end{minipage}
	\begin{minipage}{0.33\hsize}
\begin{flushleft}
\raisebox{\height}{\includegraphics[trim=2.0cm 1.8cm 0cm 1.5cm, align=c, scale=\chordsize, vmargin=0mm]{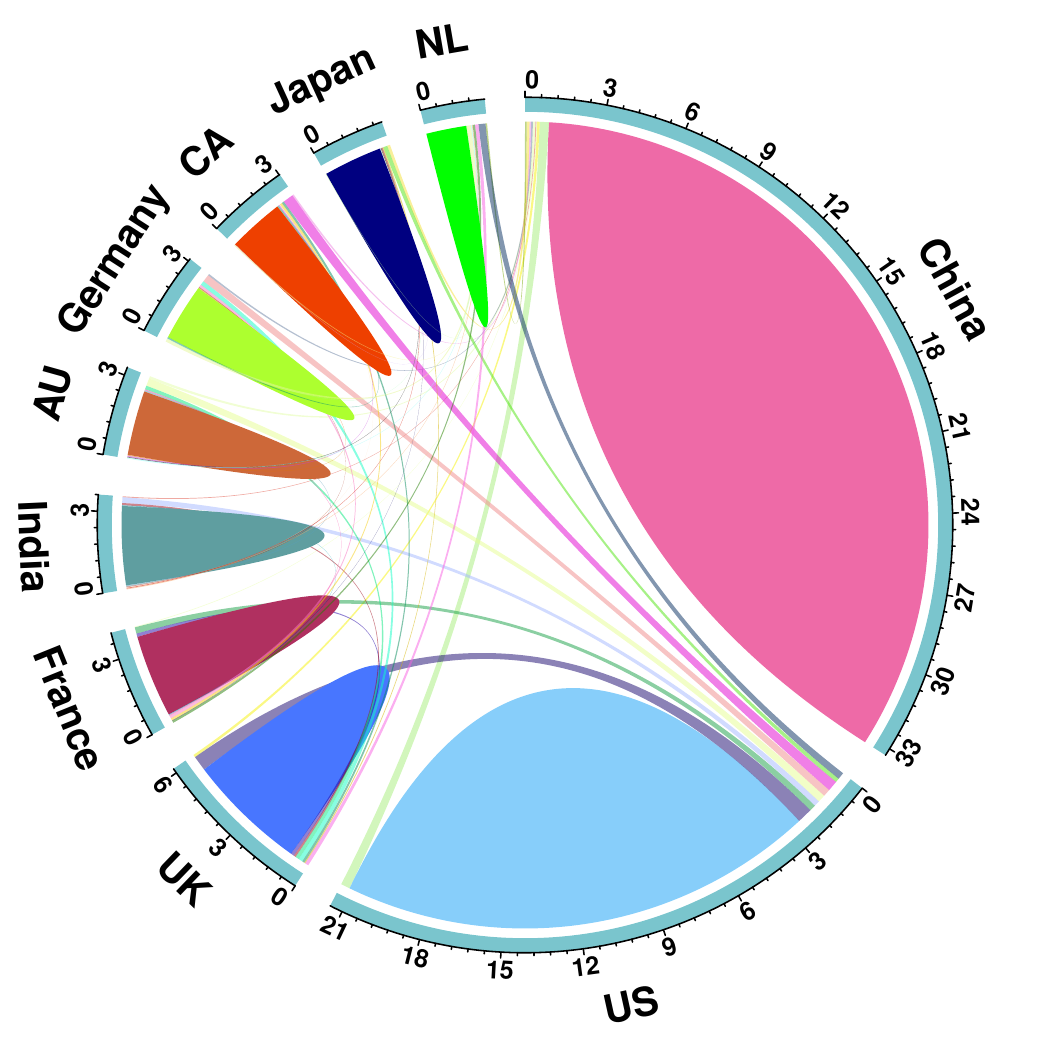}}
\end{flushleft}
	\end{minipage}
	\begin{minipage}{0.33\hsize}
\begin{flushleft}
\raisebox{\height}{\includegraphics[trim=2.0cm 1.8cm 0cm 1.5cm, align=c, scale=\chordsize, vmargin=0mm]{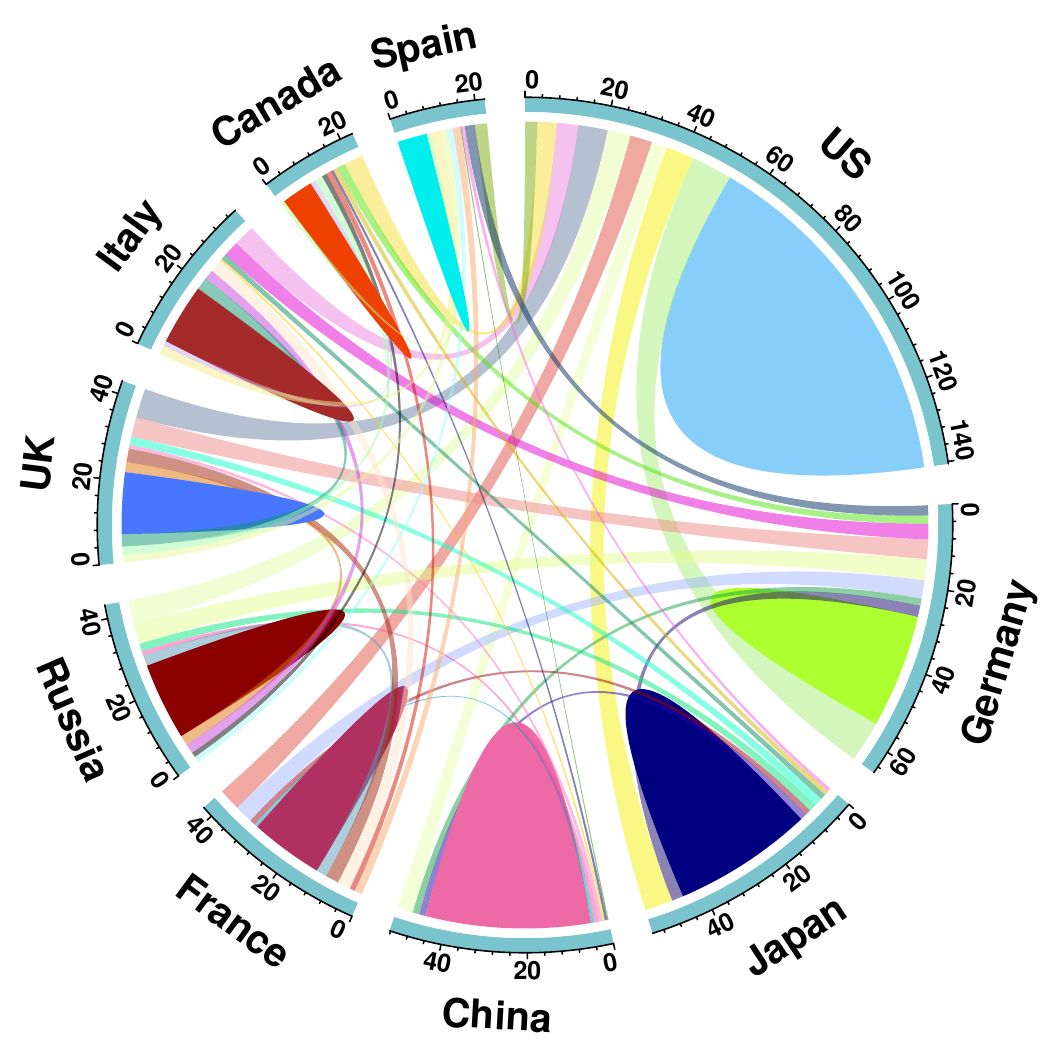}}
\end{flushleft}
	\end{minipage}
    \end{tabular}

\vspace{1mm}
\hspace{-0.5cm}{\marrow}\quad\dotfill
\vspace{4mm}

    \begin{tabular}{c}
    \begin{minipage}{0.03\hsize}
\begin{flushleft}
    \hspace{-0.7cm}\rotatebox{90}{\period{2}{1991--2000}}
\end{flushleft}
	\end{minipage}
	\begin{minipage}{0.33\hsize}
\begin{flushleft}
\raisebox{\height}{\includegraphics[trim=2.0cm 1.8cm 0cm 1.5cm, align=c, scale=\chordsize, vmargin=0mm]{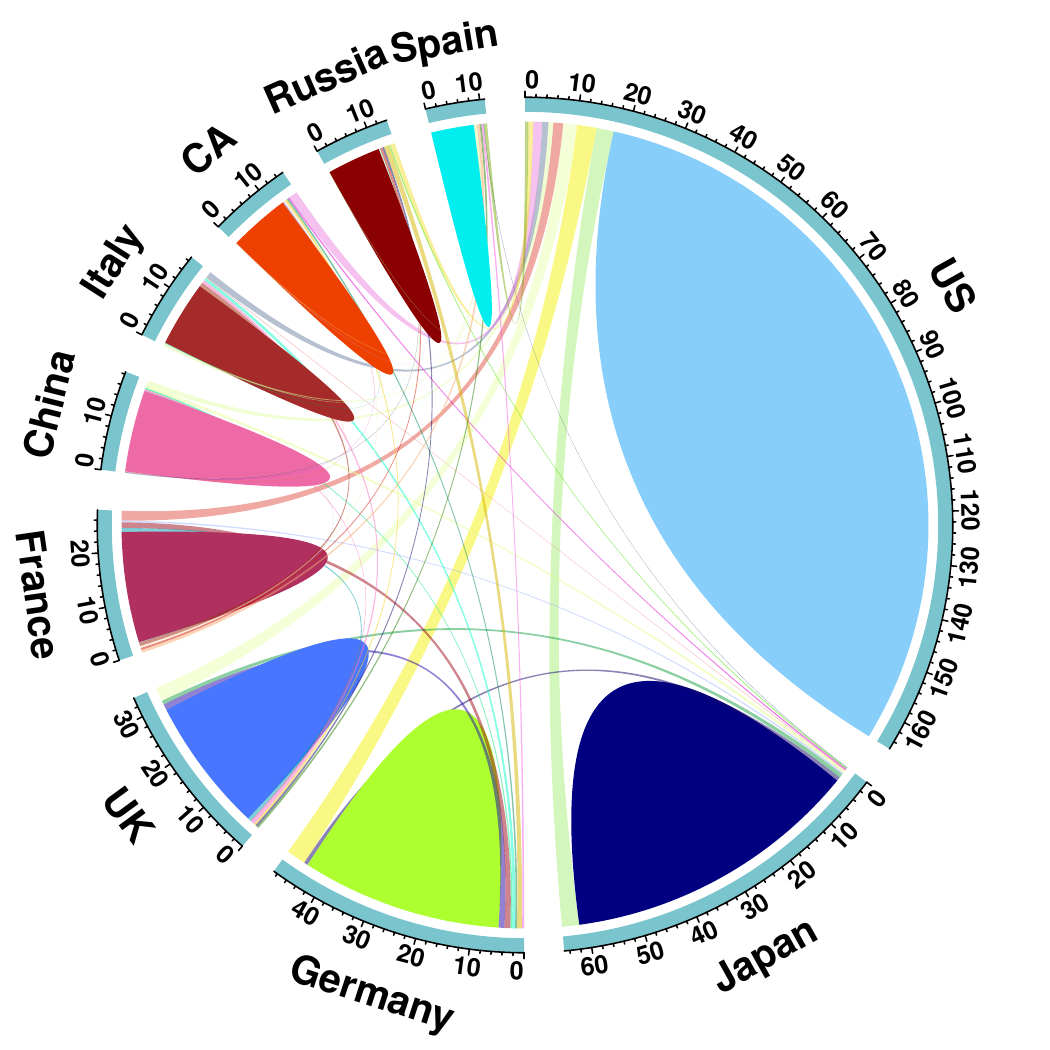}}
\end{flushleft}
    \end{minipage}
	\begin{minipage}{0.33\hsize}
\begin{flushleft}
\raisebox{\height}{\includegraphics[trim=2.0cm 1.8cm 0cm 1.5cm, align=c, scale=\chordsize, vmargin=0mm]{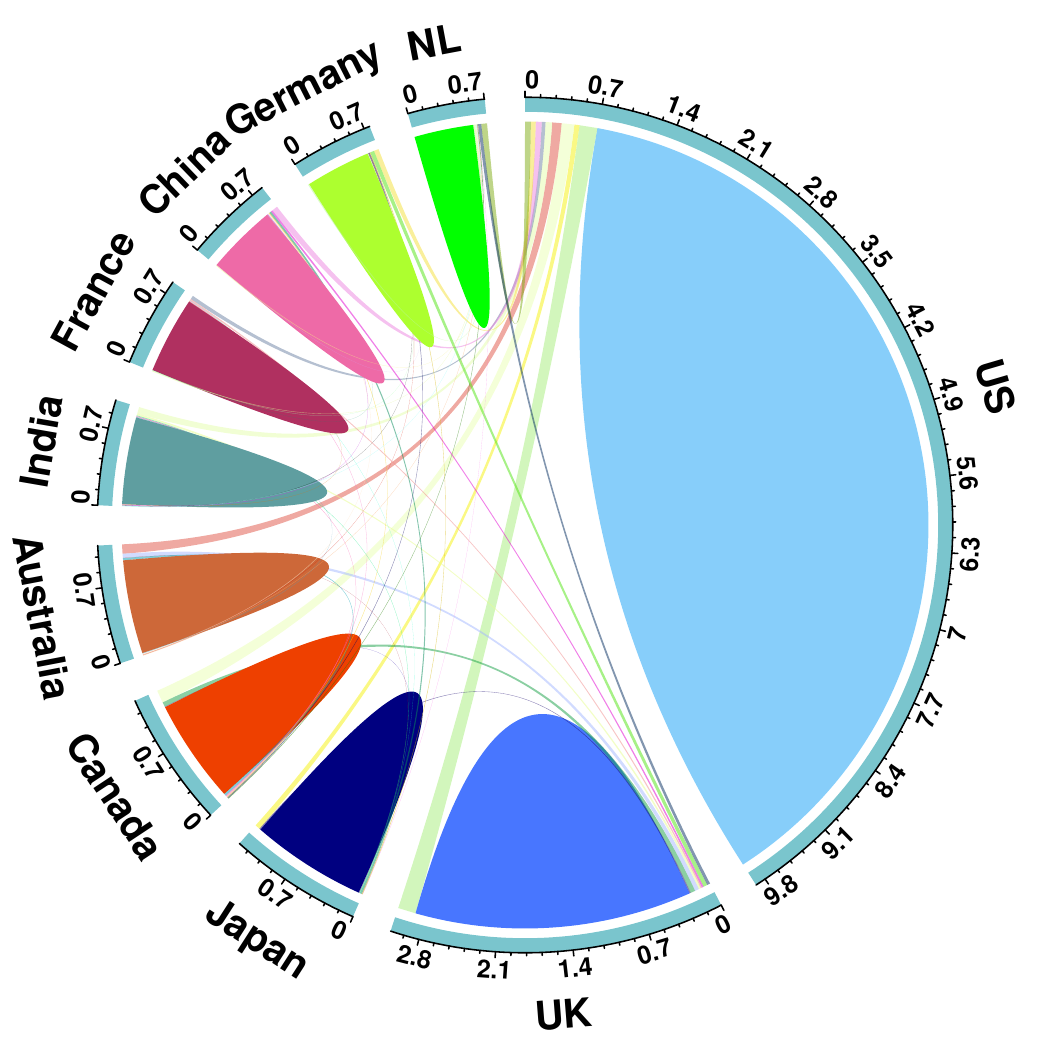}}
\end{flushleft}
	\end{minipage}
	\begin{minipage}{0.33\hsize}
\begin{flushleft}
\raisebox{\height}{\includegraphics[trim=2.0cm 1.8cm 0cm 1.5cm, align=c, scale=\chordsize, vmargin=0mm]{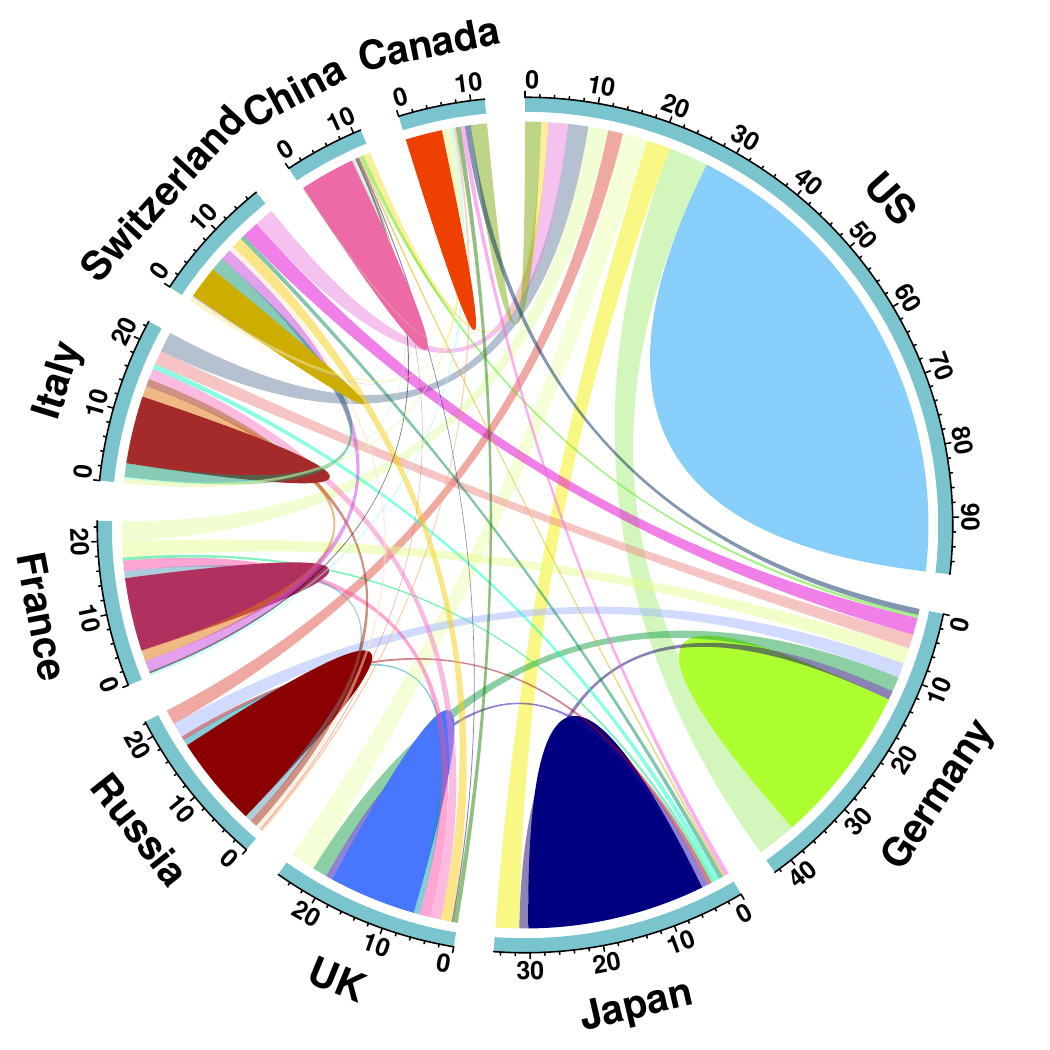}}
\end{flushleft}
	\end{minipage}
    \end{tabular}

\vspace{1mm}
\hspace{-0.5cm}{\marrow}\quad\dotfill
\vspace{4mm}

    \begin{tabular}{c}
    \begin{minipage}{0.03\hsize}
\begin{flushleft}
    \hspace{-0.7cm}\rotatebox{90}{\period{1}{1971--1990}}
\end{flushleft}
	\end{minipage}
	\begin{minipage}{0.33\hsize}
\begin{flushleft}
\raisebox{\height}{\includegraphics[trim=2.0cm 1.8cm 0cm 1.5cm, align=c, scale=\chordsize, vmargin=0mm]{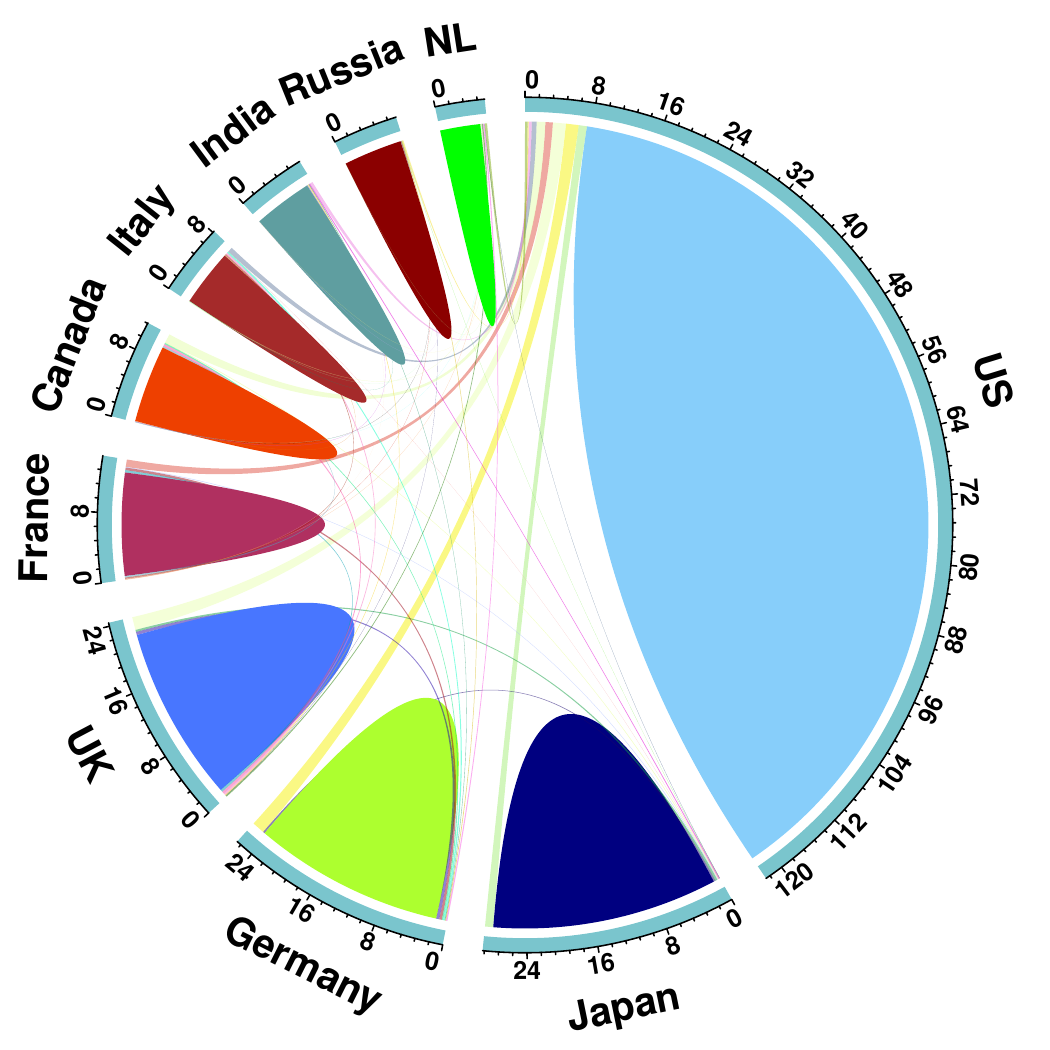}}
\end{flushleft}
    \end{minipage}
	\begin{minipage}{0.33\hsize}
\begin{flushleft}
\raisebox{\height}{\includegraphics[trim=2.0cm 1.8cm 0cm 1.5cm, align=c, scale=\chordsize, vmargin=0mm]{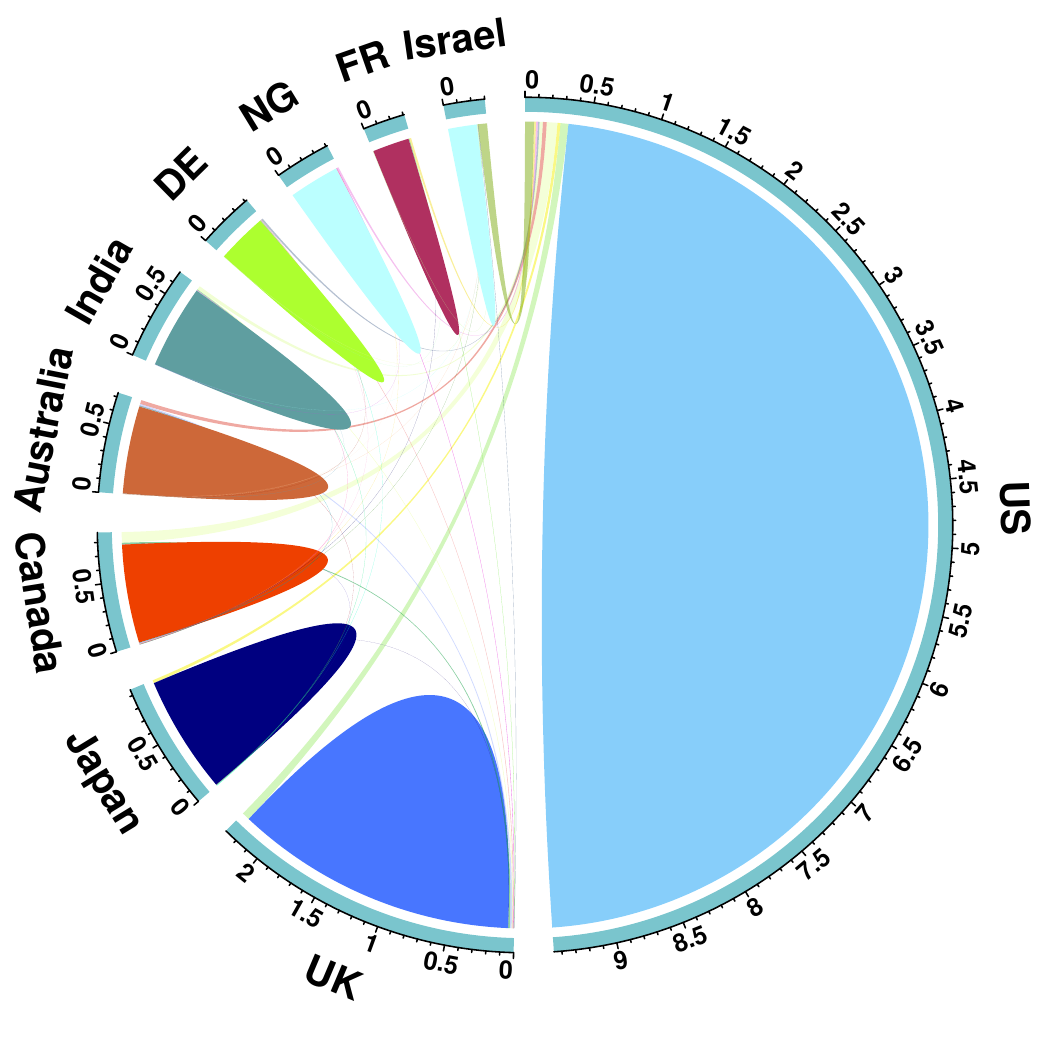}}
\end{flushleft}
	\end{minipage}
	\begin{minipage}{0.33\hsize}
\begin{flushleft}
\raisebox{\height}{\includegraphics[trim=2.0cm 1.8cm 0cm 1.5cm, align=c, scale=\chordsize, vmargin=0mm]{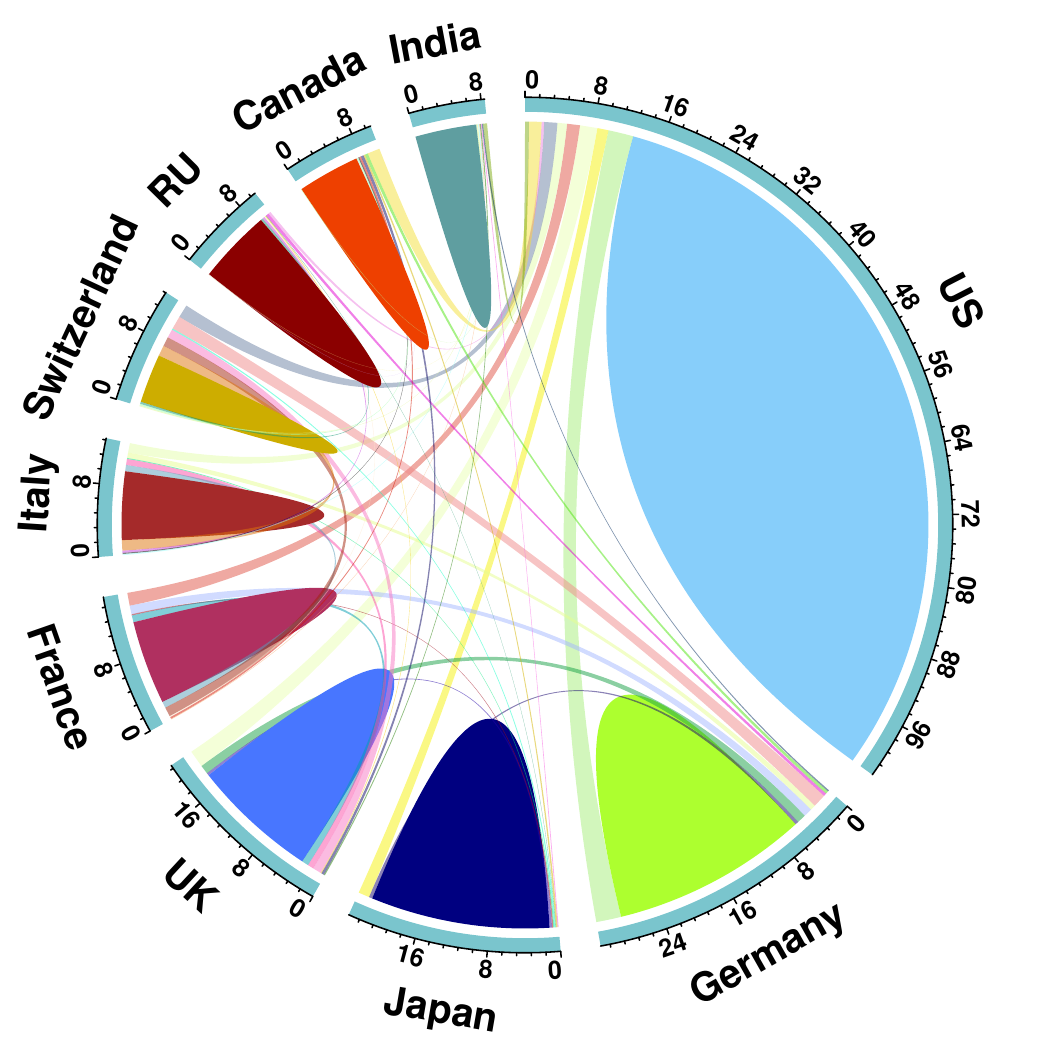}}
\end{flushleft}
	\end{minipage}
    \end{tabular}
\end{subfigure}
\vspace{2.5mm}
\caption{\textbf{Changes in bilateral relationships over time.}
The number of works is displayed in thousands.}
\label{fig:chord_2}
\end{figure}
}
\afterpage{\clearpage%
\begin{figure}[htp]\ContinuedFloat
\centering
\begin{subfigure}{1.0\textwidth}
\vspace{-0.5cm}
    \begin{tabular}{c}
    \begin{minipage}{0.03\hsize}
\begin{flushleft}
    \hspace{-0.7cm}\rotatebox{90}{\period{4}{2011--2020}}
\end{flushleft}
	\end{minipage}
	\begin{minipage}{0.33\hsize}
\begin{flushleft}
\raisebox{-0.0cm}{\hspace{-6mm}\small\textrm{\textbf{(d)~ \aerospace}}}\\[6mm]
\raisebox{\height}{\includegraphics[trim=2.0cm 1.8cm 0cm 1.5cm, align=c, scale=\chordsize, vmargin=0mm]{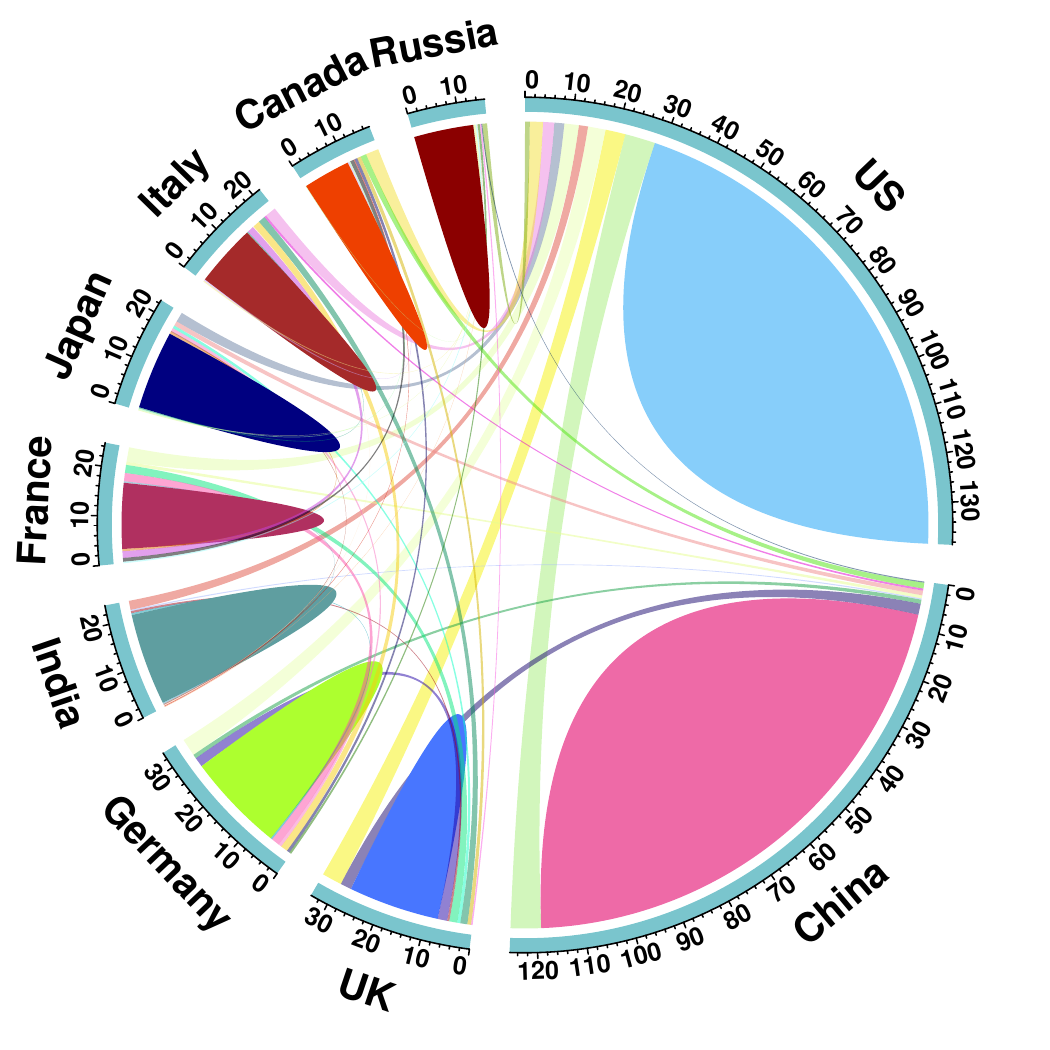}}
\end{flushleft}
    \end{minipage}
	\begin{minipage}{0.33\hsize}
\begin{flushleft}
\raisebox{-0.0cm}{\hspace{-6mm}\small\textrm{\textbf{(e)~ \nuclear}}}\\[6mm]
\raisebox{\height}{\includegraphics[trim=2.0cm 1.8cm 0cm 1.5cm, align=c, scale=\chordsize, vmargin=0mm]{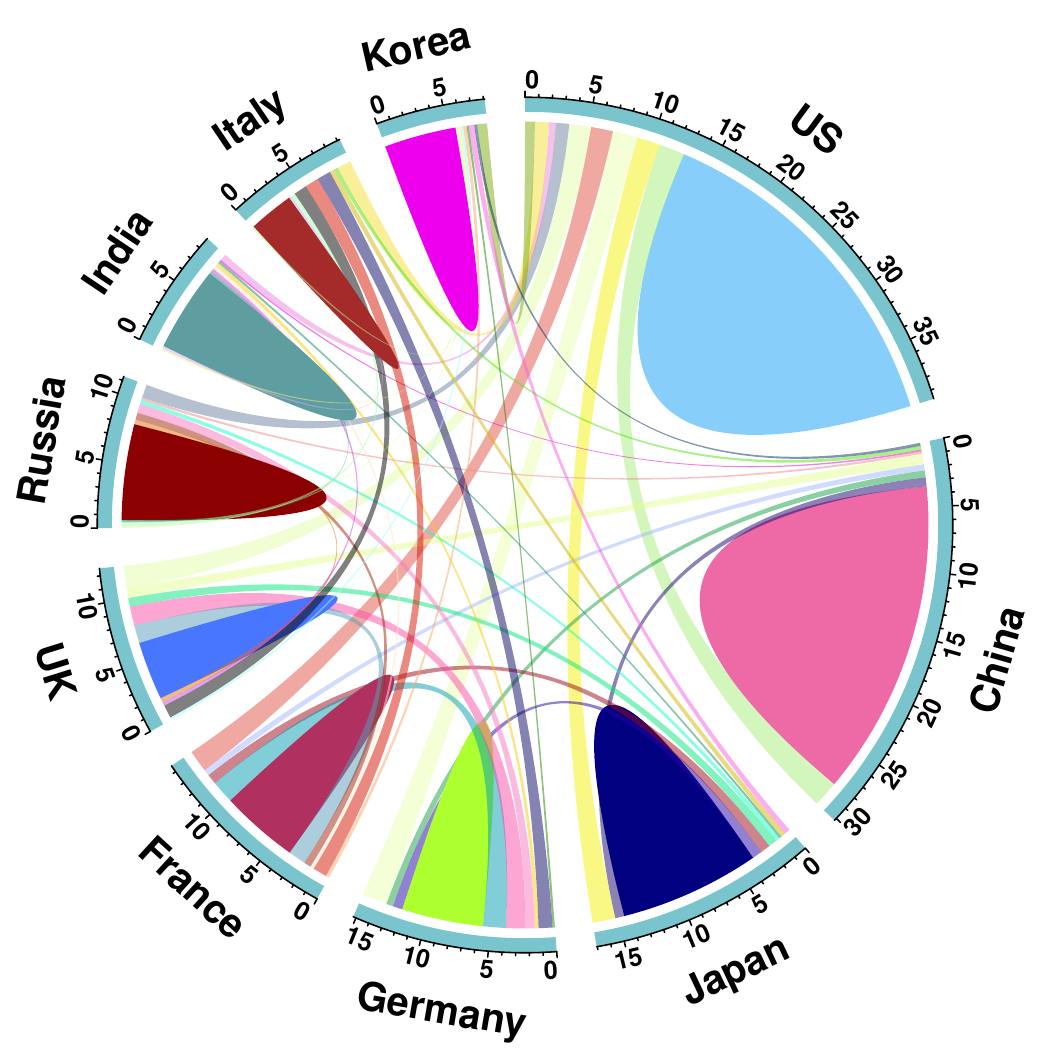}}
\end{flushleft}
	\end{minipage}
	\begin{minipage}{0.33\hsize}
\begin{flushleft}
\raisebox{-0.0cm}{\hspace{-6mm}\small\textrm{\textbf{(f)~ \marine}}}\\[6mm]
\raisebox{\height}{\includegraphics[trim=2.0cm 1.8cm 0cm 1.5cm, align=c, scale=\chordsize, vmargin=0mm]{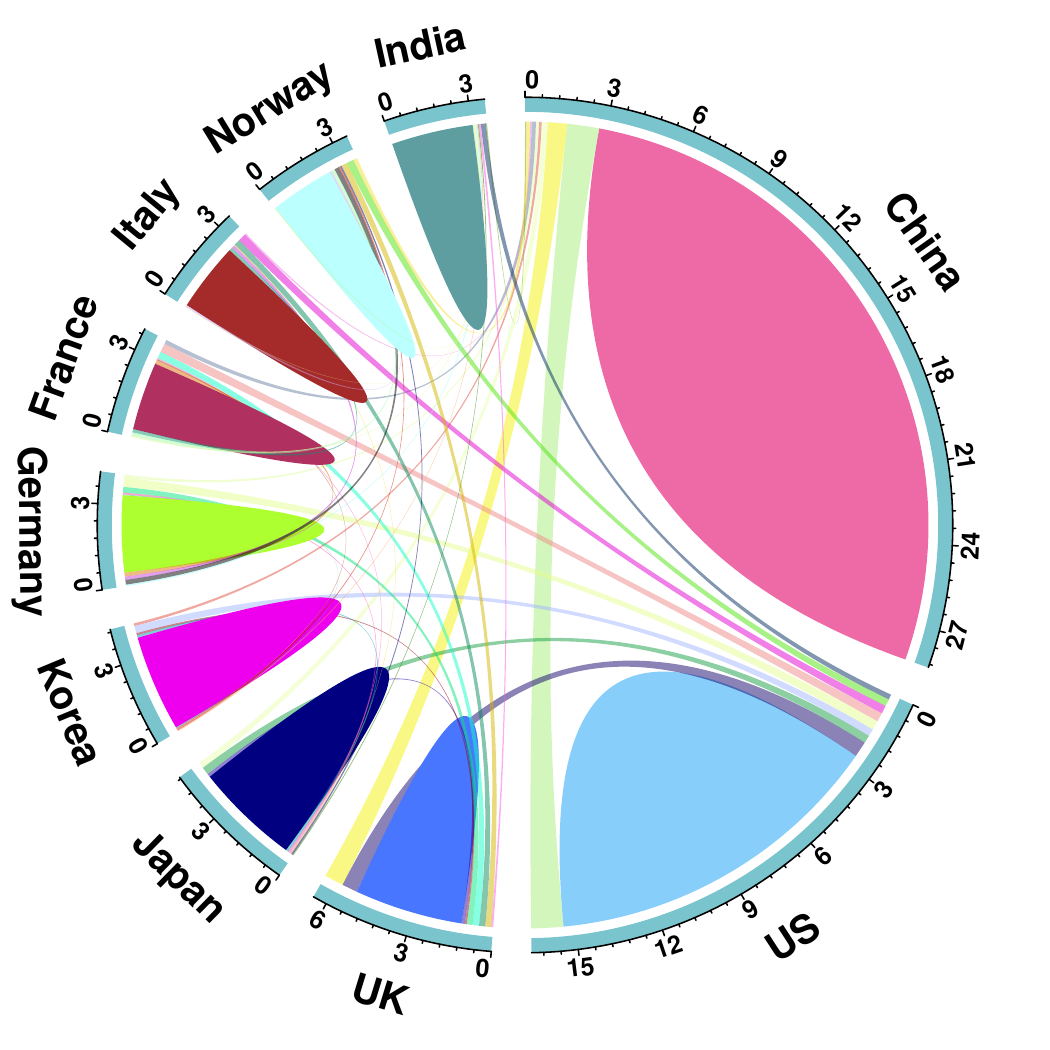}}
\end{flushleft}
	\end{minipage}
    \end{tabular}

\vspace{1mm}
\hspace{-0.5cm}{\marrow}\quad\dotfill
\vspace{4mm}

    \begin{tabular}{c}
    \begin{minipage}{0.03\hsize}
\begin{flushleft}
    \hspace{-0.7cm}\rotatebox{90}{\period{3}{2001--2010}}
\end{flushleft}
	\end{minipage}
	\begin{minipage}{0.33\hsize}
\begin{flushleft}
\raisebox{\height}{\includegraphics[trim=2.0cm 1.8cm 0cm 1.5cm, align=c, scale=\chordsize, vmargin=0mm]{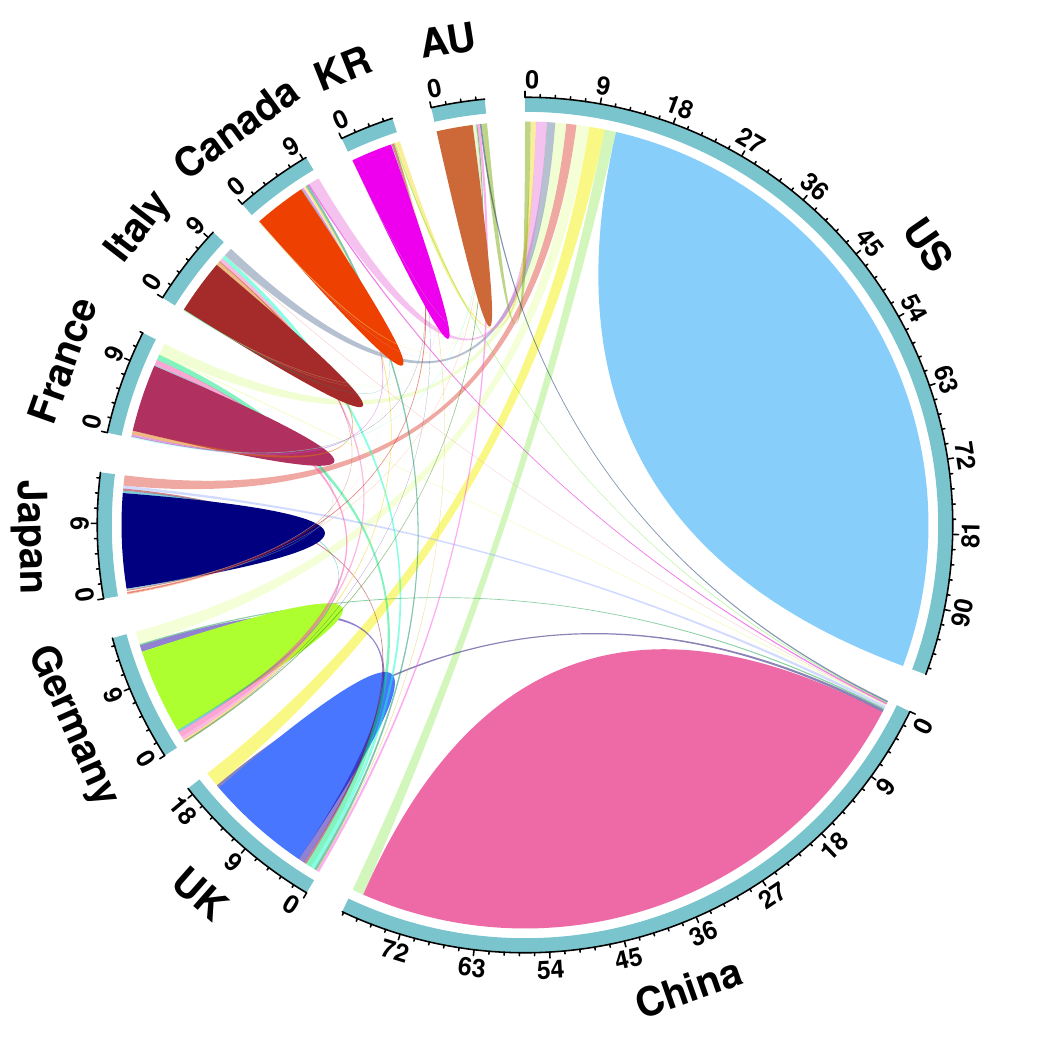}}
\end{flushleft}
    \end{minipage}
	\begin{minipage}{0.33\hsize}
\begin{flushleft}
\raisebox{\height}{\includegraphics[trim=2.0cm 1.8cm 0cm 1.5cm, align=c, scale=\chordsize, vmargin=0mm]{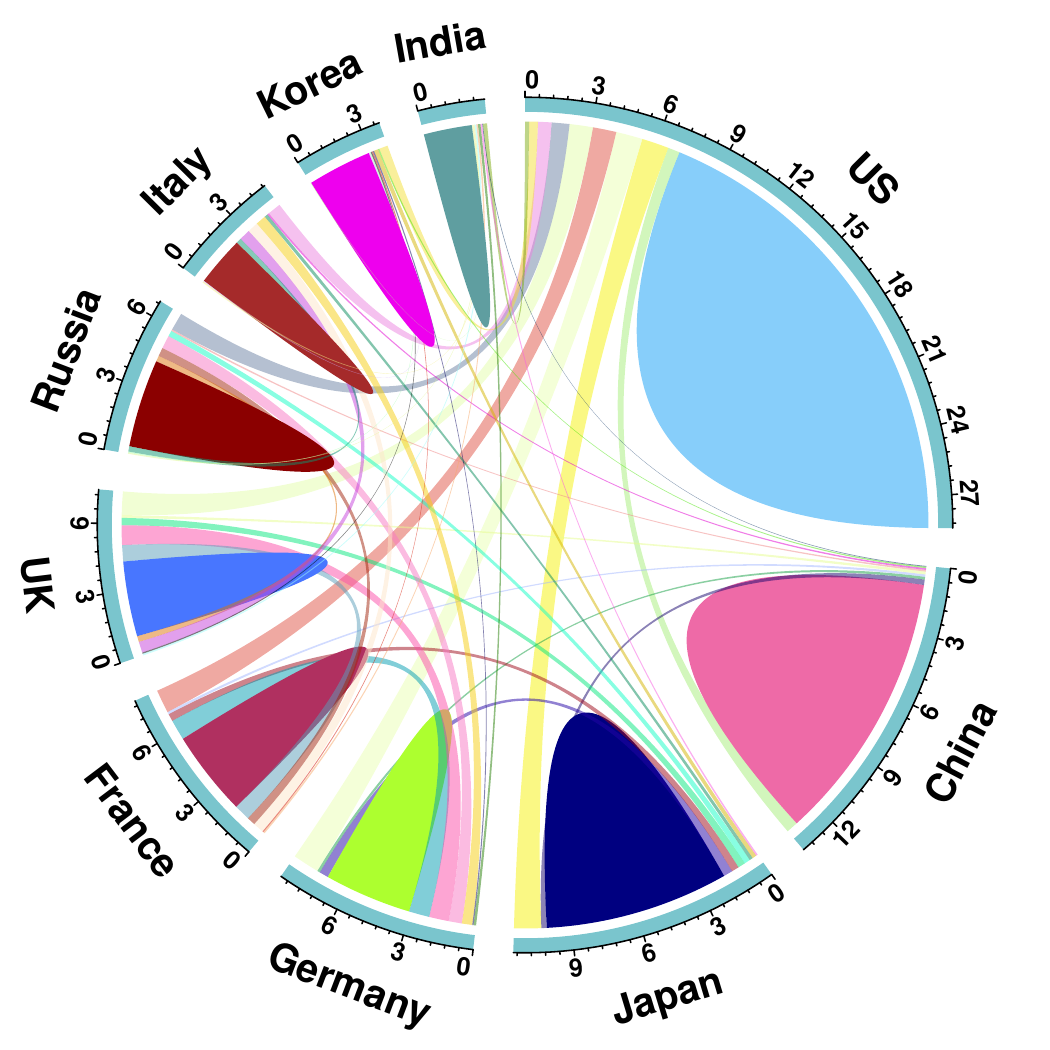}}
\end{flushleft}
	\end{minipage}
	\begin{minipage}{0.33\hsize}
\begin{flushleft}
\raisebox{\height}{\includegraphics[trim=2.0cm 1.8cm 0cm 1.5cm, align=c, scale=\chordsize, vmargin=0mm]{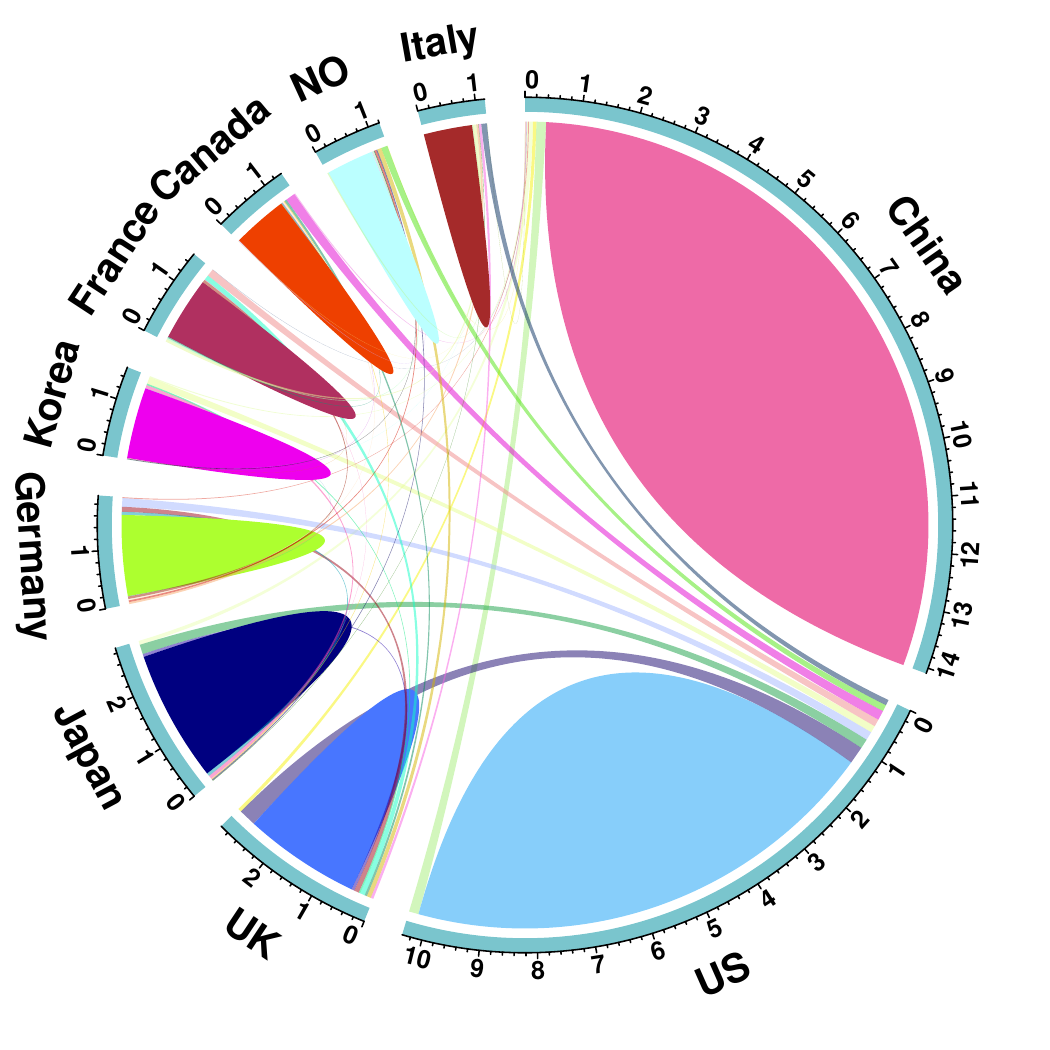}}
\end{flushleft}
	\end{minipage}
    \end{tabular}

\vspace{1mm}
\hspace{-0.5cm}{\marrow}\quad\dotfill
\vspace{4mm}

    \begin{tabular}{c}
    \begin{minipage}{0.03\hsize}
\begin{flushleft}
    \hspace{-0.7cm}\rotatebox{90}{\period{2}{1991--2000}}
\end{flushleft}
	\end{minipage}
	\begin{minipage}{0.33\hsize}
\begin{flushleft}
\raisebox{\height}{\includegraphics[trim=2.0cm 1.8cm 0cm 1.5cm, align=c, scale=\chordsize, vmargin=0mm]{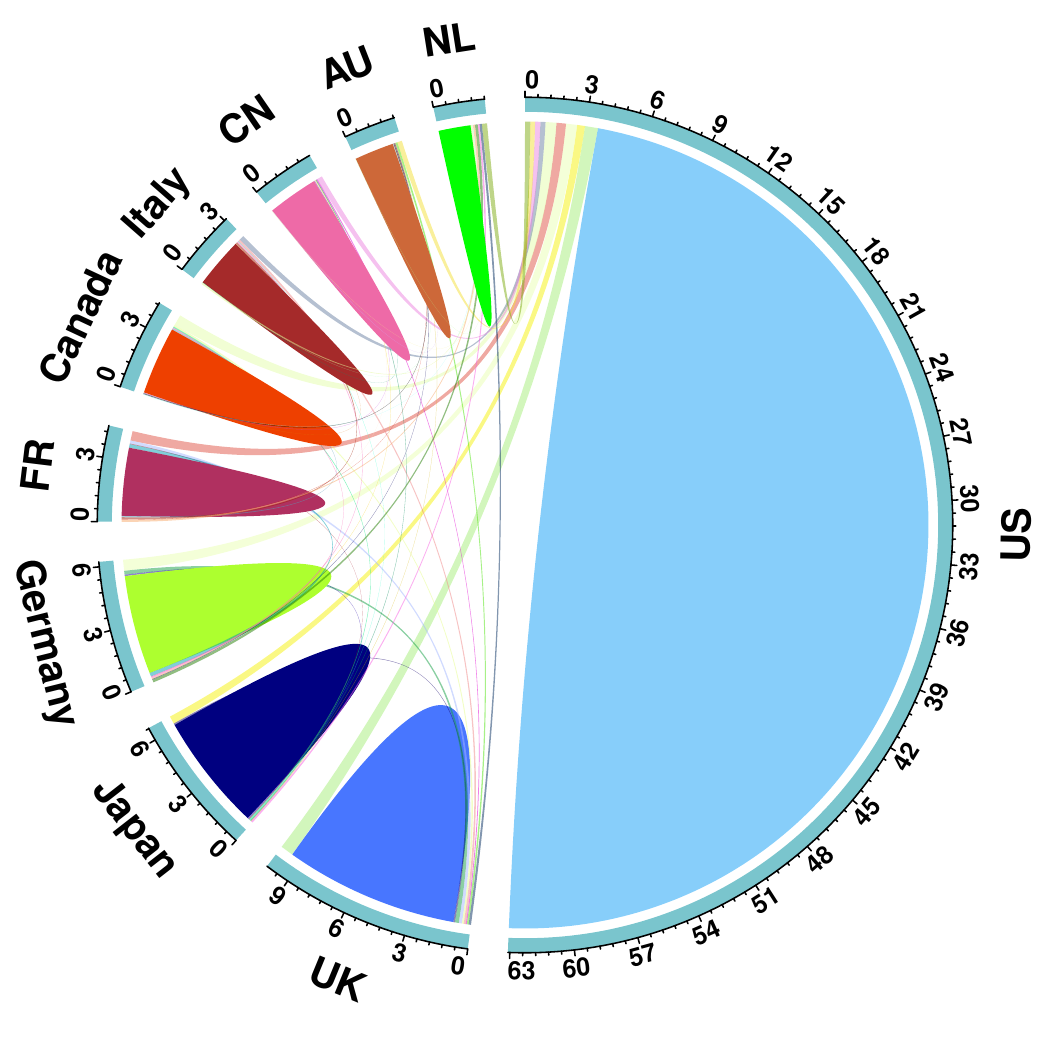}}
\end{flushleft}
    \end{minipage}
	\begin{minipage}{0.33\hsize}
\begin{flushleft}
\raisebox{\height}{\includegraphics[trim=2.0cm 1.8cm 0cm 1.5cm, align=c, scale=\chordsize, vmargin=0mm]{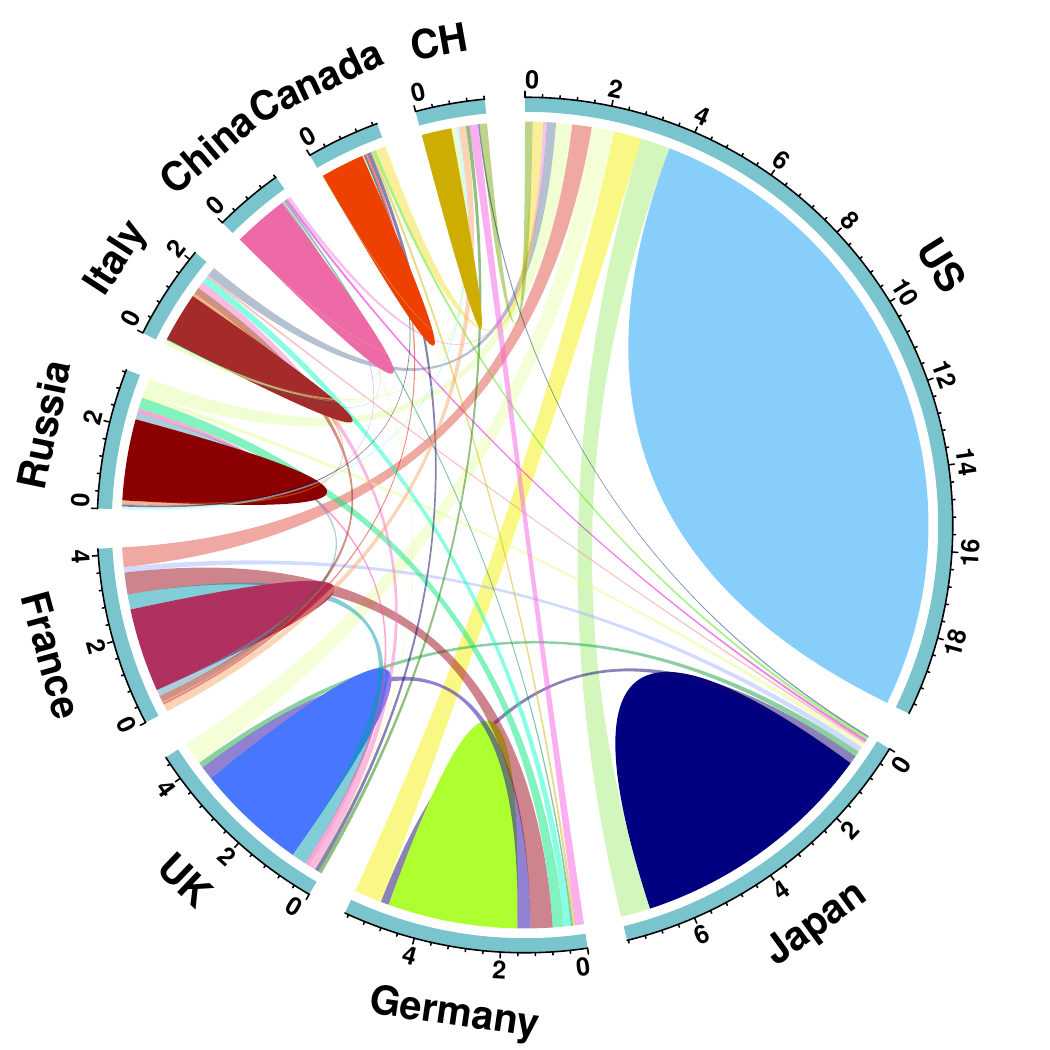}}
\end{flushleft}
	\end{minipage}
	\begin{minipage}{0.33\hsize}
\begin{flushleft}
\raisebox{\height}{\includegraphics[trim=2.0cm 1.8cm 0cm 1.5cm, align=c, scale=\chordsize, vmargin=0mm]{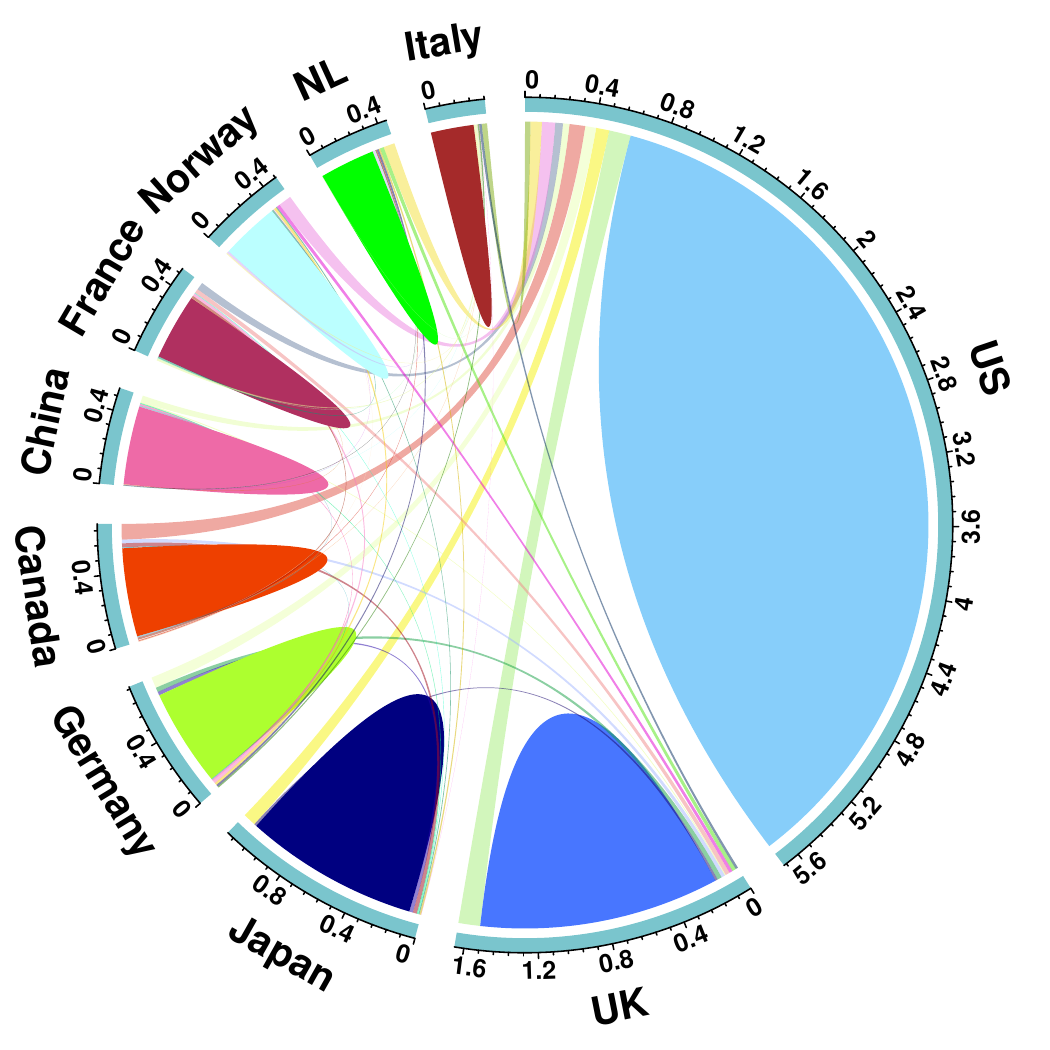}}
\end{flushleft}
	\end{minipage}
    \end{tabular}

\vspace{1mm}
\hspace{-0.5cm}{\marrow}\quad\dotfill
\vspace{4mm}

    \begin{tabular}{c}
    \begin{minipage}{0.03\hsize}
\begin{flushleft}
    \hspace{-0.7cm}\rotatebox{90}{\period{1}{1971--1990}}
\end{flushleft}
	\end{minipage}
	\begin{minipage}{0.33\hsize}
\begin{flushleft}
\raisebox{\height}{\includegraphics[trim=2.0cm 1.8cm 0cm 1.5cm, align=c, scale=\chordsize, vmargin=0mm]{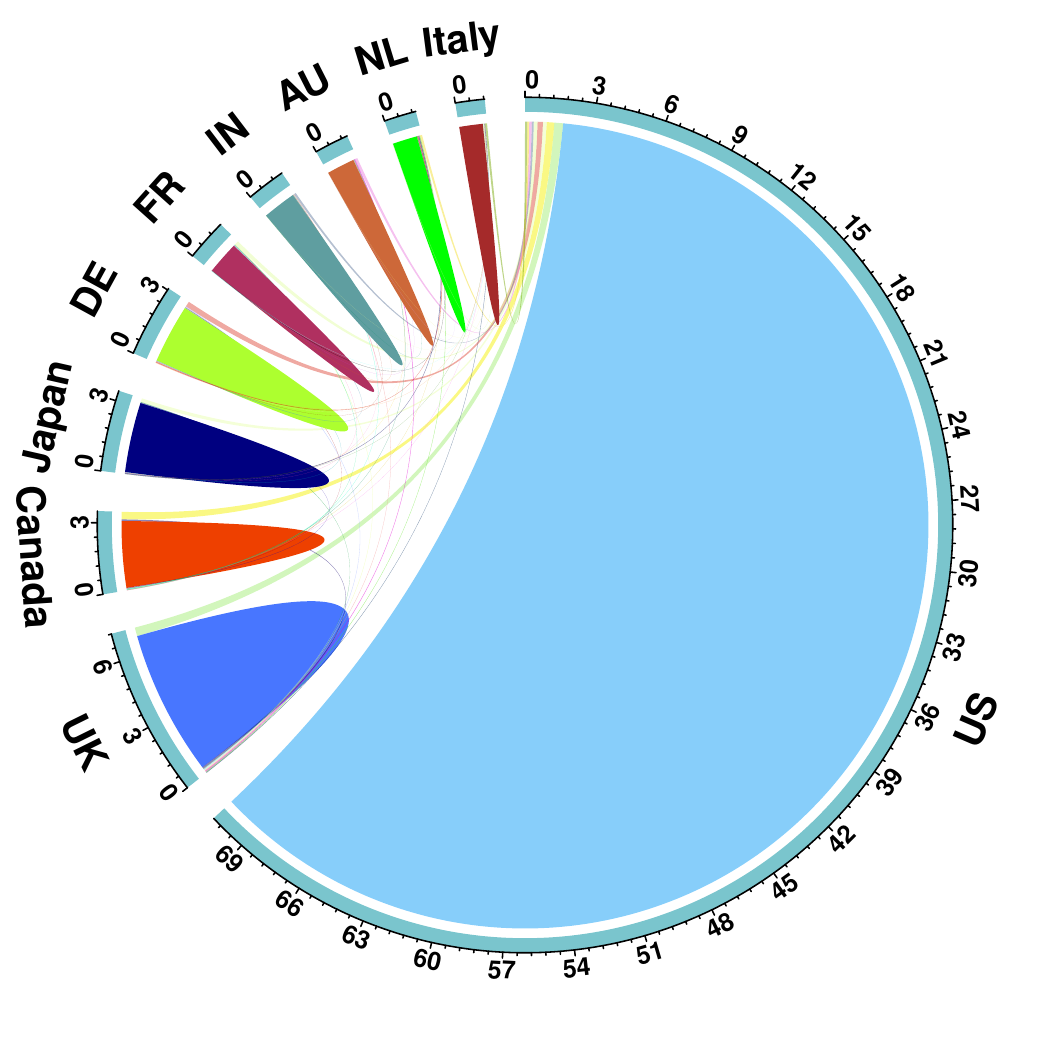}}
\end{flushleft}
    \end{minipage}
	\begin{minipage}{0.33\hsize}
\begin{flushleft}
\raisebox{\height}{\includegraphics[trim=2.0cm 1.8cm 0cm 1.5cm, align=c, scale=\chordsize, vmargin=0mm]{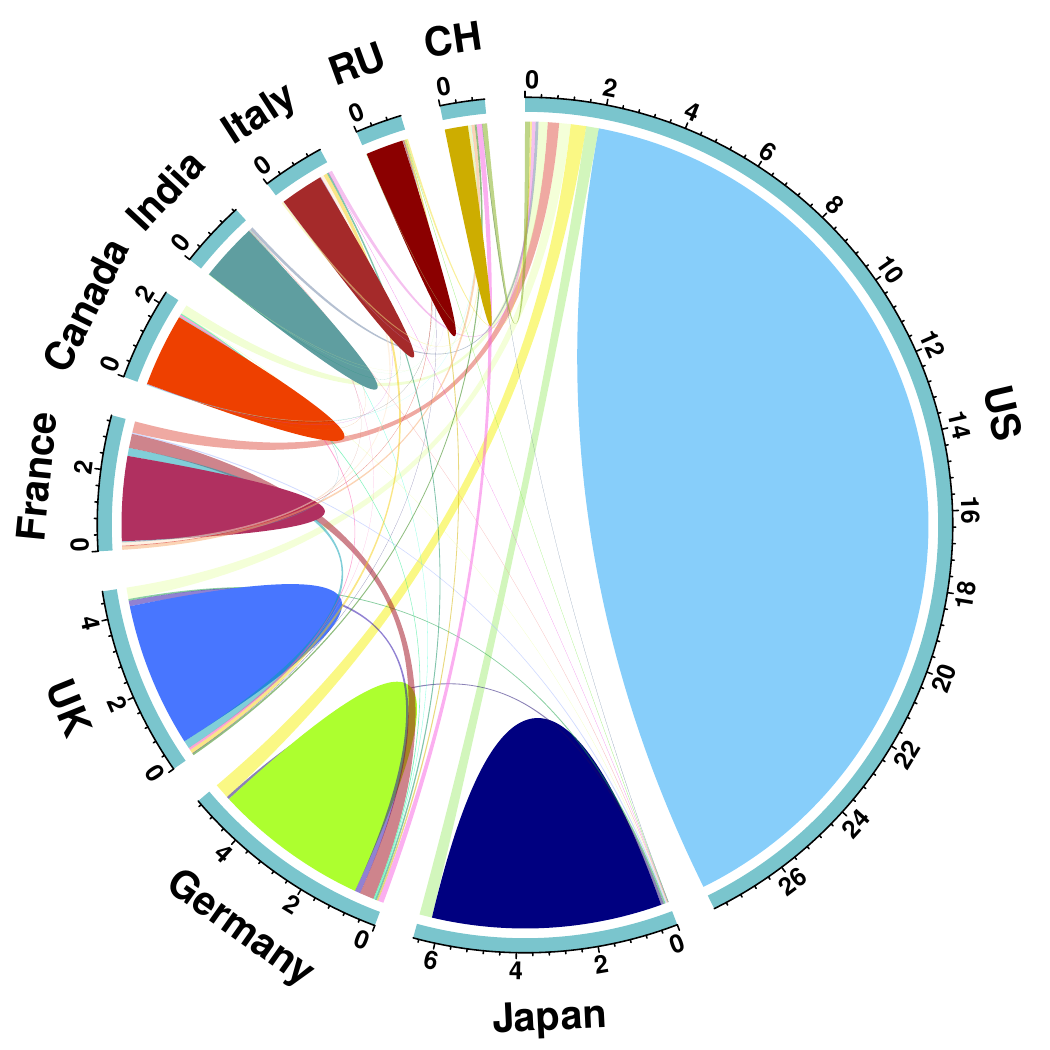}}
\end{flushleft}
	\end{minipage}
	\begin{minipage}{0.33\hsize}
\begin{flushleft}
\raisebox{\height}{\includegraphics[trim=2.0cm 1.8cm 0cm 1.5cm, align=c, scale=\chordsize, vmargin=0mm]{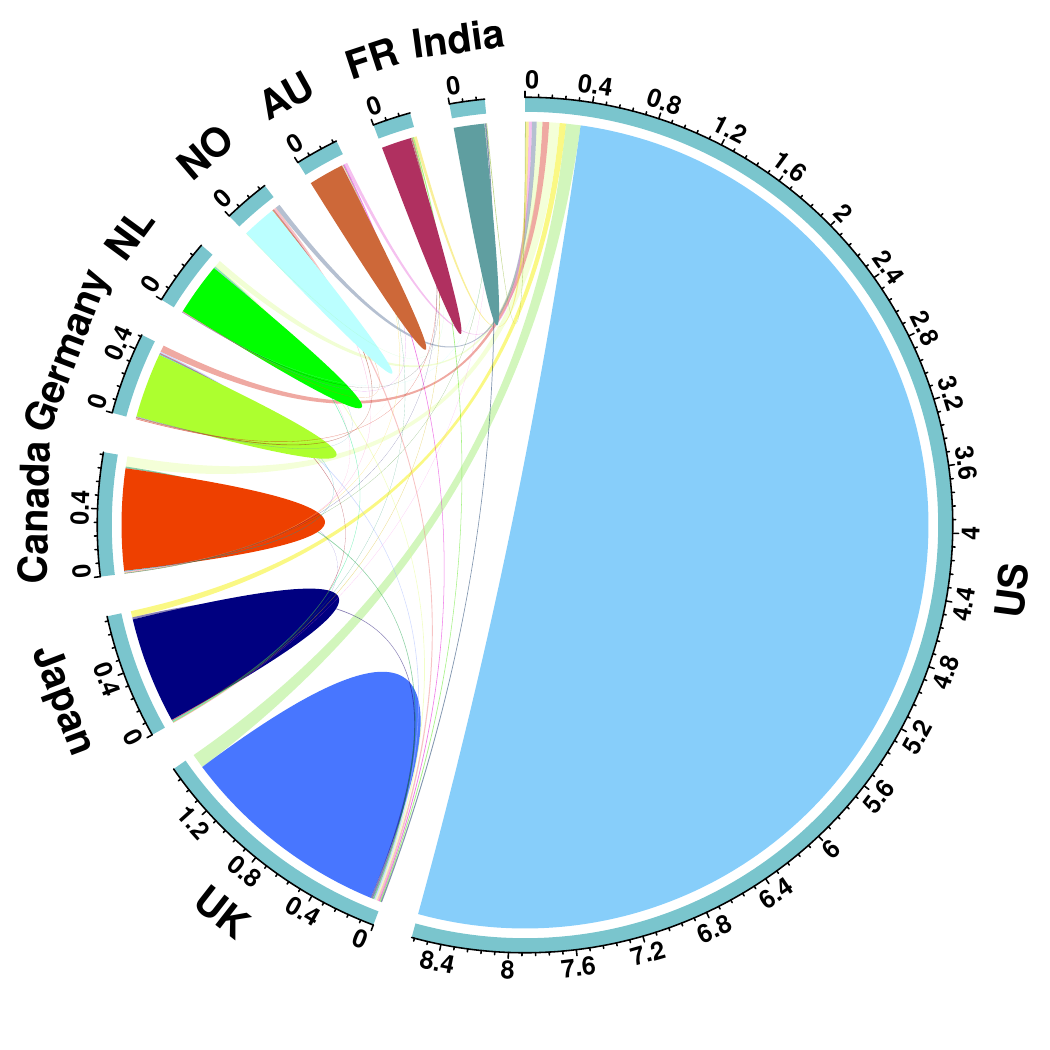}}
\end{flushleft}
	\end{minipage}
    \end{tabular}
\end{subfigure}
\vspace{2.5mm}
\caption{\textbf{Changes in bilateral relationships over time. \emph{(Cont.)}}
The number of works is displayed in thousands.}
\label{fig:chord_3}
\end{figure}
}
\afterpage{\clearpage%
\begin{figure}[htp]\ContinuedFloat
\centering
\begin{subfigure}{1.0\textwidth}
\vspace{-0.5cm}
    \begin{tabular}{c}
    \begin{minipage}{0.03\hsize}
\begin{flushleft}
    \hspace{-0.7cm}\rotatebox{90}{\period{4}{2011--2020}}
\end{flushleft}
	\end{minipage}
	\begin{minipage}{0.33\hsize}
\begin{flushleft}
\raisebox{-0.0cm}{\hspace{-6mm}\small\textrm{\textbf{(g)~ \neuro}}}\\[6mm]
\raisebox{\height}{\includegraphics[trim=2.0cm 1.8cm 0cm 1.5cm, align=c, scale=\chordsize, vmargin=0mm]{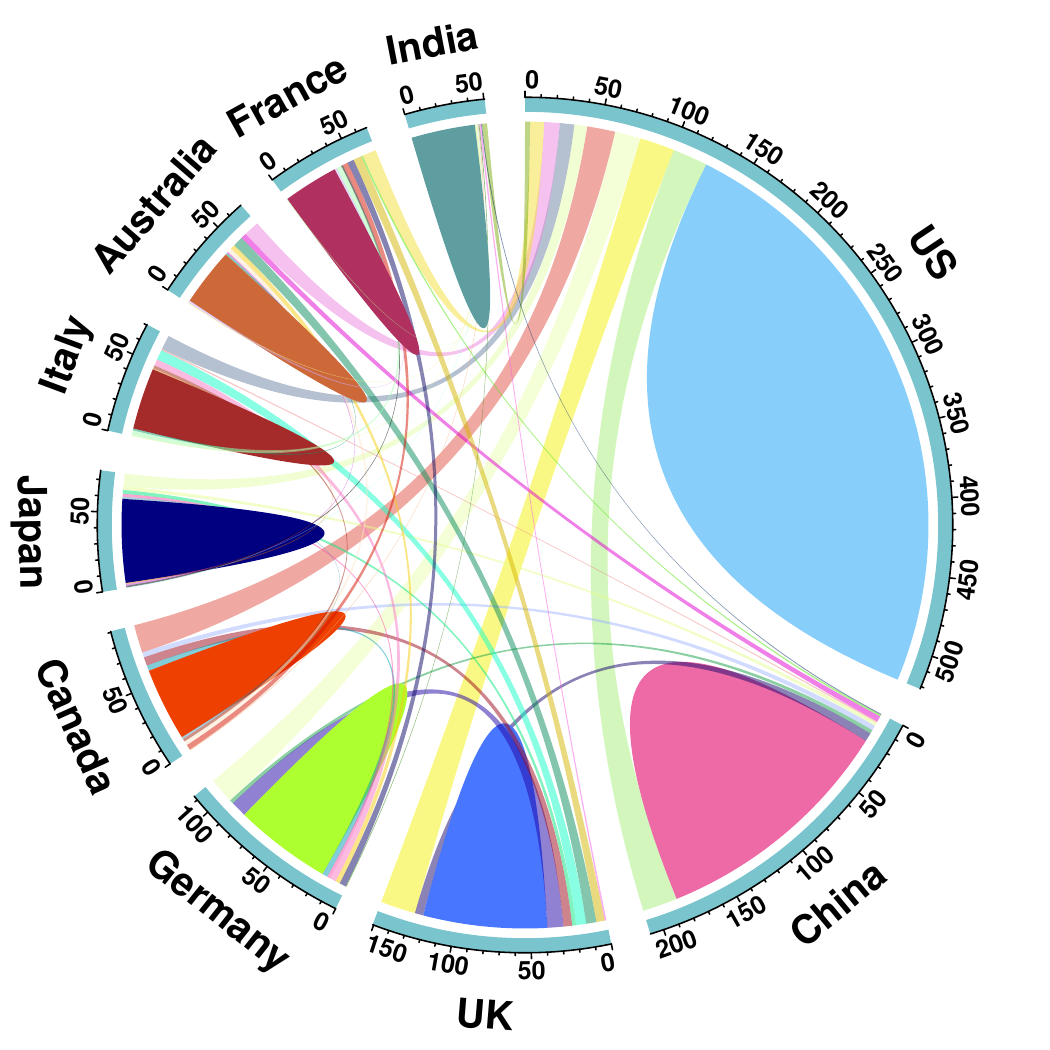}}
\end{flushleft}
    \end{minipage}
	\begin{minipage}{0.33\hsize}
\begin{flushleft}
\raisebox{-0.0cm}{\hspace{-6mm}\small\textrm{\textbf{(h)~ \condensed}}}\\[6mm]
\raisebox{\height}{\includegraphics[trim=2.0cm 1.8cm 0cm 1.5cm, align=c, scale=\chordsize, vmargin=0mm]{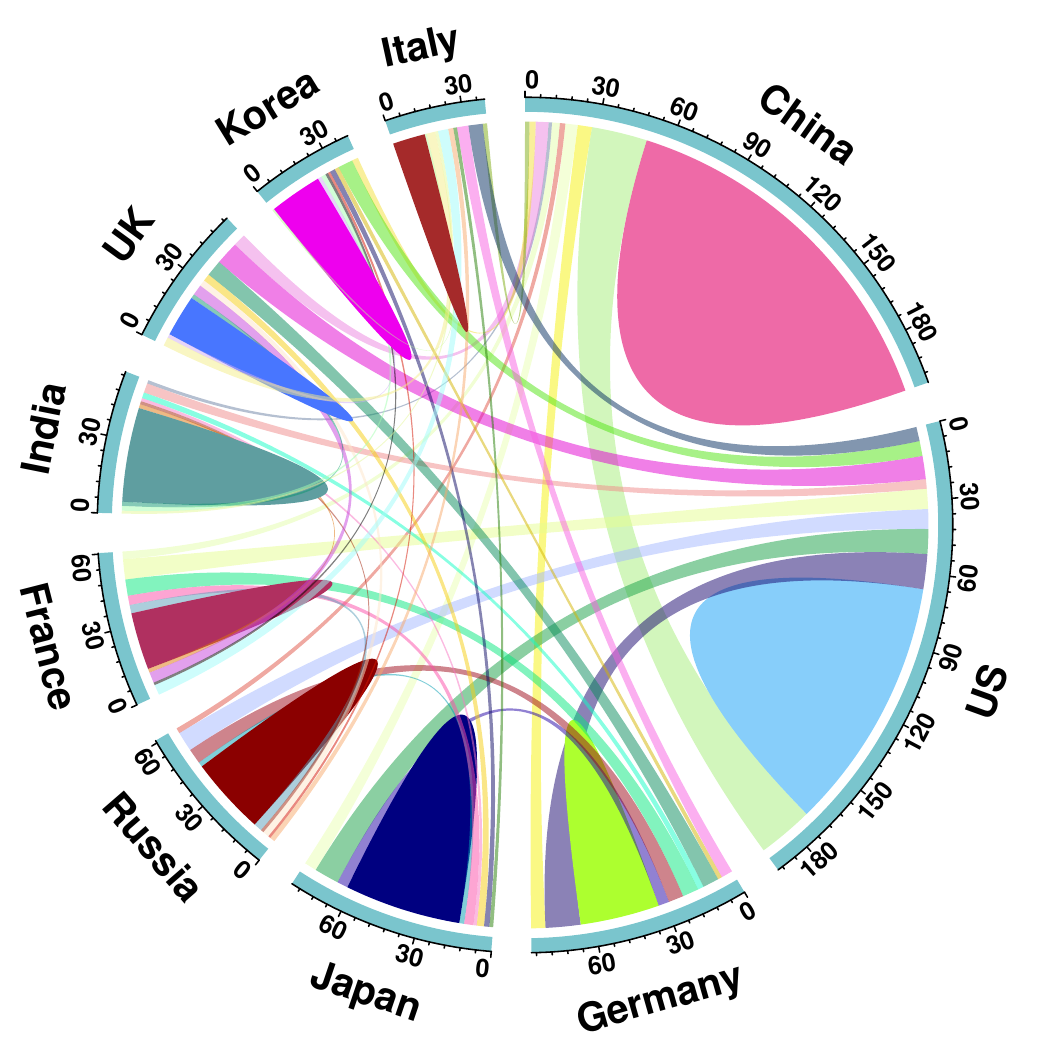}}
\end{flushleft}
	\end{minipage}
	\begin{minipage}{0.33\hsize}
\begin{flushleft}
\raisebox{-0.0cm}{\hspace{-6mm}\small\textrm{\textbf{(i)~ \envi}}}\\[6mm]
\raisebox{\height}{\includegraphics[trim=2.0cm 1.8cm 0cm 1.5cm, align=c, scale=\chordsize, vmargin=0mm]{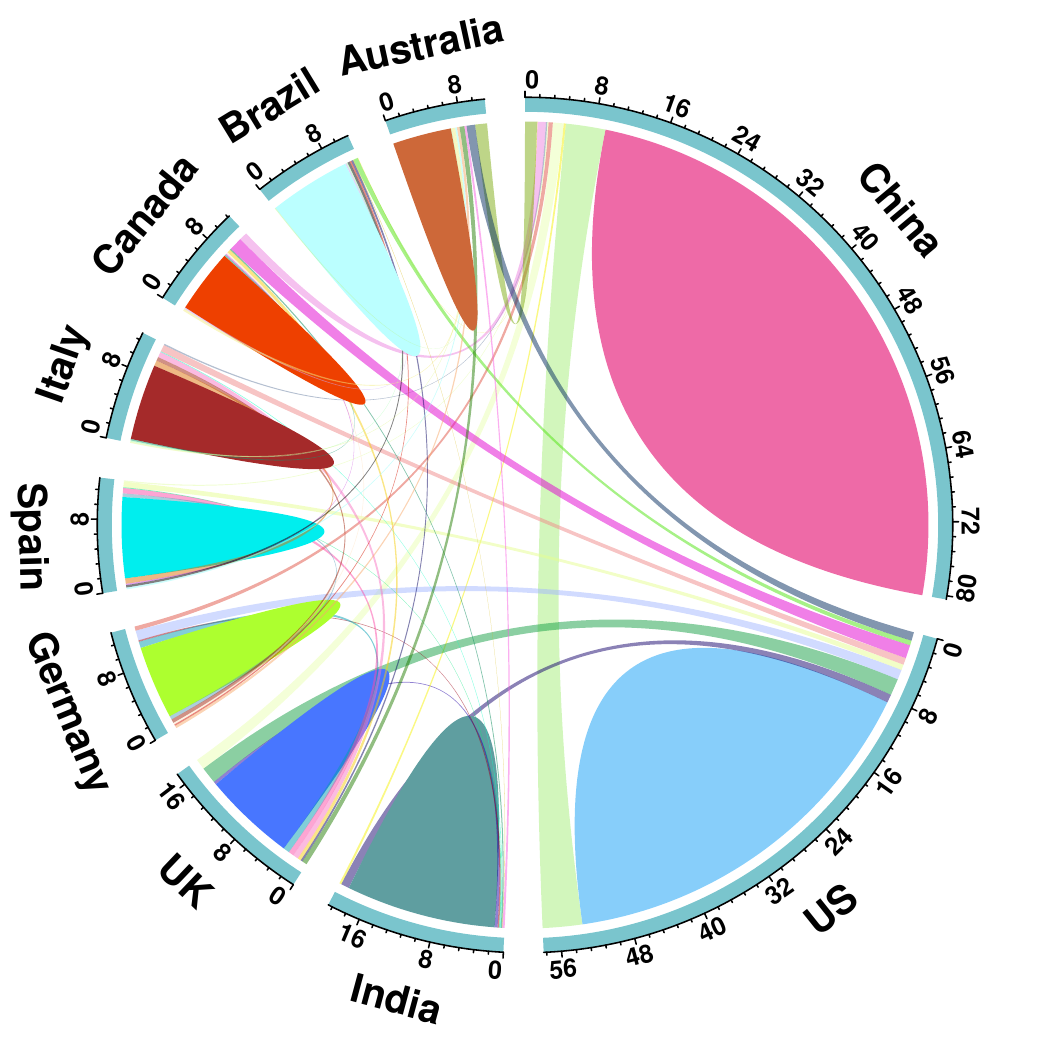}}
\end{flushleft}
	\end{minipage}
    \end{tabular}

\vspace{1mm}
\hspace{-0.5cm}{\marrow}\quad\dotfill
\vspace{4mm}

    \begin{tabular}{c}
    \begin{minipage}{0.03\hsize}
\begin{flushleft}
    \hspace{-0.7cm}\rotatebox{90}{\period{3}{2001--2010}}
\end{flushleft}
	\end{minipage}
	\begin{minipage}{0.33\hsize}
\begin{flushleft}
\raisebox{\height}{\includegraphics[trim=2.0cm 1.8cm 0cm 1.5cm, align=c, scale=\chordsize, vmargin=0mm]{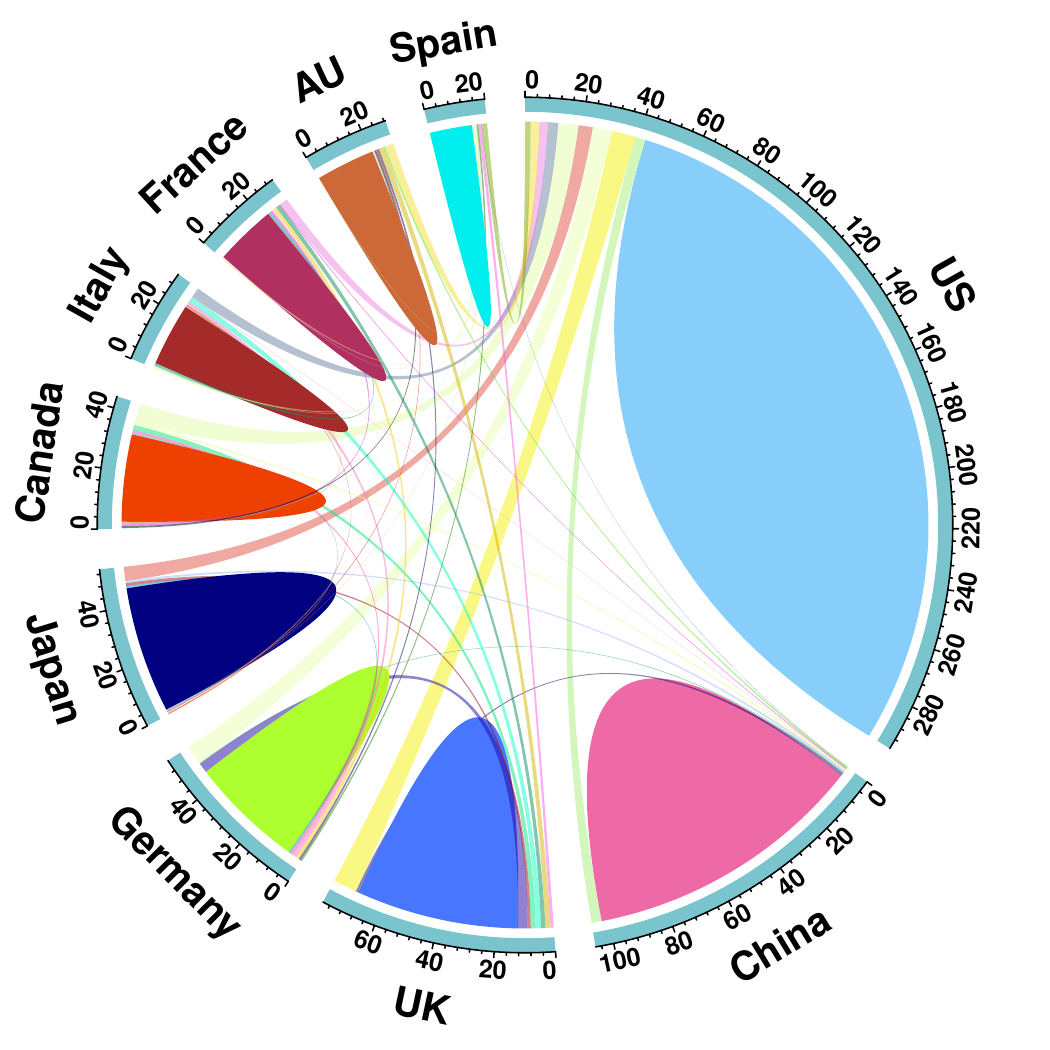}}
\end{flushleft}
    \end{minipage}
	\begin{minipage}{0.33\hsize}
\begin{flushleft}
\raisebox{\height}{\includegraphics[trim=2.0cm 1.8cm 0cm 1.5cm, align=c, scale=\chordsize, vmargin=0mm]{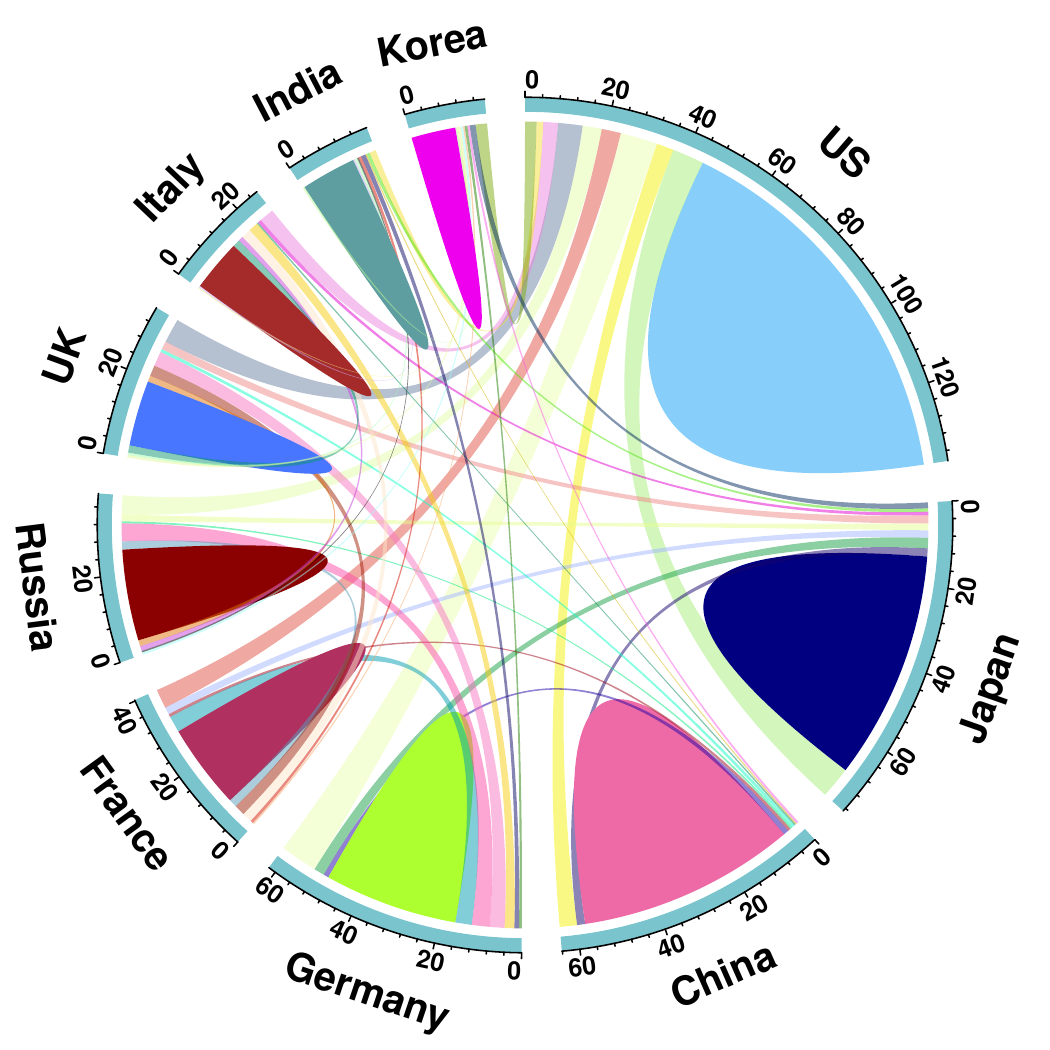}}
\end{flushleft}
	\end{minipage}
	\begin{minipage}{0.33\hsize}
\begin{flushleft}
\raisebox{\height}{\includegraphics[trim=2.0cm 1.8cm 0cm 1.5cm, align=c, scale=\chordsize, vmargin=0mm]{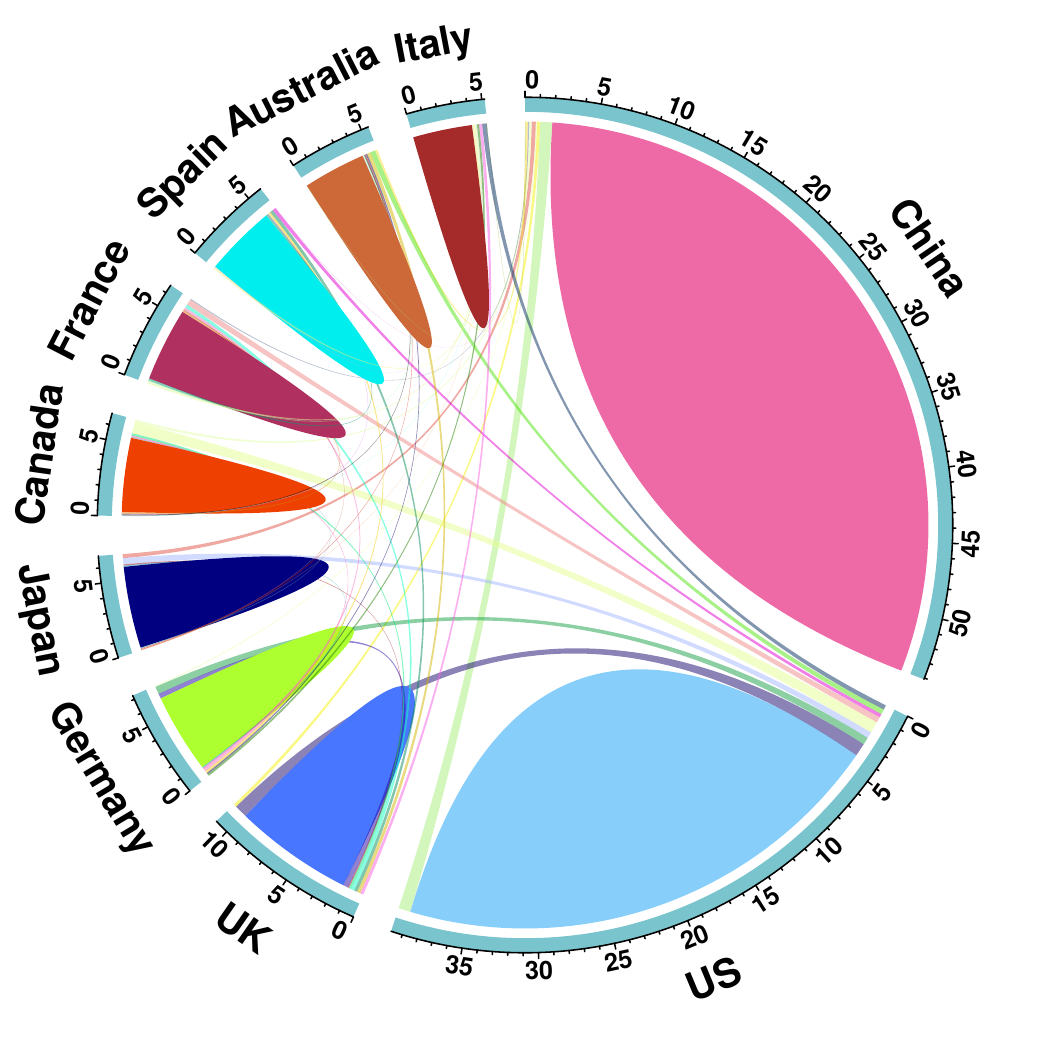}}
\end{flushleft}
	\end{minipage}
    \end{tabular}

\vspace{1mm}
\hspace{-0.5cm}{\marrow}\quad\dotfill
\vspace{4mm}

    \begin{tabular}{c}
    \begin{minipage}{0.03\hsize}
\begin{flushleft}
    \hspace{-0.7cm}\rotatebox{90}{\period{2}{1991--2000}}
\end{flushleft}
	\end{minipage}
	\begin{minipage}{0.33\hsize}
\begin{flushleft}
\raisebox{\height}{\includegraphics[trim=2.0cm 1.8cm 0cm 1.5cm, align=c, scale=\chordsize, vmargin=0mm]{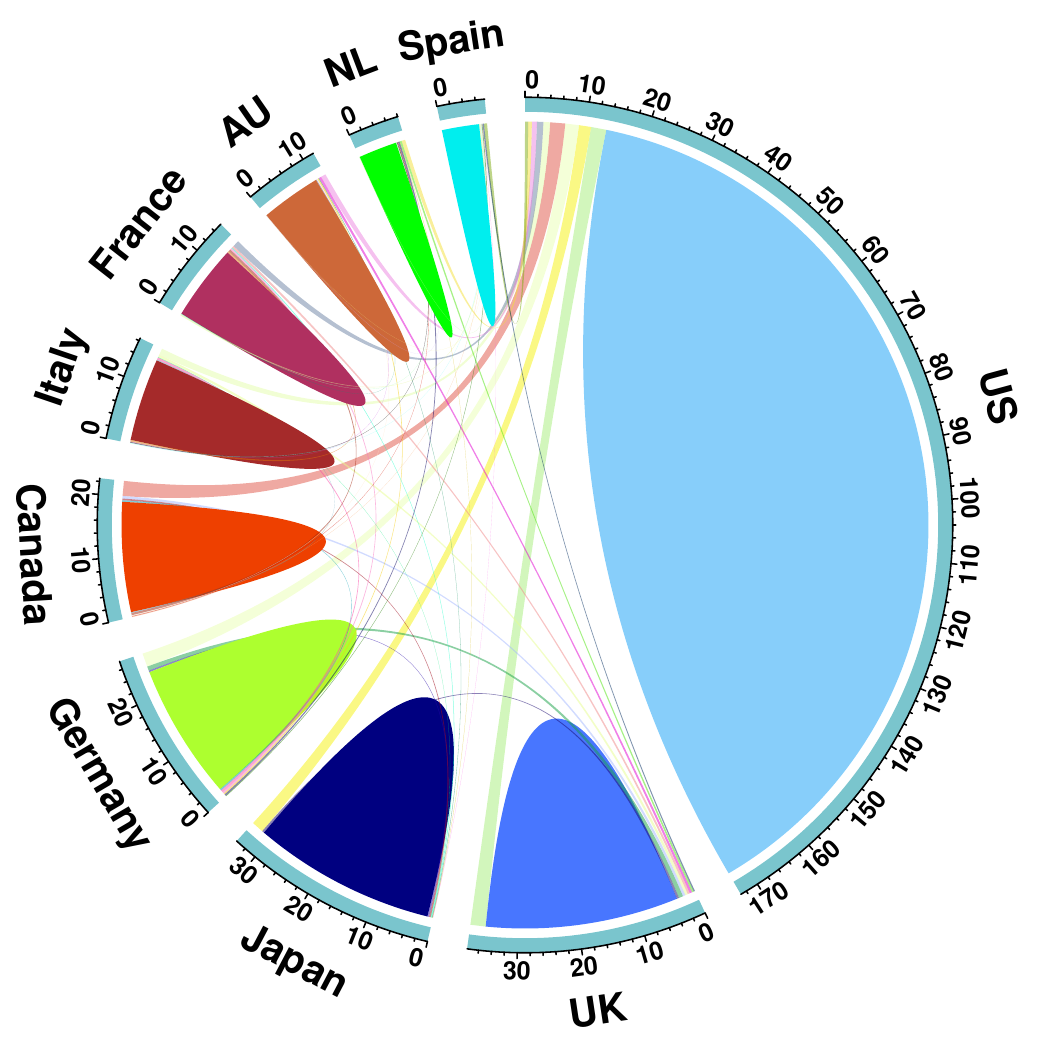}}
\end{flushleft}
    \end{minipage}
	\begin{minipage}{0.33\hsize}
\begin{flushleft}
\raisebox{\height}{\includegraphics[trim=2.0cm 1.8cm 0cm 1.5cm, align=c, scale=\chordsize, vmargin=0mm]{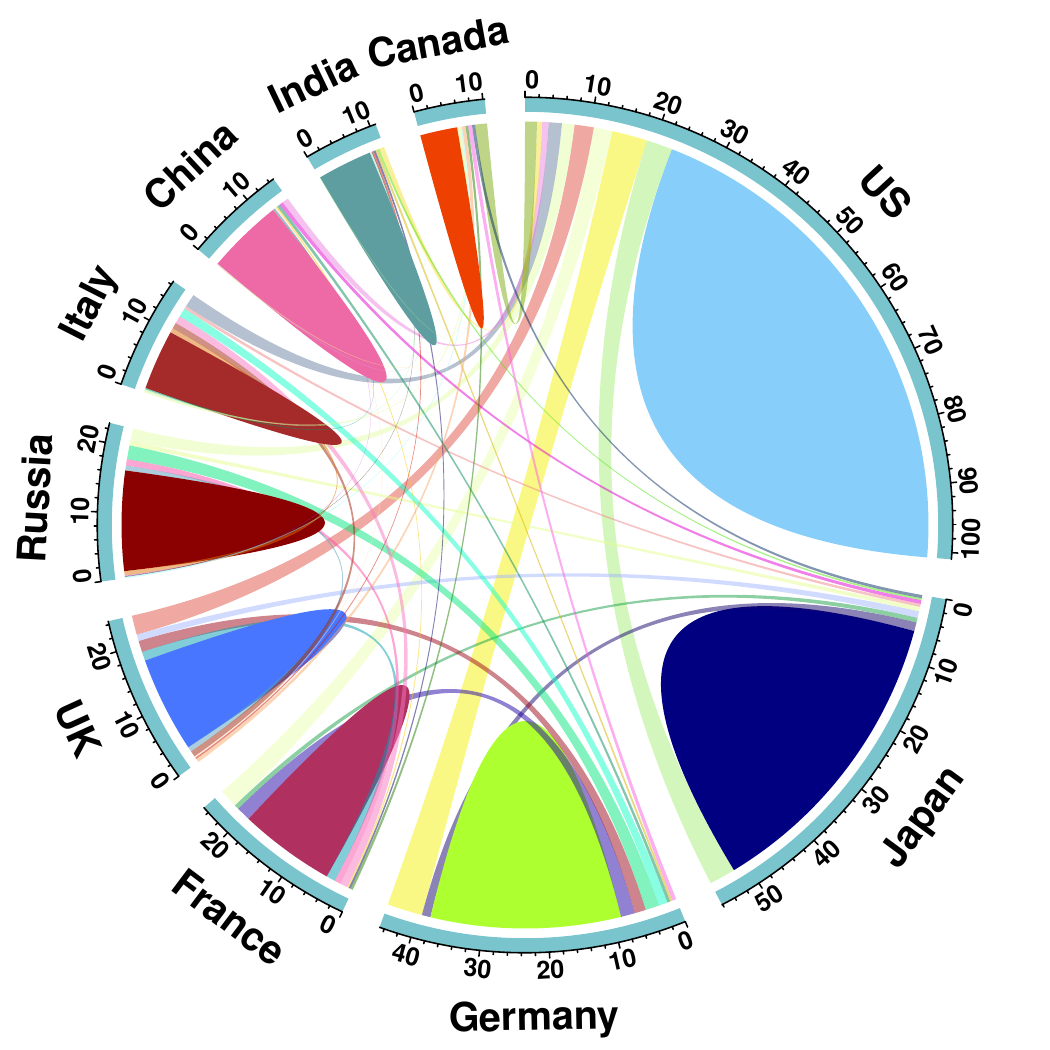}}
\end{flushleft}
	\end{minipage}
	\begin{minipage}{0.33\hsize}
\begin{flushleft}
\raisebox{\height}{\includegraphics[trim=2.0cm 1.8cm 0cm 1.5cm, align=c, scale=\chordsize, vmargin=0mm]{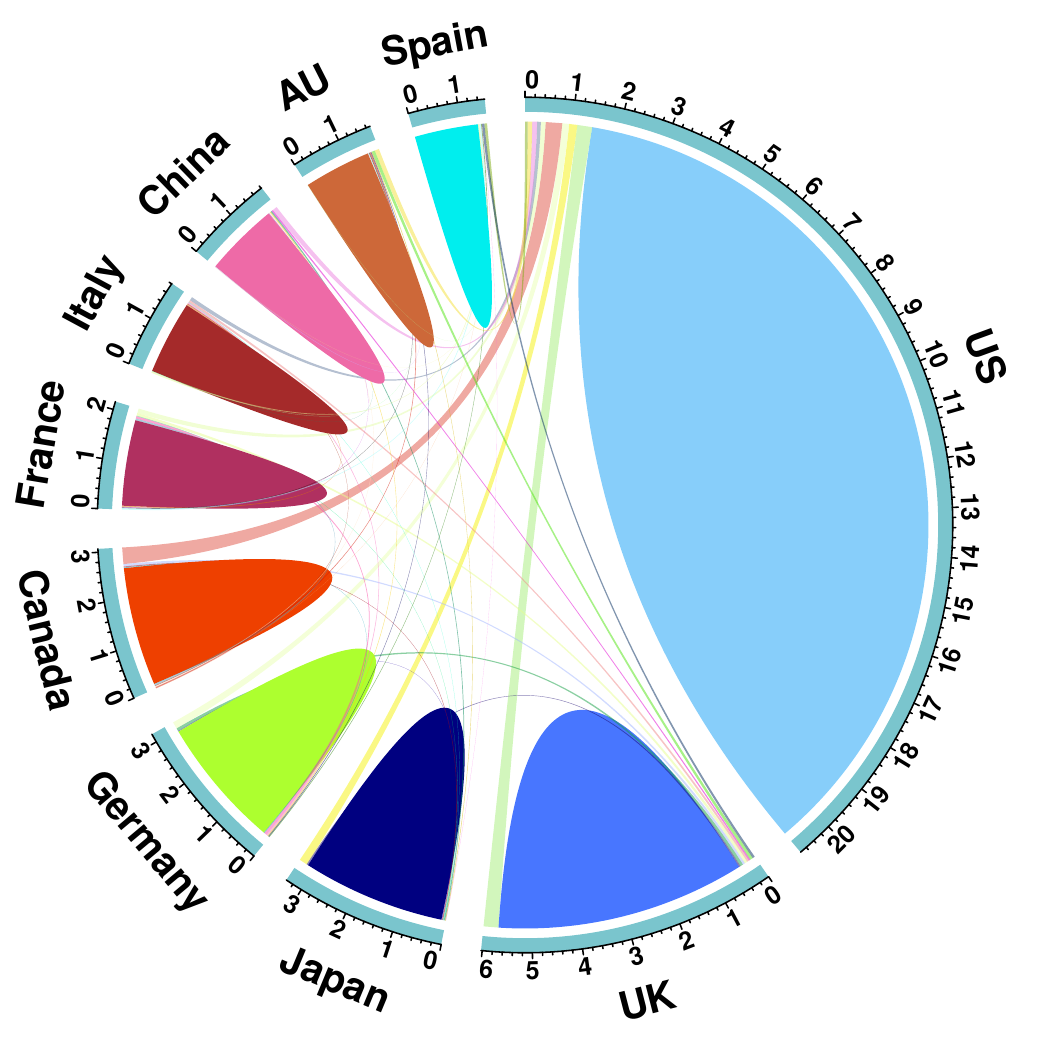}}
\end{flushleft}
	\end{minipage}
    \end{tabular}

\vspace{1mm}
\hspace{-0.5cm}{\marrow}\quad\dotfill
\vspace{4mm}

    \begin{tabular}{c}
    \begin{minipage}{0.03\hsize}
\begin{flushleft}
    \hspace{-0.7cm}\rotatebox{90}{\period{1}{1971--1990}}
\end{flushleft}
	\end{minipage}
	\begin{minipage}{0.33\hsize}
\begin{flushleft}
\raisebox{\height}{\includegraphics[trim=2.0cm 1.8cm 0cm 1.5cm, align=c, scale=\chordsize, vmargin=0mm]{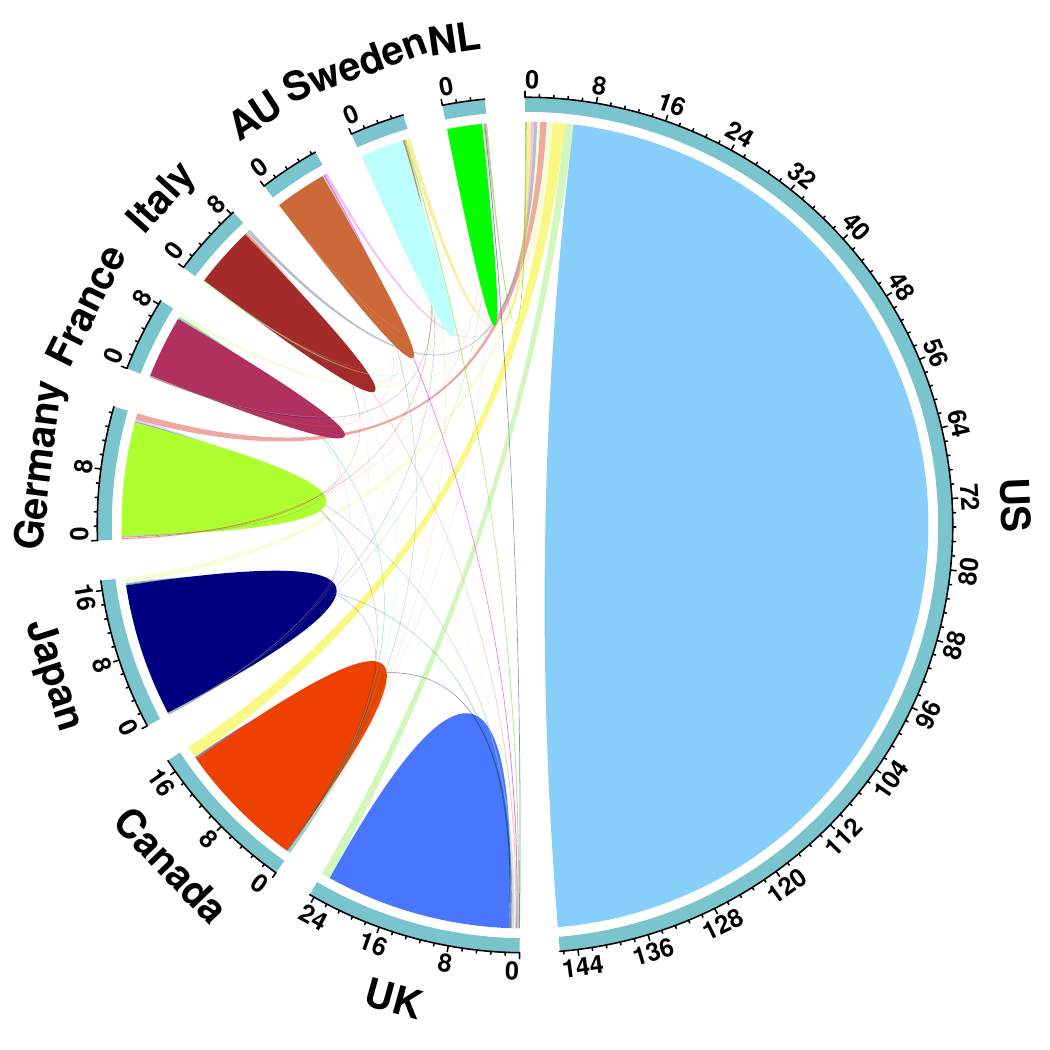}}
\end{flushleft}
    \end{minipage}
	\begin{minipage}{0.33\hsize}
\begin{flushleft}
\raisebox{\height}{\includegraphics[trim=2.0cm 1.8cm 0cm 1.5cm, align=c, scale=\chordsize, vmargin=0mm]{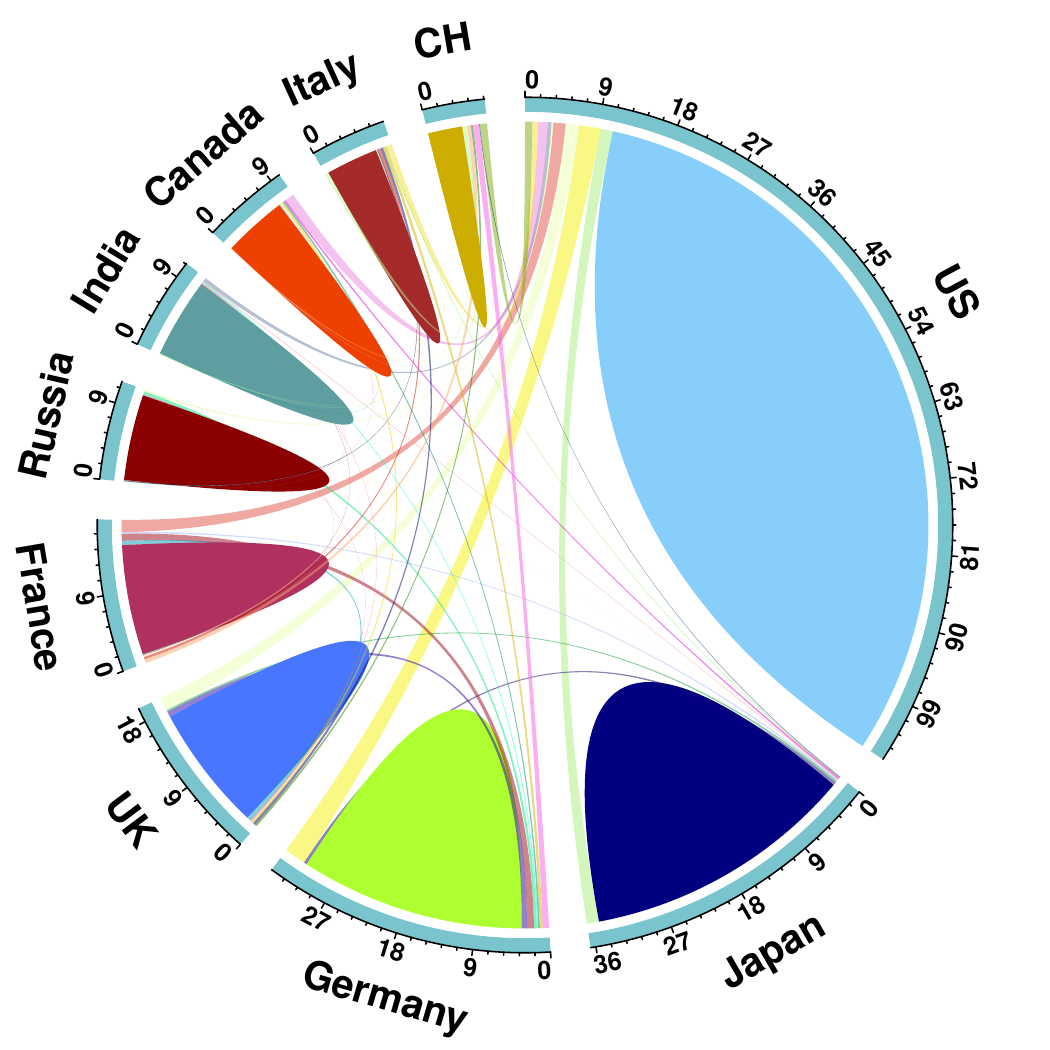}}
\end{flushleft}
	\end{minipage}
	\begin{minipage}{0.33\hsize}
\begin{flushleft}
\raisebox{\height}{\includegraphics[trim=2.0cm 1.8cm 0cm 1.5cm, align=c, scale=\chordsize, vmargin=0mm]{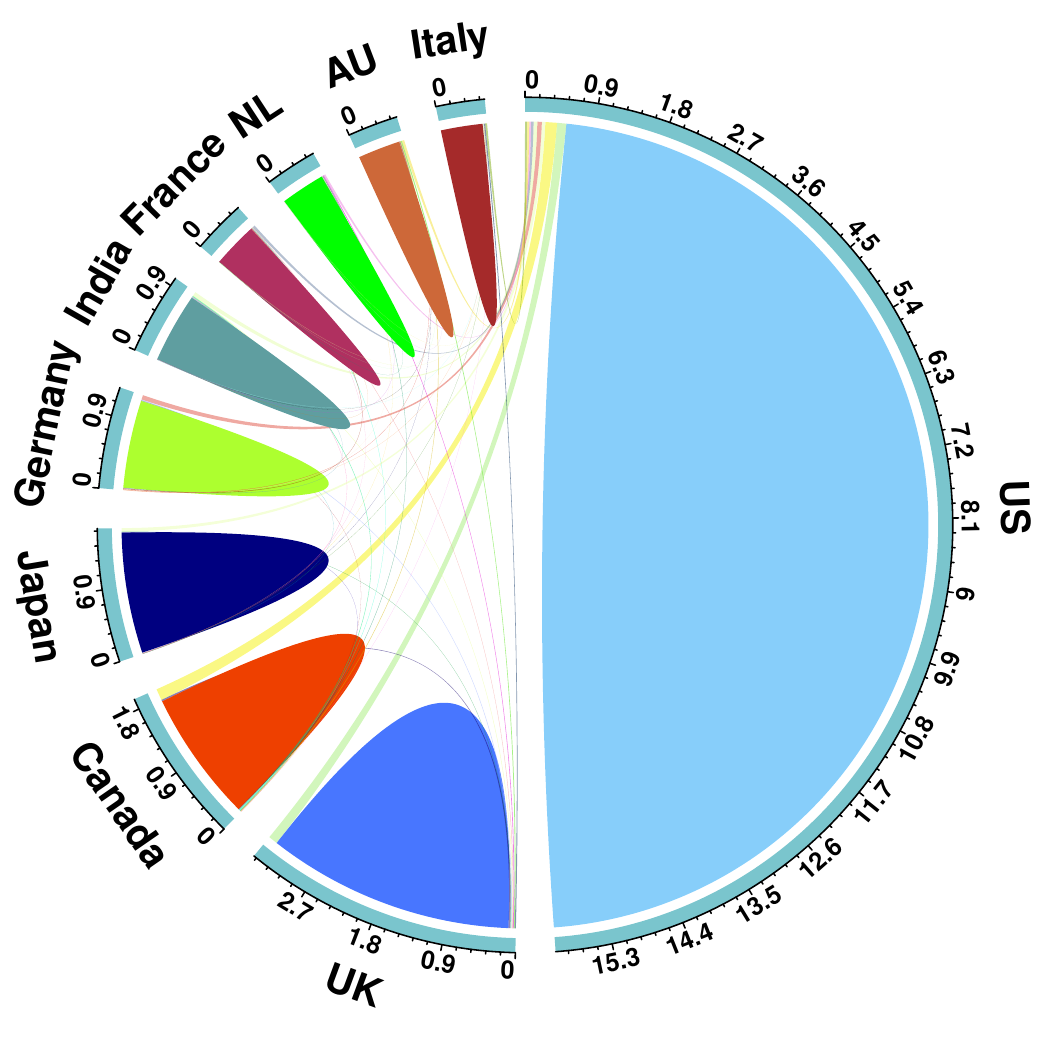}}
\end{flushleft}
	\end{minipage}
    \end{tabular}
\end{subfigure}
\vspace{2.5mm}
\caption{\textbf{Changes in bilateral relationships over time. \emph{(Cont.)}}
The number of works is displayed in thousands.}
\label{fig:chord_4}
\end{figure}
}
\afterpage{\clearpage%
\begin{figure}[htp]\ContinuedFloat
\centering
\begin{subfigure}{1.0\textwidth}
\vspace{-0.5cm}
    \begin{tabular}{c}
    \begin{minipage}{0.03\hsize}
\begin{flushleft}
    \hspace{-0.7cm}\rotatebox{90}{\period{4}{2011--2020}}
\end{flushleft}
	\end{minipage}
	\begin{minipage}{0.33\hsize}
\begin{flushleft}
\raisebox{-0.0cm}{\hspace{-6mm}\small\textrm{\textbf{(j)~ \earth}}}\\[6mm]
\raisebox{\height}{\includegraphics[trim=2.0cm 1.8cm 0cm 1.5cm, align=c, scale=\chordsize, vmargin=0mm]{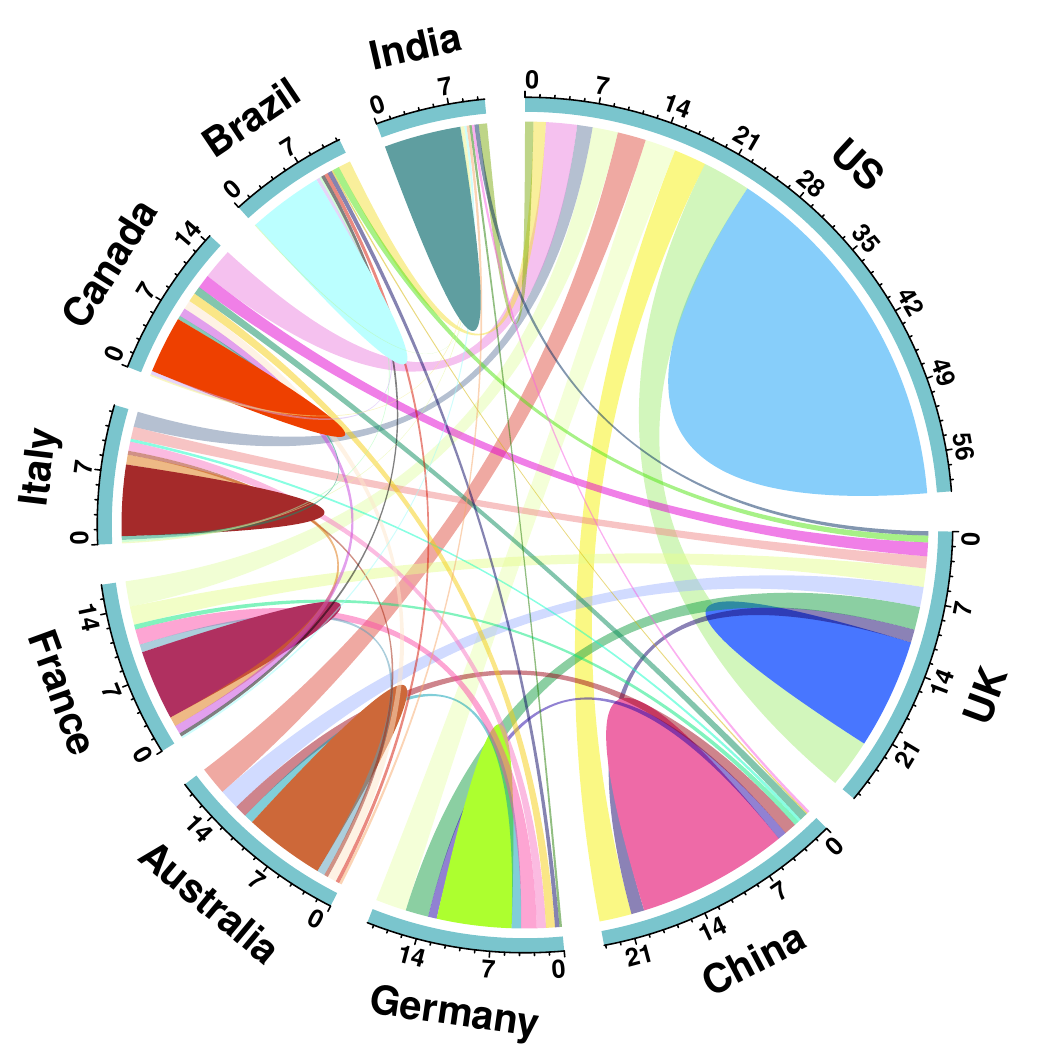}}
\end{flushleft}
    \end{minipage}
	\begin{minipage}{0.33\hsize}
\begin{flushleft}
\raisebox{-0.0cm}{\hspace{-6mm}\small\textrm{\textbf{(k)~ \astro}}}\\[6mm]
\raisebox{\height}{\includegraphics[trim=2.0cm 1.8cm 0cm 1.5cm, align=c, scale=\chordsize, vmargin=0mm]{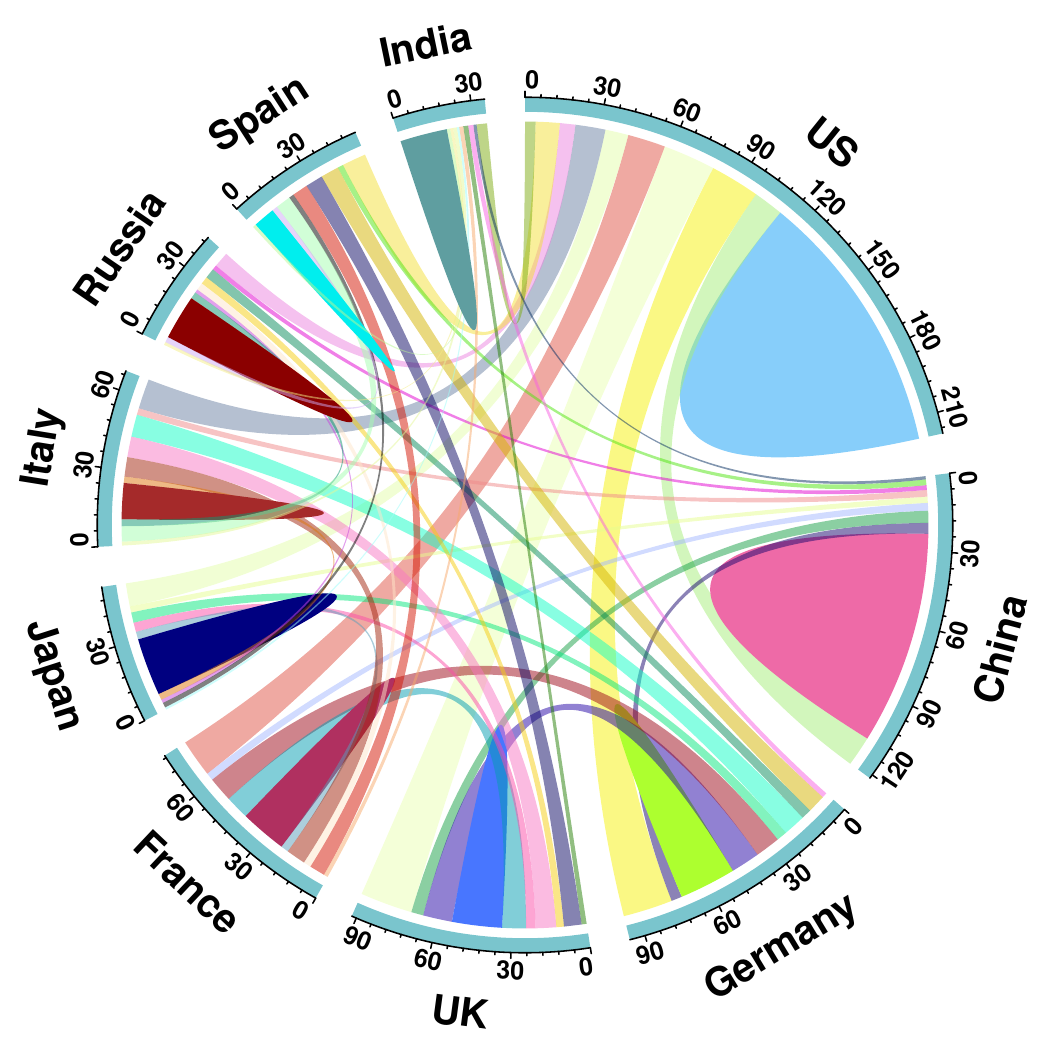}}
\end{flushleft}
	\end{minipage}
	\begin{minipage}{0.33\hsize}
\begin{flushleft}
\raisebox{-0.0cm}{\hspace{-6mm}\small\textrm{\textbf{(l)~ \math}}}\\[6mm]
\raisebox{\height}{\includegraphics[trim=2.0cm 1.8cm 0cm 1.5cm, align=c, scale=\chordsize, vmargin=0mm]{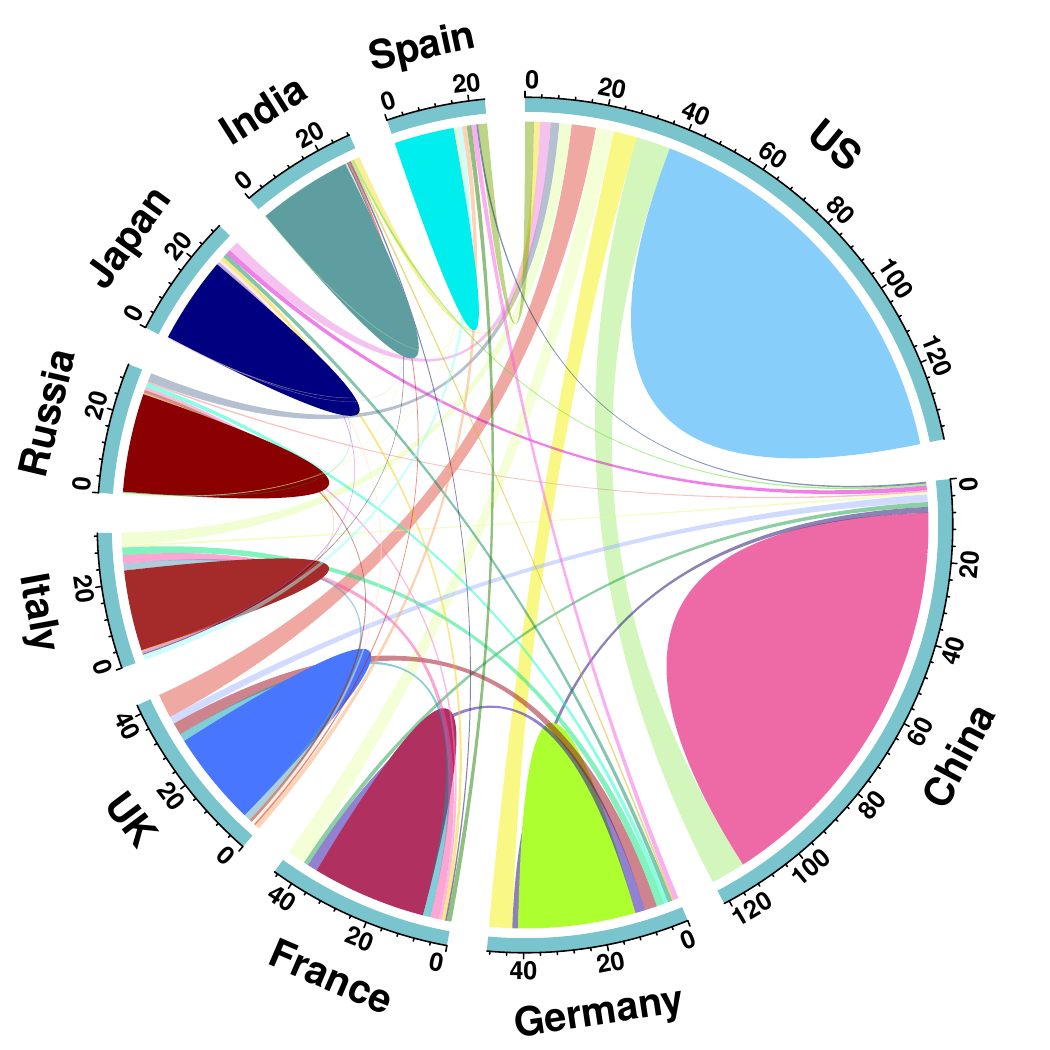}}
\end{flushleft}
	\end{minipage}
    \end{tabular}

\vspace{1mm}
\hspace{-0.5cm}{\marrow}\quad\dotfill
\vspace{4mm}

    \begin{tabular}{c}
    \begin{minipage}{0.03\hsize}
\begin{flushleft}
    \hspace{-0.7cm}\rotatebox{90}{\period{3}{2001--2010}}
\end{flushleft}
	\end{minipage}
	\begin{minipage}{0.33\hsize}
\begin{flushleft}
\raisebox{\height}{\includegraphics[trim=2.0cm 1.8cm 0cm 1.5cm, align=c, scale=\chordsize, vmargin=0mm]{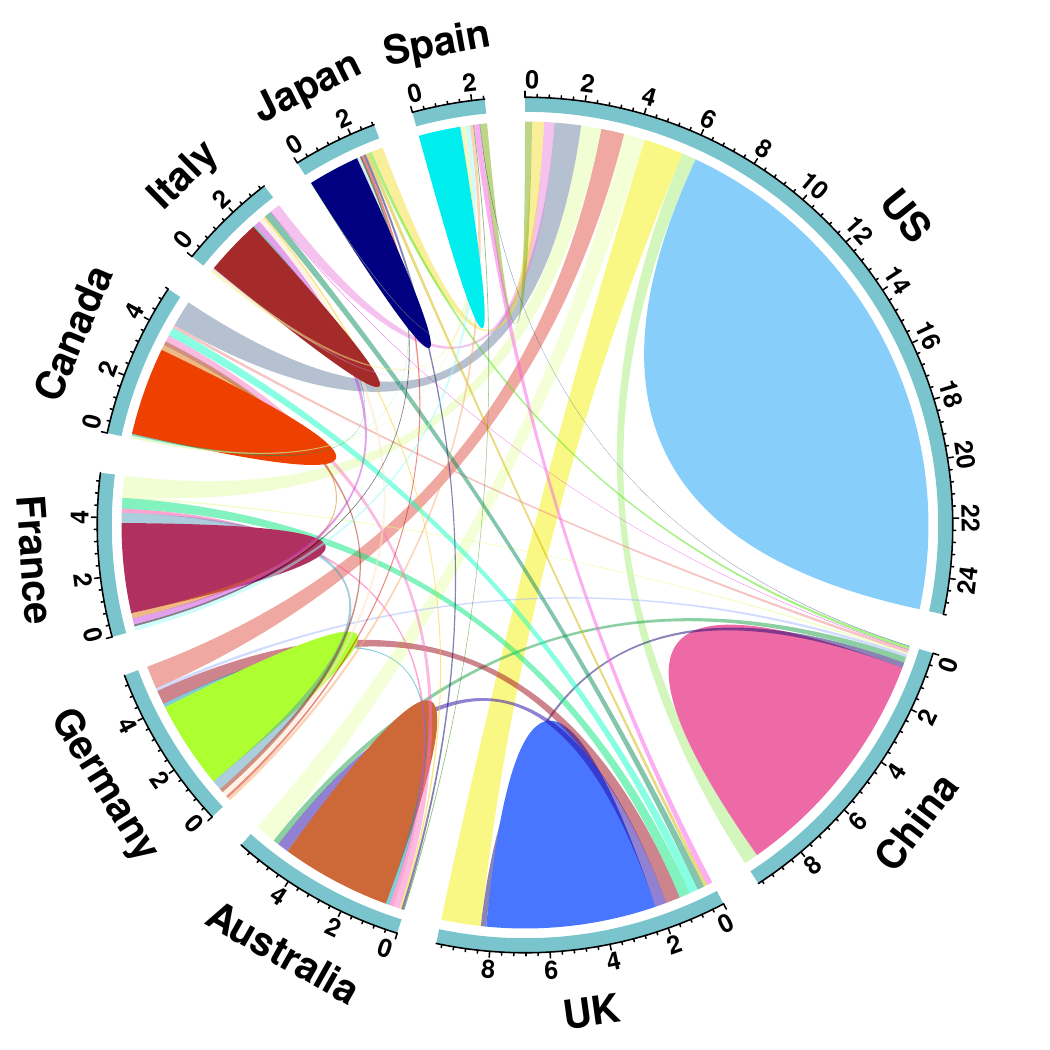}}
\end{flushleft}
    \end{minipage}
	\begin{minipage}{0.33\hsize}
\begin{flushleft}
\raisebox{\height}{\includegraphics[trim=2.0cm 1.8cm 0cm 1.5cm, align=c, scale=\chordsize, vmargin=0mm]{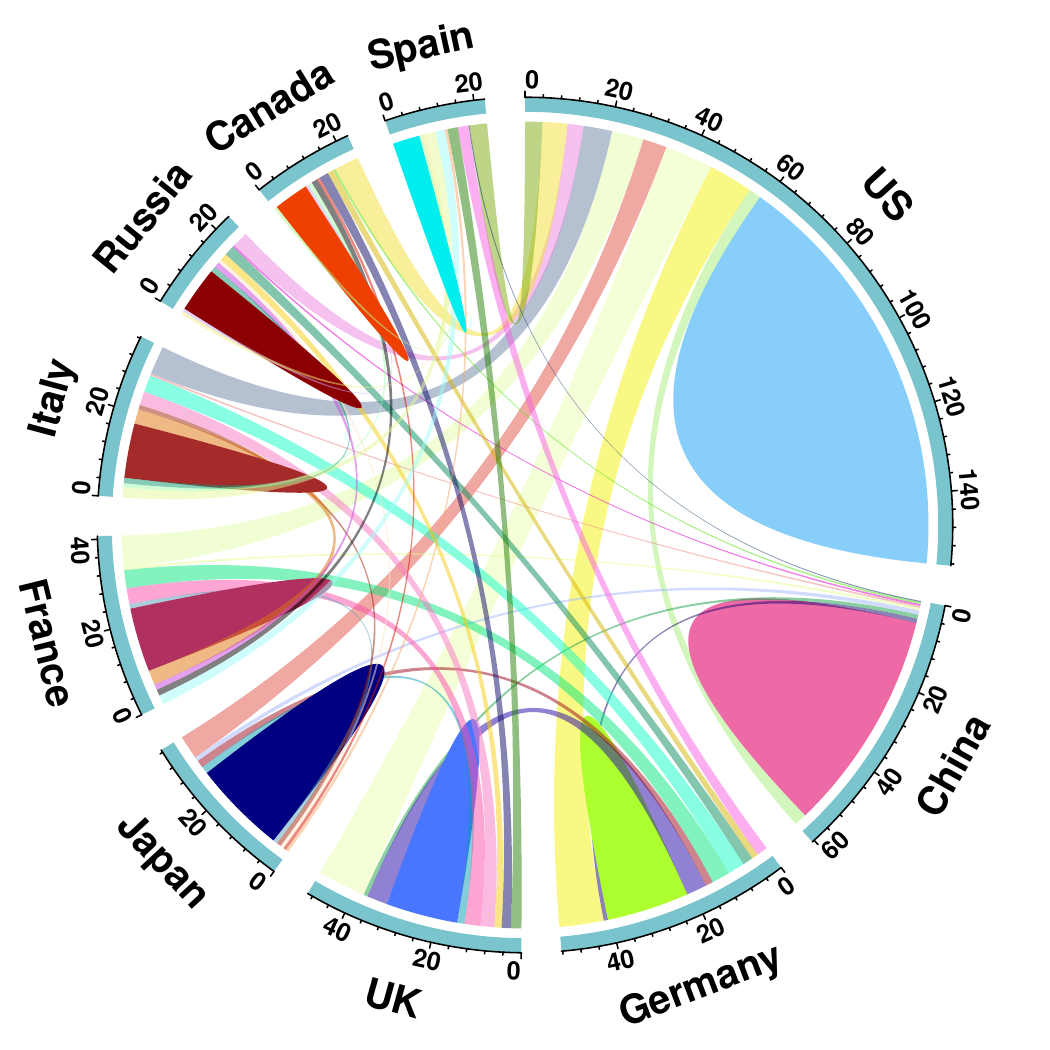}}
\end{flushleft}
	\end{minipage}
	\begin{minipage}{0.33\hsize}
\begin{flushleft}
\raisebox{\height}{\includegraphics[trim=2.0cm 1.8cm 0cm 1.5cm, align=c, scale=\chordsize, vmargin=0mm]{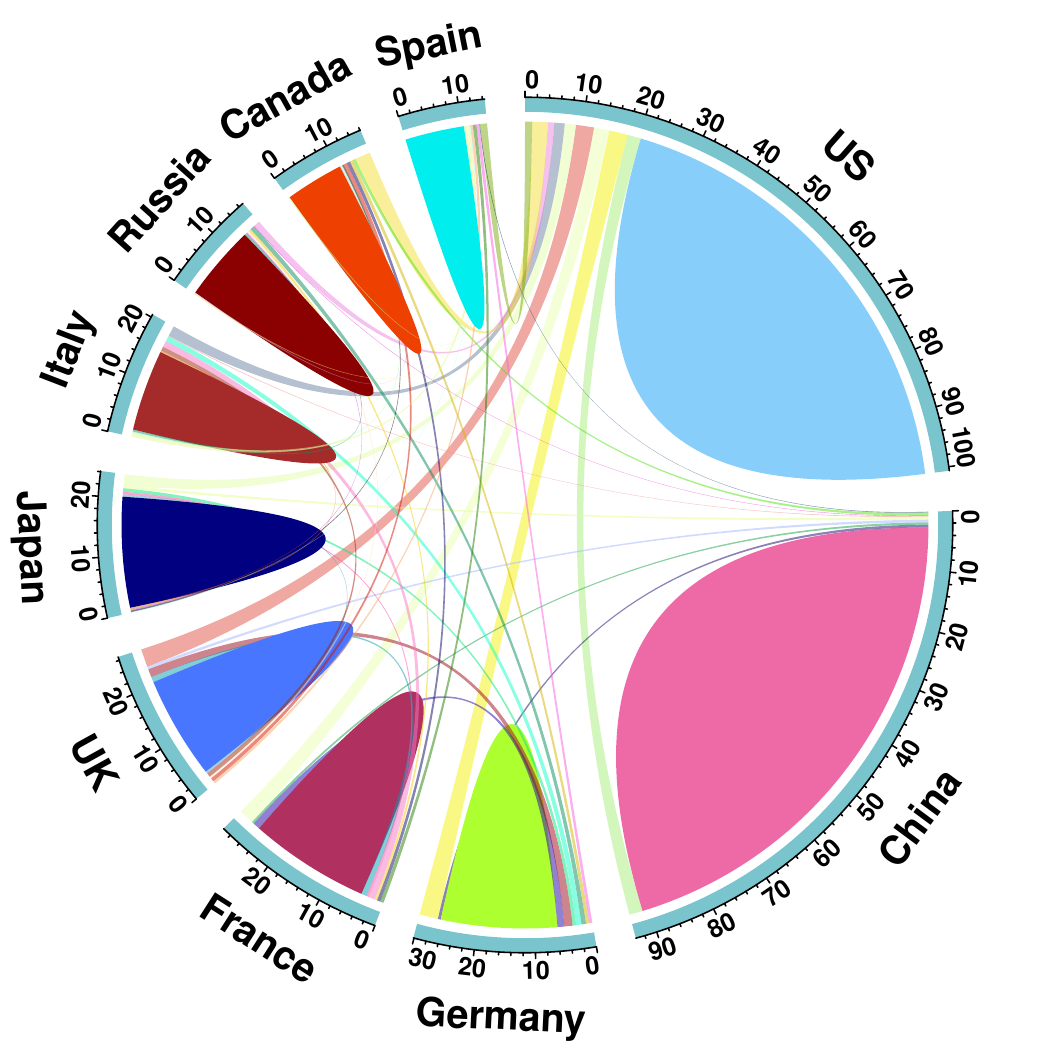}}
\end{flushleft}
	\end{minipage}
    \end{tabular}

\vspace{1mm}
\hspace{-0.5cm}{\marrow}\quad\dotfill
\vspace{4mm}

    \begin{tabular}{c}
    \begin{minipage}{0.03\hsize}
\begin{flushleft}
    \hspace{-0.7cm}\rotatebox{90}{\period{2}{1991--2000}}
\end{flushleft}
	\end{minipage}
	\begin{minipage}{0.33\hsize}
\begin{flushleft}
\raisebox{\height}{\includegraphics[trim=2.0cm 1.8cm 0cm 1.5cm, align=c, scale=\chordsize, vmargin=0mm]{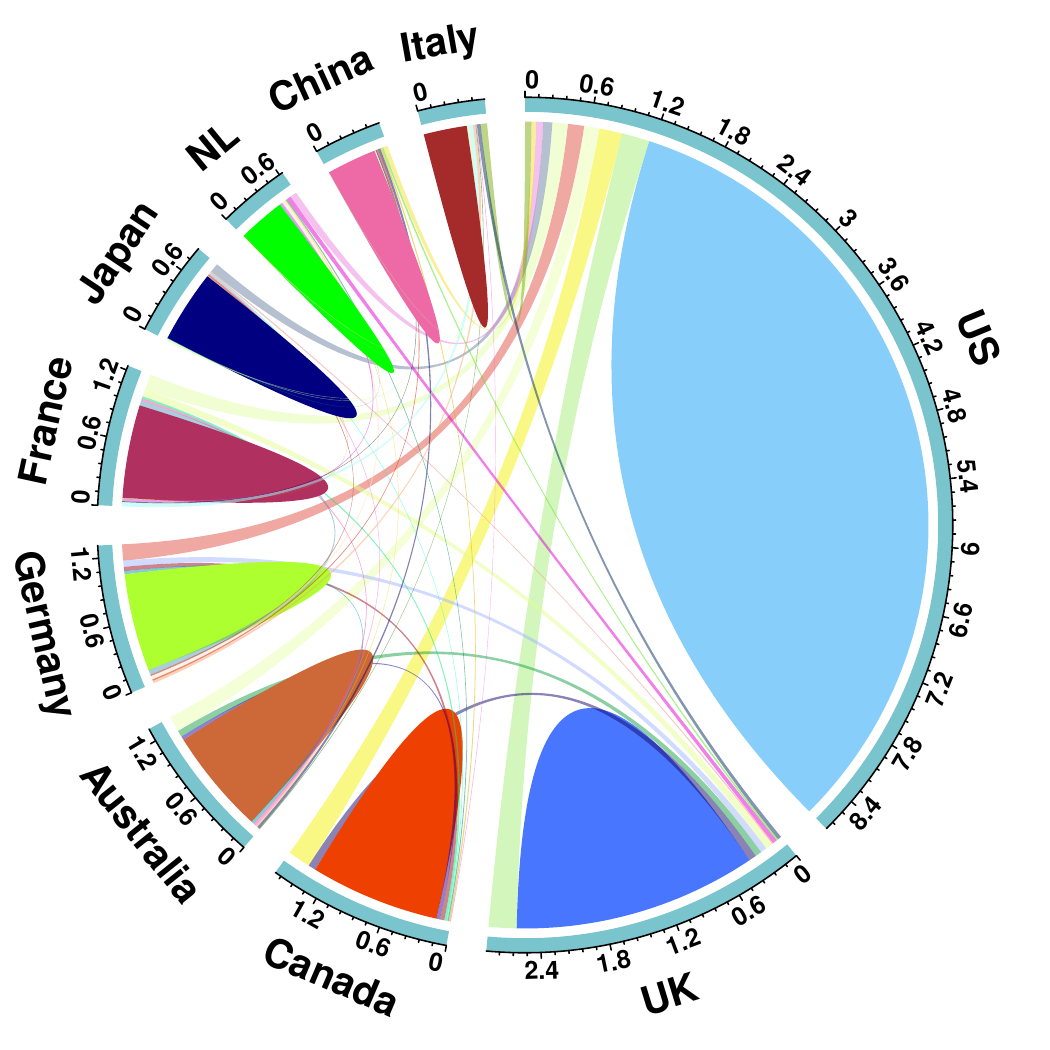}}
\end{flushleft}
    \end{minipage}
	\begin{minipage}{0.33\hsize}
\begin{flushleft}
\raisebox{\height}{\includegraphics[trim=2.0cm 1.8cm 0cm 1.5cm, align=c, scale=\chordsize, vmargin=0mm]{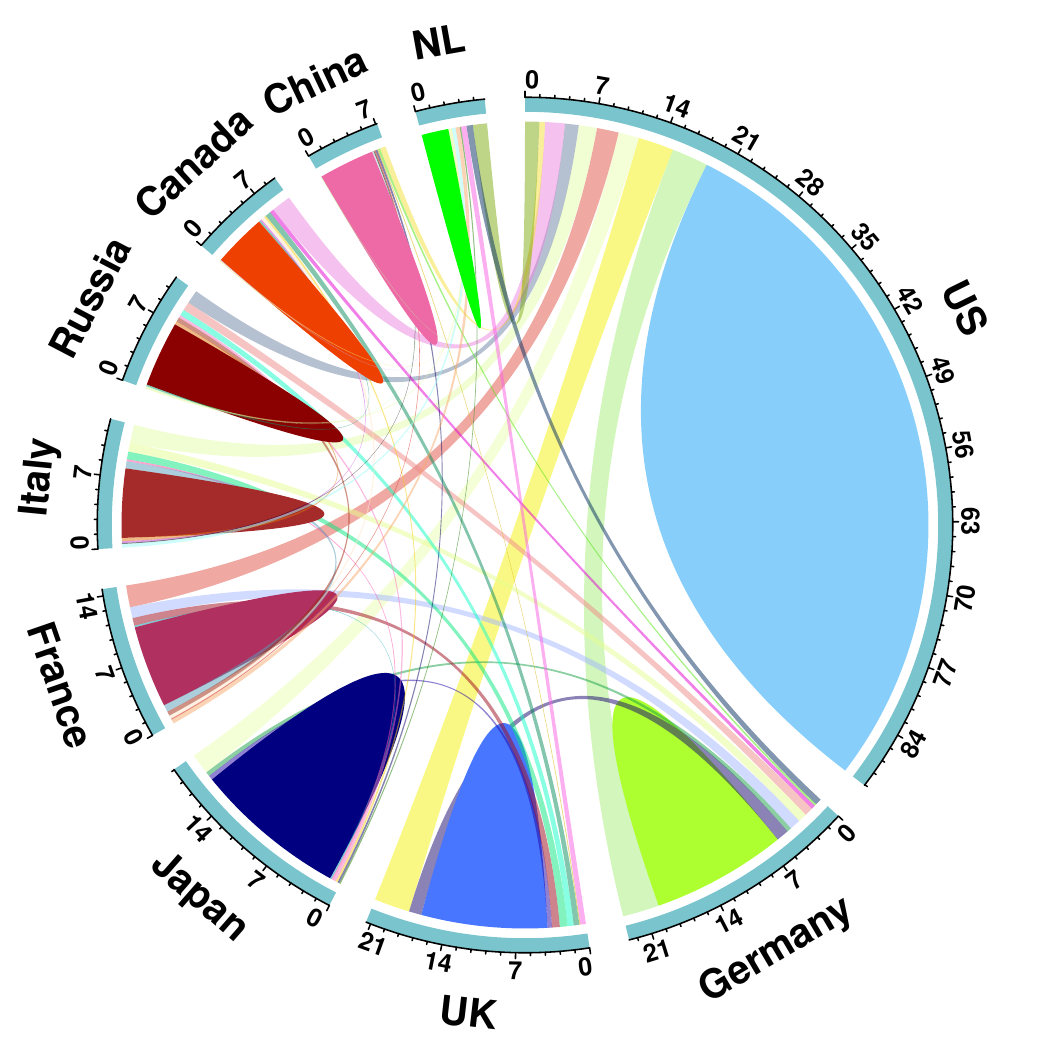}}
\end{flushleft}
	\end{minipage}
	\begin{minipage}{0.33\hsize}
\begin{flushleft}
\raisebox{\height}{\includegraphics[trim=2.0cm 1.8cm 0cm 1.5cm, align=c, scale=\chordsize, vmargin=0mm]{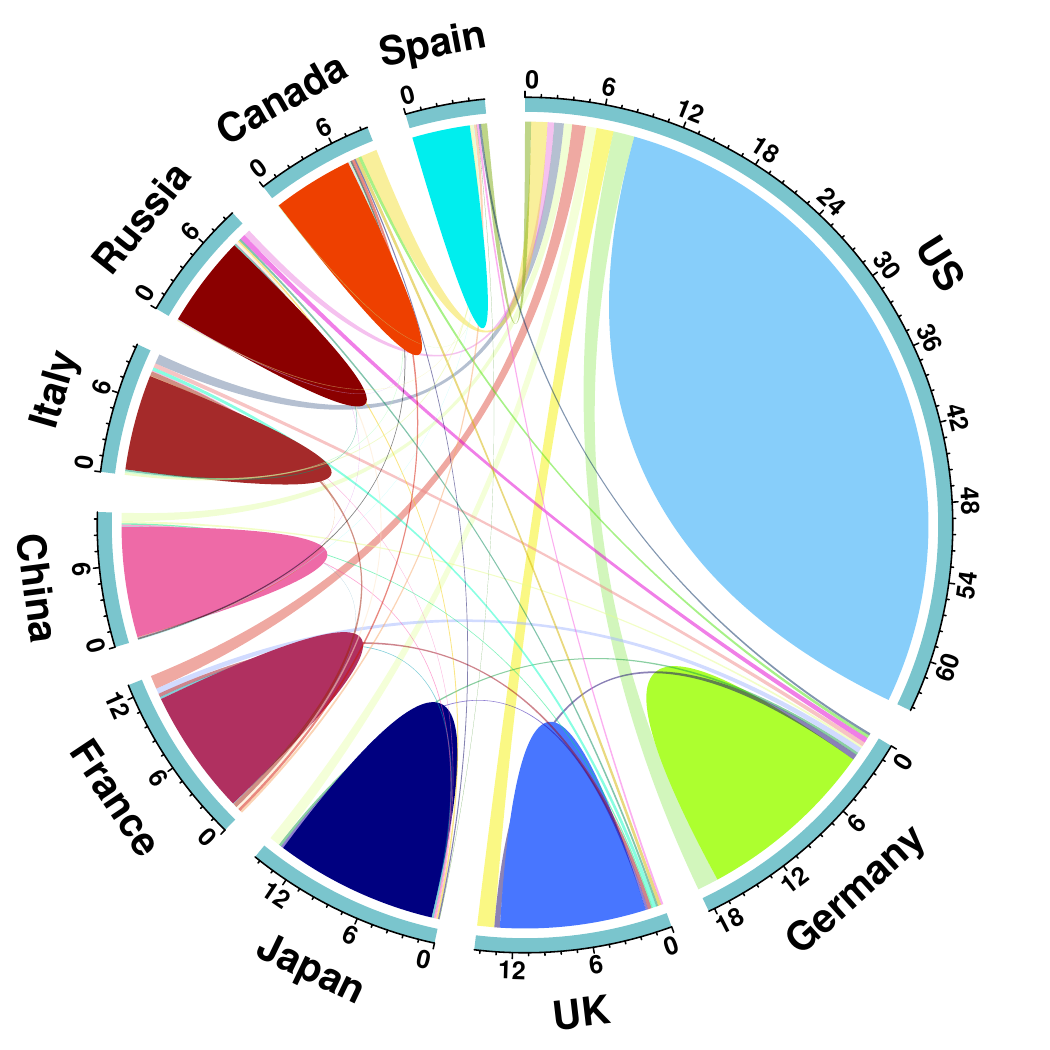}}
\end{flushleft}
	\end{minipage}
    \end{tabular}

\vspace{1mm}
\hspace{-0.5cm}{\marrow}\quad\dotfill
\vspace{4mm}

    \begin{tabular}{c}
    \begin{minipage}{0.03\hsize}
\begin{flushleft}
    \hspace{-0.7cm}\rotatebox{90}{\period{1}{1971--1990}}
\end{flushleft}
	\end{minipage}
	\begin{minipage}{0.33\hsize}
\begin{flushleft}
\raisebox{\height}{\includegraphics[trim=2.0cm 1.8cm 0cm 1.5cm, align=c, scale=\chordsize, vmargin=0mm]{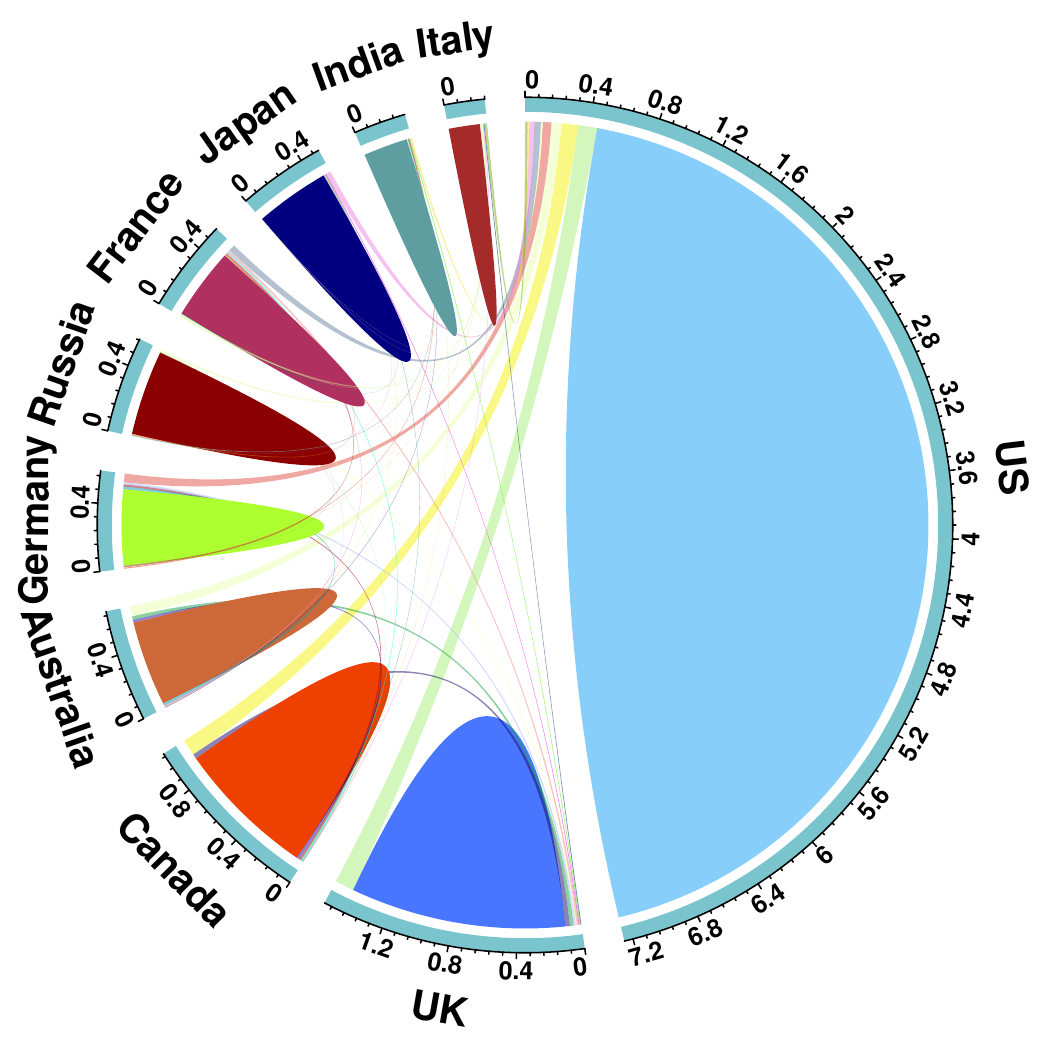}}
\end{flushleft}
    \end{minipage}
	\begin{minipage}{0.33\hsize}
\begin{flushleft}
\raisebox{\height}{\includegraphics[trim=2.0cm 1.8cm 0cm 1.5cm, align=c, scale=\chordsize, vmargin=0mm]{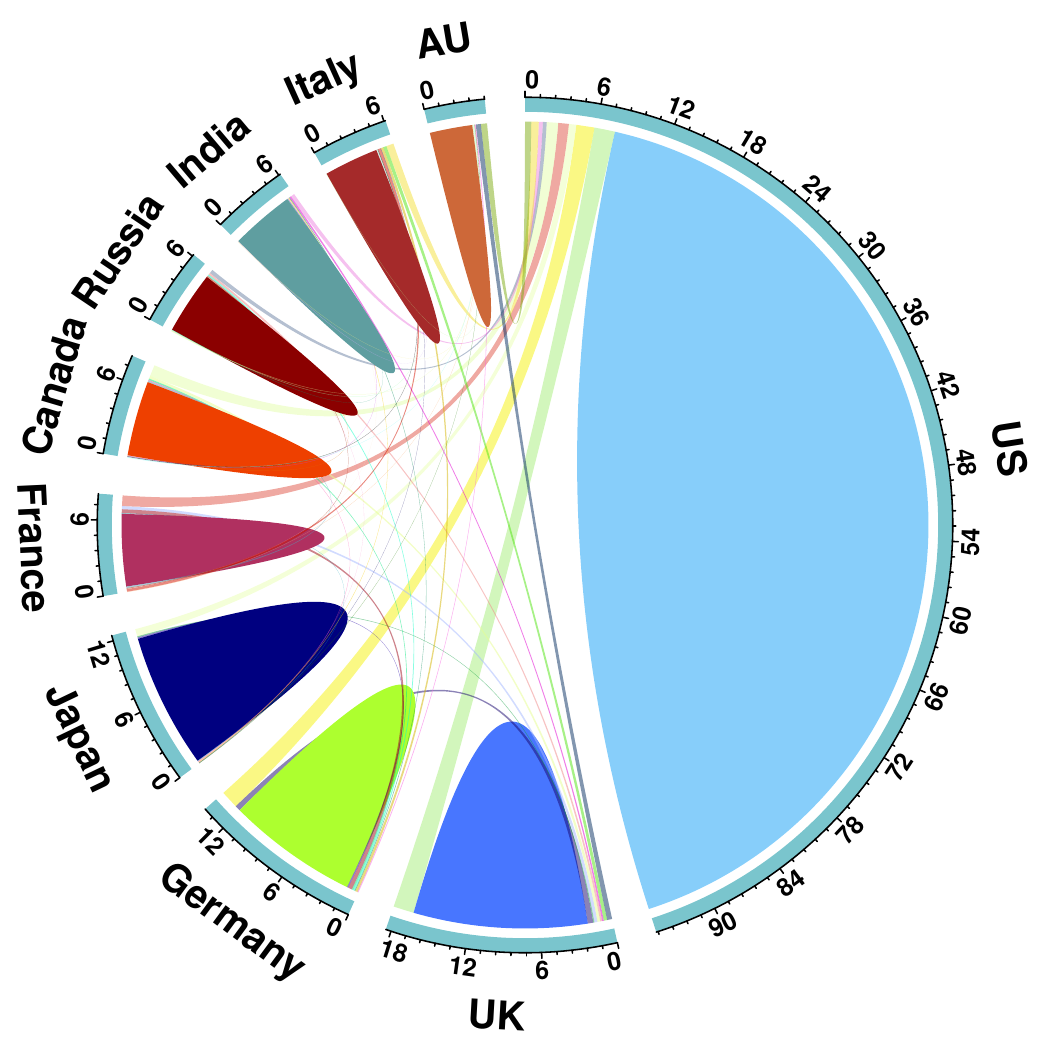}}
\end{flushleft}
	\end{minipage}
	\begin{minipage}{0.33\hsize}
\begin{flushleft}
\raisebox{\height}{\includegraphics[trim=2.0cm 1.8cm 0cm 1.5cm, align=c, scale=\chordsize, vmargin=0mm]{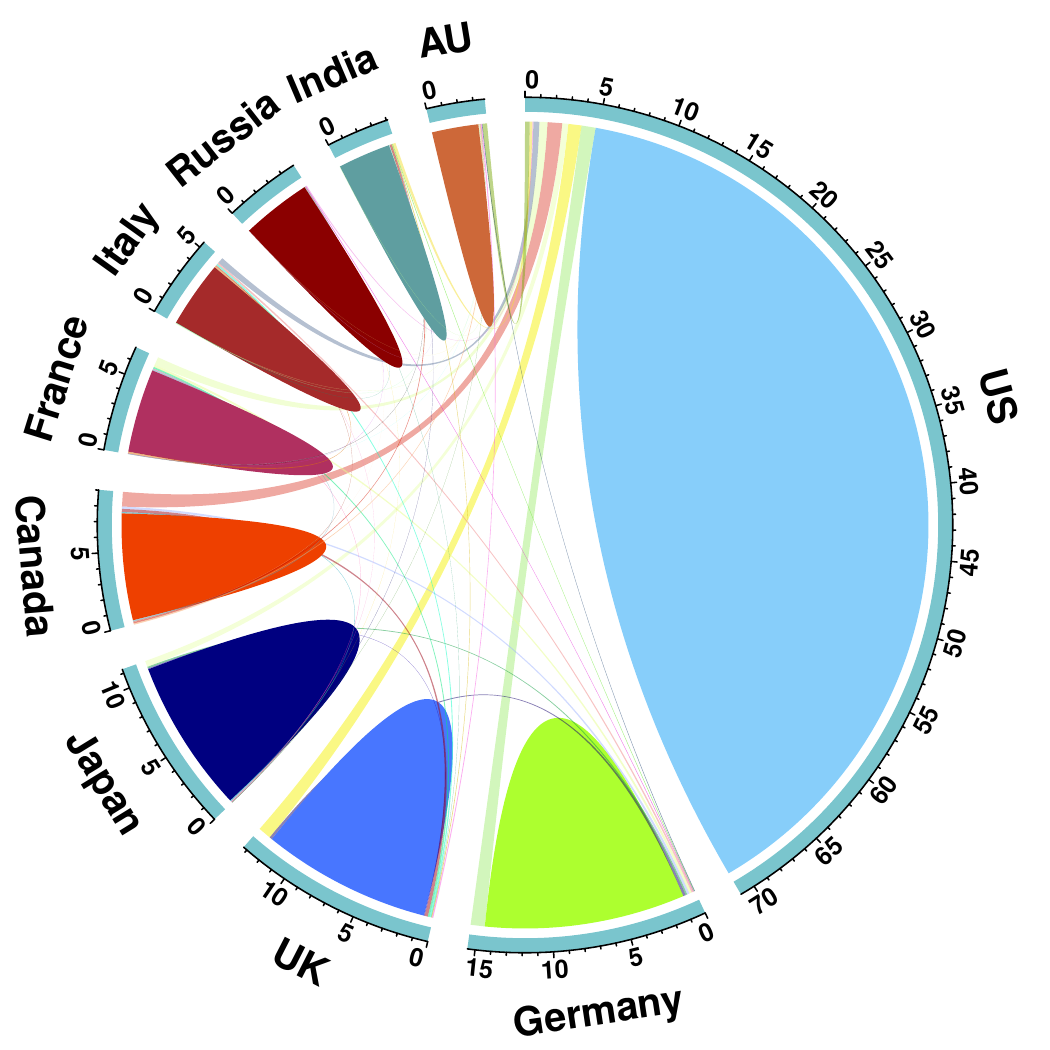}}
\end{flushleft}
	\end{minipage}
    \end{tabular}
\end{subfigure}
\vspace{2.5mm}
\caption{\textbf{Changes in bilateral relationships over time. \emph{(Cont.)}}
The number of works is displayed in thousands.}
\label{fig:chord_5}
\end{figure}
}

\afterpage{\clearpage%
\begin{figure}[htp]
\centering
\begin{subfigure}{1.0\textwidth}
\vspace{-0.5cm}
    \begin{tabular}{c}
    \begin{minipage}{0.03\hsize}
\begin{flushleft}
    \hspace{-0.7cm}\rotatebox{90}{\period{4}{2011--2020}}
\end{flushleft}
	\end{minipage}
	\begin{minipage}{0.33\hsize}
\begin{flushleft}
\raisebox{-0.0cm}{\hspace{-0mm}\small\textrm{\textbf{(a)~ \nano}}}\\[6mm]
\raisebox{\height}{\includegraphics[trim=2.0cm 1.8cm 0cm 1.5cm, align=c, scale=\csize, vmargin=0mm]{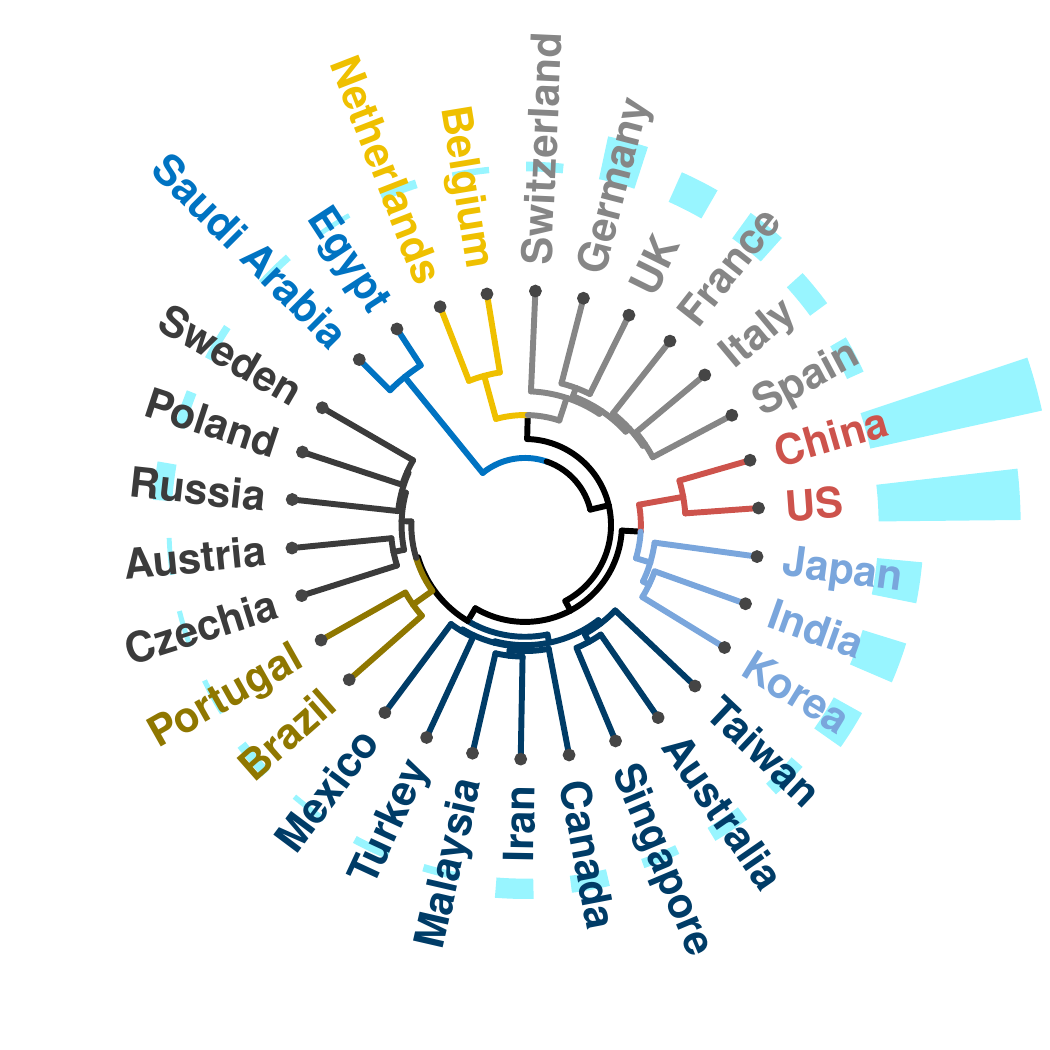}}
\end{flushleft}
    \end{minipage}
	\begin{minipage}{0.33\hsize}
\begin{flushleft}
\raisebox{-0.0cm}{\hspace{-0mm}\small\textrm{\textbf{(b)~ \agri}}}\\[6mm]
\raisebox{\height}{\includegraphics[trim=2.0cm 1.8cm 0cm 1.5cm, align=c, scale=\csize, vmargin=0mm]{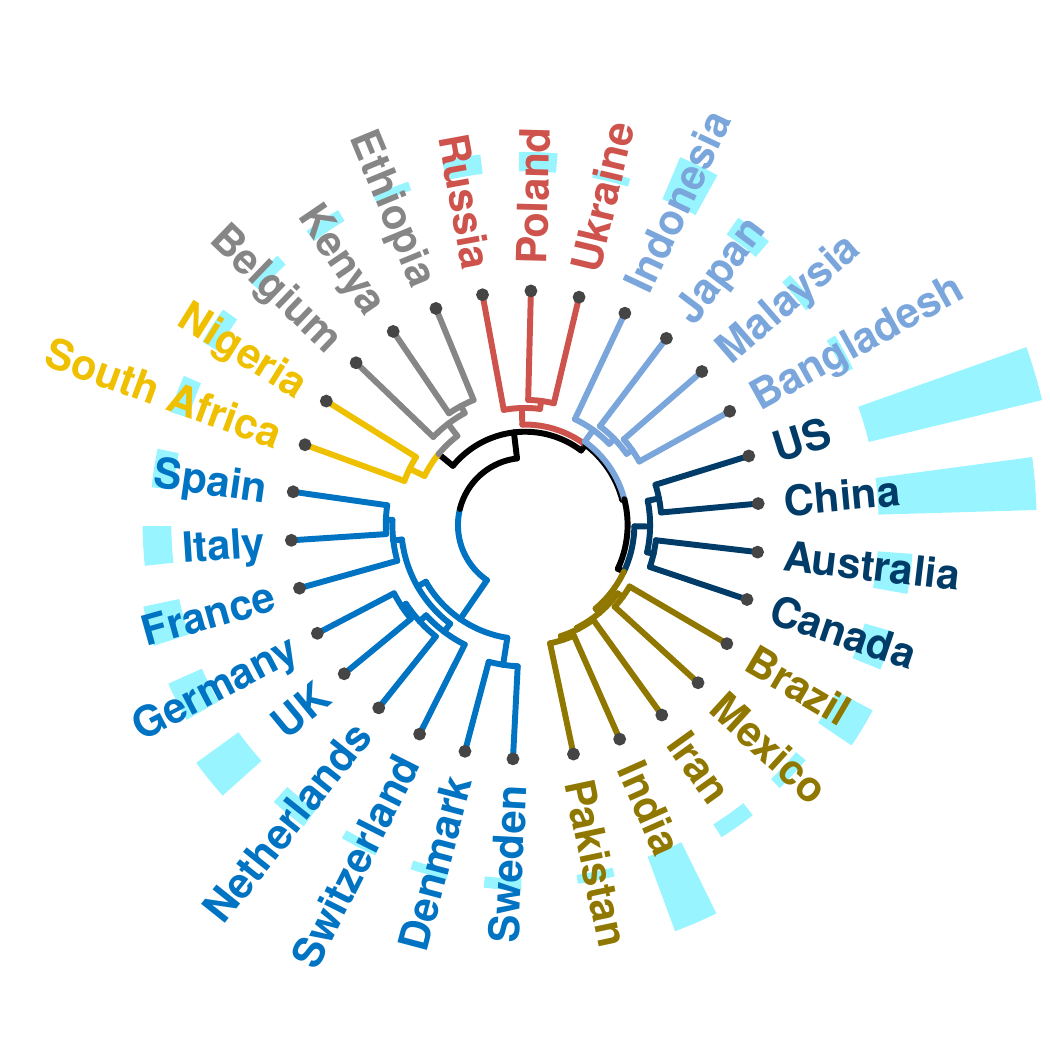}}
\end{flushleft}
	\end{minipage}
	\begin{minipage}{0.33\hsize}
\begin{flushleft}
\raisebox{-0.0cm}{\hspace{-0mm}\small\textrm{\textbf{(c)~ \particle}}}\\[6mm]
\raisebox{\height}{\includegraphics[trim=2.0cm 1.8cm 0cm 1.5cm, align=c, scale=\csize, vmargin=0mm]{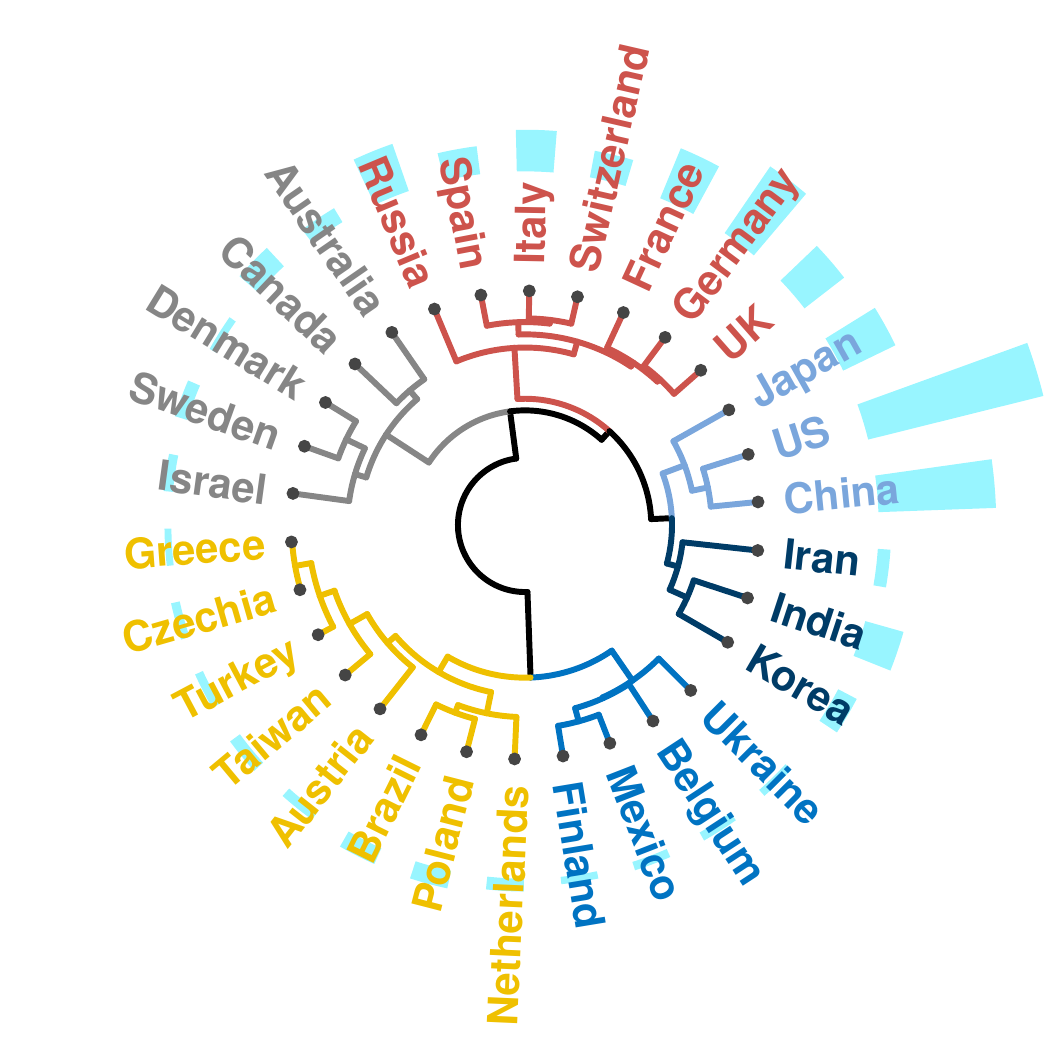}}
\end{flushleft}
	\end{minipage}
    \end{tabular}

\hspace{-0.5cm}{\marrow}\quad\dotfill

    \begin{tabular}{c}
    \begin{minipage}{0.03\hsize}
\begin{flushleft}
    \hspace{-0.7cm}\rotatebox{90}{\period{3}{2001--2010}}
\end{flushleft}
	\end{minipage}
	\begin{minipage}{0.33\hsize}
\begin{flushleft}
\raisebox{\height}{\includegraphics[trim=2.0cm 1.8cm 0cm 1.5cm, align=c, scale=\csize, vmargin=0mm]{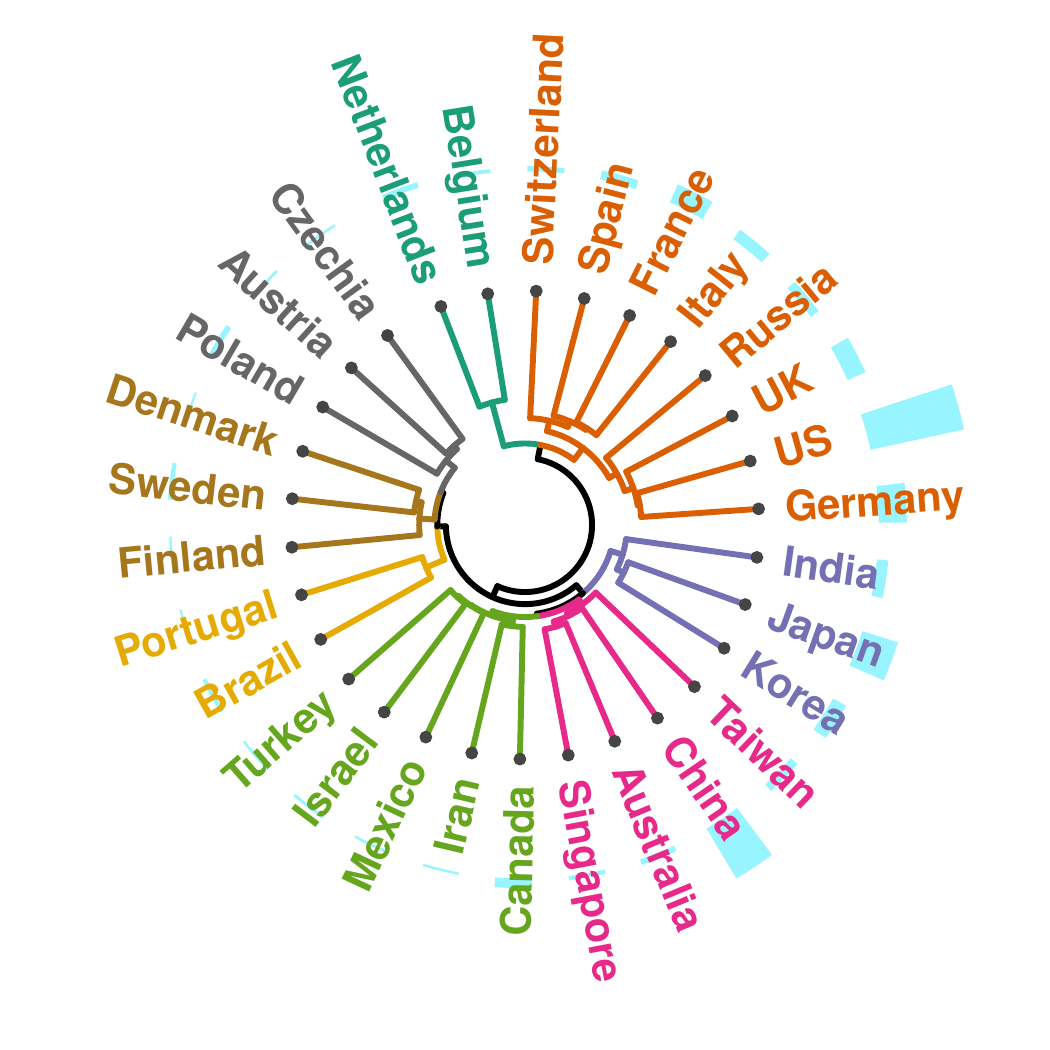}}
\end{flushleft}
    \end{minipage}
	\begin{minipage}{0.33\hsize}
\begin{flushleft}
\raisebox{\height}{\includegraphics[trim=2.0cm 1.8cm 0cm 1.5cm, align=c, scale=\csize, vmargin=0mm]{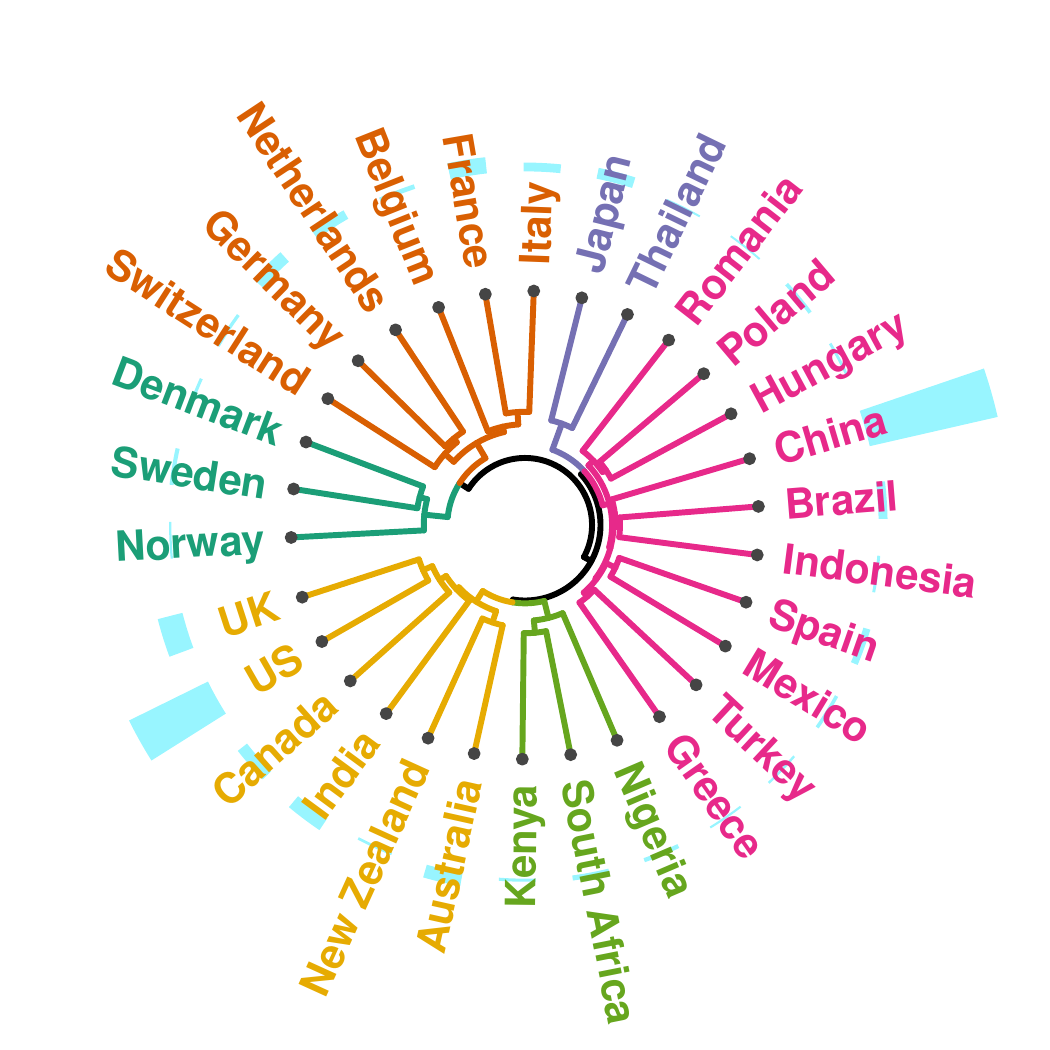}}
\end{flushleft}
	\end{minipage}
	\begin{minipage}{0.33\hsize}
\begin{flushleft}
\raisebox{\height}{\includegraphics[trim=2.0cm 1.8cm 0cm 1.5cm, align=c, scale=\csize, vmargin=0mm]{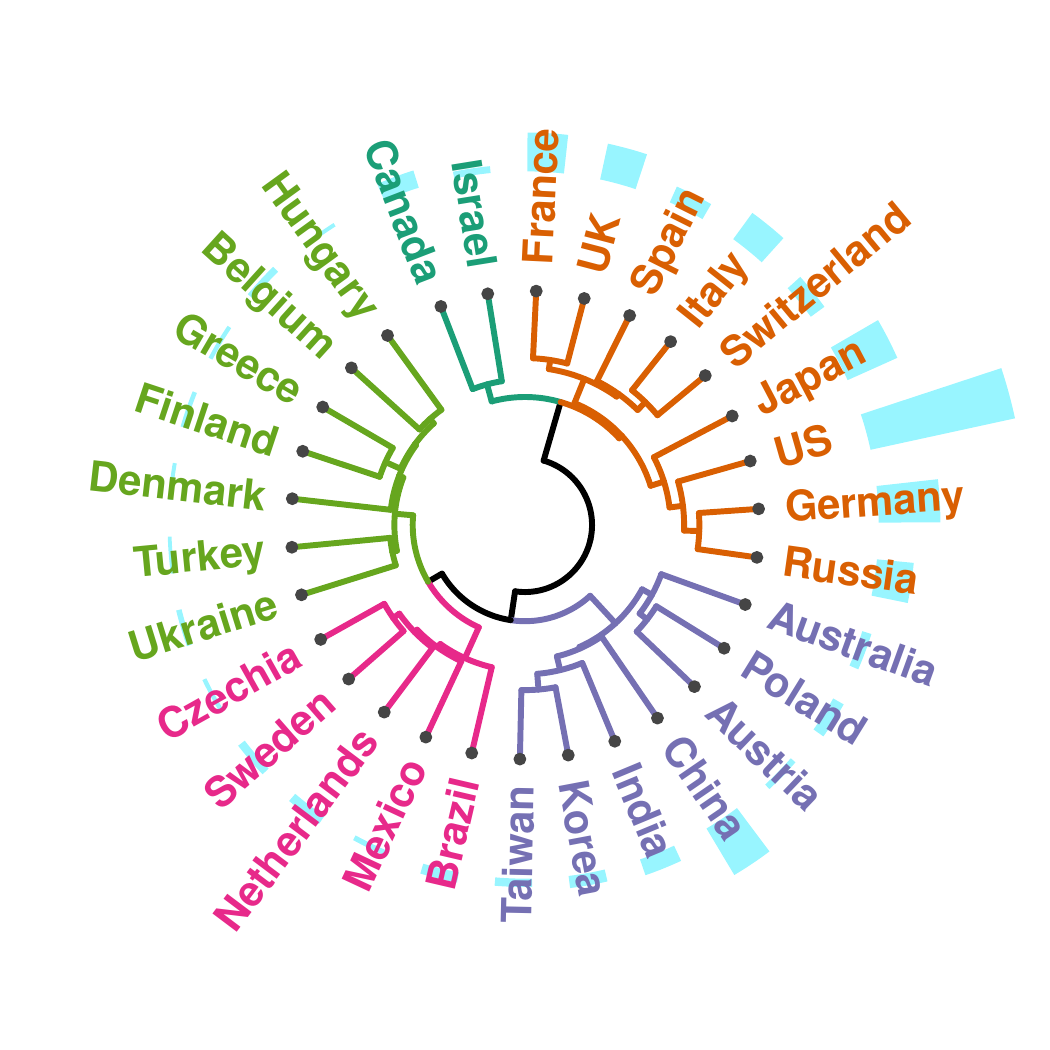}}
\end{flushleft}
	\end{minipage}
    \end{tabular}

\hspace{-0.5cm}{\marrow}\quad\dotfill

    \begin{tabular}{c}
    \begin{minipage}{0.03\hsize}
\begin{flushleft}
    \hspace{-0.7cm}\rotatebox{90}{\period{2}{1991--2000}}
\end{flushleft}
	\end{minipage}
	\begin{minipage}{0.33\hsize}
\begin{flushleft}
\raisebox{\height}{\includegraphics[trim=2.0cm 1.8cm 0cm 1.5cm, align=c, scale=\csize, vmargin=0mm]{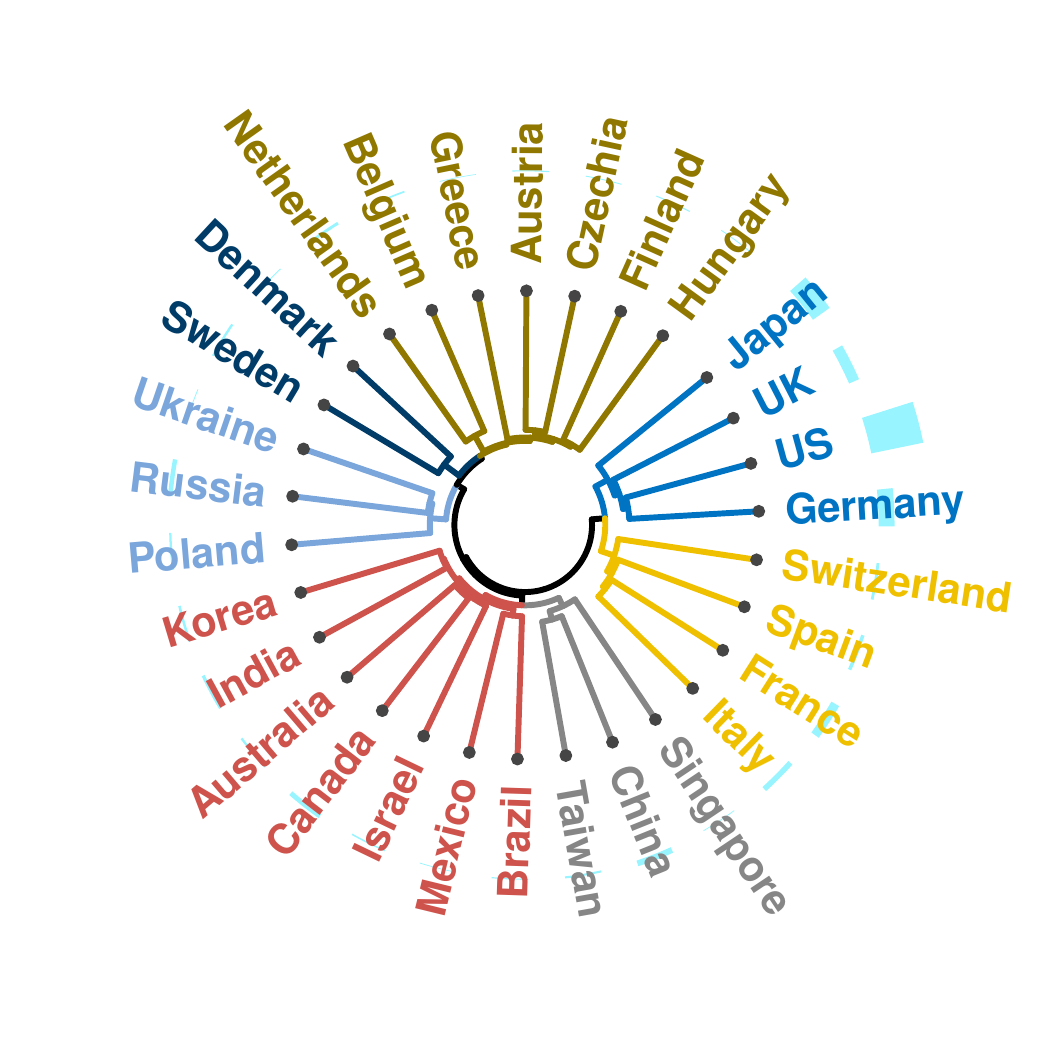}}
\end{flushleft}
    \end{minipage}
	\begin{minipage}{0.33\hsize}
\begin{flushleft}
\raisebox{\height}{\includegraphics[trim=2.0cm 1.8cm 0cm 1.5cm, align=c, scale=\csize, vmargin=0mm]{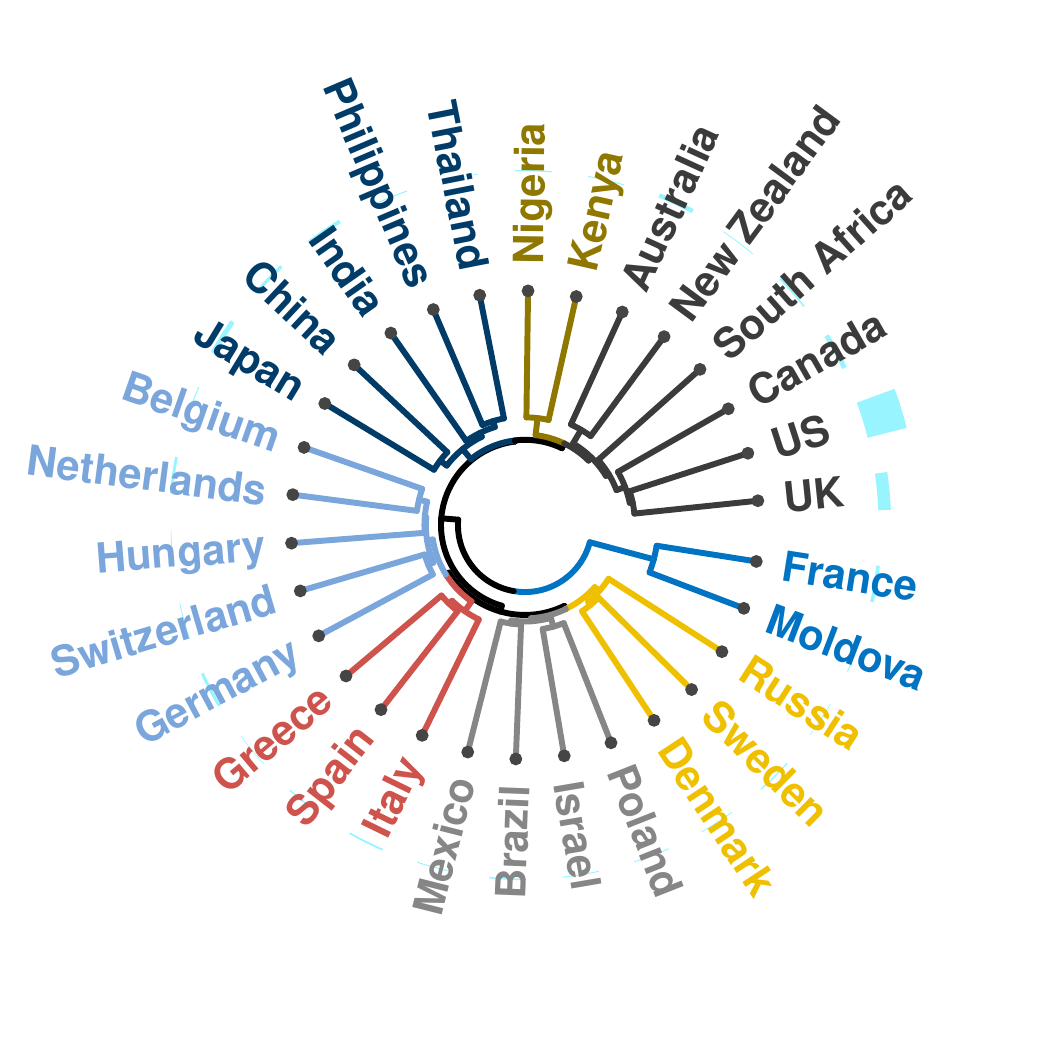}}
\end{flushleft}
	\end{minipage}
	\begin{minipage}{0.33\hsize}
\begin{flushleft}
\raisebox{\height}{\includegraphics[trim=2.0cm 1.8cm 0cm 1.5cm, align=c, scale=\csize, vmargin=0mm]{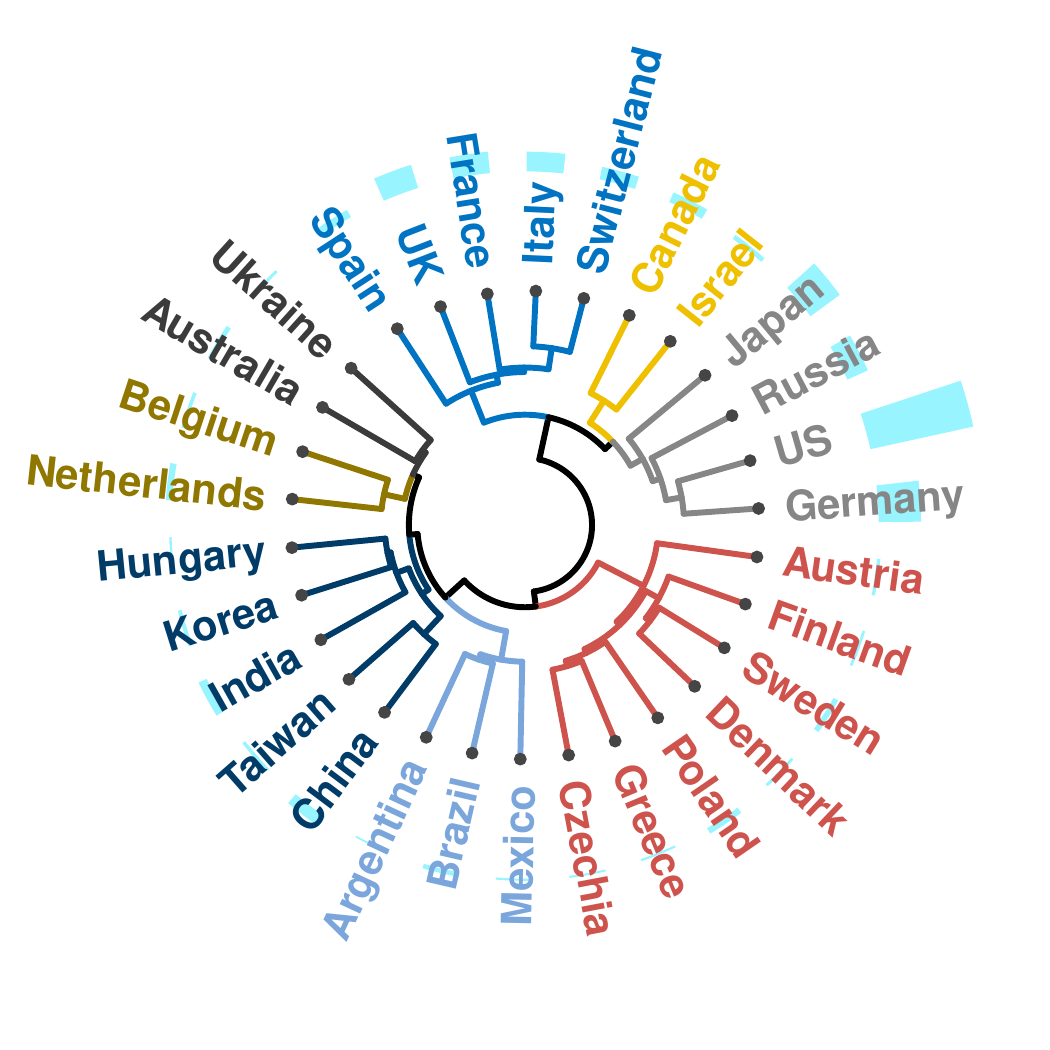}}
\end{flushleft}
	\end{minipage}
    \end{tabular}

\hspace{-0.5cm}{\marrow}\quad\dotfill

    \begin{tabular}{c}
    \begin{minipage}{0.03\hsize}
\begin{flushleft}
    \hspace{-0.7cm}\rotatebox{90}{\period{1}{1971--1990}}
\end{flushleft}
	\end{minipage}
	\begin{minipage}{0.33\hsize}
\begin{flushleft}
\raisebox{\height}{\includegraphics[trim=2.0cm 1.8cm 0cm 1.5cm, align=c, scale=\csize, vmargin=0mm]{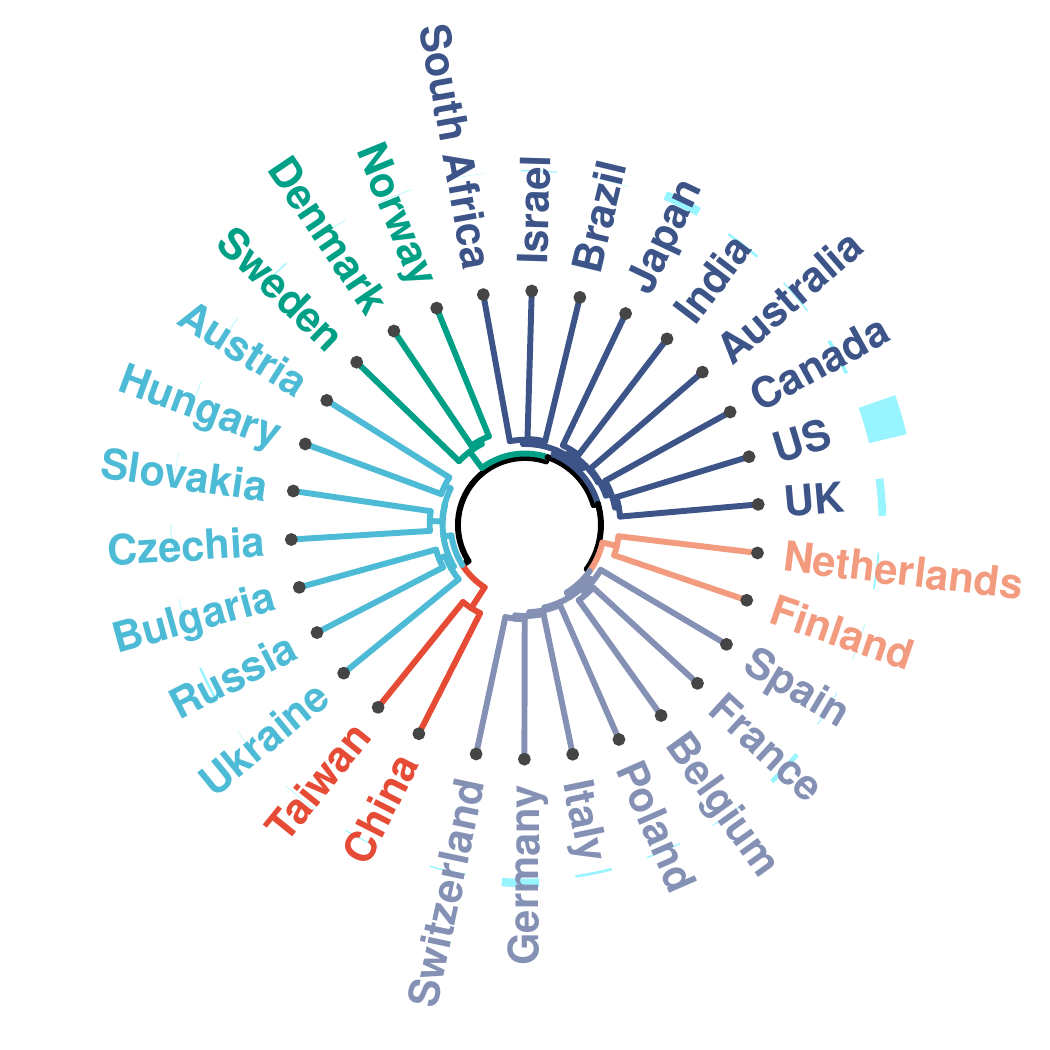}}
\end{flushleft}
    \end{minipage}
	\begin{minipage}{0.33\hsize}
\begin{flushleft}
\raisebox{\height}{\includegraphics[trim=2.0cm 1.8cm 0cm 1.5cm, align=c, scale=\csize, vmargin=0mm]{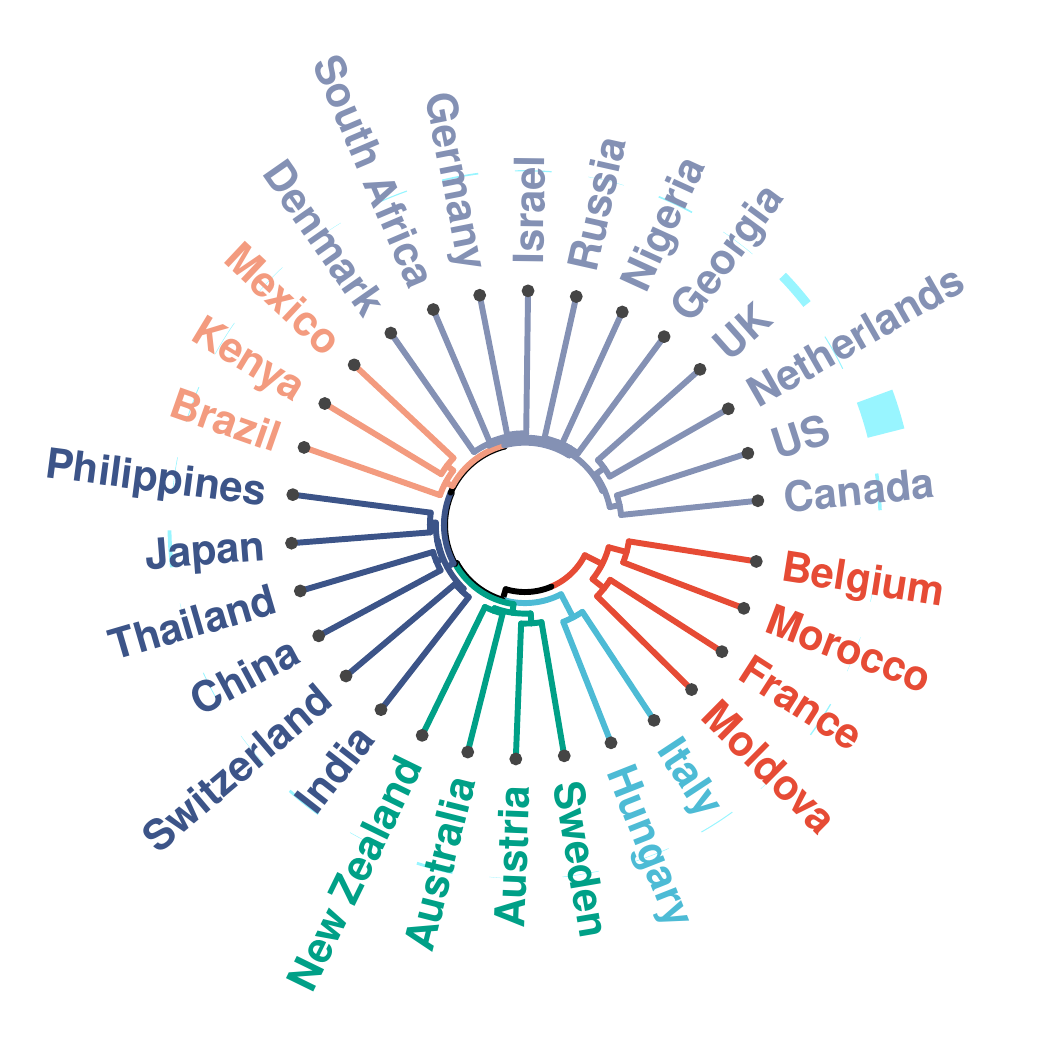}}
\end{flushleft}
	\end{minipage}
	\begin{minipage}{0.33\hsize}
\begin{flushleft}
\raisebox{\height}{\includegraphics[trim=2.0cm 1.8cm 0cm 1.5cm, align=c, scale=\csize, vmargin=0mm]{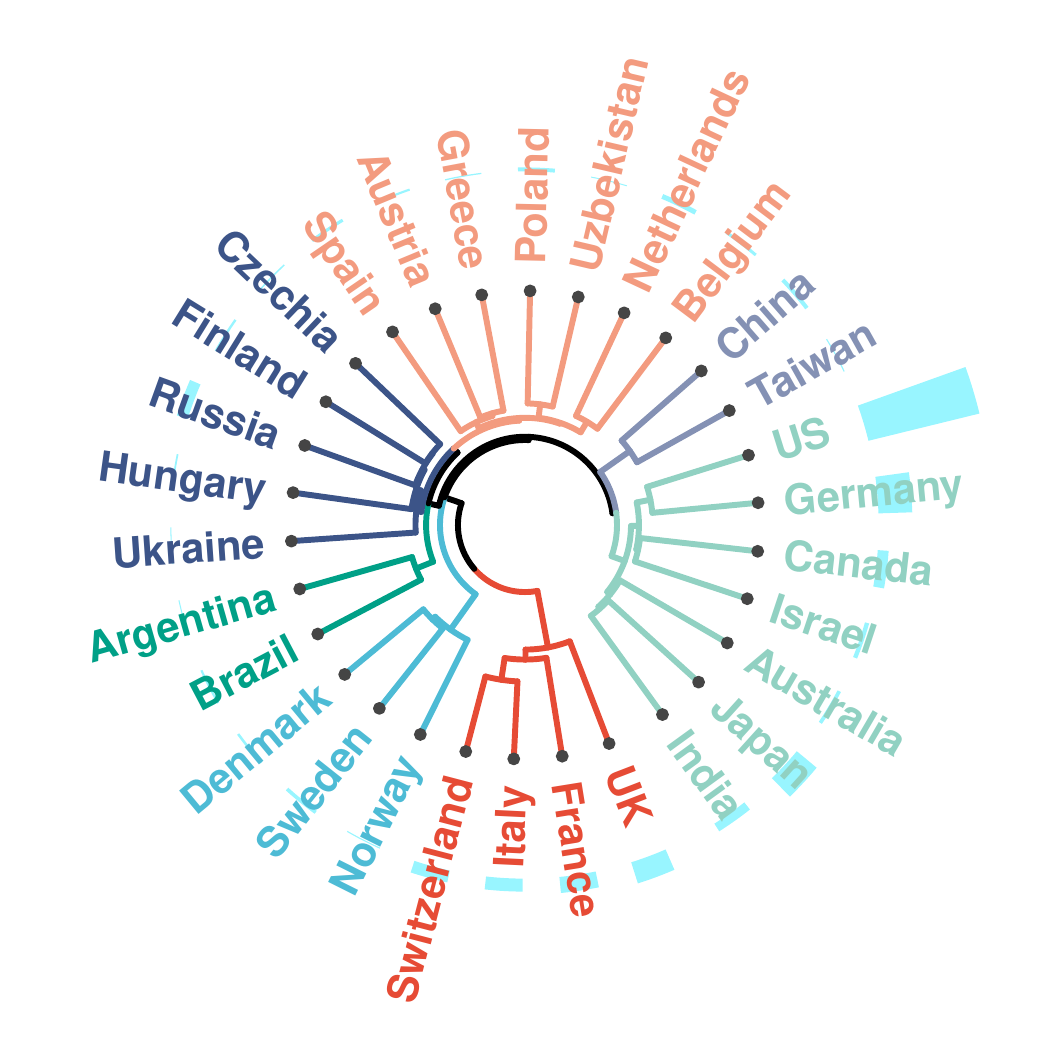}}
\end{flushleft}
	\end{minipage}
    \end{tabular}
\end{subfigure}
\vspace{2.5mm}
\caption{\textbf{Evolution of international research collaboration clusters.}}
\label{fig:cdend_2}
\end{figure}
}
\afterpage{\clearpage%
\begin{figure}[htp]\ContinuedFloat
\centering
\begin{subfigure}{1.0\textwidth}
\vspace{-0.5cm}
    \begin{tabular}{c}
    \begin{minipage}{0.03\hsize}
\begin{flushleft}
    \hspace{-0.7cm}\rotatebox{90}{\period{4}{2011--2020}}
\end{flushleft}
	\end{minipage}
	\begin{minipage}{0.33\hsize}
\begin{flushleft}
\raisebox{-0.0cm}{\hspace{-0mm}\small\textrm{\textbf{(d)~ \aerospace}}}\\[6mm]
\raisebox{\height}{\includegraphics[trim=2.0cm 1.8cm 0cm 1.5cm, align=c, scale=\csize, vmargin=0mm]{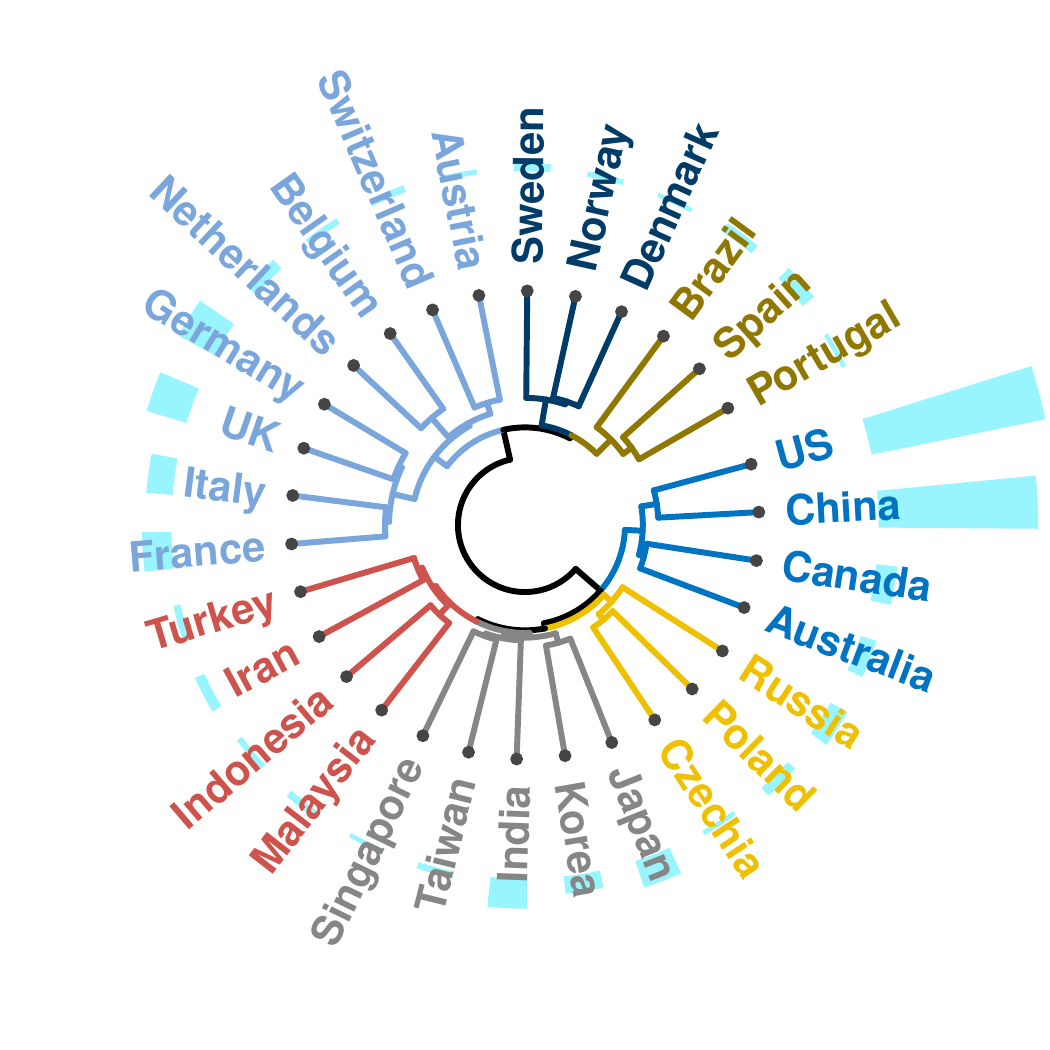}}
\end{flushleft}
    \end{minipage}
	\begin{minipage}{0.33\hsize}
\begin{flushleft}
\raisebox{-0.0cm}{\hspace{-0mm}\small\textrm{\textbf{(e)~ \nuclear}}}\\[6mm]
\raisebox{\height}{\includegraphics[trim=2.0cm 1.8cm 0cm 1.5cm, align=c, scale=\csize, vmargin=0mm]{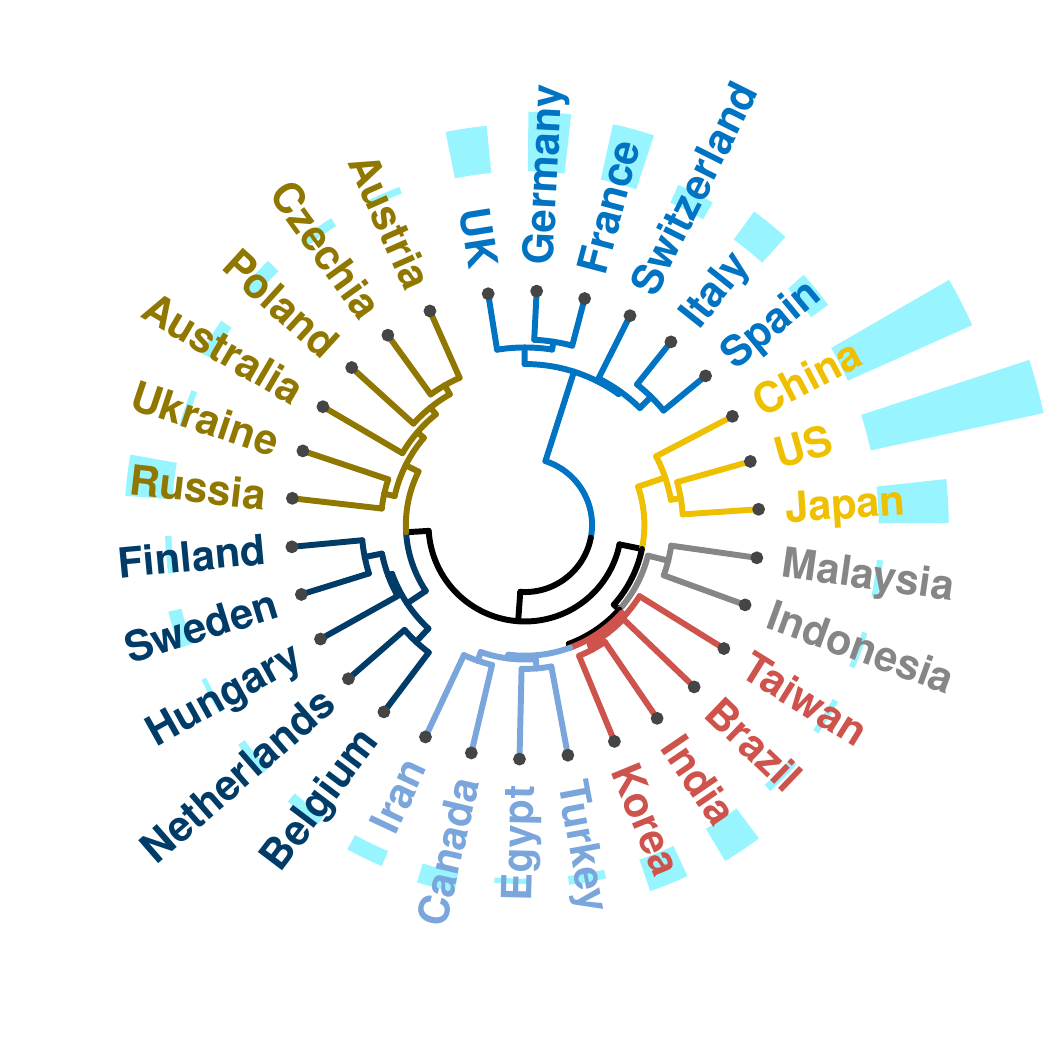}}
\end{flushleft}
	\end{minipage}
	\begin{minipage}{0.33\hsize}
\begin{flushleft}
\raisebox{-0.0cm}{\hspace{-0mm}\small\textrm{\textbf{(f)~ \marine}}}\\[6mm]
\raisebox{\height}{\includegraphics[trim=2.0cm 1.8cm 0cm 1.5cm, align=c, scale=\csize, vmargin=0mm]{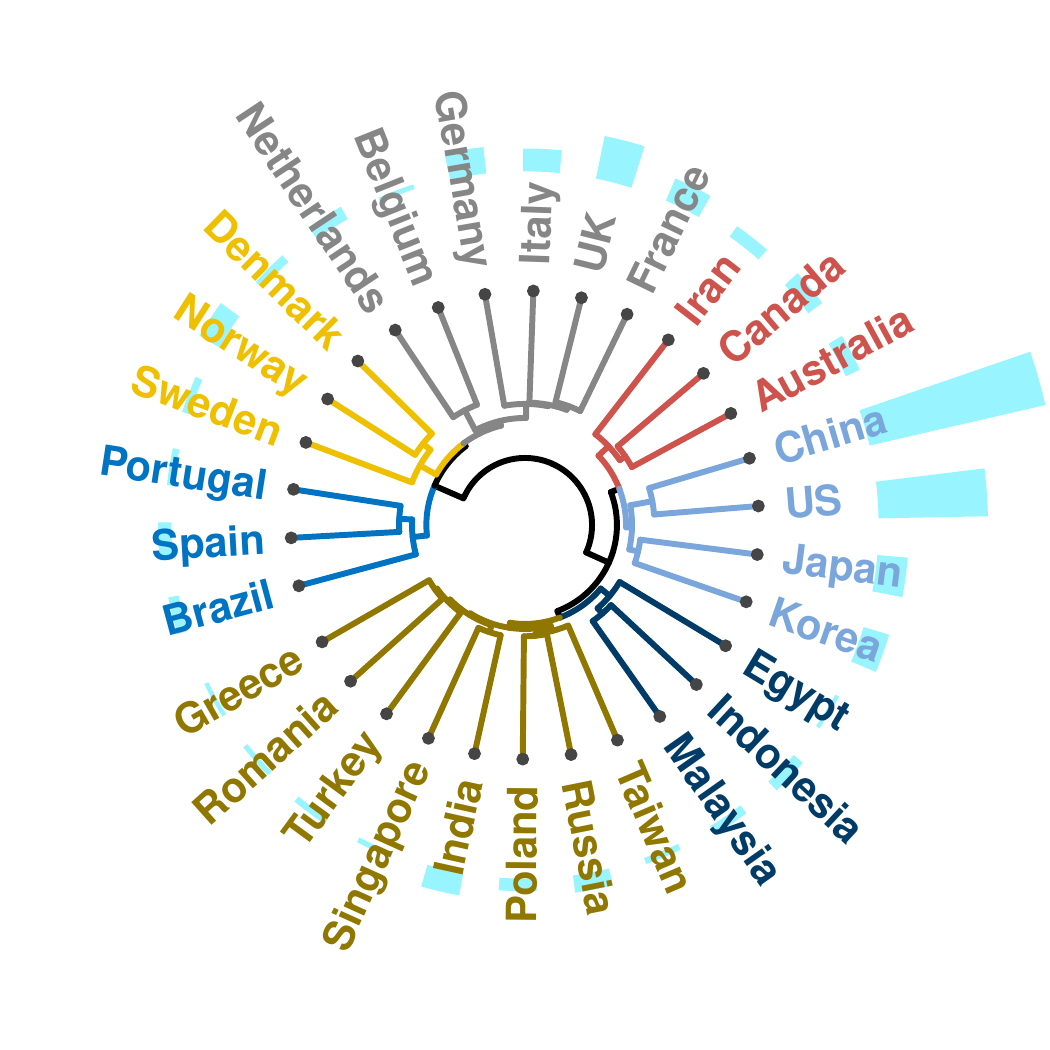}}
\end{flushleft}
	\end{minipage}
    \end{tabular}

\hspace{-0.5cm}{\marrow}\quad\dotfill

    \begin{tabular}{c}
    \begin{minipage}{0.03\hsize}
\begin{flushleft}
    \hspace{-0.7cm}\rotatebox{90}{\period{3}{2001--2010}}
\end{flushleft}
	\end{minipage}
	\begin{minipage}{0.33\hsize}
\begin{flushleft}
\raisebox{\height}{\includegraphics[trim=2.0cm 1.8cm 0cm 1.5cm, align=c, scale=\csize, vmargin=0mm]{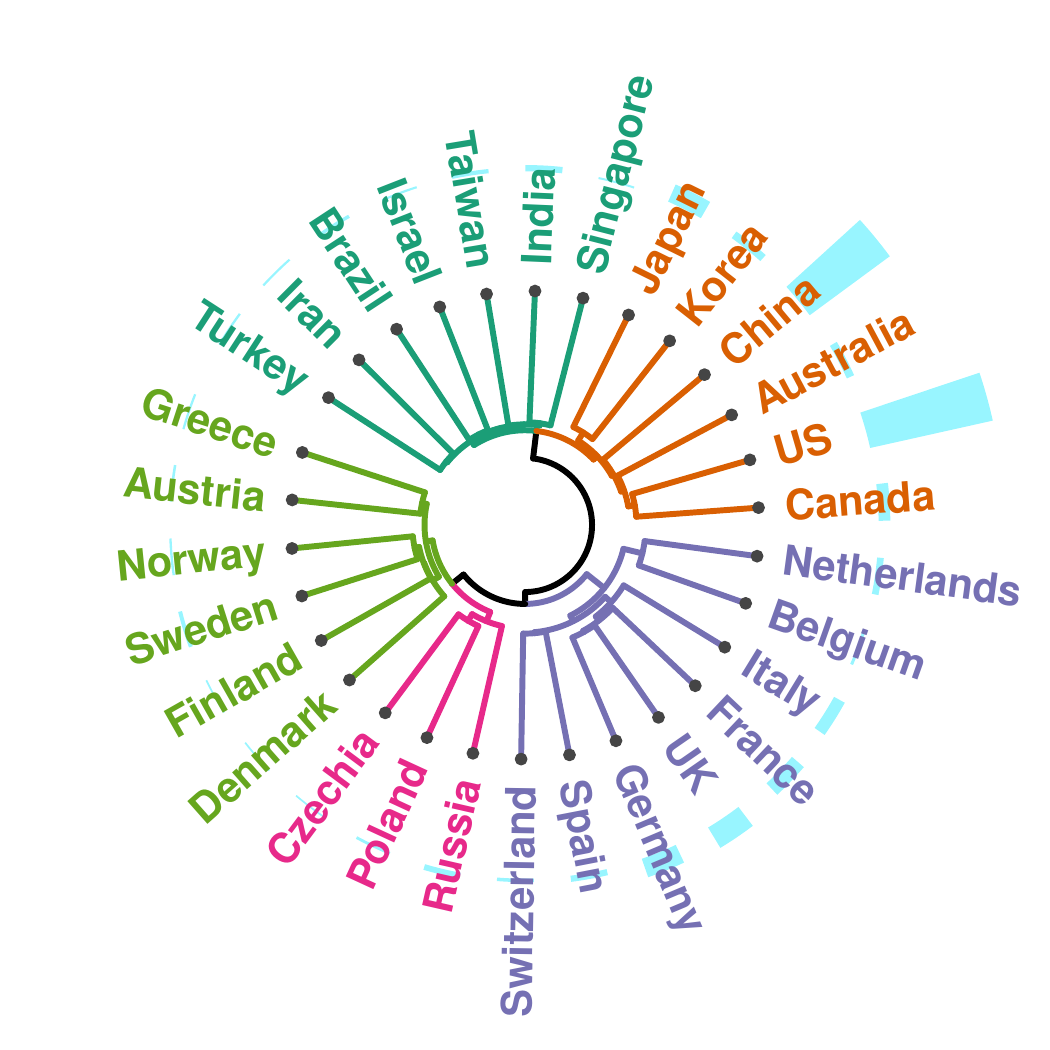}}
\end{flushleft}
    \end{minipage}
	\begin{minipage}{0.33\hsize}
\begin{flushleft}
\raisebox{\height}{\includegraphics[trim=2.0cm 1.8cm 0cm 1.5cm, align=c, scale=\csize, vmargin=0mm]{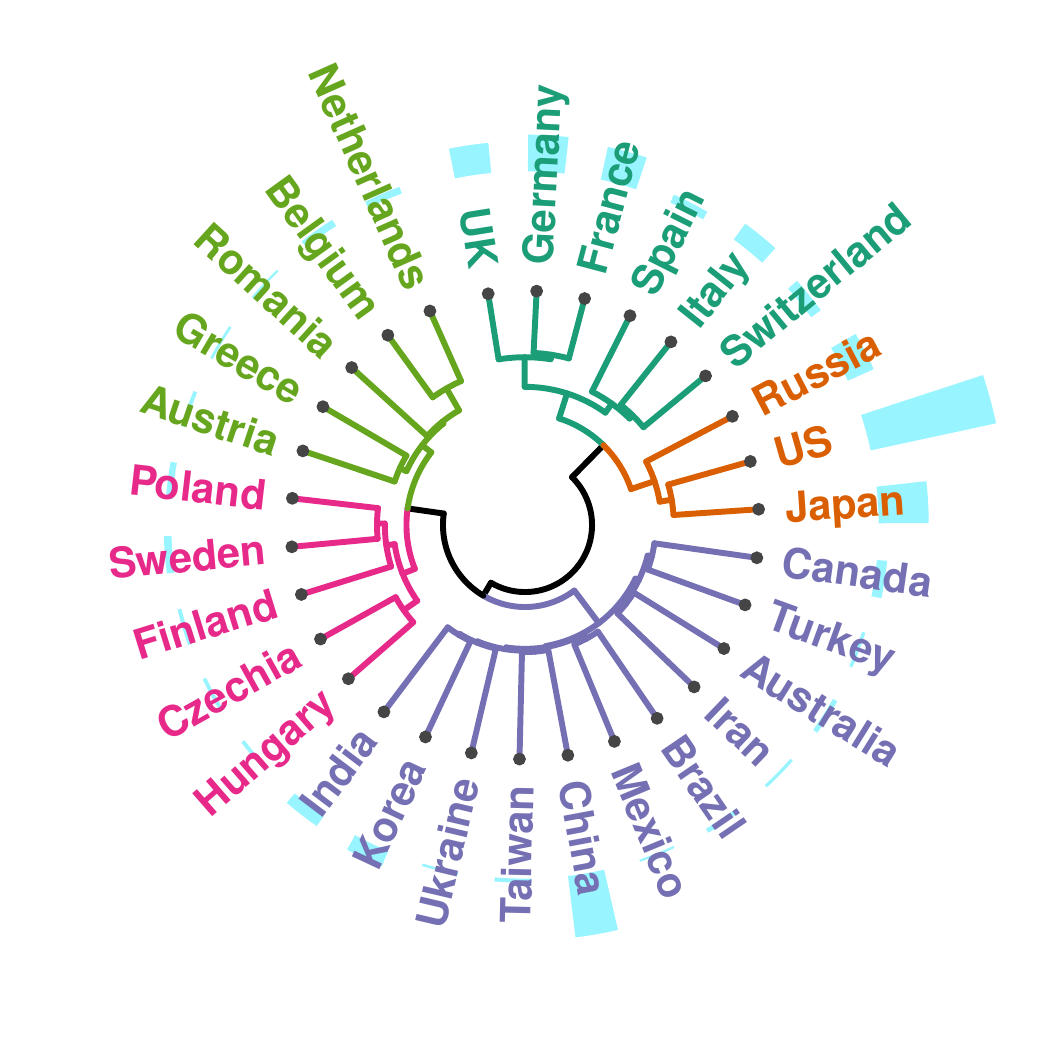}}
\end{flushleft}
	\end{minipage}
	\begin{minipage}{0.33\hsize}
\begin{flushleft}
\raisebox{\height}{\includegraphics[trim=2.0cm 1.8cm 0cm 1.5cm, align=c, scale=\csize, vmargin=0mm]{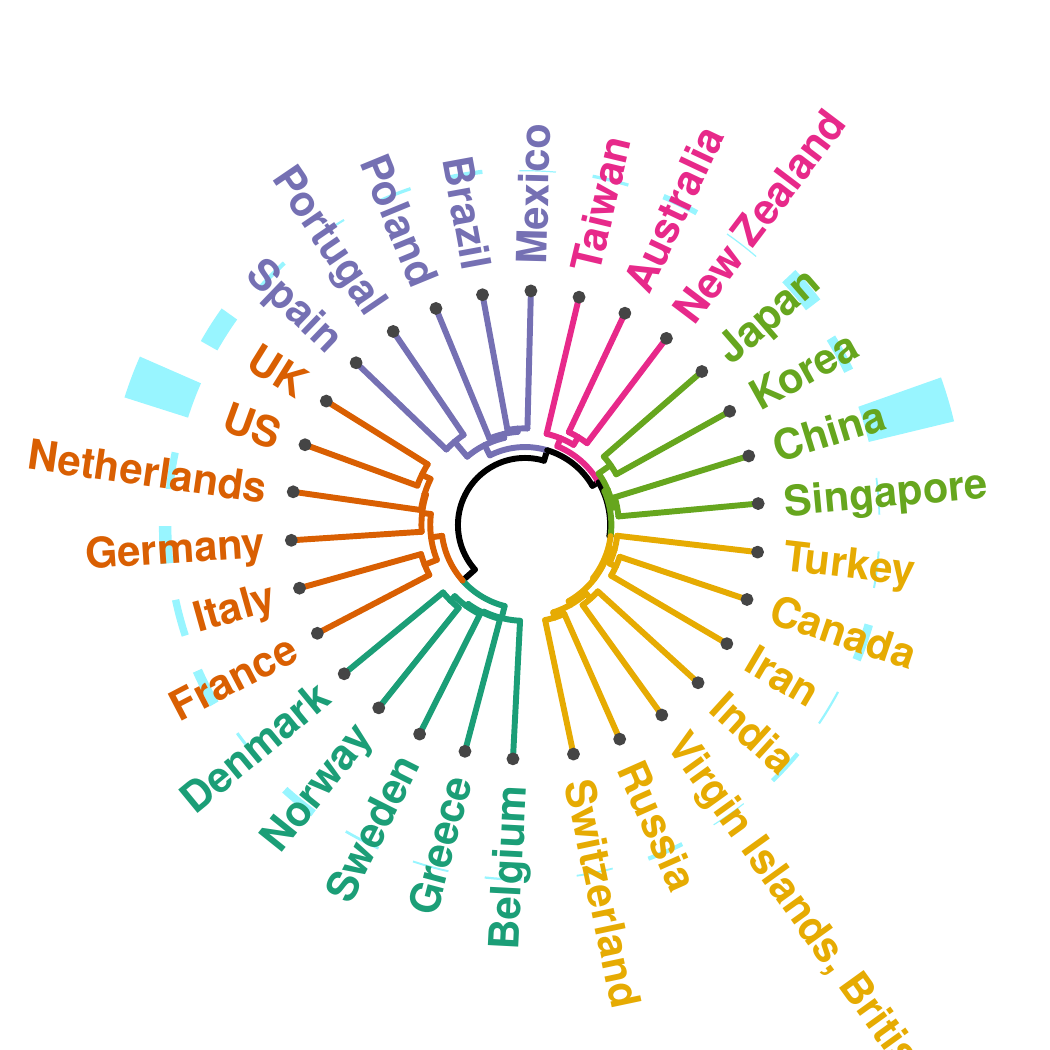}}
\end{flushleft}
	\end{minipage}
    \end{tabular}

\hspace{-0.5cm}{\marrow}\quad\dotfill

    \begin{tabular}{c}
    \begin{minipage}{0.03\hsize}
\begin{flushleft}
    \hspace{-0.7cm}\rotatebox{90}{\period{2}{1991--2000}}
\end{flushleft}
	\end{minipage}
	\begin{minipage}{0.33\hsize}
\begin{flushleft}
\raisebox{\height}{\includegraphics[trim=2.0cm 1.8cm 0cm 1.5cm, align=c, scale=\csize, vmargin=0mm]{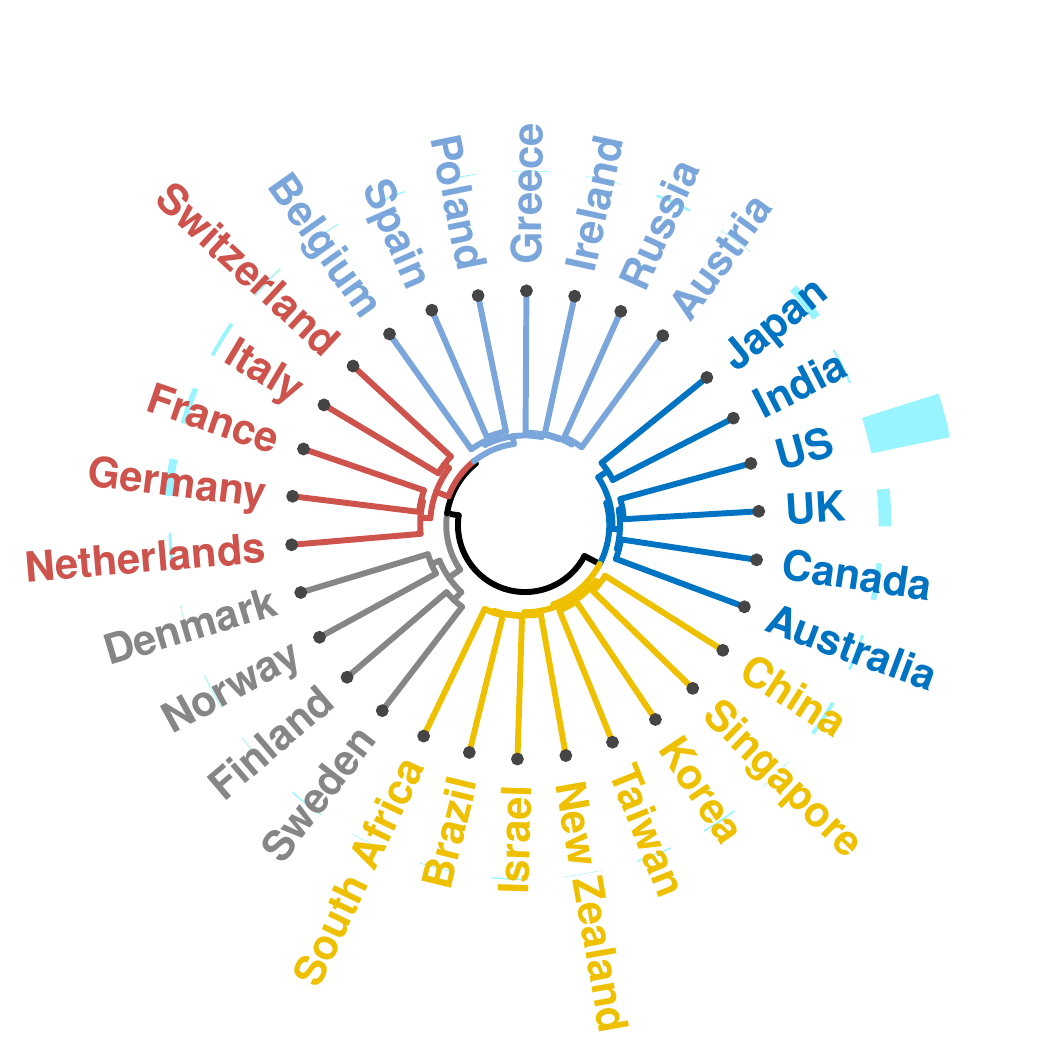}}
\end{flushleft}
    \end{minipage}
	\begin{minipage}{0.33\hsize}
\begin{flushleft}
\raisebox{\height}{\includegraphics[trim=2.0cm 1.8cm 0cm 1.5cm, align=c, scale=\csize, vmargin=0mm]{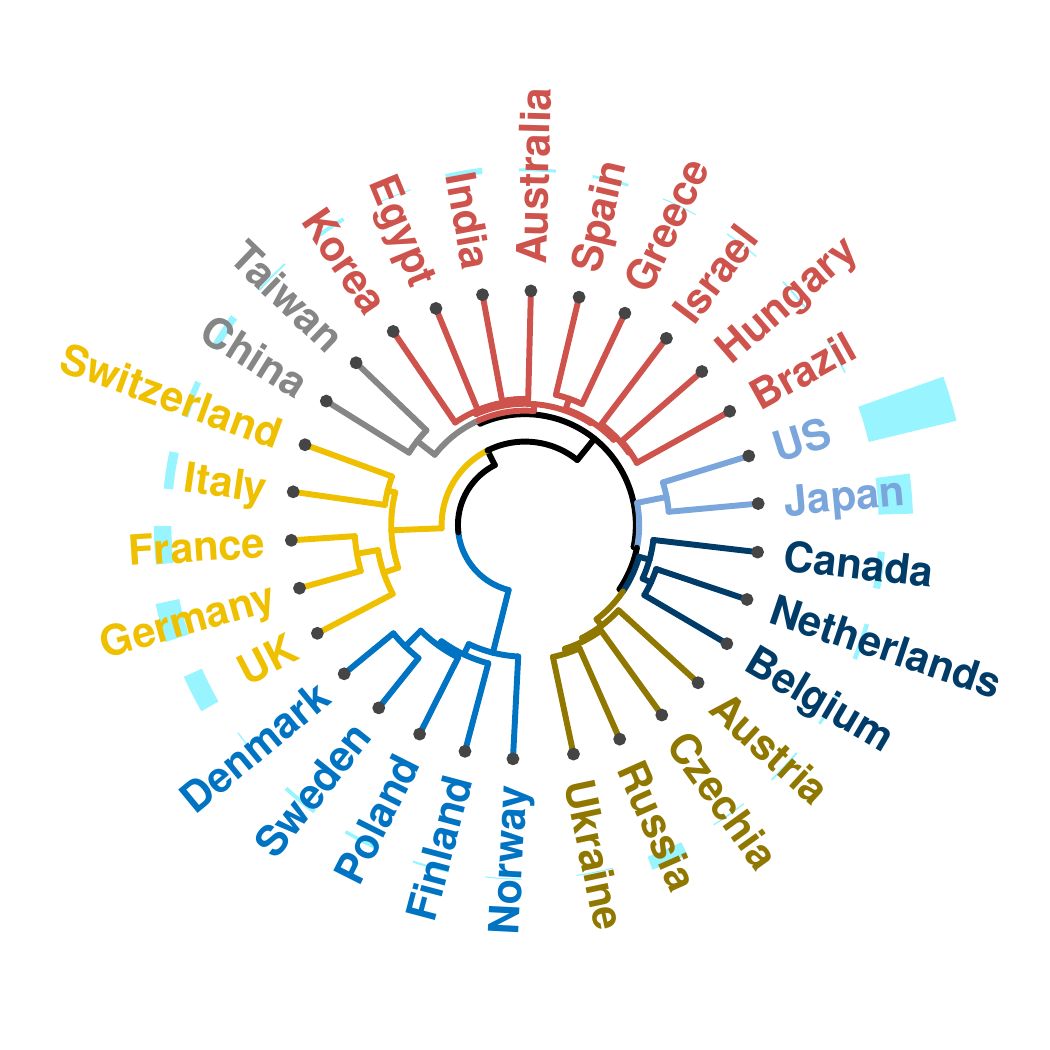}}
\end{flushleft}
	\end{minipage}
	\begin{minipage}{0.33\hsize}
\begin{flushleft}
\raisebox{\height}{\includegraphics[trim=2.0cm 1.8cm 0cm 1.5cm, align=c, scale=\csize, vmargin=0mm]{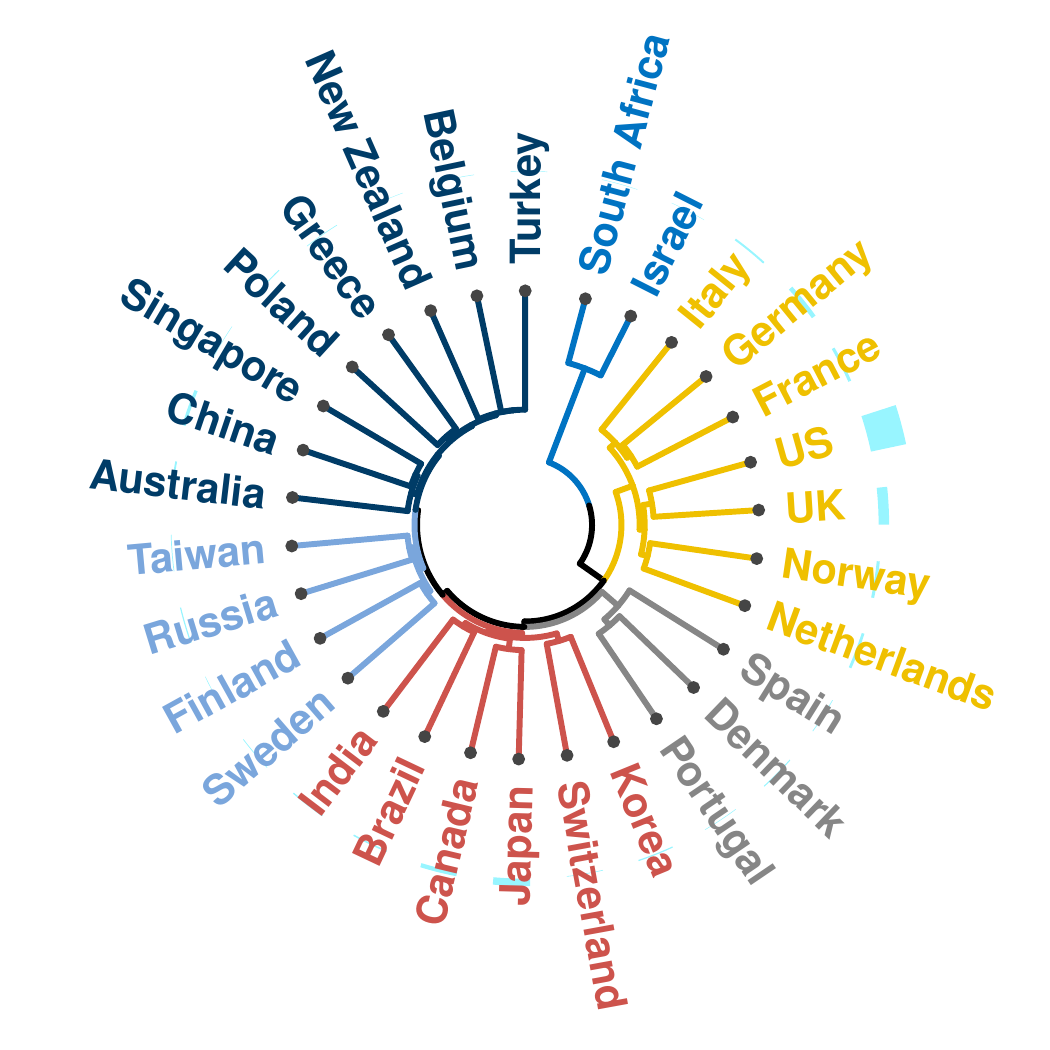}}
\end{flushleft}
	\end{minipage}
    \end{tabular}

\hspace{-0.5cm}{\marrow}\quad\dotfill

    \begin{tabular}{c}
    \begin{minipage}{0.03\hsize}
\begin{flushleft}
    \hspace{-0.7cm}\rotatebox{90}{\period{1}{1971--1990}}
\end{flushleft}
	\end{minipage}
	\begin{minipage}{0.33\hsize}
\begin{flushleft}
\raisebox{\height}{\includegraphics[trim=2.0cm 1.8cm 0cm 1.5cm, align=c, scale=\csize, vmargin=0mm]{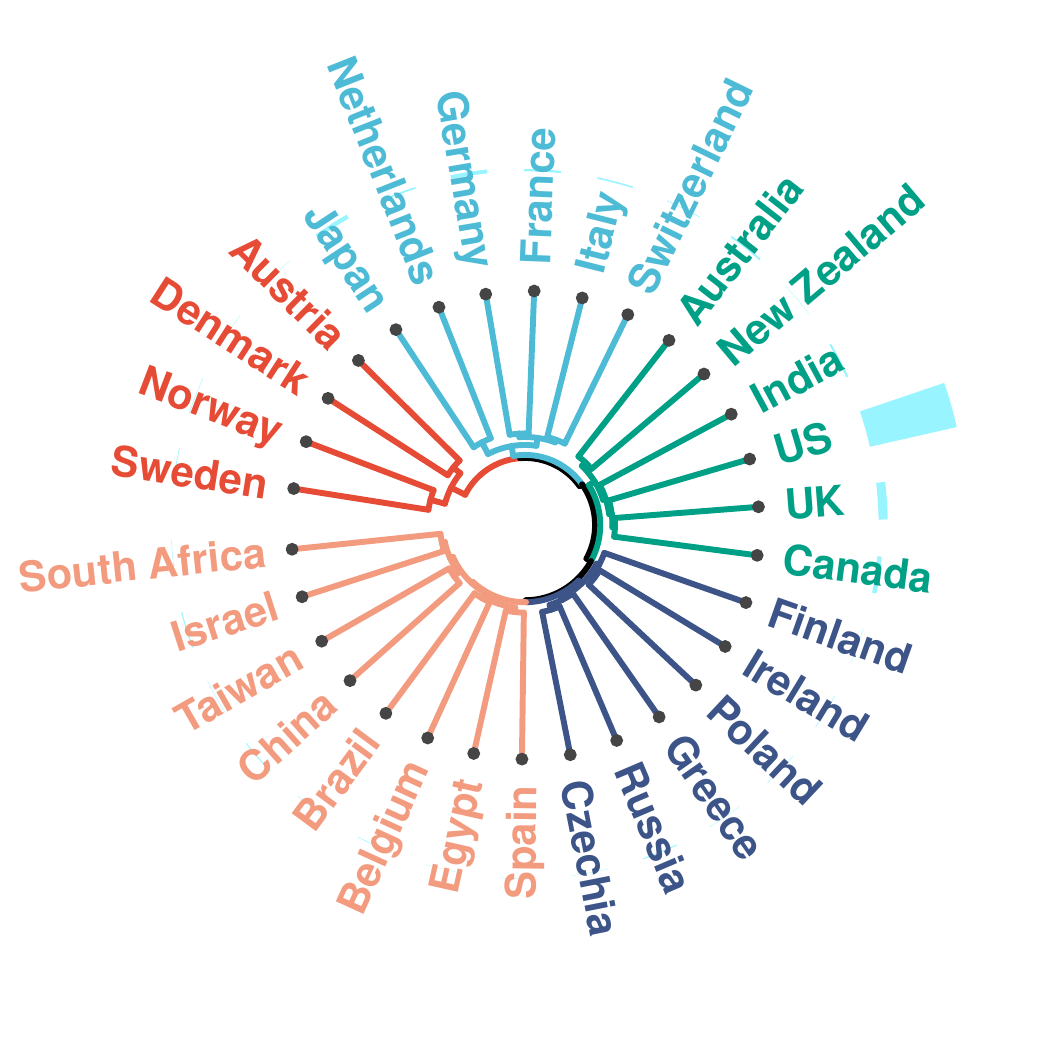}}
\end{flushleft}
    \end{minipage}
	\begin{minipage}{0.33\hsize}
\begin{flushleft}
\raisebox{\height}{\includegraphics[trim=2.0cm 1.8cm 0cm 1.5cm, align=c, scale=\csize, vmargin=0mm]{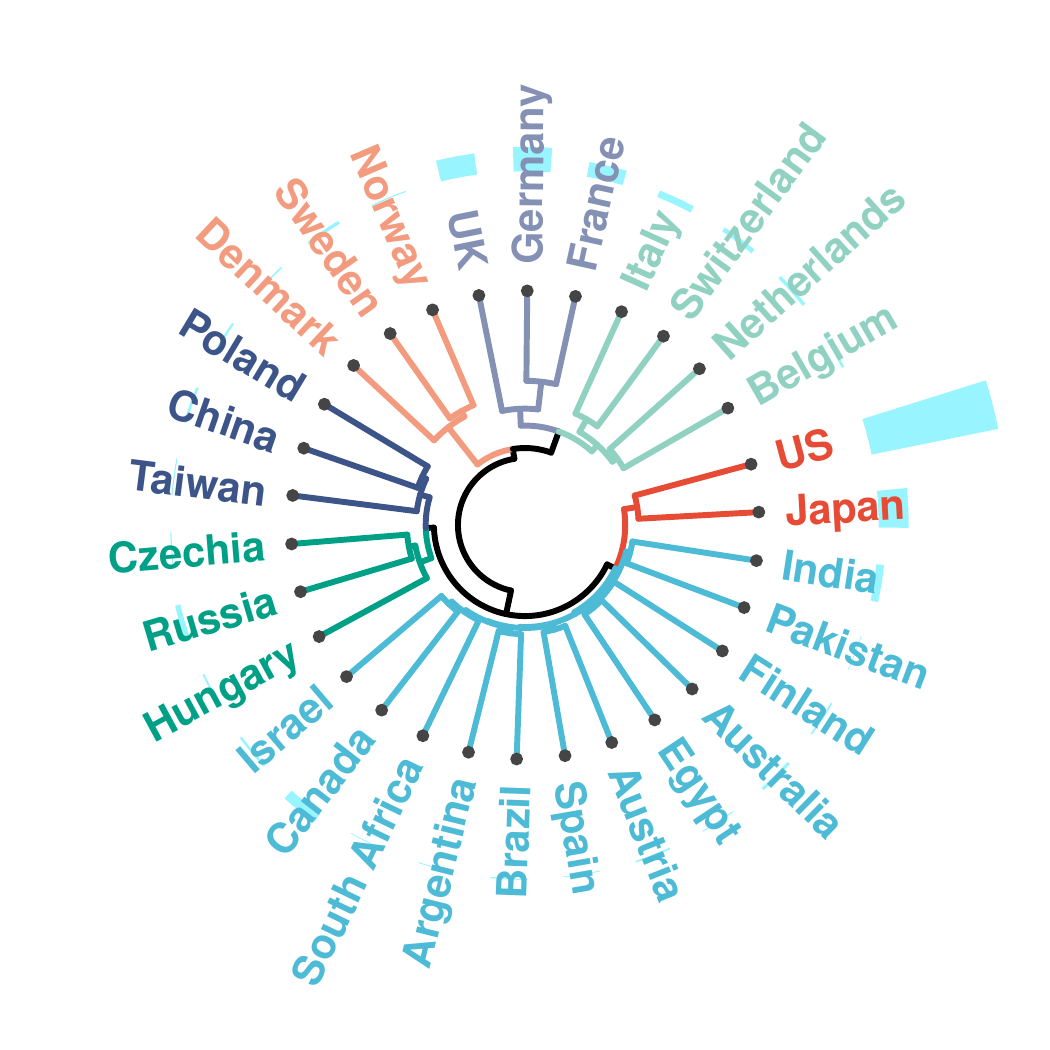}}
\end{flushleft}
	\end{minipage}
	\begin{minipage}{0.33\hsize}
\begin{flushleft}
\raisebox{\height}{\includegraphics[trim=2.0cm 1.8cm 0cm 1.5cm, align=c, scale=\csize, vmargin=0mm]{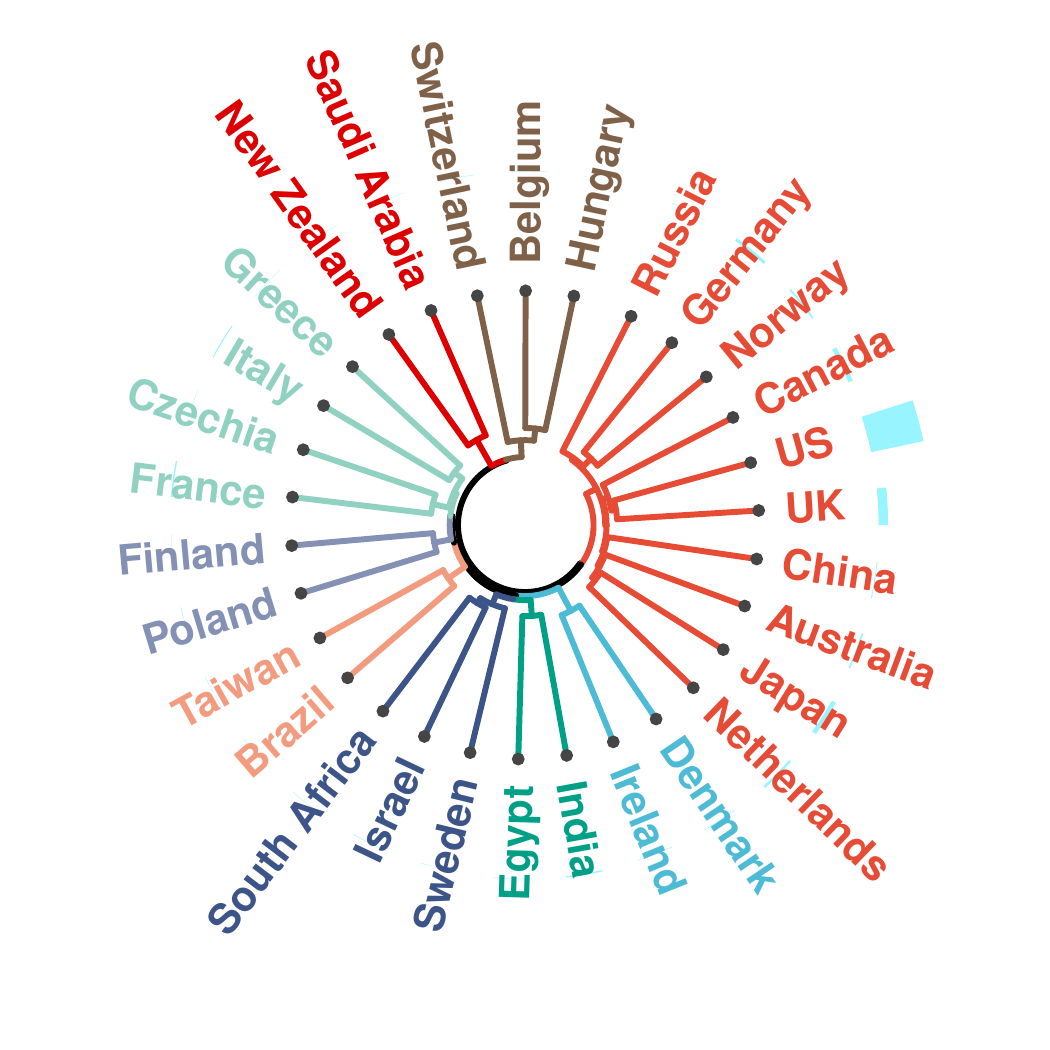}}
\end{flushleft}
	\end{minipage}
    \end{tabular}
\end{subfigure}
\vspace{2.5mm}
\caption{\textbf{Evolution of international research collaboration clusters. \emph{(Cont.)}}}
\label{fig:cdend_3}
\end{figure}
}
\afterpage{\clearpage%
\begin{figure}[htp]\ContinuedFloat
\centering
\begin{subfigure}{1.0\textwidth}
\vspace{-0.5cm}
    \begin{tabular}{c}
    \begin{minipage}{0.03\hsize}
\begin{flushleft}
    \hspace{-0.7cm}\rotatebox{90}{\period{4}{2011--2020}}
\end{flushleft}
	\end{minipage}
	\begin{minipage}{0.33\hsize}
\begin{flushleft}
\raisebox{-0.0cm}{\hspace{-0mm}\small\textrm{\textbf{(g)~ \neuro}}}\\[6mm]
\raisebox{\height}{\includegraphics[trim=2.0cm 1.8cm 0cm 1.5cm, align=c, scale=\csize, vmargin=0mm]{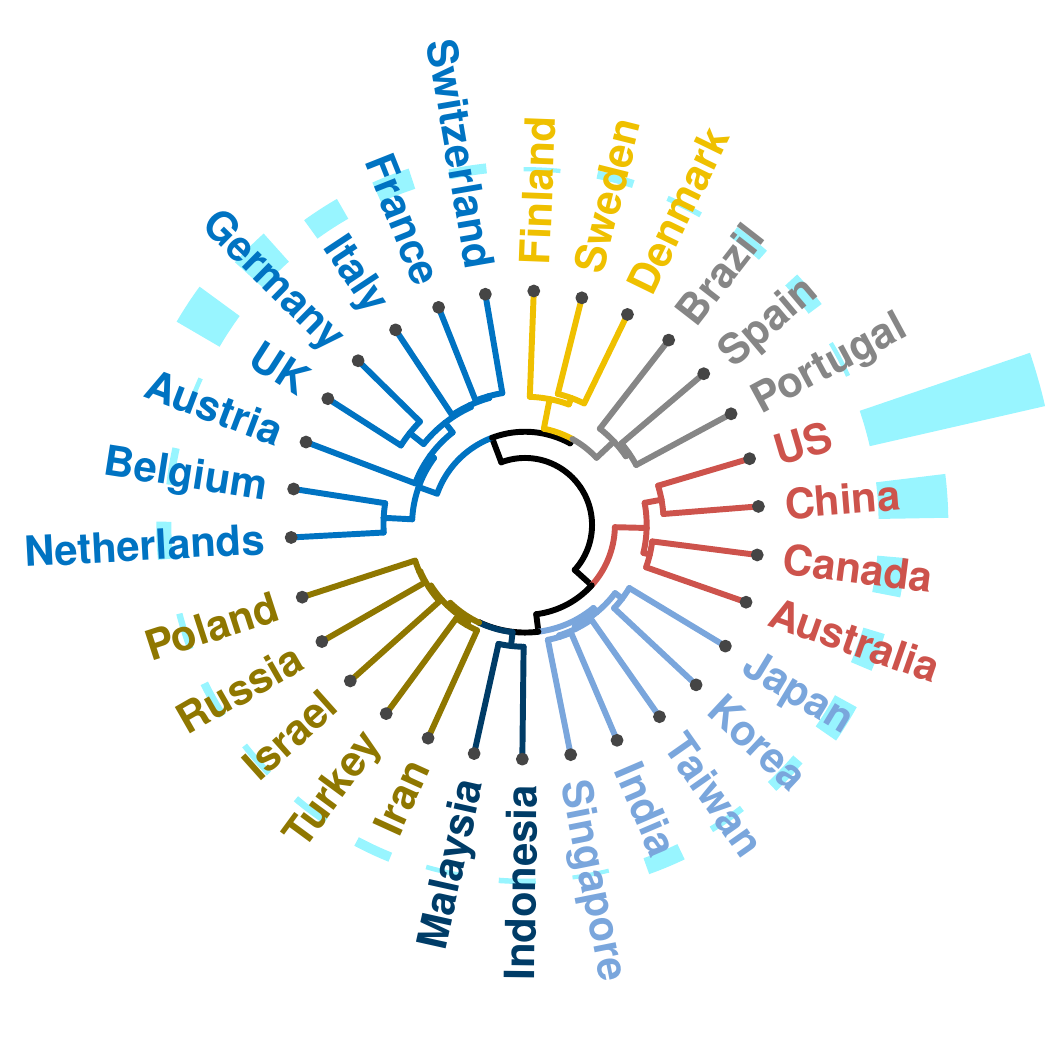}}
\end{flushleft}
    \end{minipage}
	\begin{minipage}{0.33\hsize}
\begin{flushleft}
\raisebox{-0.0cm}{\hspace{-0mm}\small\textrm{\textbf{(h)~ \condensed}}}\\[6mm]
\raisebox{\height}{\includegraphics[trim=2.0cm 1.8cm 0cm 1.5cm, align=c, scale=\csize, vmargin=0mm]{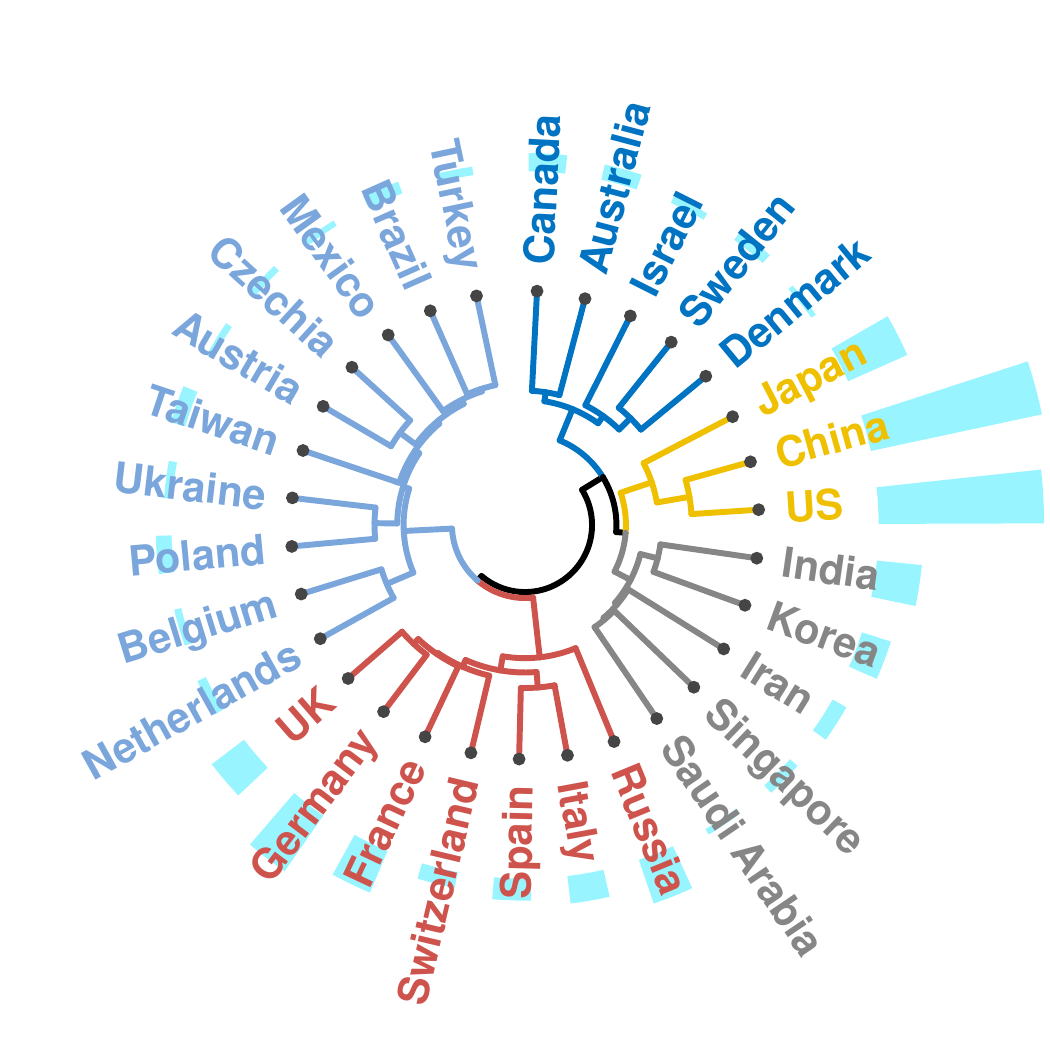}}
\end{flushleft}
	\end{minipage}
	\begin{minipage}{0.33\hsize}
\begin{flushleft}
\raisebox{-0.0cm}{\hspace{-0mm}\small\textrm{\textbf{(i)~ \envi}}}\\[6mm]
\raisebox{\height}{\includegraphics[trim=2.0cm 1.8cm 0cm 1.5cm, align=c, scale=\csize, vmargin=0mm]{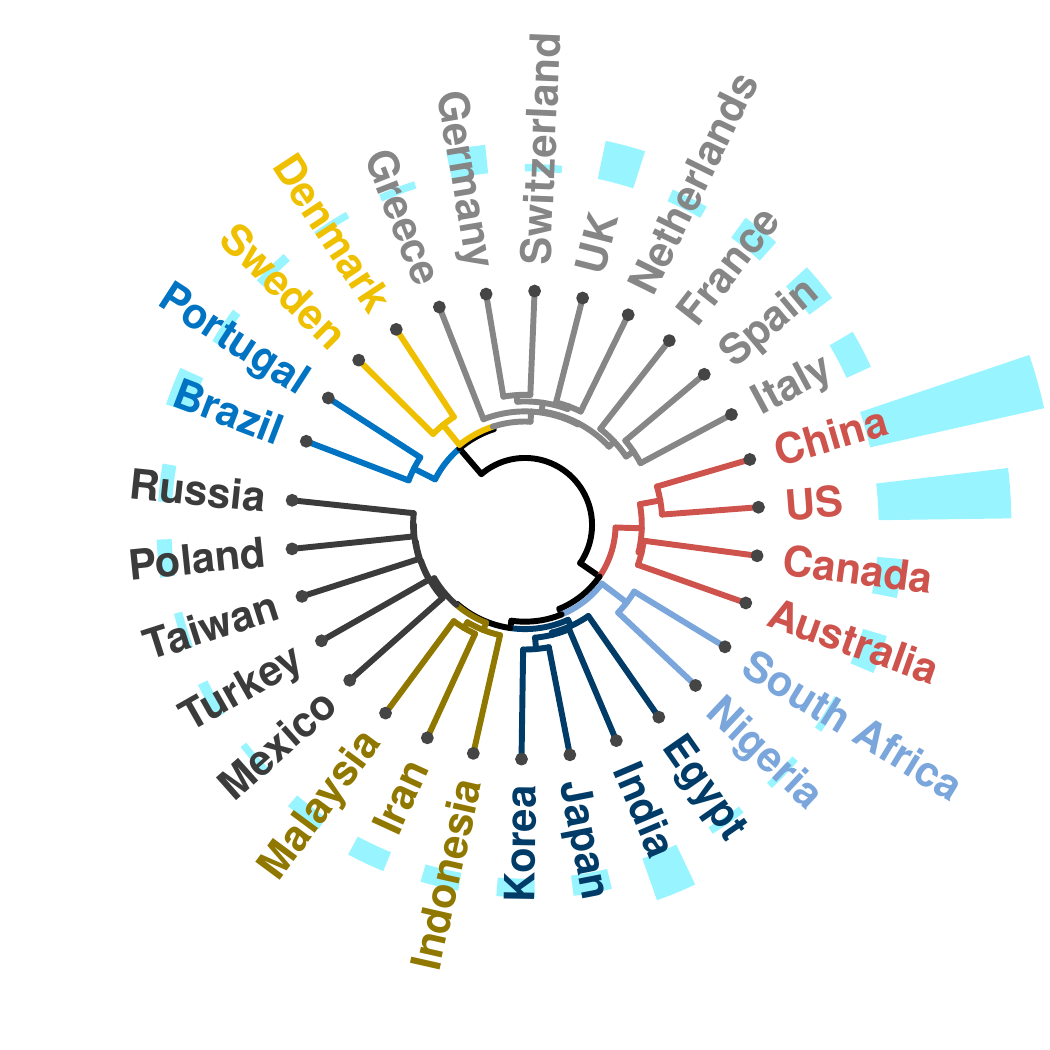}}
\end{flushleft}
	\end{minipage}
    \end{tabular}

\hspace{-0.5cm}{\marrow}\quad\dotfill

    \begin{tabular}{c}
    \begin{minipage}{0.03\hsize}
\begin{flushleft}
    \hspace{-0.7cm}\rotatebox{90}{\period{3}{2001--2010}}
\end{flushleft}
	\end{minipage}
	\begin{minipage}{0.33\hsize}
\begin{flushleft}
\raisebox{\height}{\includegraphics[trim=2.0cm 1.8cm 0cm 1.5cm, align=c, scale=\csize, vmargin=0mm]{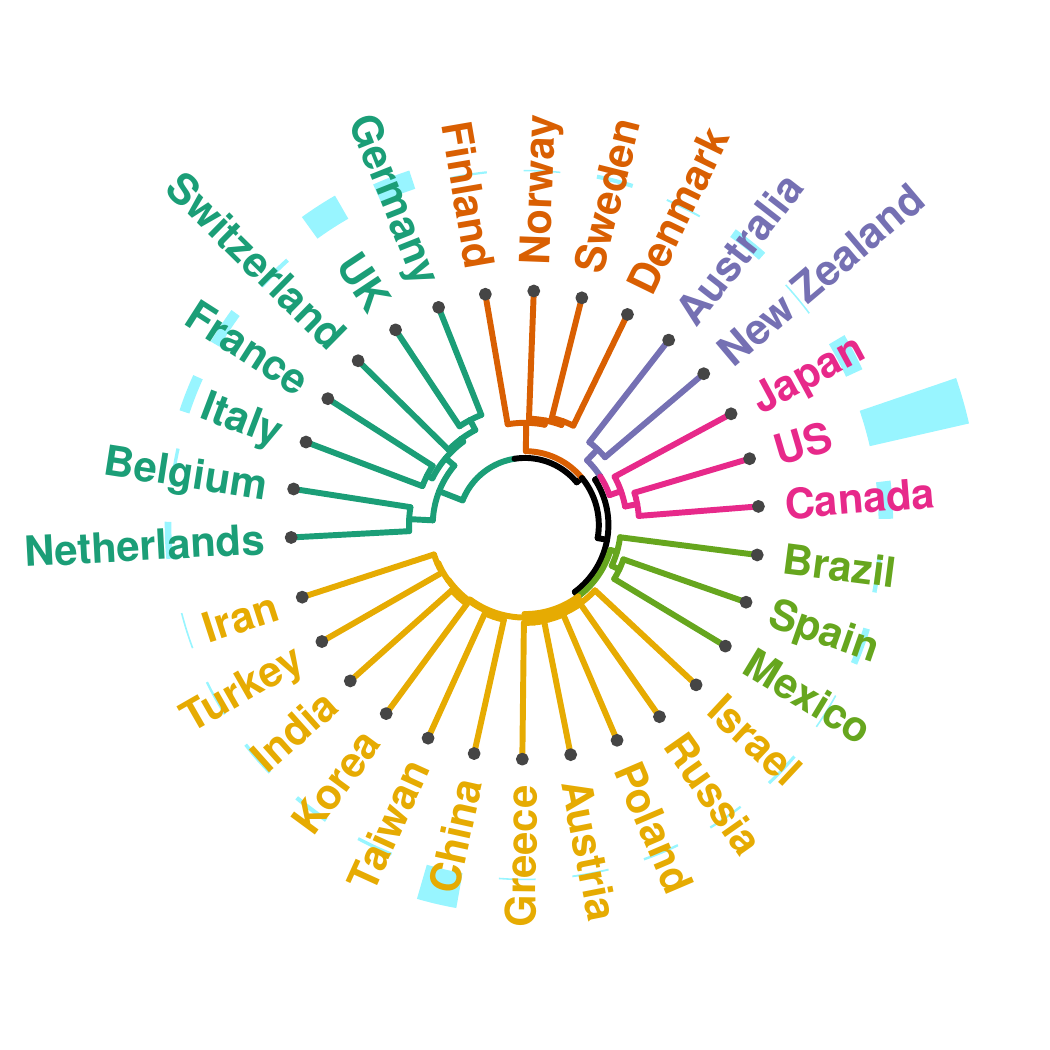}}
\end{flushleft}
    \end{minipage}
	\begin{minipage}{0.33\hsize}
\begin{flushleft}
\raisebox{\height}{\includegraphics[trim=2.0cm 1.8cm 0cm 1.5cm, align=c, scale=\csize, vmargin=0mm]{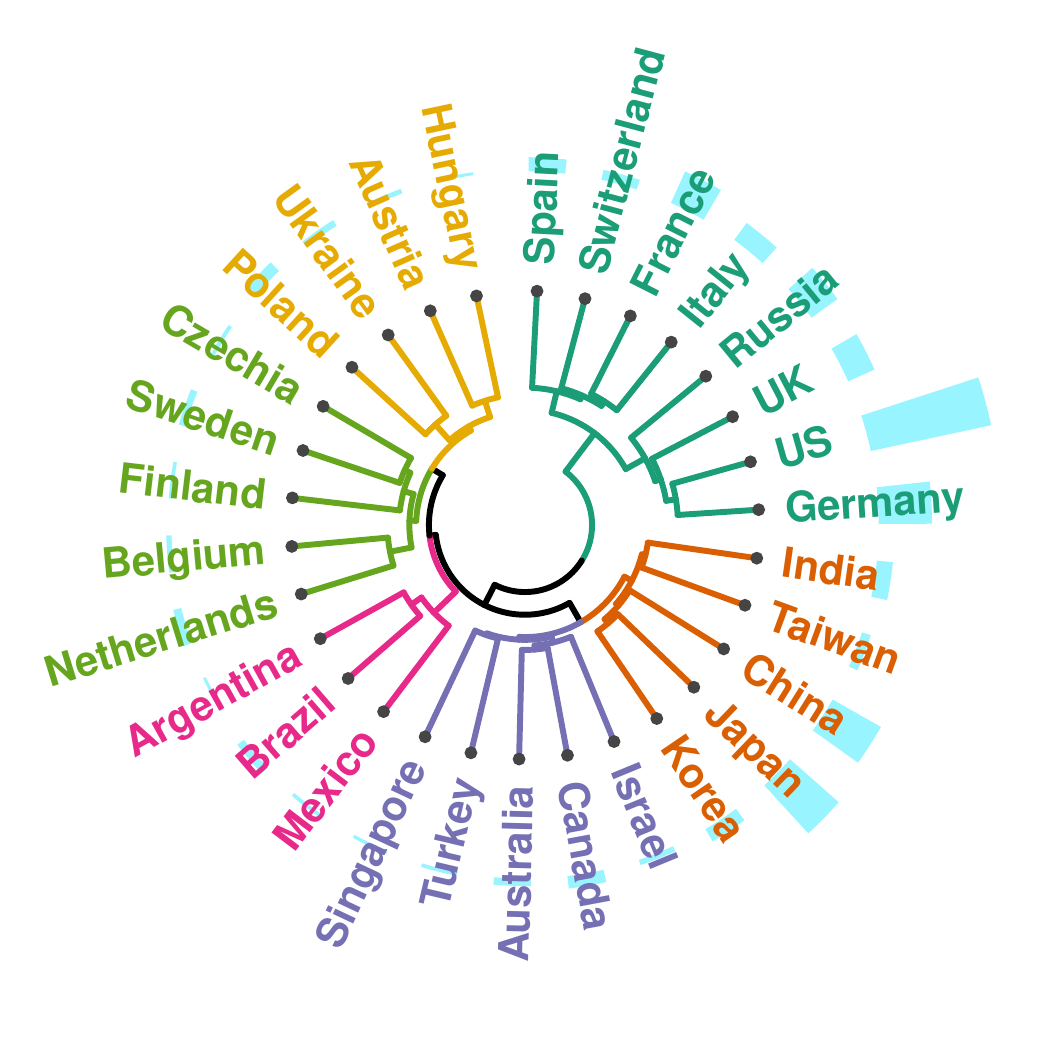}}
\end{flushleft}
	\end{minipage}
	\begin{minipage}{0.33\hsize}
\begin{flushleft}
\raisebox{\height}{\includegraphics[trim=2.0cm 1.8cm 0cm 1.5cm, align=c, scale=\csize, vmargin=0mm]{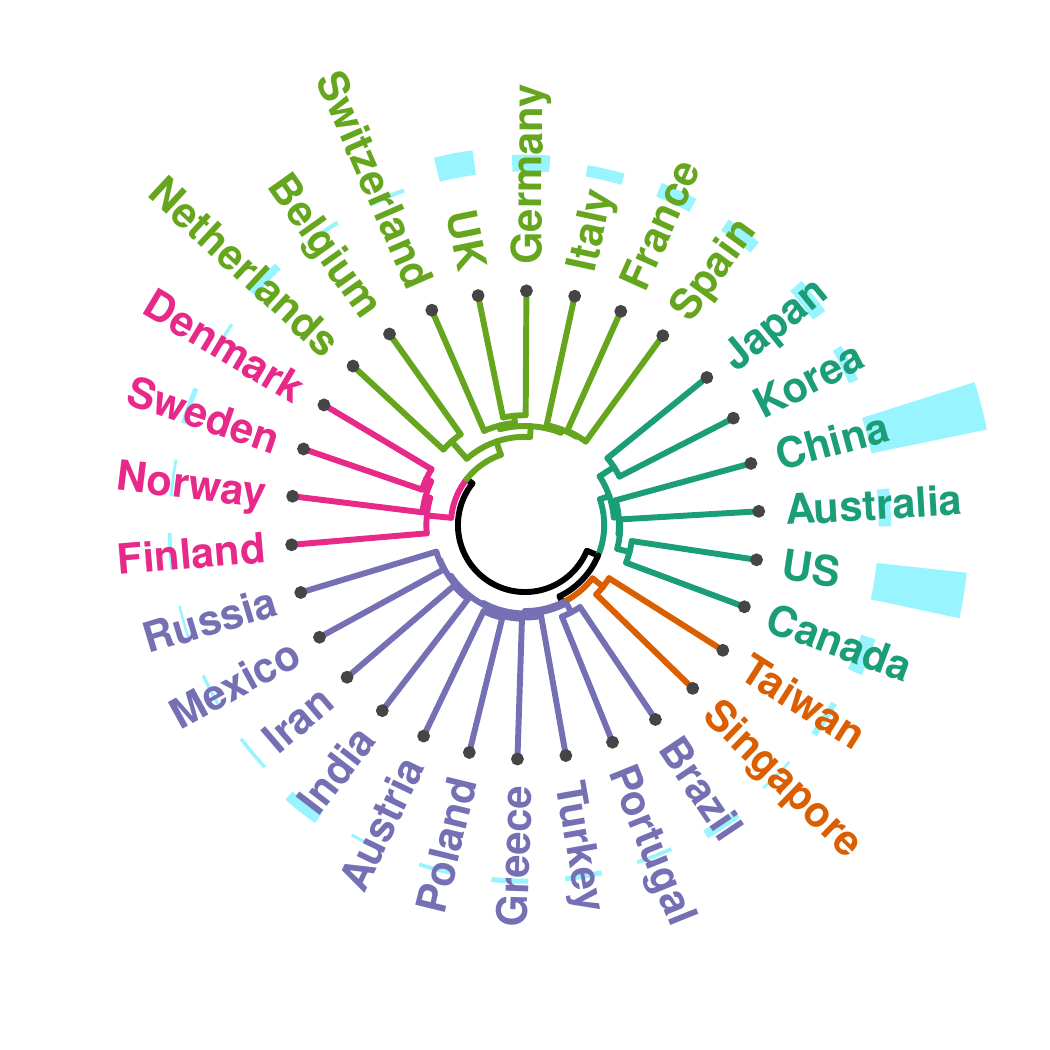}}
\end{flushleft}
	\end{minipage}
    \end{tabular}

\hspace{-0.5cm}{\marrow}\quad\dotfill

    \begin{tabular}{c}
    \begin{minipage}{0.03\hsize}
\begin{flushleft}
    \hspace{-0.7cm}\rotatebox{90}{\period{2}{1991--2000}}
\end{flushleft}
	\end{minipage}
	\begin{minipage}{0.33\hsize}
\begin{flushleft}
\raisebox{\height}{\includegraphics[trim=2.0cm 1.8cm 0cm 1.5cm, align=c, scale=\csize, vmargin=0mm]{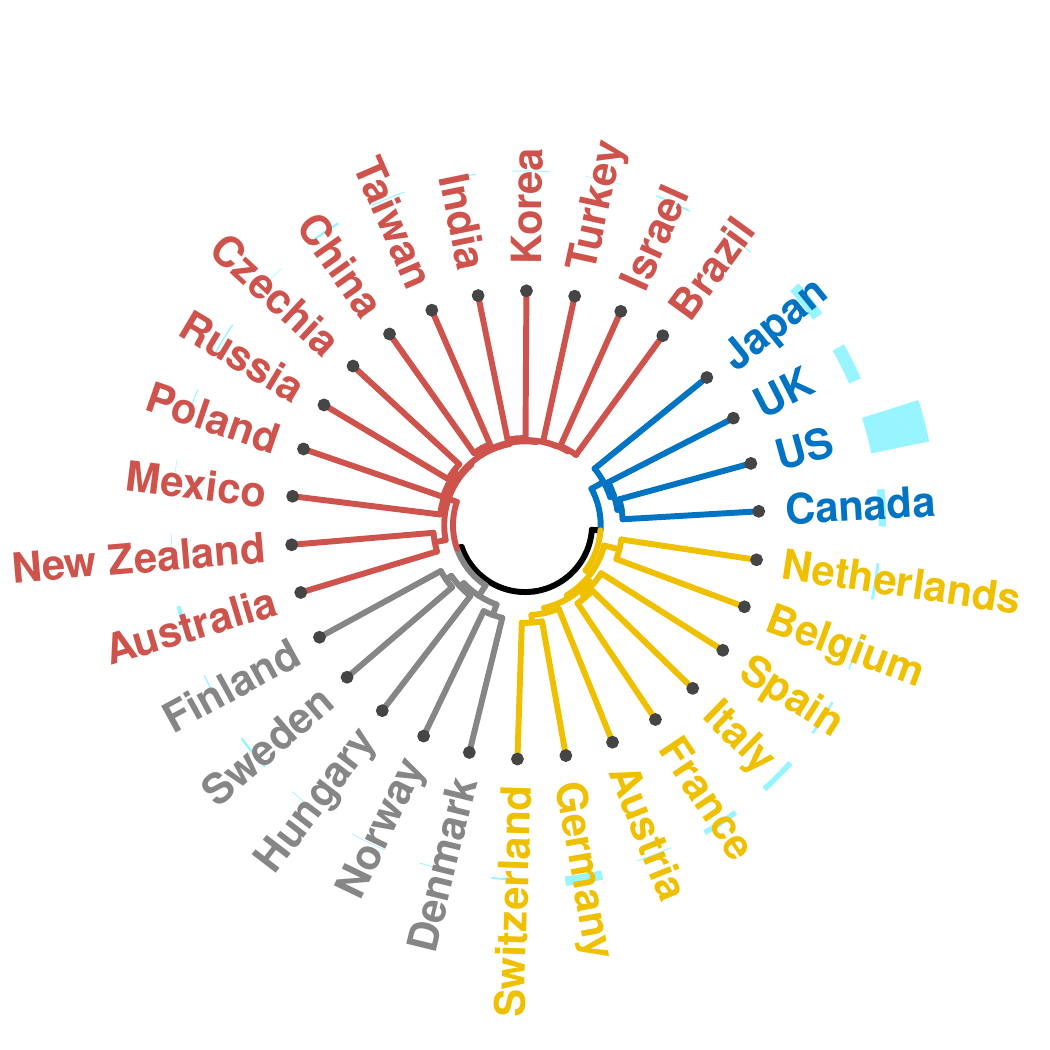}}
\end{flushleft}
    \end{minipage}
	\begin{minipage}{0.33\hsize}
\begin{flushleft}
\raisebox{\height}{\includegraphics[trim=2.0cm 1.8cm 0cm 1.5cm, align=c, scale=\csize, vmargin=0mm]{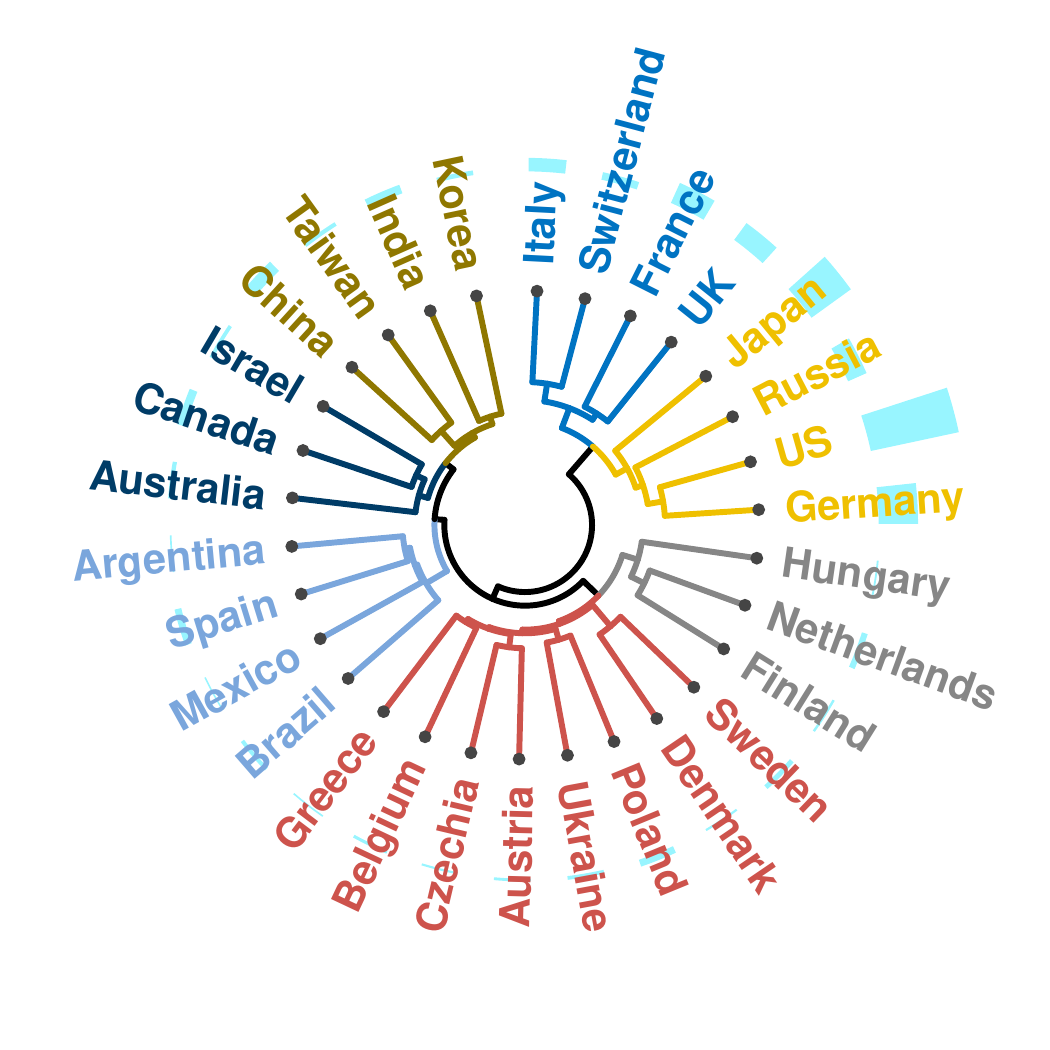}}
\end{flushleft}
	\end{minipage}
	\begin{minipage}{0.33\hsize}
\begin{flushleft}
\raisebox{\height}{\includegraphics[trim=2.0cm 1.8cm 0cm 1.5cm, align=c, scale=\csize, vmargin=0mm]{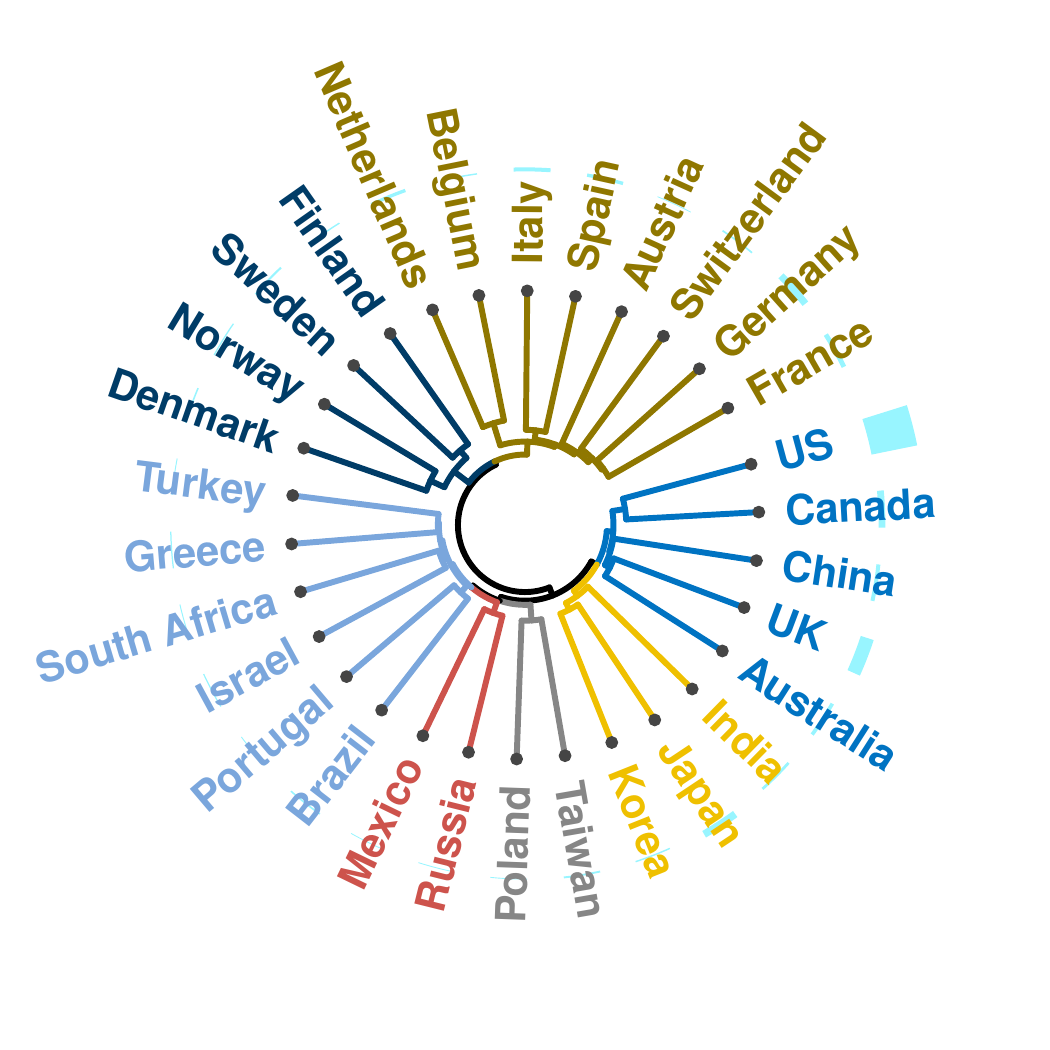}}
\end{flushleft}
	\end{minipage}
    \end{tabular}

\hspace{-0.5cm}{\marrow}\quad\dotfill

    \begin{tabular}{c}
    \begin{minipage}{0.03\hsize}
\begin{flushleft}
    \hspace{-0.7cm}\rotatebox{90}{\period{1}{1971--1990}}
\end{flushleft}
	\end{minipage}
	\begin{minipage}{0.33\hsize}
\begin{flushleft}
\raisebox{\height}{\includegraphics[trim=2.0cm 1.8cm 0cm 1.5cm, align=c, scale=\csize, vmargin=0mm]{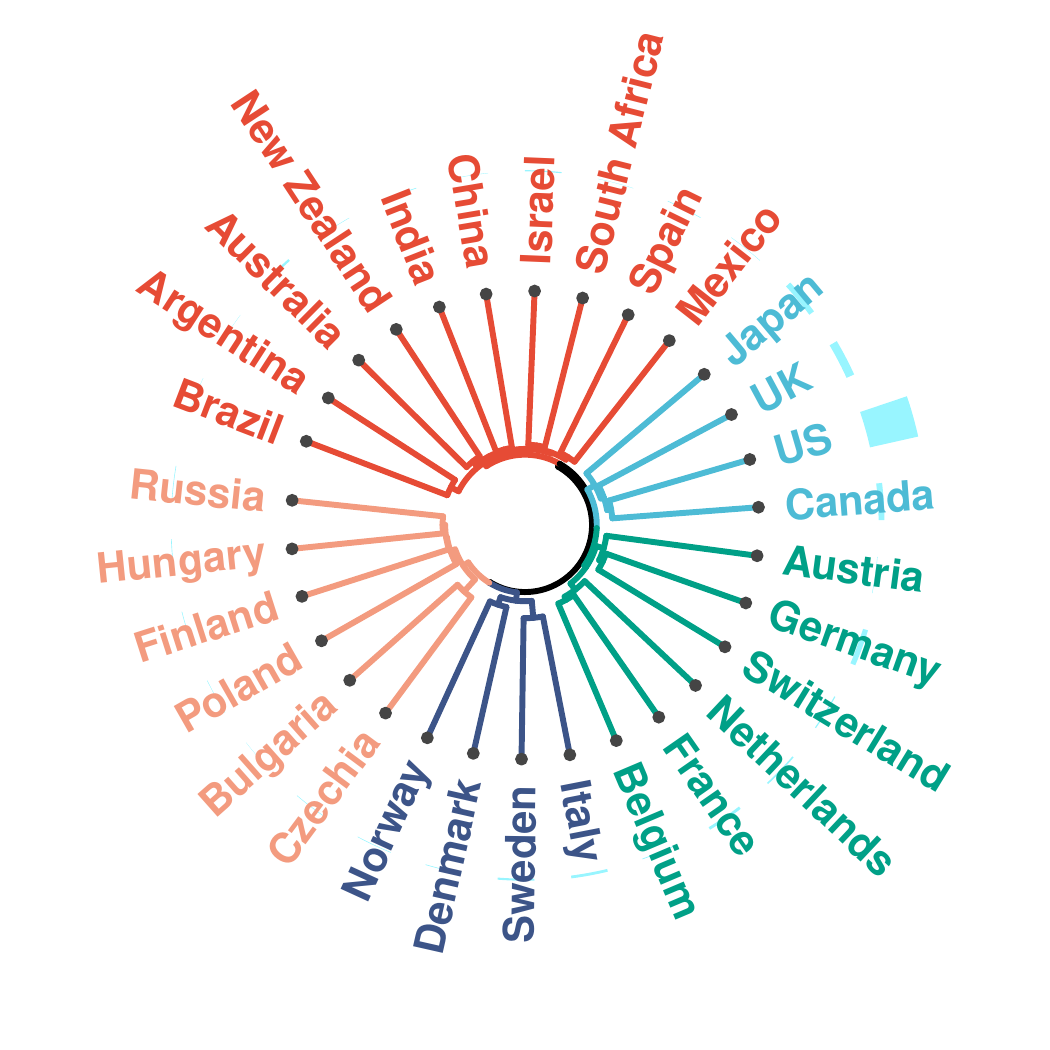}}
\end{flushleft}
    \end{minipage}
	\begin{minipage}{0.33\hsize}
\begin{flushleft}
\raisebox{\height}{\includegraphics[trim=2.0cm 1.8cm 0cm 1.5cm, align=c, scale=\csize, vmargin=0mm]{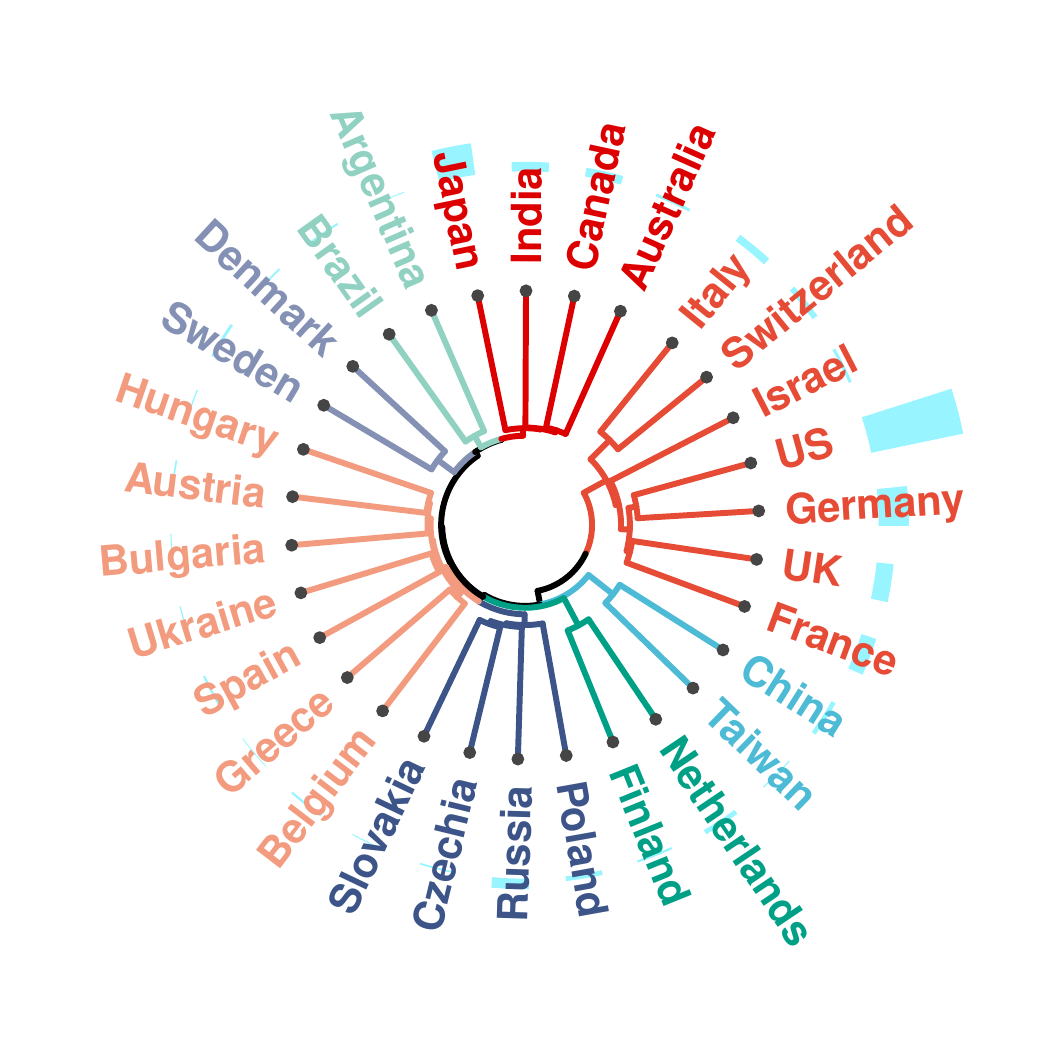}}
\end{flushleft}
	\end{minipage}
	\begin{minipage}{0.33\hsize}
\begin{flushleft}
\raisebox{\height}{\includegraphics[trim=2.0cm 1.8cm 0cm 1.5cm, align=c, scale=\csize, vmargin=0mm]{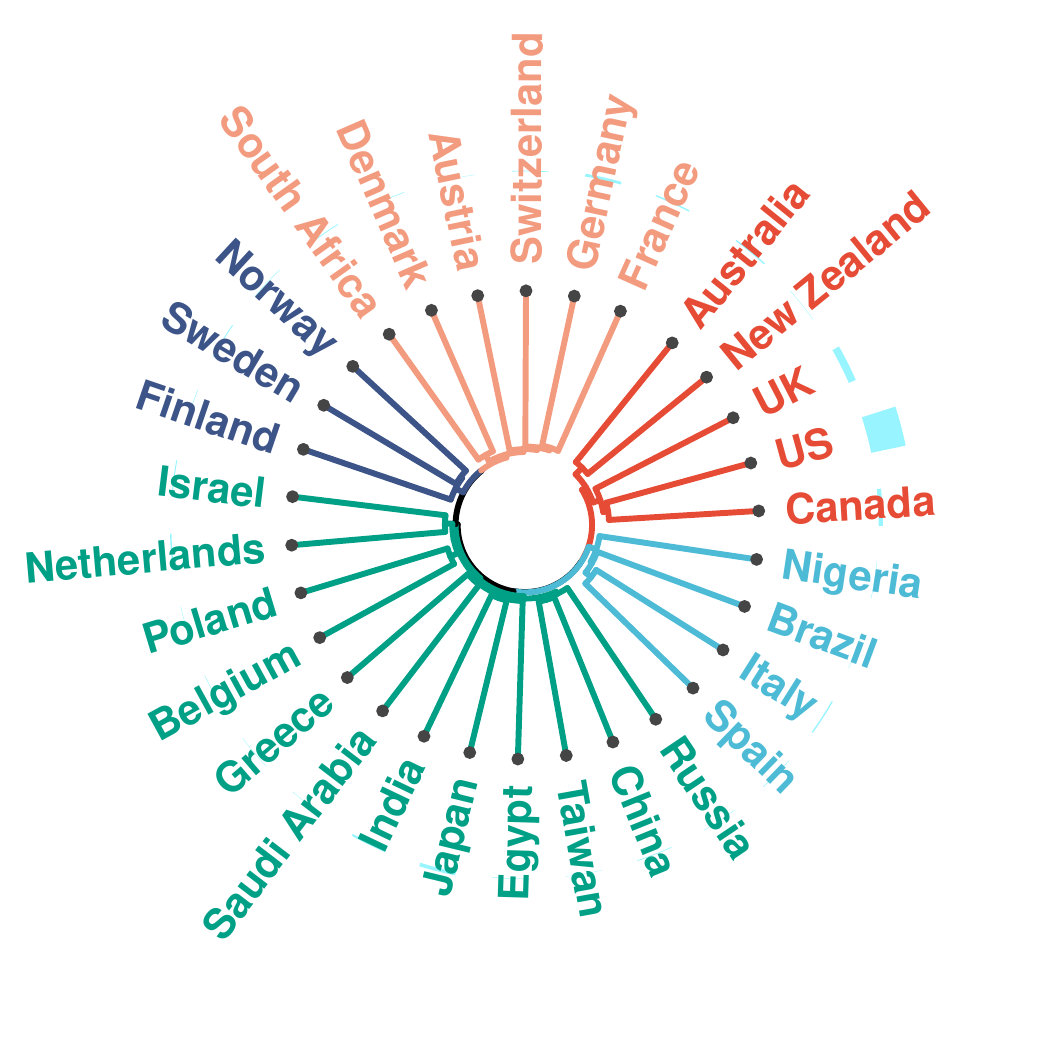}}
\end{flushleft}
	\end{minipage}
    \end{tabular}
\end{subfigure}
\vspace{2.5mm}
\caption{\textbf{Evolution of international research collaboration clusters. \emph{(Cont.)}}}
\label{fig:cdend_4}
\end{figure}
}
\afterpage{\clearpage%
\begin{figure}[htp]\ContinuedFloat
\centering
\begin{subfigure}{1.0\textwidth}
\vspace{-0.5cm}
    \begin{tabular}{c}
    \begin{minipage}{0.03\hsize}
\begin{flushleft}
    \hspace{-0.7cm}\rotatebox{90}{\period{4}{2011--2020}}
\end{flushleft}
	\end{minipage}
	\begin{minipage}{0.33\hsize}
\begin{flushleft}
\raisebox{-0.0cm}{\hspace{-0mm}\small\textrm{\textbf{(j)~ \earth}}}\\[6mm]
\raisebox{\height}{\includegraphics[trim=2.0cm 1.8cm 0cm 1.5cm, align=c, scale=\csize, vmargin=0mm]{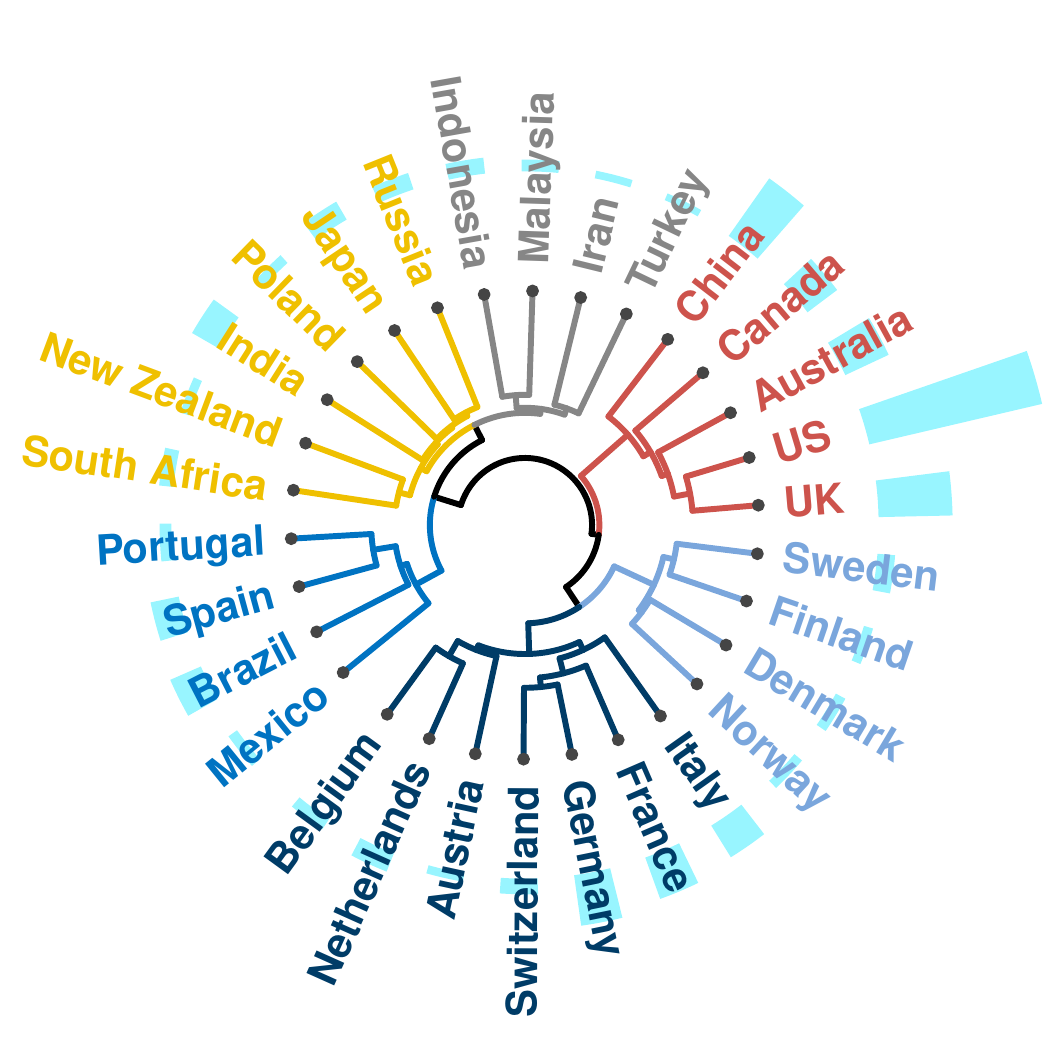}}
\end{flushleft}
    \end{minipage}
	\begin{minipage}{0.33\hsize}
\begin{flushleft}
\raisebox{-0.0cm}{\hspace{-0mm}\small\textrm{\textbf{(k)~ \astro}}}\\[6mm]
\raisebox{\height}{\includegraphics[trim=2.0cm 1.8cm 0cm 1.5cm, align=c, scale=\csize, vmargin=0mm]{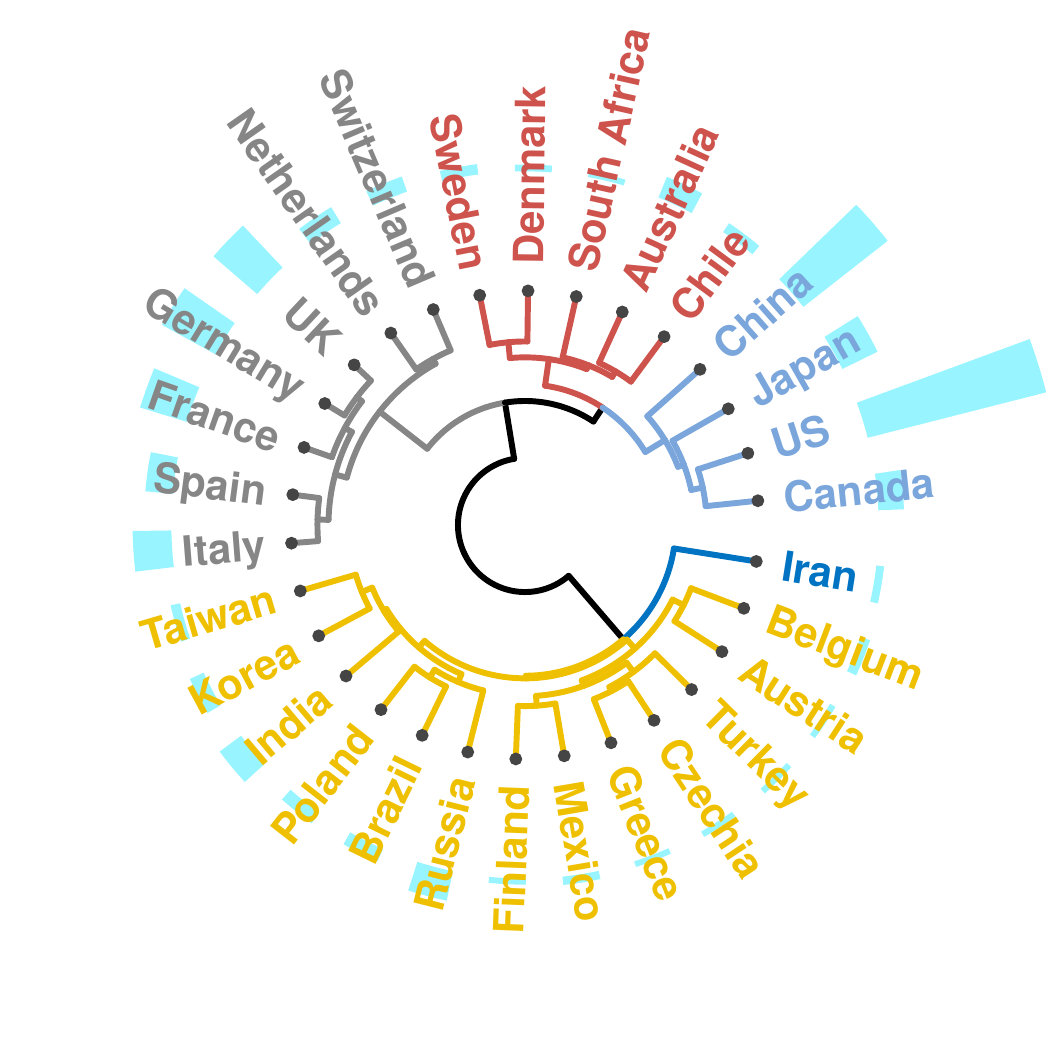}}
\end{flushleft}
	\end{minipage}
	\begin{minipage}{0.33\hsize}
\begin{flushleft}
\raisebox{-0.0cm}{\hspace{-0mm}\small\textrm{\textbf{(l)~ \math}}}\\[6mm]
\raisebox{\height}{\includegraphics[trim=2.0cm 1.8cm 0cm 1.5cm, align=c, scale=\csize, vmargin=0mm]{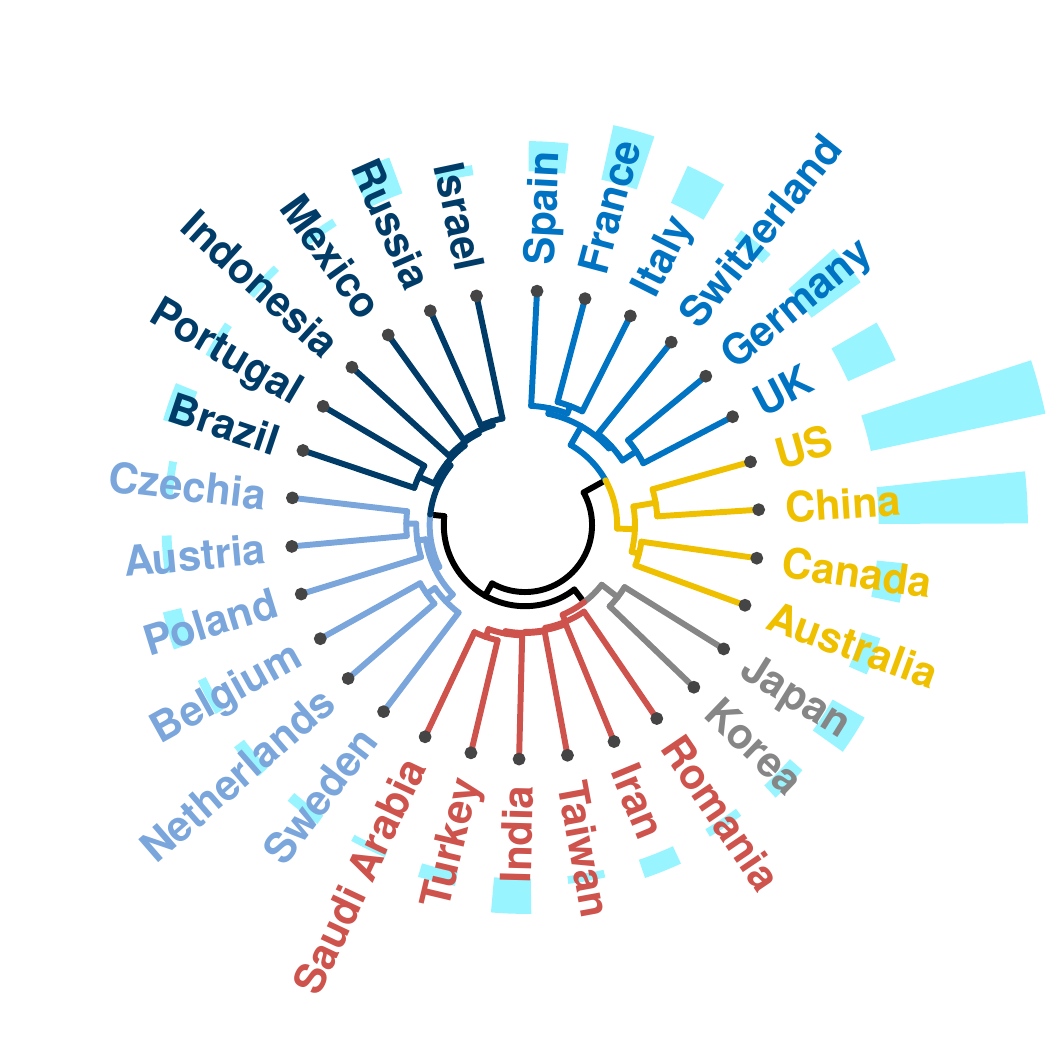}}
\end{flushleft}
	\end{minipage}
    \end{tabular}

\hspace{-0.5cm}{\marrow}\quad\dotfill

    \begin{tabular}{c}
    \begin{minipage}{0.03\hsize}
\begin{flushleft}
    \hspace{-0.7cm}\rotatebox{90}{\period{3}{2001--2010}}
\end{flushleft}
	\end{minipage}
	\begin{minipage}{0.33\hsize}
\begin{flushleft}
\raisebox{\height}{\includegraphics[trim=2.0cm 1.8cm 0cm 1.5cm, align=c, scale=\csize, vmargin=0mm]{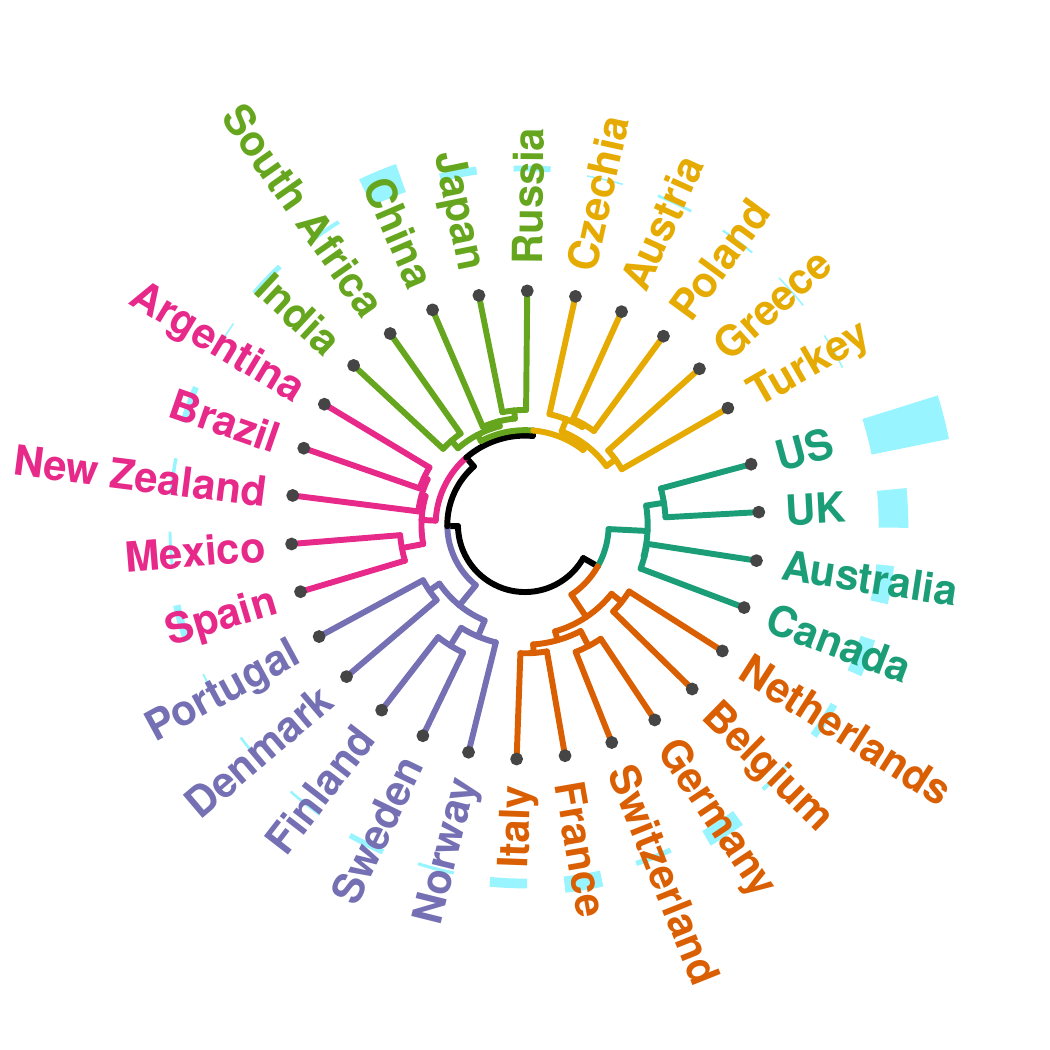}}
\end{flushleft}
    \end{minipage}
	\begin{minipage}{0.33\hsize}
\begin{flushleft}
\raisebox{\height}{\includegraphics[trim=2.0cm 1.8cm 0cm 1.5cm, align=c, scale=\csize, vmargin=0mm]{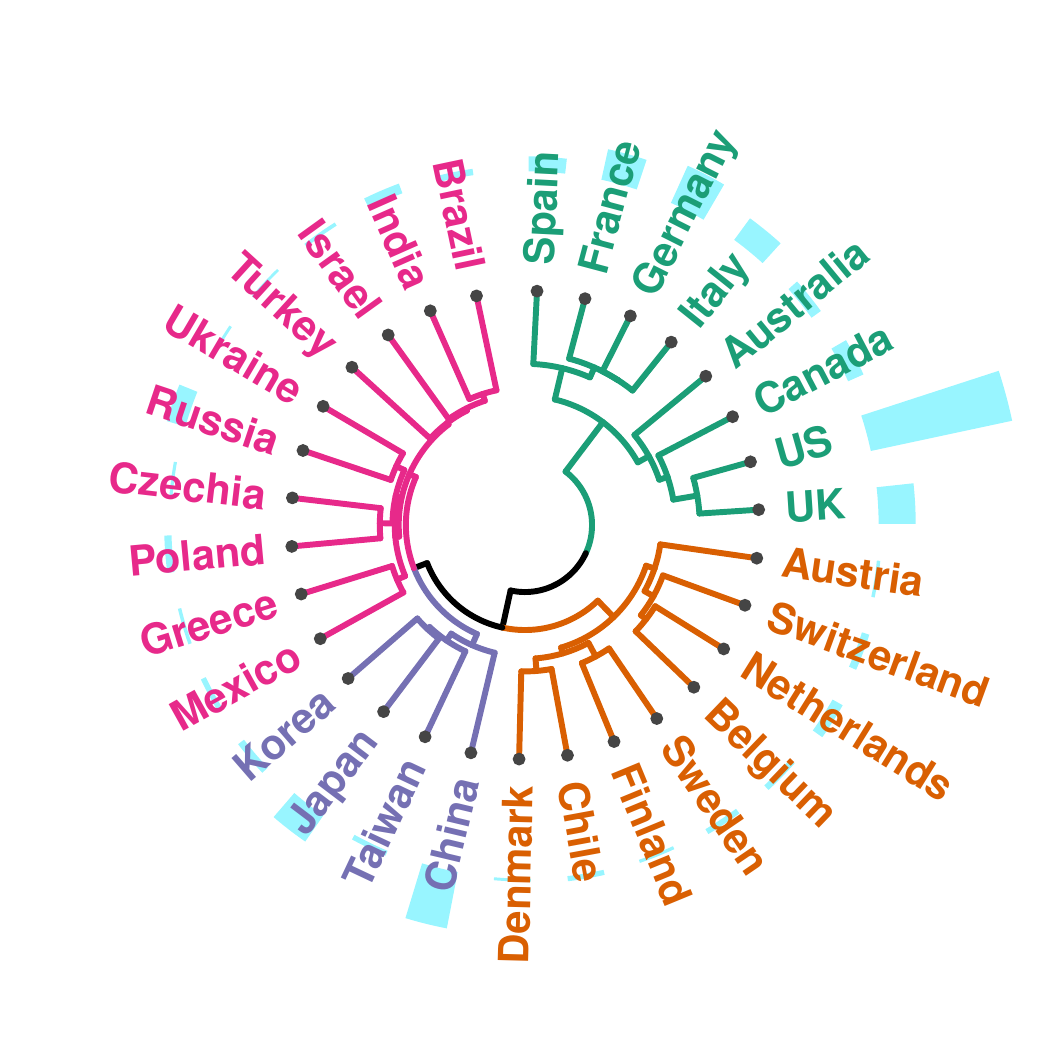}}
\end{flushleft}
	\end{minipage}
	\begin{minipage}{0.33\hsize}
\begin{flushleft}
\raisebox{\height}{\includegraphics[trim=2.0cm 1.8cm 0cm 1.5cm, align=c, scale=\csize, vmargin=0mm]{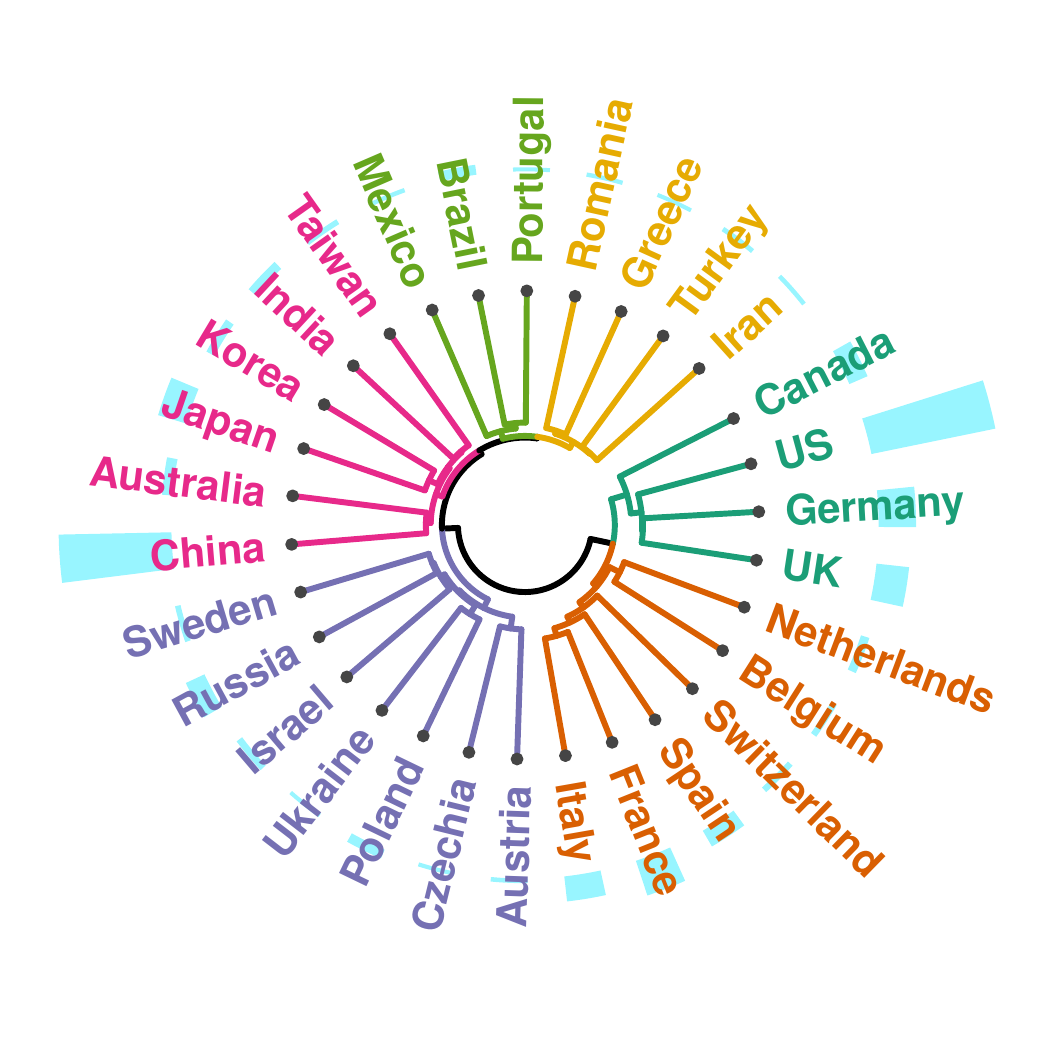}}
\end{flushleft}
	\end{minipage}
    \end{tabular}

\hspace{-0.5cm}{\marrow}\quad\dotfill

    \begin{tabular}{c}
    \begin{minipage}{0.03\hsize}
\begin{flushleft}
    \hspace{-0.7cm}\rotatebox{90}{\period{2}{1991--2000}}
\end{flushleft}
	\end{minipage}
	\begin{minipage}{0.33\hsize}
\begin{flushleft}
\raisebox{\height}{\includegraphics[trim=2.0cm 1.8cm 0cm 1.5cm, align=c, scale=\csize, vmargin=0mm]{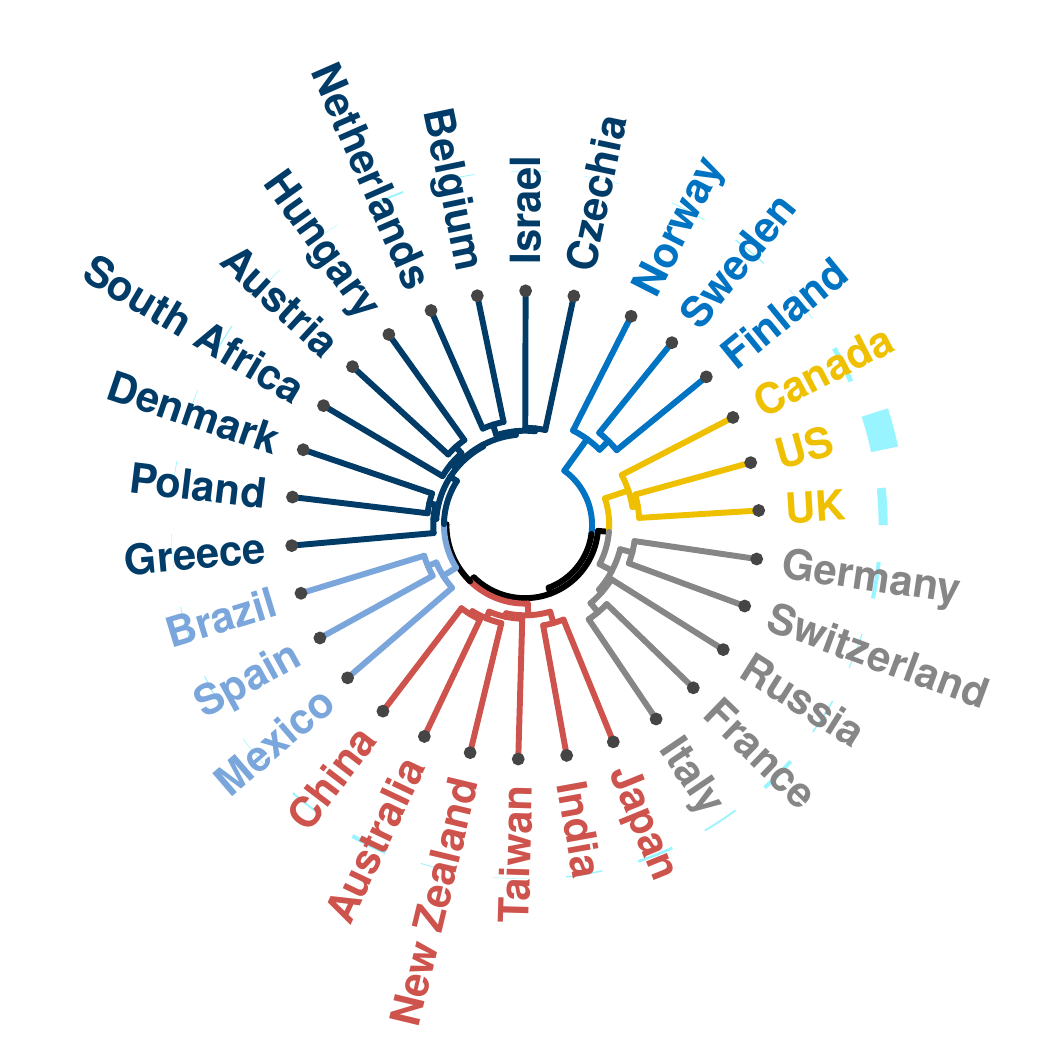}}
\end{flushleft}
    \end{minipage}
	\begin{minipage}{0.33\hsize}
\begin{flushleft}
\raisebox{\height}{\includegraphics[trim=2.0cm 1.8cm 0cm 1.5cm, align=c, scale=\csize, vmargin=0mm]{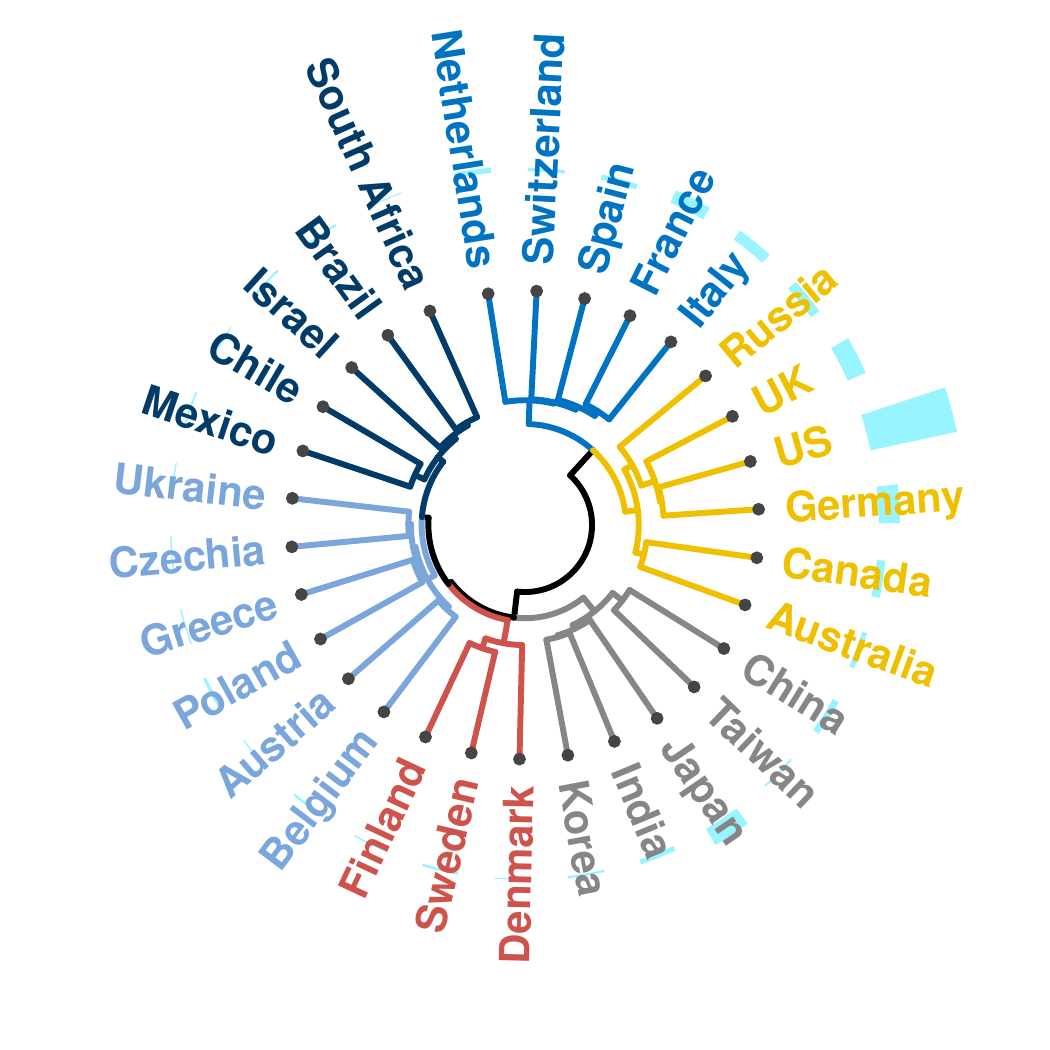}}
\end{flushleft}
	\end{minipage}
	\begin{minipage}{0.33\hsize}
\begin{flushleft}
\raisebox{\height}{\includegraphics[trim=2.0cm 1.8cm 0cm 1.5cm, align=c, scale=\csize, vmargin=0mm]{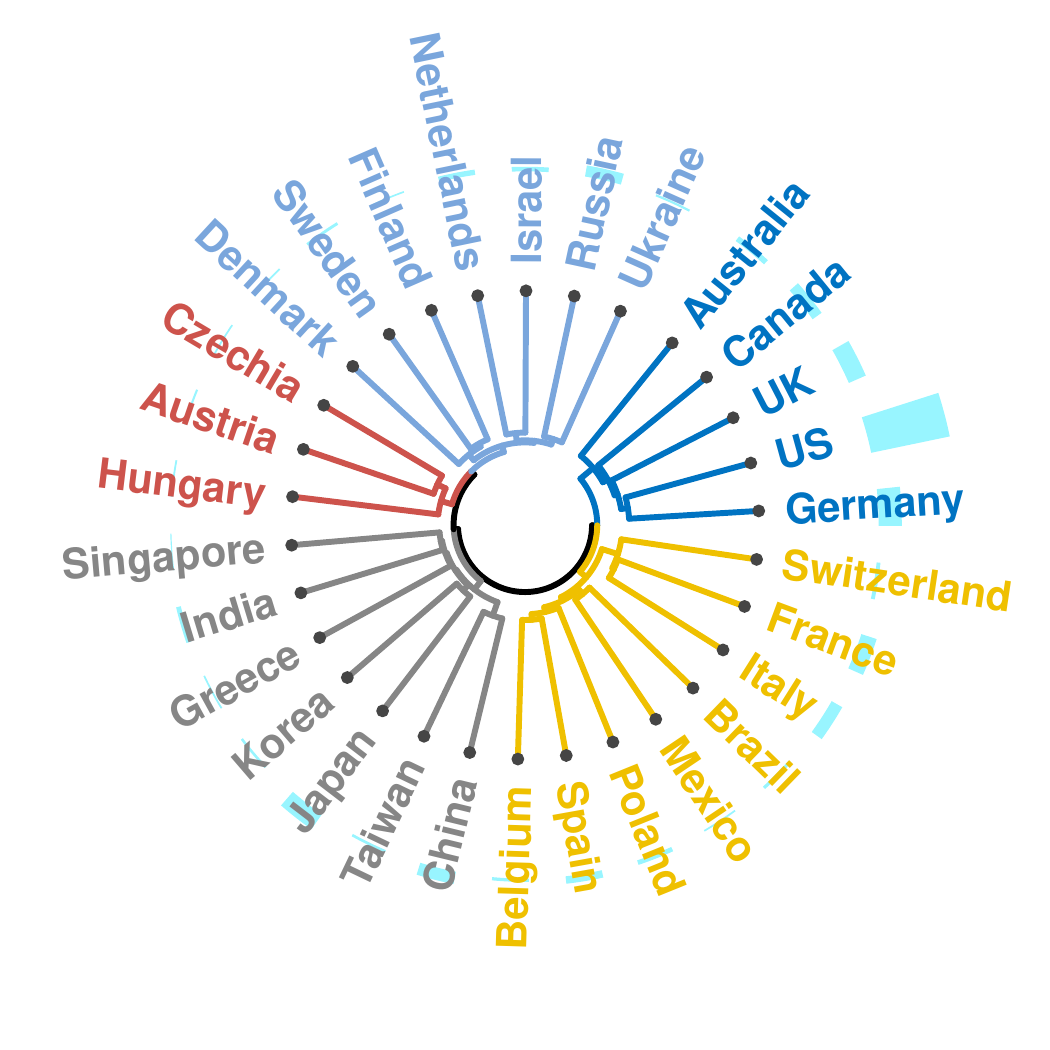}}
\end{flushleft}
	\end{minipage}
    \end{tabular}

\hspace{-0.5cm}{\marrow}\quad\dotfill

    \begin{tabular}{c}
    \begin{minipage}{0.03\hsize}
\begin{flushleft}
    \hspace{-0.7cm}\rotatebox{90}{\period{1}{1971--1990}}
\end{flushleft}
	\end{minipage}
	\begin{minipage}{0.33\hsize}
\begin{flushleft}
\raisebox{\height}{\includegraphics[trim=2.0cm 1.8cm 0cm 1.5cm, align=c, scale=\csize, vmargin=0mm]{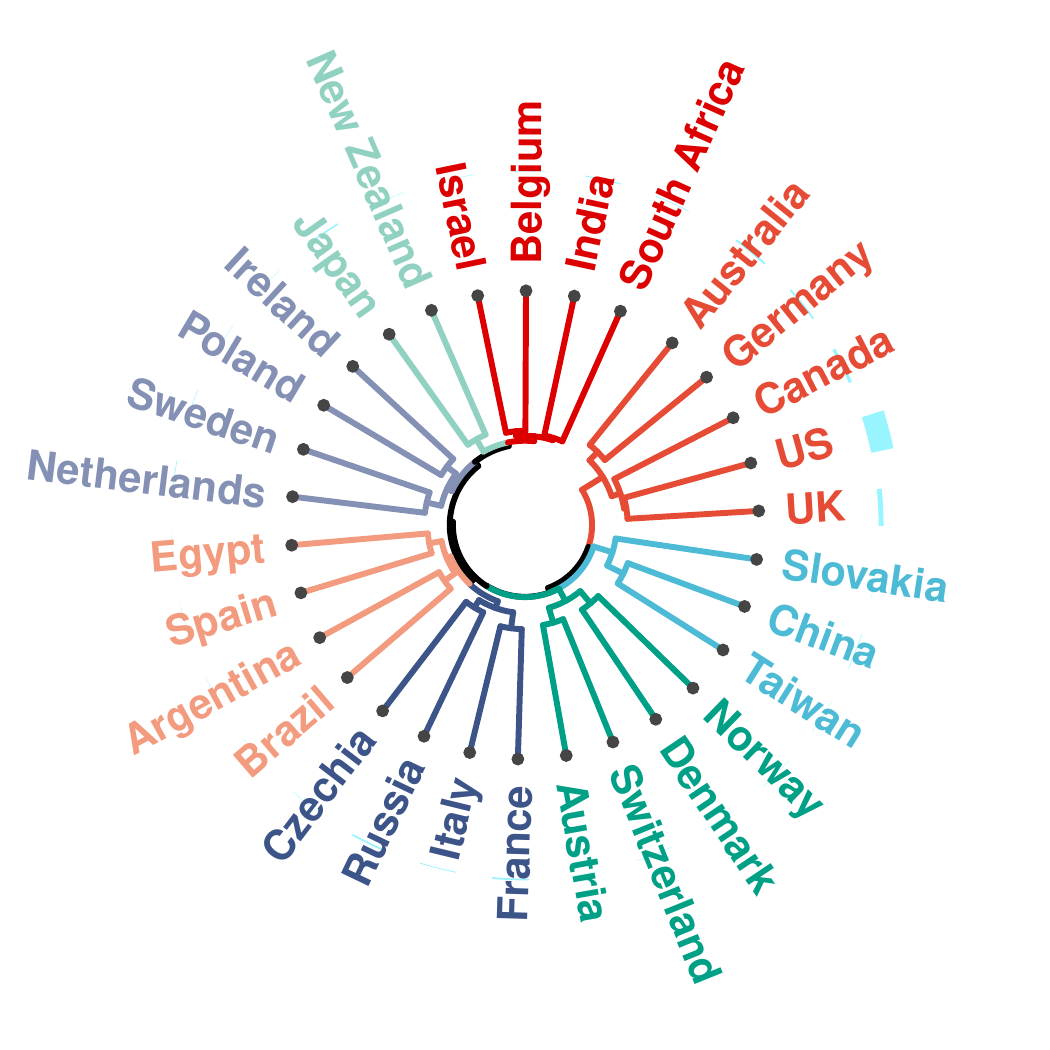}}
\end{flushleft}
    \end{minipage}
	\begin{minipage}{0.33\hsize}
\begin{flushleft}
\raisebox{\height}{\includegraphics[trim=2.0cm 1.8cm 0cm 1.5cm, align=c, scale=\csize, vmargin=0mm]{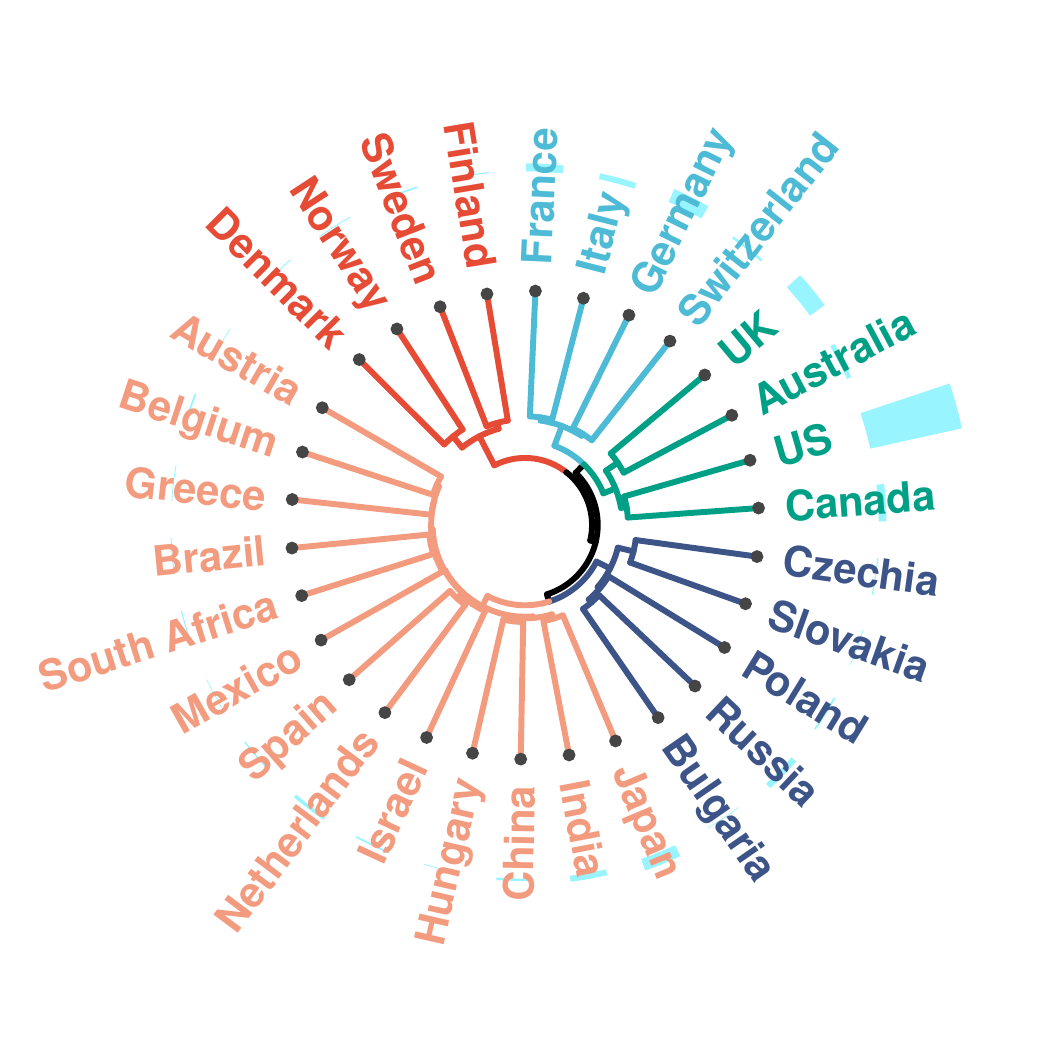}}
\end{flushleft}
	\end{minipage}
	\begin{minipage}{0.33\hsize}
\begin{flushleft}
\raisebox{\height}{\includegraphics[trim=2.0cm 1.8cm 0cm 1.5cm, align=c, scale=\csize, vmargin=0mm]{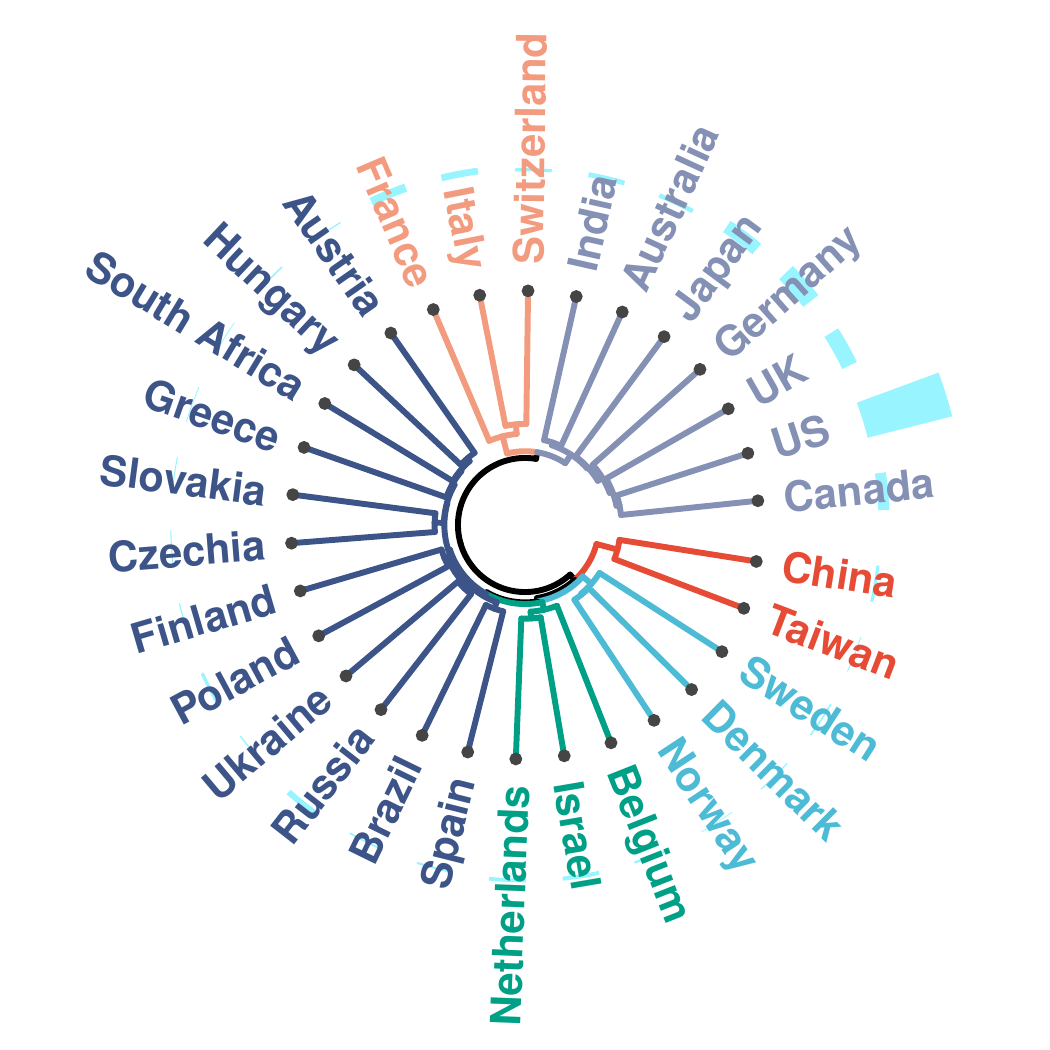}}
\end{flushleft}
	\end{minipage}
    \end{tabular}
\end{subfigure}
\vspace{2.5mm}
\caption{\textbf{Evolution of international research collaboration clusters. \emph{(Cont.)}}}
\label{fig:cdend_5}
\end{figure}
}

\afterpage{\clearpage%
\begin{figure}[!tp]
\centering
\vspace{-0.5cm}
\noindent
\includegraphics[align=c, scale=0.95, vmargin=1mm]{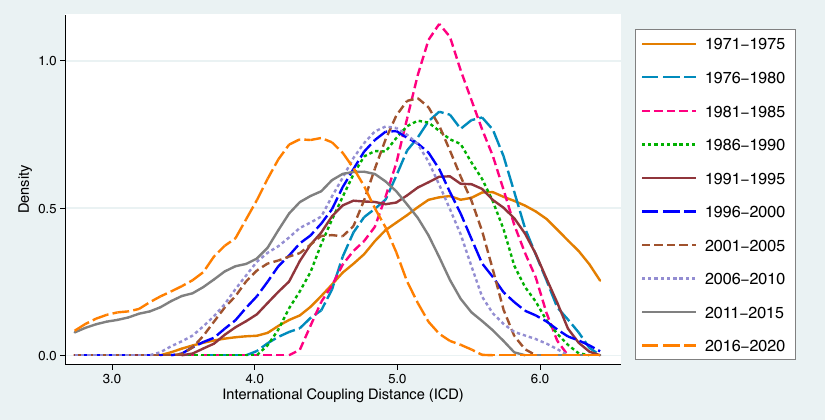}
\caption{\textbf{Change in the estimated kernel density of the International Coupling Distance (ICD).}
}
\label{fig:kdensity_ICD}
\vspace{5mm}
\end{figure}
\quad\\
\vfill
}

\afterpage{\clearpage%
\begin{figure}[htp]
\centering
\begin{subfigure}{1.0\textwidth}
\vspace{-0.5cm}
    \begin{tabular}{c}
    \begin{minipage}{0.03\hsize}
\begin{flushleft}
    \hspace{0.7cm}\rotatebox{0}{\period{4}{2011--2020}}
\end{flushleft}
	\end{minipage}
	\begin{minipage}{0.5\hsize}
\begin{flushright}
\raisebox{\height}{\includegraphics[trim=2.0cm 1.8cm 0cm 1.5cm, align=c, scale=\chordsize, vmargin=0mm]{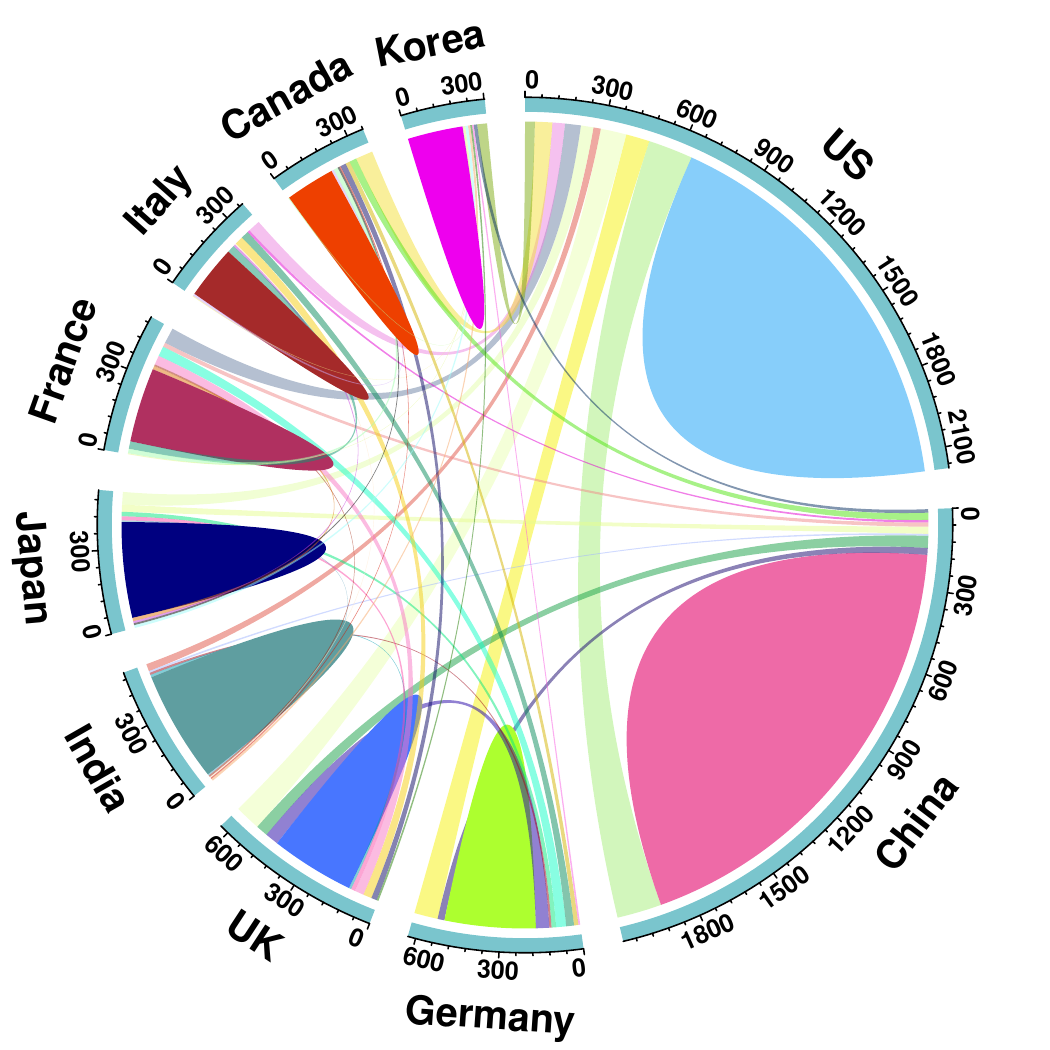}}
\end{flushright}
    \end{minipage}
	\begin{minipage}{0.5\hsize}
\begin{center}
\raisebox{\height}{\includegraphics[trim=2.0cm 1.8cm 0cm 1.5cm, align=c, scale=\csize, vmargin=0mm]{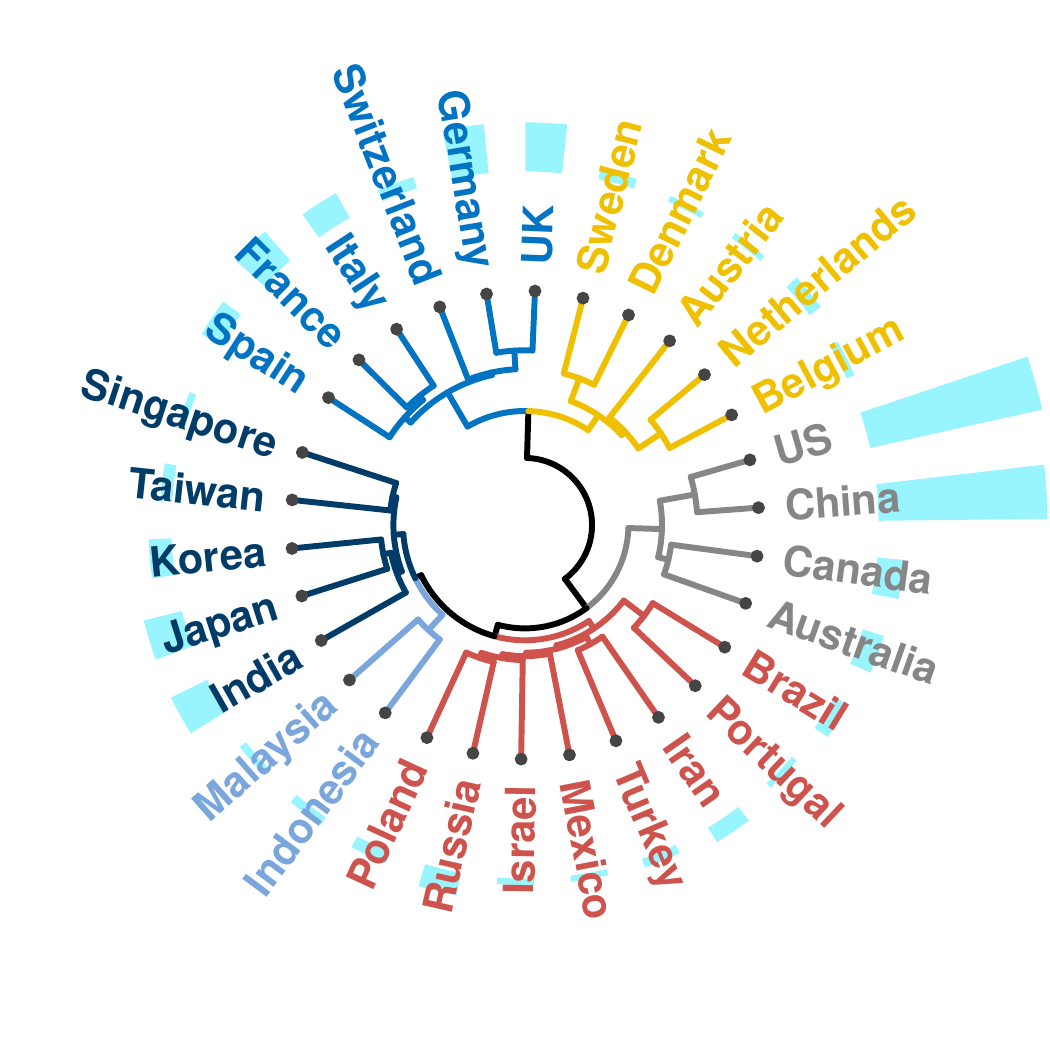}}
\end{center}
	\end{minipage}
    \end{tabular}

\vspace{0mm}
\hspace{0.5cm}{\marrow}\quad\dotfill
\vspace{2mm}

    \begin{tabular}{c}
    \begin{minipage}{0.03\hsize}
\begin{flushleft}
    \hspace{0.7cm}\rotatebox{0}{\period{3}{2001--2010}}
\end{flushleft}
	\end{minipage}
	\begin{minipage}{0.5\hsize}
\begin{flushright}
\raisebox{\height}{\includegraphics[trim=2.0cm 1.8cm 0cm 1.5cm, align=c, scale=\chordsize, vmargin=0mm]{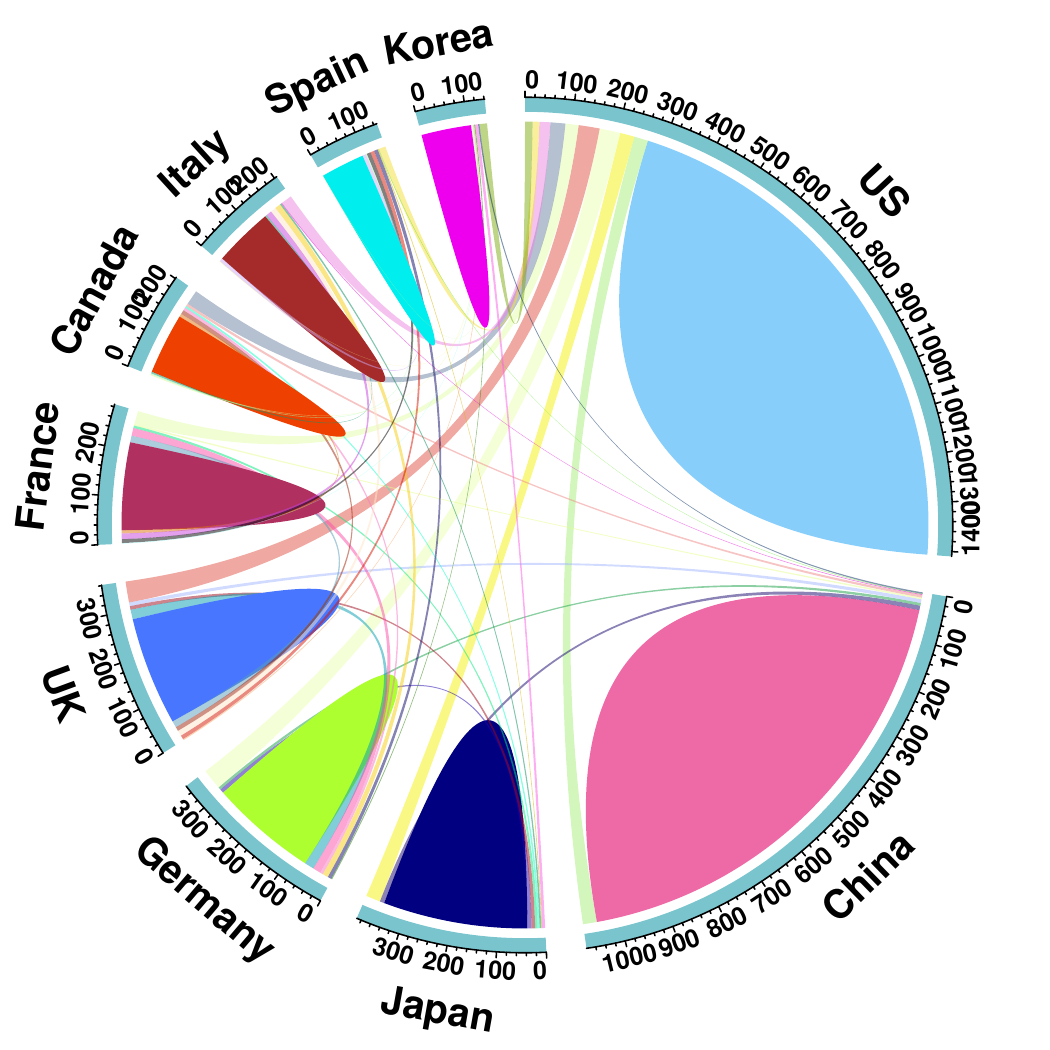}}
\end{flushright}
    \end{minipage}
	\begin{minipage}{0.5\hsize}
\begin{center}
\raisebox{\height}{\includegraphics[trim=2.0cm 1.8cm 0cm 1.5cm, align=c, scale=\csize, vmargin=0mm]{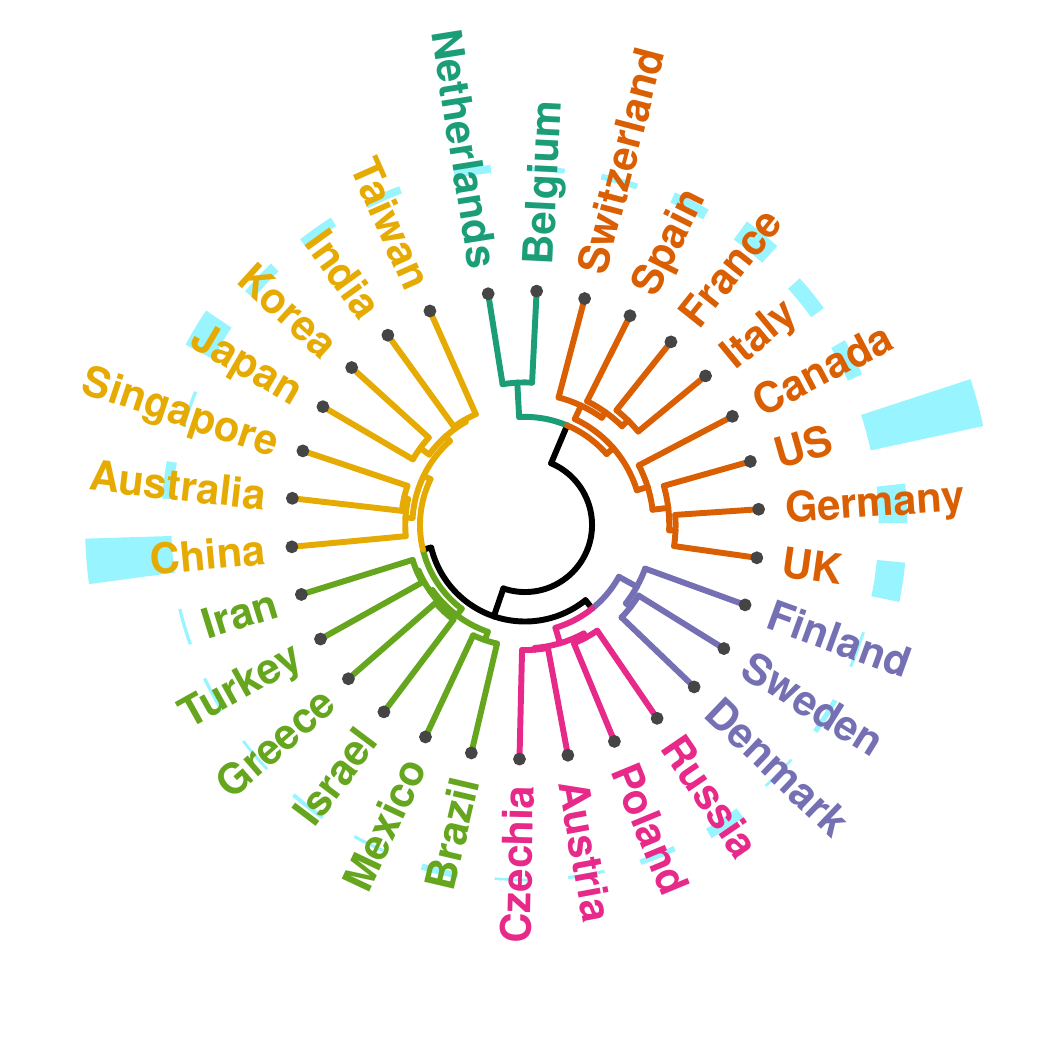}}
\end{center}
	\end{minipage}
    \end{tabular}

\vspace{0mm}
\hspace{0.5cm}{\marrow}\quad\dotfill
\vspace{2mm}

    \begin{tabular}{c}
    \begin{minipage}{0.03\hsize}
\begin{flushleft}
    \hspace{0.7cm}\rotatebox{0}{\period{2}{1991--2000}}
\end{flushleft}
	\end{minipage}
	\begin{minipage}{0.5\hsize}
\begin{flushright}
\raisebox{\height}{\includegraphics[trim=2.0cm 1.8cm 0cm 1.5cm, align=c, scale=\chordsize, vmargin=0mm]{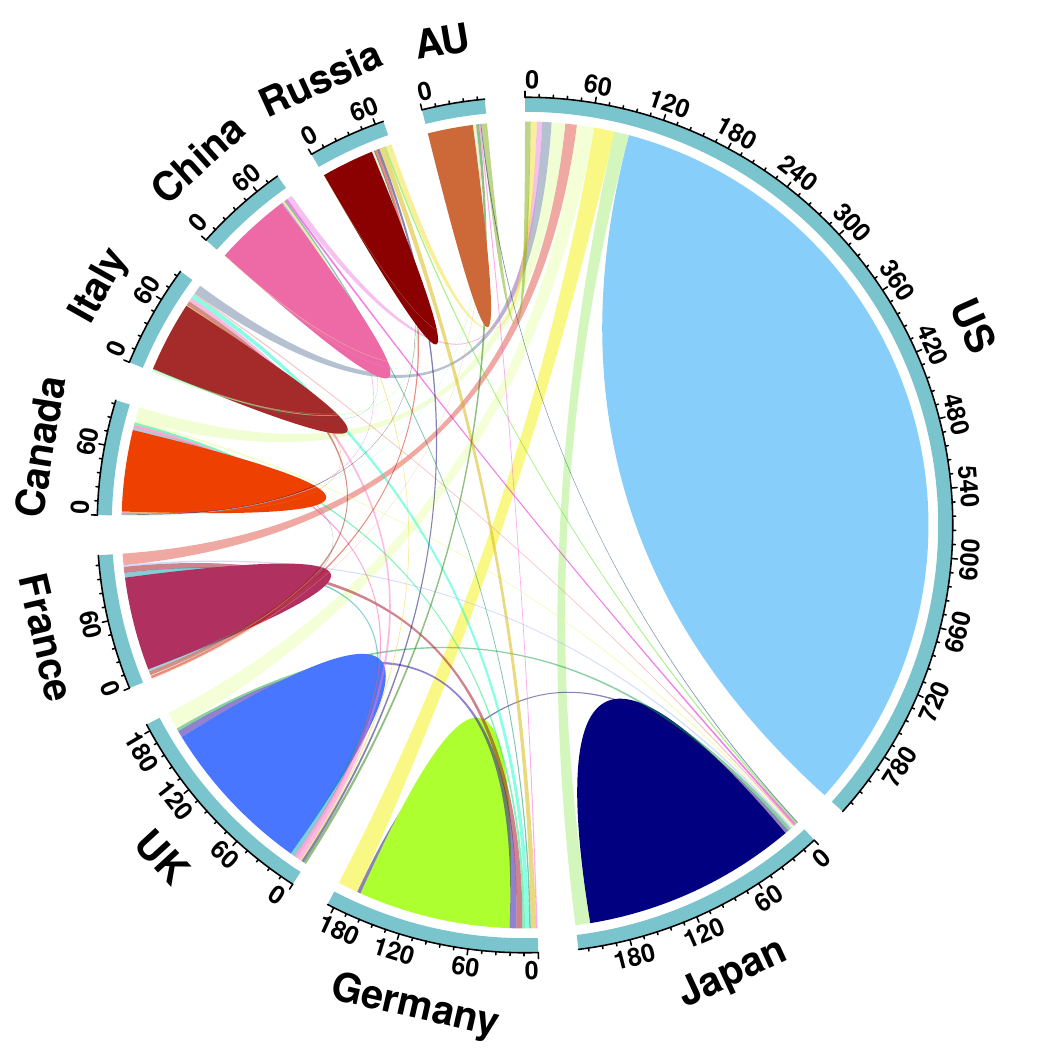}}
\end{flushright}
    \end{minipage}
	\begin{minipage}{0.5\hsize}
\begin{center}
\raisebox{\height}{\includegraphics[trim=2.0cm 1.8cm 0cm 1.5cm, align=c, scale=\csize, vmargin=0mm]{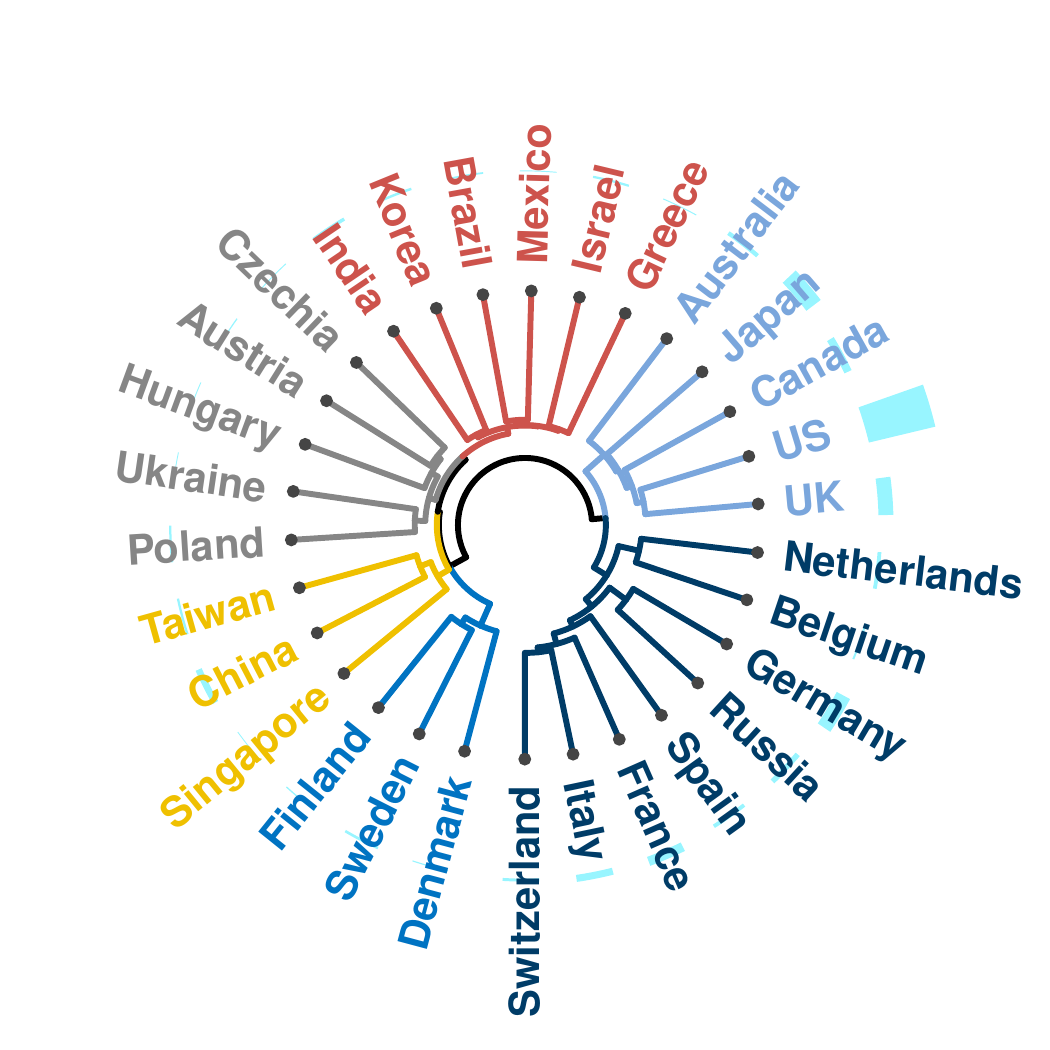}}
\end{center}
	\end{minipage}
    \end{tabular}

\vspace{0mm}
\hspace{0.5cm}{\marrow}\quad\dotfill
\vspace{2mm}

    \begin{tabular}{c}
    \begin{minipage}{0.03\hsize}
\begin{flushleft}
    \hspace{0.7cm}\rotatebox{0}{\period{1}{1971--1990}}
\end{flushleft}
	\end{minipage}
	\begin{minipage}{0.5\hsize}
\begin{flushright}
\raisebox{\height}{\includegraphics[trim=2.0cm 1.8cm 0cm 1.5cm, align=c, scale=\chordsize, vmargin=0mm]{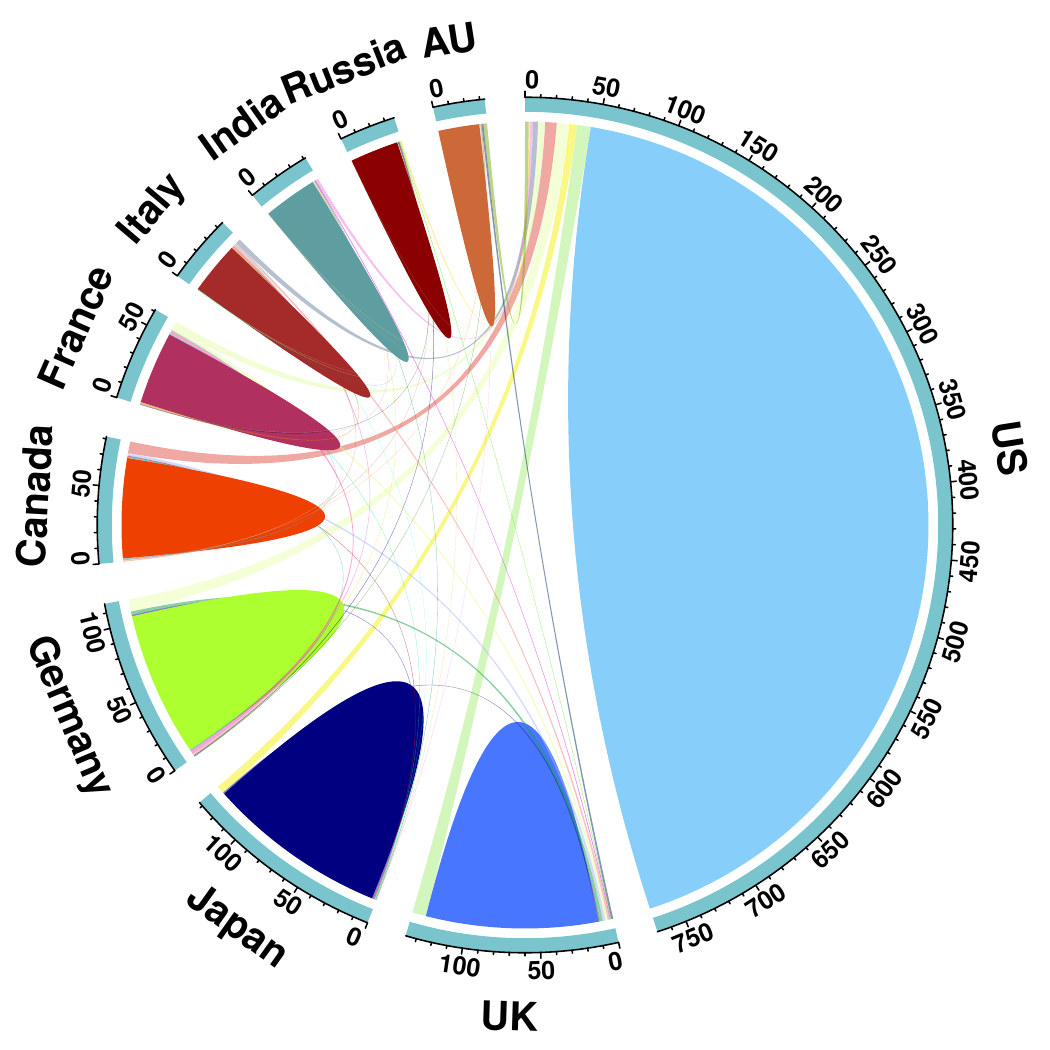}}
\end{flushright}
    \end{minipage}
	\begin{minipage}{0.5\hsize}
\begin{center}
\raisebox{\height}{\includegraphics[trim=2.0cm 1.8cm 0cm 1.5cm, align=c, scale=\csize, vmargin=0mm]{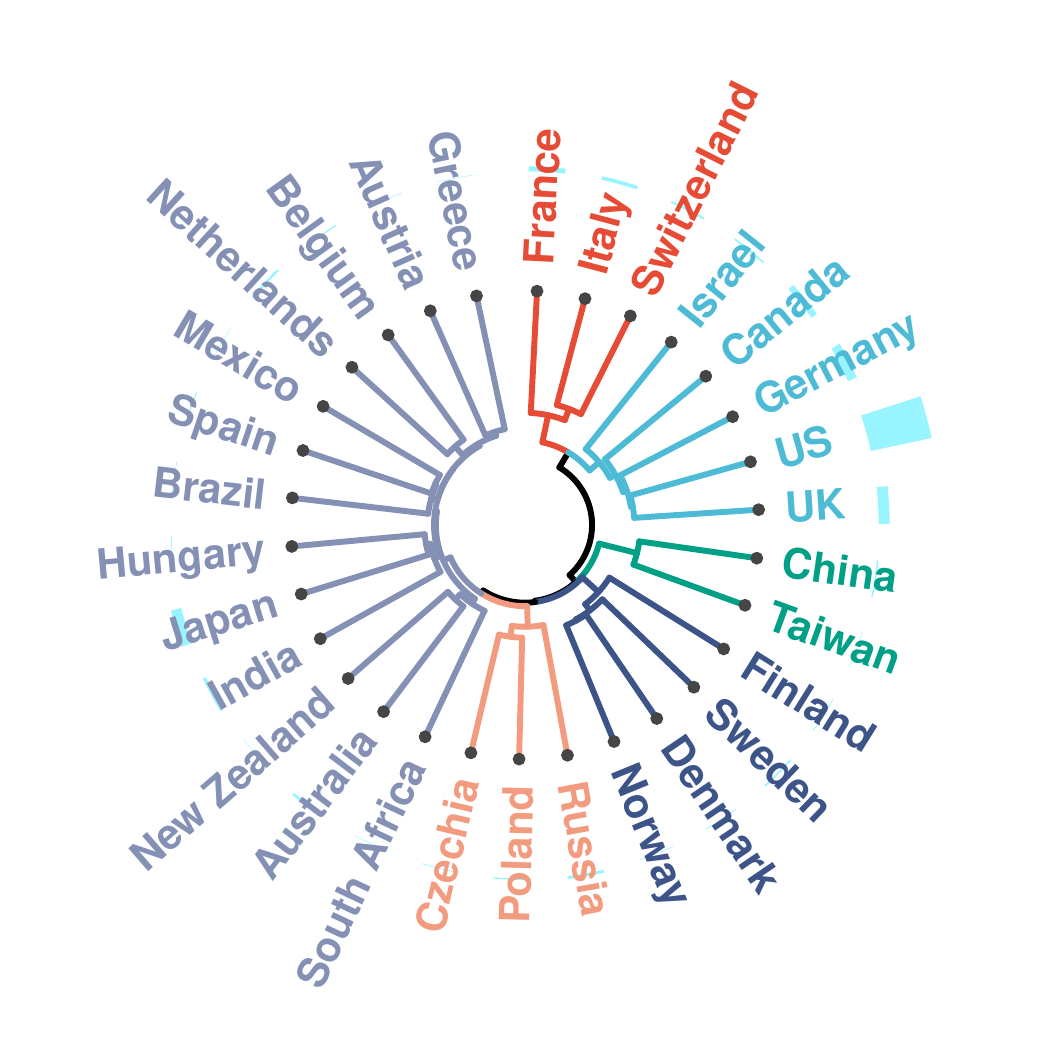}}
\end{center}
	\end{minipage}
    \end{tabular}
\end{subfigure}
\vspace{0mm}
\caption{\textbf{Changes in bilateral relationships over time and the evolution of international research collaboration clusters for the aggregate field of \q{Natural Sciences}.}
The number of works is displayed in thousands.}
\label{fig:chord_cdend_allf}
\end{figure}
}

\addcontentsline{toc}{subsection}{Supplementary Table}

\afterpage{\clearpage%
\begin{table}[h]
\centering
\vspace{-0.5cm}
\caption{\textbf{Numbers of clusters by discipline and period.}
The number of clusters is systematically derived with the threshold value of the coupling heights set commonly to all disciplines and periods.}
\label{tab:ncluster}
{\footnotesize
{\setlength{\tabcolsep}{0.5em}
\begin{tabular}{rl@{\hspace{1.2cm}}ccccccccccccccc}\\[-2mm] \toprule[1pt] \\[-4.2mm]
\multicolumn{2}{c}{R\&D discipline}	&\rn{1}. 1971--1990	&	&\rn{2}. 1991--2000	&	&\rn{3}. 2001--2010	&	&\rn{4}. 2011--2020 \\[-0.2mm] \midrule[0.3pt]
{1.} & \!\!\ai 				& {4} & & {4} & & {5} & & {7} \\
{2.} & \!\!\quantum 		& {7} & & {6} & & {5} & & {7} \\
{3.} & \!\!\bio 			& {6} & & {6} & & {5} & & {6} \\
{4.} & \!\!\nano 			& {6} & & {7} & & {8} & & {8} \\
{5.} & \!\!\agri 			& {6} & & {8} & & {6} & & {7} \\
{6.} & \!\!\particle 		& {7} & & {8} & & {5} & & {6} \\
{7.} & \!\!\aerospace 		& {5} & & {5} & & {5} & & {7} \\
{8.} & \!\!\nuclear 		& {7} & & {7} & & {5} & & {7} \\
{9.} & \!\!\marine 			& {9} & & {6} & & {6} & & {7} \\
{10.} & \!\!\neuro 			& {5} & & {4} & & {6} & & {7} \\
{11.} & \!\!\condensed 		& {8} & & {7} & & {6} & & {5} \\
{12.} & \!\!\envi			& {5} & & {7} & & {5} & & {8} \\
{13.} & \!\!\earth 			& {8} & & {6} & & {6} & & {6} \\
{14.} & \!\!\astro 			& {5} & & {6} & & {4} & & {5} \\
{15.} & \!\!\math 			& {6} & & {5} & & {6} & & {6} \\[0mm] \bottomrule[1pt] \\
\end{tabular}}
}
\vspace{5mm}
\end{table}
\quad\\
\vfill
}


\afterpage{\clearpage%
\renewcommand\refname{\fontsize{14}{15}\selectfont References for Supplementary Materials}

\setlength{\bibsep}{0\baselineskip plus 0.2\baselineskip}

}

\end{document}